%% file: Thesis.tex
\documentclass[
12pt, 
a4paper, 
onecolumn,
oneside, 
hidelinks,
table]{book}

\input{structure.tex}
\usepackage[backend=biber,
 giveninits=true,
 isbn=true,
 url=true,
 doi=true,
 style=ieee,
 citestyle=numeric-comp,
 dashed=false,
 maxnames = 5]{biblatex}
{%
  \fancyhf{}
  \fancyhead[R]{\nouppercase{\leftmark}}
  \fancyfoot[C]{\thepage} 
  \renewcommand{\headrulewidth}{0pt}
  \renewcommand{\footrulewidth}{0pt}
}

\input{./tex/custom_commands}
\input{./tex/acronyms_definitions.tex}

\begin{document}
\frontmatter
\thispagestyle{empty}
\pagenumbering{alph}

%

\input{./tex/title.tex}

\pagenumbering{roman}
\setstretch{1.18}
\input{./tex/abstract.tex}
\setstretch{1.5}
\input{./tex/publications.tex}

\input{./tex/table_of_content.tex}
\doublespacing

\adjustmtc

\glsaddall

\clearpage


\printglossary[style=super, type=\acronymtype, nonumberlist]


\onehalfspace
\raggedbottom


\mainmatter
\doublespacing
\pagenumbering{arabic}
\pagestyle{fancy}{%
    \fancyhf{}
    \fancyhead[L,C]{}
    \fancyhead[R]{\nouppercase{\leftmark}}
    \fancyfoot[L]{}
    \fancyfoot[C]{\thepage}
    \fancyfoot[R]{}
    \renewcommand{\headrulewidth}{0pt}
    \renewcommand{\footrulewidth}{0pt}
}

\thispagestyle{fancy}
\input{./tex/Chapter1-Intro/main.tex}

\input{./tex/Chapter2-LitRev/main.tex}

\input{./tex/Chapter3-Smart-Infer/main.tex}

\input{./tex/Chapter4-Obscure-Infer/main.tex}

\input{./tex/Chapter5-Scalable-Infer/main.tex}
\input{./tex/Chapter6-Resiliency/main.tex}

\input{./tex/Chapter7-Conclusion/main.tex}

\backmatter
\renewcommand{\refname}{References} 
\appto{\bibsetup}{\sloppy}
\apptocmd{\sloppy}{\hbadness 10000\relax}{}{}
\begin{singlespace}
\setlength\bibitemsep{10pt}   
\printbibliography[heading=bibintoc, title={References}]
\end{singlespace}
\end{document}

%% file: structure.tex
\usepackage[columnsep=2cm,
 left=1.575in,
 right=0.78in,
 top=1.18in,
 bottom=1.18in]{geometry} 

 \usepackage{relsize}
 \usepackage{marvosym}
 \usepackage{makecell}
 \setcellgapes{4pt}

 \usepackage{hhline}
 \usepackage{fancybox}
 \usepackage{adjustbox}
 \usepackage{amsmath}
 \usepackage{amsthm}
 \usepackage{tikz}
 \usepackage{algpseudocode}
 \usepackage{subfigure}
 \usepackage{array}
 \usepackage{soul}
 \usepackage{balance}
 \usepackage{xspace}
 \usepackage{graphicx}
 \usepackage{multirow}
 \usepackage{url}
 \usepackage{colortbl}
 \usepackage{enumitem}
 \usetikzlibrary{arrows}
 \usetikzlibrary{shapes}
 \usepackage{tcolorbox}
 \usepackage{comment}
 \usepackage{morewrites}
 \usepackage{amssymb}
 \usepackage{pifont}

\usepackage[T1]{fontenc}
\usepackage[utf8]{inputenc}

\usepackage{sectsty}

\usepackage[titles,subfigure]{tocloft} 

\usepackage{textcomp}
\usepackage{gensymb}
\usepackage{enumitem}
\usepackage{graphicx}
\usepackage[colorinlistoftodos]{todonotes}
\usepackage{fancyhdr}
\usepackage{setspace}
\usepackage{url}
\usepackage{verbatim}
\usepackage[bottom]{footmisc}
\usepackage[english]{babel}
\usepackage{multirow}

\usepackage[nottoc, numbib]{tocbibind}

\usepackage{amsfonts}
\usepackage{amssymb}
\usepackage{longtable,tabularx}
\usepackage{csquotes}
\usepackage{siunitx}
\usepackage{booktabs}

\usepackage{moreverb}
\usepackage{bm}
\usepackage{algorithm}
\usepackage{minitoc}

\usepackage{multicol}
\usepackage{hyperref}
\usepackage{cleveref}
\usepackage{bookmark}
\usepackage{xcolor}
\usepackage{pdfpages}
\usepackage[titletoc]{appendix}
\usepackage{mathtools}
\usepackage{etoolbox}
\usepackage[htt]{hyphenat}
\usepackage{soul}

\usepackage{lipsum}

\usepackage{filecontents}
\usepackage{balance}
\usepackage{xcolor}

\usepackage{caption}

\usepackage[utf8]{inputenc}
\usepackage[english]{babel}

\usepackage[acronym, nogroupskip, nopostdot]{glossaries} 
\newglossary[slg]{symbol}{sld}{sdn}{Nomenclature}
\makeglossaries
\glstoctrue
\usepackage{afterpage}

\usepackage{emptypage}
    
\let\Algorithm\algorithm
\renewcommand\algorithm[1][]{\Algorithm[#1]\setstretch{1.4}}

\usepackage{lscape}
\usepackage{array}
\usepackage{multirow}
\usepackage{rotating}
\PassOptionsToPackage{table,xcdraw}{xcolor}

\renewcommand{\texttt}[1]{%
	\begingroup
	\ttfamily
	\begingroup\lccode`~=`/\lowercase{\endgroup\def~}{/\discretionary{}{}{}}%
	\begingroup\lccode`~=`[\lowercase{\endgroup\def~}{[\discretionary{}{}{}}%
	\begingroup\lccode`~=`.\lowercase{\endgroup\def~}{.\discretionary{}{}{}}%
	\catcode`/=\active\catcode`[=\active\catcode`.=\active
	\scantokens{#1\noexpand}%
	\endgroup
}

\newboolean{rev}
\setboolean{rev}{true}
\newcommand{\setrevtrue}{\color{blue}}
\newcommand{\setrevfalse}{}
\newcommand{\rev}{\ifthenelse{\boolean{rev}}{\setrevtrue}             
	{\setrevfalse}}

%% file: tex/custom_commands.tex
\definecolor{yellow}{rgb}{1, 1, 0}
\definecolor{yellowgreen}{rgb}{0.71, 1, 0}
\definecolor{lightgreen}{rgb}{0.48, 1, 0}
\definecolor{green}{rgb}{0, 1, 0}
\newcommand{\low}{\cellcolor{yellow}}
\newcommand{\midL}{\cellcolor{yellowgreen}} 
\newcommand{\midH}{\cellcolor{lightgreen}} 
\newcommand{\high}{\cellcolor{green}}

\theoremstyle{definition}

\definecolor{GrayCell}{HTML}{DDDDDD}
\definecolor{LightGrayCell}{HTML}{EEEEEE}
\definecolor{RedCell}{HTML}{FF0000}
\definecolor{GreenCell}{HTML}{32CB00}
\definecolor{YellowCell}{HTML}{F8FF00}
\definecolor{WhiteCell}{HTML}{FFFFFF}
\newcolumntype{L}[1]{>{\raggedright\let\newline\\\arraybackslash\hspace{0pt}}m{#1}}
\newcolumntype{C}[1]{>{\centering\let\newline\\\arraybackslash\hspace{0pt}}m{#1}}
\newcolumntype{R}[1]{>{\raggedleft\let\newline\\\arraybackslash\hspace{0pt}}m{#1}}

\newcolumntype{M}[1]{%
	>{\begin{turn}{90}\begin{minipage}{#1}%
				\raggedright\hspace{0pt}}l%
			<{\end{minipage}\end{turn}}}

\newcommand{\myverb}{\fontsize{11}{48}\usefont{OT1}{lmtt}{b}{n}\noindent }

\newcommand{\ie}{{\em i.e., }}
\newcommand{\eg}{{\em e.g., }}

\newcommand{\ciaoo}[1]{{\setlength\fboxrule{0pt}\fbox{\tcbox[colframe=black,colback=white,shrink tight,boxrule=0.8pt,extrude by=1mm]{\small #1}}}}

\newcommand{\Null}{\textit{Null}}
\DeclareMathOperator{\score}{score}
\DeclareMathOperator{\pred}{pred}
\DeclareMathOperator{\aand}{and}

\algdef{SE}[SUBALG]{Indent}{EndIndent}{}{\algorithmicend\ }%
\algtext*{Indent}
\algtext*{EndIndent}

\newcommand\nd{\textsuperscript{nd}\xspace}

\newcommand\nth{\textsuperscript{th}\xspace} 
\newcommand\nst{\textsuperscript{st}\xspace}


%% file: tex/acronyms_definitions.tex
\newacronym{iot}{IoT}{Internet of Things}
\newacronym{it}{IT}{Information Technology}
\newacronym{iana}{IANA}{Internet Assigned Numbers Authority}
\newacronym{ip}{IP}{Internet Protocol}
\newacronym{ipfix}{IPFIX}{Internet Protocol Flow Information Export}
\newacronym{ai}{AI}{Artifical Intelligence}
\newacronym{aml}{AML}{Adversarial Machine Learning}
\newacronym{aws}{AWS}{Amazon Web Services}
\newacronym{ddos}{DDoS}{Distributed Denial of Service}
\newacronym{dos}{DoS}{Denial of Service}
\newacronym{ccdf}{CCDF}{Complementary Cumulative Distribution Function }
\newacronym{cdn}{CDN}{Content Delivery Network}
\newacronym{cpu}{CPU}{Central Processing Unit}
\newacronym{cve}{CVE}{Common Vulnerability Exposure}
\newacronym{csv}{CSV}{Comma Separated Values}
\newacronym{oui}{OUI}{Organizational Unique Identifier}
\newacronym{pcap}{PCAP}{Packet Capture}
\newacronym{ttl}{TTL}{Time To Live}
\newacronym{dns}{DNS}{Domain Name System}
\newacronym{mdns}{mDNS}{Multicast Domain Name System}
\newacronym{dhcp}{DHCP}{Dynamic Host Configuration Protocol}
\newacronym{http}{HTTP}{Hyper Text Transfer Protocol}
\newacronym{html}{HTML}{Hyper Text Markup Language}
\newacronym{ftp}{FTP}{File Transfer Protocol}
\newacronym{rtsp}{RTSP}{Real Time Streaming Protocol}
\newacronym{tls}{TLS}{Transport Layer Security}
\newacronym{sdn}{SDN}{Software Defined Networking}
\newacronym{mud}{MUD}{Manufacturer Usage Description}
\newacronym{mqtt}{MQTT}{Message Queuing Telemetry Transport}
\newacronym{ntp}{NTP}{Network Time Protocol}
\newacronym{nvd}{NVD}{National Vulnerability Database}
\newacronym{nat}{NAT}{Network Address Translation}
\newacronym{isp}{ISP}{Internet Service Provider}
\newacronym{gan}{GAN}{Generative Adversarial Networks}
\newacronym{ssdp}{SSDP}{Simple Service Discovery Protocol}
\newacronym{arp}{ARP}{Address Resolution Protocol}
\newacronym{mac}{MAC}{Medium Access Control}
\newacronym{ml}{ML}{Machine Learning}
\newacronym{svm}{SVM}{Support Vector Machine}
\newacronym{upnp}{UPnP}{Universal Plug and Play}
\newacronym{icmp}{ICMP}{Internet Control Message Protocol}
\newacronym{igmp}{IGMP}{Internet Group Management Protocol}
\newacronym{snmp}{SNMP}{Simple Network Management Protocol}
\newacronym{tcp}{TCP}{Transmission Control Protocol}
\newacronym{tcam}{TCAM}{Ternary Content-Addressable Memory}
\newacronym{vlan}{VLAN}{Virtual Local Area Network}
\newacronym{udp}{UDP}{User Datagram Protocol}
\newacronym{ndr}{NDR}{Network Detection and Response}
\newacronym{ids}{IDS}{Intrusion Detection System}
\newacronym{span}{SPAN}{Switched Port Analyzer}

%% file: tex/title.tex
\begin{titlepage}
\thispagestyle{empty}
\begin{center}



\Huge \textbf{Managing Networked IoT Assets Using Practical and Scalable \\Traffic Inference}\\
\vspace{1.5cm}
\huge \textbf{Arman Pashamokhtari}\\
\vspace{2cm}
\large A dissertation submitted in fulfillment\\of the requirements for the degree of\\
\vspace{0.5cm}
\large \textbf{Doctor of Philosophy}\\
\vspace{1cm}
\includegraphics[width=0.35\columnwidth]{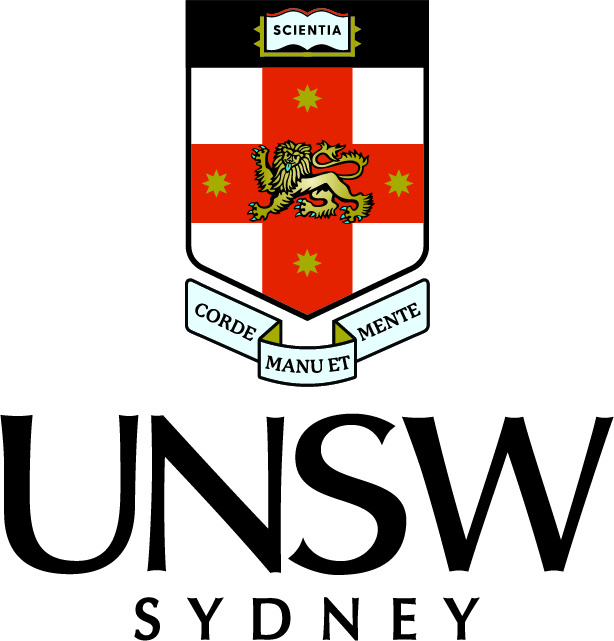}\\
\vspace{0.5cm}
\large School of Electrical Engineering and Telecommunications\\
\vspace{0.4cm}
\large Faculty of Engineering\\
\vspace{0.4cm}
\large The University of New South Wales\\
\vspace{1cm}
\large {January 2023}
\end{center}

\end{titlepage}

%% file: tex/abstract.tex
\newgeometry{top=-1cm}
\chapter*{Abstract}
\bookmarksetupnext{level=section}
\addcontentsline{toc}{chapter}{Abstract}
\onehalfspace
\bookmarksetupnext{level=section}
\thispagestyle{plain}
\adjustmtc
\vspace{-1cm}

The Internet has recently witnessed unprecedented growth of a class of connected assets called the Internet of Things (IoT). Due to relatively immature manufacturing processes and limited computing resources, IoTs have inadequate device-level security measures, exposing the Internet to various cyber risks. Therefore, network-level security has been considered a practical and scalable approach for securing IoTs, but this cannot be employed without discovering the connected devices and characterizing their behavior. Prior research leveraged predictable patterns in IoT network traffic to develop inference models. However, they fall short of expectations in addressing practical challenges, preventing them from being deployed in production settings. This thesis identifies four practical challenges and develops techniques to address them which can help secure businesses and protect user privacy against growing cyber threats.

My {\textbf{first contribution}} balances prediction gains against computing costs of traffic features for IoT traffic classification and monitoring. I develop a method to find the best set of specialized models for multi-view classification that can reach an average accuracy of 99\%, \ie a similar accuracy compared to existing works but reducing the cost by a factor of 6. I develop a hierarchy of one-class models per asset class, each at certain granularity, to progressively monitor IoT traffic. My {\textbf{second contribution}} addresses the challenges of measurement costs and data quality. I develop an inference method that uses stochastic and deterministic modeling to predict IoT devices in home networks from opaque and coarse-grained IPFIX flow data. Evaluations show that false positive rates can be reduced by 75\% compared to related work without significantly affecting true positives. My {\textbf{third contribution}} focuses on the challenge of concept drifts by analyzing over six million flow records collected from 12 real home networks. I develop several inference strategies and compare their performance under concept drift, particularly when labeled data is unavailable in the testing phase. Finally, my {\textbf{fourth contribution}} studies the resilience of machine learning models against adversarial attacks with a specific focus on decision tree-based models. I develop methods to quantify the vulnerability of a given decision tree-based model against data-driven adversarial attacks and refine vulnerable decision trees, making them robust against 92\% of adversarial attacks.

\cleardoublepage
\restoregeometry

%% file: tex/publications.tex
\chapter*{List of Publications} %

\bookmarksetupnext{level=section}
\addcontentsline{toc}{chapter}{List of Publications}
\bookmarksetupnext{level=section}
\thispagestyle{plain}

\vspace{-10mm}
During the course of my Ph.D. candidature, a number of publications have been made based on the work presented here and are listed below for reference.

\noindent{\underline{\large Journal Publications}}
\begin{enumerate}

\vspace{-5mm}
\item {\textbf{A. Pashamokhtari}, N. Okui, M. Nakahara, A. Kubota, G. Batista, H. Habibi Gharakheili, ``Dynamic Inference from IoT Traffic Flows under Concept Drifts in Residential ISP Networks
	'', to appear in \textit{IEEE Internet of Things Journal (IoT-J)}, Apr 2023.}

\item {\textbf{A. Pashamokhtari}, G. Batista, H. Habibi Gharakheili, ``Efficient IoT Traffic Inference: from Multi-View Classification to Progressive Monitoring'', Under review at \textit{ACM Transactions on Internet of Things (TIOT)}.}

\item \textbf{A. Pashamokhtari}, N. Okui, Y. Miyake, M. Nakahara, and H. Habibi Gharakheili, ``Combining Stochastic and Deterministic Modeling of IPFIX Records to Infer Connected IoT Devices in Residential ISP Networks'', \textit{IEEE Internet of Things Journal (IoT-J)}, 10(6):5128-5145, Nov 2022.

\item \textbf{A. Pashamokhtari}, G. Batista, H. Habibi Gharakheili, ``AdIoTack: Quantifying and Refining Resilience of Decision Tree Ensemble Inference Models Against Adversarial Volumetric Attacks on IoT Networks'', \textit{Elsevier Computers \& Security}, 120(1):1-16, Sep 2022.

\end{enumerate}

\newpage
\noindent{\underline{\large Conference Publications}}
\begin{enumerate}
\setcounter{enumi}{4}

\vspace{-5mm}

\item  \textbf{A. Pashamokhtari}, N. Okui, Y. Miyake, M. Nakahara, H. Habibi Gharakheili, ``Inferring Connected IoT Devices from IPFIX Records in Residential ISP Networks'', \emph{IEEE LCN}, Virtual, Edmonton, Canada, Oct 2021.

\item  \textbf{A. Pashamokhtari}, H. Habibi Gharakheili, and V. Sivaraman, ``Progressive Monitoring of IoT Networks Using SDN and Cost-Effective Traffic Signatures'', \emph{ACM Workshop on Emerging Technologies for Security in IoT (ETSecIoT)}, Sydney, Australia, Apr 2020. \textbf{[Best paper award]}

\item  \textbf{A. Pashamokhtari}, ``PhD Forum Abstract: Dynamic Inference on IoT Network Traffic using Programmable Telemetry and Machine Learning'', \emph{IEEE/ACM IPSN}, Sydney, Australia, Apr 2020.

\end{enumerate}

Additionally, the following paper was published based on my summer internship at CyAmast:
\begin{enumerate}
	\setcounter{enumi}{7}
	
	\item  \textbf{A. Pashamokhtari}, A. Sivanathan, A. Hamza and H. Habibi Gharakheili, ``PicP-MUD: Profiling Information Content of Payloads in MUD Flows for IoT Devices'', \emph{IEEE WoWMoM workshop on Smart Computing for Smart Cities}, Virtual, Belfast, United Kingdom, Jun 2022. \textbf{[Runner's up paper award]}
	
\end{enumerate}

\vspace{5mm}
\noindent{\underline{\large Patents}}
\begin{enumerate}
	\setcounter{enumi}{8}

\item \textbf{A. Pashamokhtari}, A. Sivanathan, A. Hamza, H. Habibi Gharakheili, ``Apparatus and Process for Monitoring Network Traffic of Networked Assets'', \emph{AU Provisional Patent No. 2022901601}, filed Jun 2022.

\end{enumerate}

%% file: tex/table_of_content.tex
\dominitoc
\clearpage
\tableofcontents
\adjustmtc
\cleardoublepage

\clearpage
\listoffigures
\adjustmtc
\cleardoublepage

\clearpage
\listoftables
\adjustmtc

%% file: tex/Chapter1-Intro/main.tex
\chapter{Introduction}\label{chapter:Intro}


\vspace{8mm}
The Internet continues to evolve, empowering users, businesses, and governments to benefit from online interactions and controls. Over the past decade, the Internet's growth has made connected devices more popular, especially Internet-of-things (IoT) devices such as cameras, screens, printers, medical devices,  or industrial logical controllers. IoTs are often resource-constrained devices designed for a limited set of tasks. They are increasingly being adopted in various domains like industry, healthcare, and smart homes. Example use cases for IoT devices are sensors to collect data, wearable devices, and indoor/outdoor cameras. According to \cite{iot_market_2022}, there were about 12.2 billion connected IoT devices around the globe at the beginning of 2022. The same source forecasts that this number will peak at 27 billion devices by 2025. The growth of IoT devices is significantly larger than that of other computer devices. For example, the number of smartphones was about 6.2 billion by 2022 and is expected to be around 7.2 billion by 2025 \cite{statista_phone}, which is almost four times less than the projected figure for IoT devices.

\vspace{8mm}
Despite their benefits to individuals and organizations, IoT devices are prone to cyber attacks more than Information Technology (IT) devices like servers, computers, or even smartphones \cite{Palo2020,cyberreport}. This is primarily attributed to the fact that IoT devices come with relatively constrained computing resources and hence unlike IT devices, lack appropriate security measures embedded. Additionally, the diversity of immature IoT manufacturers with low-security standards and policies results in IoT devices emerging with insufficient device-level security \cite{Harkin2022}. An example of weak device-level security is using default credentials. The default username and passwords of different IoT device models and manufacturers reported publicly \cite{defaultpass}, along with insecure authentication \cite{Kumar2019, Anand2021}, are among key drivers for compromising IoT devices and/or enabling Internet-scale cyber attacks \cite{Palo2020}. Therefore, network operators need a network-level approach to better manage and/or augment the security of connected IoT devices. However, current solutions used for IT devices are not readily applicable to IoT devices because of the heterogeneity of IoTs \cite{Palo2020}. This makes most connected IoT devices remain relatively unmanaged, exposing networks (and, to some extent, the Internet) to huge risks.

As highlighted by cyber industry reports \cite{Palo2020,cyberreport}, the foundational step to secure IoT networks is obtaining rich ``visibility'' into connected IoT devices. Visibility can be viewed from different aspects: (1) Assessing the potential risks of IoT devices in a network -- \eg unencrypted traffic exchange, open transport-layer services, and default credentials are some of the common risks associated with IoT devices, (2) Identifying/classifying IoT devices connected to networks -- \eg determining their role (camera, health sensor, printer), manufacturer, operating system, behavioral profile to maintain an accurate asset inventory; or, (3) Continuously/periodically monitoring the network behavior of IoT devices and flagging changes or deviations from an expected norm. Visibility can be best achieved by modeling activity patterns in network traffic of IoT devices, given their limited, identifiable, and repeatable communications and activities on the network, as opposed to IT devices displaying unbounded network behaviors driven by their users' activities. Traffic inference can vary in multiple ways. Inference models may consume traffic features of varying resolutions, costs (measurement and computing), and prediction powers. Also, their internal decision-making logic may differ depending on data availability and quality.

IoT traffic inference has particularly interested researchers for the past several years. However, existing methods are yet to be production-ready as they do not address some practical challenges. In this thesis, I identify and address four crucial challenges when inference models are applied in real practice. These challenges are as follows:

(1) \textit{Balancing traffic measurement and feature computation costs against prediction power:} Inputs of network-based inference models are features and/or signals extracted from network traffic. Traffic measurement (the process of collecting network traffic) incurs computing costs depending on how and where on the network the measurement is performed. For example, measuring traffic from several distributed access switches is more challenging and expensive than measuring traffic from a single vantage point (an aggregation switch). Subsequently, computing features from the measured traffic is relatively expensive, depending on the intended features. For example, features that need to be extracted from individual packet contents will require somewhat deeper inspection, which is more expensive than features computed from network flows (aggregate of packets). A practical inference model must have a reasonable balance between its prediction quality and computing costs.

(2) \textit{Data quality:} The performance of data-driven inference models is tightly bound to the quality of the data they consume. In traffic inference, data quality can be impacted by factors like resolution and transparency. Coarse-grained (low-resolution) data records like statistical features (\eg average packet/byte count) computed from flows are less predictive than fine-grained (high-resolution) data records like those extracted from individual packets (\eg TCP SYN signatures) or even time-series signals. Less transparent refers to scenarios where for some reason, the measured network traffic is relatively opaque. For example, when the traffic is measured after it crosses a middlebox (\eg a NAT-enabled router), there will be no access to the local traffic of IoT devices as well as their identity (MAC/IP address). The lack of high-quality data makes it challenging to make an accurate inference.

(3) \textit{Concept drifts:} The network behavior of IoT devices may change from time to time or even from network to network due to reasons like software/firmware updates, modifying configurations, varying user behaviors, or different compositions of connected devices. Changes (even subtle ones) in network behaviors can impact inference quality as the distribution of trained data evolves temporally and spatially. A common solution to manage this issue is having a mechanism (automatic and/or manual) to re-train inference models on new data. However, model re-training is nontrivial and may be infeasible in certain scenarios, as obtaining fresh ``labeled" data is challenging and expensive. Collecting labeled data is a time-consuming and costly part of modeling which cannot be done very frequently (weekly or even monthly) at scale. Thus, it is necessary to develop some techniques to cope with concept drifts without necessarily re-training the inference models.

(4) \textit {Resiliency of learning-based models}: Machine learning algorithms have proven to be effective in network traffic modeling \cite{Thangavelu2019,Marchal19,arunan19,Nizar2019,Yang2019,Kotak2021,Xiaobo2022,Meidan2020,Doshi2018,Hasan2019,Eirini2019,Enkhtur2021}. However, they are shown \cite{Barreno2010} to be vulnerable to adversarial attacks which target the integrity of these models, impacting their prediction performance and make them less reliable in critical environments. Adversarial attacks have been studied for years in the context of computer vision. However, network traffic differs from images and hence demands adapted techniques to combat adversarial data-driven attacks. Plus, in the computer vision context, the primary focus of adversarial attack research is mostly on neural network models \cite{Caminero2019,Abusnaina2019,Ferdowsi2019,Ibitoye2019,Goodfellow2020}, leaving other models like decision trees relatively under-explored.

\vspace{-8mm}
\section{Thesis Contributions}
\vspace{-8mm}
This thesis makes the following four contributions:

My first contribution, presented in \cite{SmartInfer2022}, addresses the challenge of measurement and computing costs, \ie balancing the cost versus accuracy of inference models. For this contribution, I consider an enterprise network of IoT devices where IoT network traffic can be measured transparently. I develop a generic inference architecture consisting of (a) multi-view classification and (b) progressive monitoring phases. Given transparent traffic, I systematically quantify the efficacy of various traffic features using four metrics, namely cost, accuracy, availability, and frequency. With several features to infer from, I develop an optimization problem to find the optimal selection of features for a multi-view classification scheme that reduces the processing costs by a factor of $6$ without affecting the accuracy. For the monitoring phase, I develop a progressive system with three granularity levels that help reduce the computation cost and increase the monitoring system's explainability. For this, I develop a refined version of a convex hull to train one-class models (per device type) that yield high true positive and low false positive rates of $94$\% and $5$\% on average, respectively.

My second contribution, presented in \cite{ArmanIoTJ2022}, examines the challenge of data quality. For this, I consider the problem of inferring IoT devices in residential networks where we have partial visibility of their network traffic, primarily due to managing the cost of measurement at the scale of ISPs. For this contribution, I develop a stochastic machine learning-based model that can accurately detect IoT devices in each home network using coarse-grained flow-based features with an average accuracy of $96$\%. I also develop a deterministic model using cloud services that IoT devices communicate with. I demonstrate that combining stochastic and deterministic models can significantly reduce false positives by $75$\%.

\vspace{-4mm}
My third contribution, presented in \cite{ConceptDriftExt2022}, addresses the challenge of concept drift. It considers a scenario of an ISP  network, similar to the second contribution, detecting IoT devices connected to residential networks. It focuses on the challenge of concept drifts. For this contribution, I begin by highlighting and quantifying concept drifts using more than six million flow records collected from 12 real home networks. Following that, I develop two inference models (global versus contextualized) and compare their performance under concept drifts. I develop a method to manage concept drifts that dynamically applies the best model (selected from a set) to labeled network traffic of unseen homes during the testing phase, yielding better performance in $20$\% of cases when the labels are available for the testing data (ideal but unrealistic settings). Finally, I develop a method to automatically select the best model without needing labels of unseen data (a realistic inference) and show that it can achieve $94$\% of the ideal model's accuracy.

\vspace{-3mm}
My fourth contribution, presented in \cite{AdIoTack2022}, studies the resiliency of decision tree-based ML models against adversarial attacks. For this contribution, first, I generate decision tree-based multi-class models like Random Forest, AdaBoost, and gradient boosting, trained on benign network patterns of IoT devices, that are capable of detecting non-adversarial volumetric attacks. Given trained models and an intended volumetric attack on a target IoT class, I develop a method that determines traffic instances (adversarial vulnerabilities) containing volumetric attacks that subvert those models. I demonstrate on a real testbed of IoT devices how those pre-computed vulnerabilities can be exploited to launch impactful volumetric attacks without being detected by the model continuously monitoring the network traffic. Lastly, I develop a technique that refines the trained decision trees model to become resilient against $92$\% of the studied adversarial attacks.

\section{Thesis Organization}

The rest of this thesis is organized as follows.

In Chapter~\ref{chapter:LiteratureReview}, I explain the state-of-the-art IoT traffic modeling approaches and discuss their objectives, traffic measurement techniques, visibility, traffic feature units, and inference models. I also review existing solutions from the viewpoints of practical challenges of data-driven models, namely concept drifts and adversarial attacks. Chapter~\ref{ch:smart infer} addresses the challenge of having a balance between prediction accuracy and computing costs. Chapter~\ref{ch:obscure infer} discusses the challenge of having an accurate inference by relying on coarse-grained and partially-transparent data, and develops two inference models: one stochastic and one deterministic, for this purpose. Chapter~\ref{ch:scalable infer} develops methods to tackle concept drifts using contextualized modeling. In Chapter~\ref{ch:secure infer}, I describe the resiliency challenges of ML models against adversarial volumetric attacks by focusing on decision tree-based models. To demonstrate this challenge, I first develop a method to automatically determine the vulnerability of decision tree-based models. I then explain an effective patching technique for making the models robust against volumetric adversarial attacks. Chapter~\ref{chapter:conclusion} concludes the thesis and notes future directions.

\section{Summary}
IoT devices are increasingly being used in different domains, but due to their limited computing resources and manufacturers’ immaturity, IoT devices expose various cyber risks. Researchers have been studying network-level inference techniques for several years. However, the existing solutions are not sufficiently scalable and practical. In this thesis, I identify and address four practical challenges of IoT traffic inference, including measurement and computing costs, data quality, concept drift, and resiliency of decision tree-based models against adversarial attacks. The methods I develop throughout this thesis can help network operators protect their IoT assets at scale. In particular, the methods discussed in Chapters~\ref{ch:obscure infer} and \ref{ch:scalable infer} enable ISPs to offer IoT security solutions for their subscribers, benefiting both the ISP (financially and cybersecurity) and end-users (cybersecurity).

%% file: tex/Chapter2-LitRev/main.tex
\chapter[Literature Review on IoT Traffic Inference]{Literature Review on \\IoT Traffic Inference}\label{chapter:LiteratureReview}


Unmanaged IoT devices are considered a security risk for network operators. As highlighted by the recent Gartner report \cite{hypeCycle} on technologies for security operations, not all of the network detection and response (NDR) tools used for IT devices are capable of covering IoT devices. This report noted that gaining visibility into the characteristics and behavior of IoT assets is essential for understanding their attack surface in order to take remedial actions. Traffic inference techniques help network operators identify IoT devices and continuously track their cyber health. Over the past few years, researchers have been developing different inference techniques. Inference methods are distinct in aspects like the traffic features they use and the decision-making models that consume those features. In this chapter, I begin by explaining the existing methods of IoT traffic inference. I next highlight some of the challenges existing works fail to address.

\begin{figure}[t!]
	\begin{center}
		\includegraphics[width=0.99\textwidth]{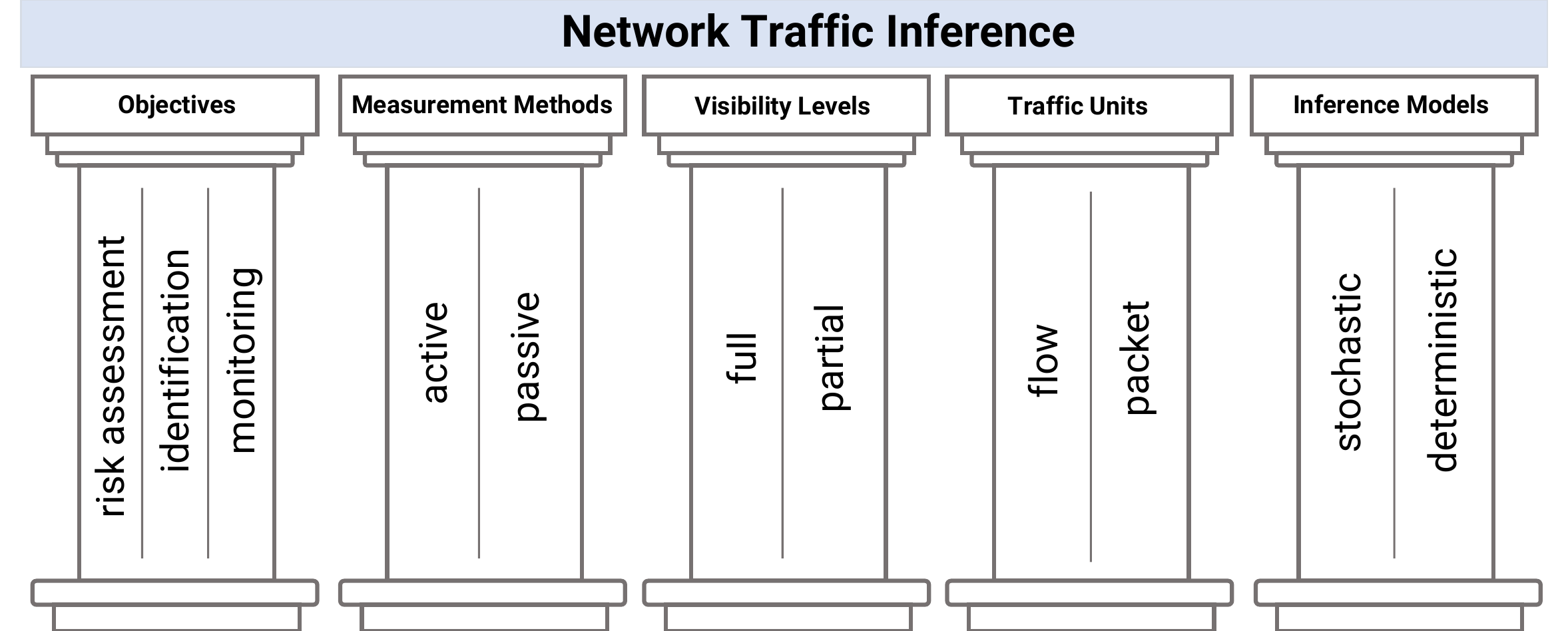}
		\caption{Taxonomy of network traffic inference techniques.}
		\label{fig:taxonomy}
	\end{center}
		\vspace{-4mm}
\end{figure}

Fig.~\ref{fig:taxonomy} shows the taxonomy of IoT traffic inference that includes five pillars: objectives, measurement methods, visibility levels, traffic units, and inference models. For each of these pillars, I have identified various  techniques that have been developed by prior work in the literature. The remainder of this chapter expands on each pillar and explains existing techniques.

\section{Inference Objectives}
Existing works on IoT traffic inference can be categorized into three groups based on their objective. \textbf{Risk assessment} focuses on finding risks, such as communicating in plaintext or authenticating with weak credentials, associated with network behaviors of IoT devices. \textbf{Identification}, also known as classification, aims to determine various information (\eg their manufacturer, make/model, role, or operating system/firmware) about IoT devices and characterize their network behavior. \textbf{Monitoring} is another objective that aims to detect changes in the behavior of IoT devices. Monitoring can be performed by either whitelisting and/or blacklisting network behaviors. The former approach involves the construction of the benign behavior for each device type (class) as a baseline (``norm'') so that any deviation from that baseline is flagged as an anomaly. In contrast, the latter approach requires a set of patterns or signatures for malicious behaviors (\eg DDoS attacks, Command-and-Control communications), and alerts will be raised if any of those patterns are found in the network traffic.

\subsection{IoT Asset Risk Assessment}
IoT devices can expose networks to a number of risks, which an attacker can exploit for different malicious purposes. Risk assessment is performed to discover those risks, guiding network operators to take appropriate actions to mitigate risks. In the literature, different risks have been studied for IoT networks and assets, listed as follows.

\textbf{Confidentiality:} Some IoT devices communicate with unencrypted data, disclosing sensitive information and violating users' privacy. Works in \cite{Loi2017, Ren2019} used Shannon's Entropy metric to distinguish encrypted from plaintext traffic. However, Shannon's Entropy falls short in differentiating encrypted from encoded (like image, audio, and video) traffic. Although encoded traffic is not human-readable, it is not as secure as encrypted traffic; hence, encoded communication should be considered medium-risk (as opposed to plaintext, which is high-risk). In one of my recent works \cite{Arman2022}, I used Shannon's Entropy, arithmetic mean, and Chi-square metrics to train a machine learning model for mapping the content type of a network flow (5-tuple) to three classes of plaintext, encrypted, and encoded contents.

\textbf{Exposed ports and weak authentication:} IoT devices often expose network services (ports) for various purposes like SSH, Telnet, or web-based configurations. Manufacturers set some default login credentials, and users rarely change them. Therefore, those IoT assets become attractive targets for attackers to compromise. Websites like \cite{defaultpass} have databases of default credentials for a number of IoT devices based on manufacturers and device models. Works in \cite{Loi2017, Kumar2019} scanned open ports of IoT devices, and for some of them, the authors were able to authorize through SSH and Telnet ports using default credentials like ``{\myverb{admin/admin}}''.
Work in \cite{Anand2021} developed a system to extract weak authentication methods and/or known passwords that networked cameras use for authentication via HTTP protocol.

\textbf{Data exfiltration:} IoT devices typically communicate with a number of remote servers maintained by their manufacturer, cloud providers (\eg Amazon AWS), common services like time synchronization (\eg {\myverb{ntp.org}}), or third-party services (\eg {\myverb{doubleclick.net}}) for other services like advertising \cite{Mandalari2021}. As highlighted by \cite{Mandalari2021,Saidi2020,Ren2019}, many IoT devices often interact with third-party  platforms, enabling those platforms to collect large-scale user data used for tracking and/or advertisement purposes \cite{Mandalari2021}. Work in \cite{Mandalari2021} developed a system to automatically identify ``non-essential'' network communications of IoT devices so that blocking them would not affect the normal operation of devices. The authors explained that limiting IoT remote communications can make them more robust against data exfiltration and enhance users' privacy protection.

\textbf{DDoS attacks:} IoT devices with weak embedded security measures not only put their users' privacy at risk but also are low-hanging fruits for cyber criminals who can weaponize them for DDoS attacks on the Internet. Malware like LizardStresser and Mirai have widely compromised thousands of IoT devices and used them for high-impact DDoS attacks by forwarding hundreds of gigabits per second of attack traffic to their victim servers \cite{ibm2017}. Another type of DDoS is reflection attacks which are launched by sending request packets with spoofed source IP addresses (of a victim server) to IoT devices, so they reply unsolicited responses to the victim server at high rates. Reflection attacks do not even need to compromise IoT devices. Work in \cite{Loi2017} reported that some IoT devices could be utilized to launch ICMP, SSDP, and SNMP reflection attacks. Authors of \cite{Lyu2017} quantified the impact of IoT reflection attacks and noted that they could be amplified to a great extent.


\vspace{-2mm}
\subsection{IoT Asset Identification}
\vspace{-2mm}
Identification (also known as classification) aims to provide information about IoT assets on a network that may come at different granularities. For example, one may want to identify groups (clusters) of IoT devices based on their roles, like cameras, printers, and building management sensors, regardless of their manufacturer and model. Another may aim for fine-grained classes, identifying the exact make and model of IoT devices like the Cisco Webex camera, Xerox printer AltaLink C8055, or Philips indoor motion detector. Asset identification is the first step for managing IoT networks \cite{hypeCycle, cyberreport}. Some of the key prior works on characterizing IoT devices from their network behavior are summarized in what follows.

IoT Sentinel \cite{Miettinen2017} developed a machine learning-based inference model to identify IoT devices in order to discover vulnerable devices by checking their characteristics against public vulnerability databases (\eg CVE \cite{cve}). The authors employed SDN-enabled gateways to isolate vulnerable devices upon detection. DEFT \cite{Thangavelu2019} uses a hierarchical inference architecture where a central inference model runs on the cloud, and local inference models are embedded in home gateways. If a local model cannot identify a device connected to its home network, the corresponding traffic data is sent to the cloud for further analysis.

AuDI \cite{Marchal19} performs identification by grouping IoT devices into abstract clusters. AuDI autonomously creates clusters with abstract labels like type 1, type 2, etc. Abstract labels (also used in \cite{Nguyen2019}) allow identifying device types without requiring a labeled dataset (unsupervised learning). However, manual analysis is needed to translate the abstract labels to actual classes (roles) to make meaningful inferences and take action.

PingPong \cite{Trimananda19} developed an inference model to detect devices' make and model as well as their operating states (turning on or off). Authors in \cite{arunan19} used a two-stage inference scheme that starts by distinguishing IoT devices from non-IoTs, followed by classifying IoT devices detected by the first stage. Work in \cite{Yang2019} developed an inference model trained on data collected by scanning the Internet. The authors used regular expression matching to label the training dataset by obtaining the device model and vendor from the headers of application protocols like HTTP and SSDP. Some other works \cite{Bezawada2018,Guo2018,Nizar2019,Meidan2017,Arunan2018,Saidi2020} also focused on determining the make and model of IoT devices from their network traffic. However, they differ from each other in terms of traffic features, visibility levels, and inference models used that will be further explained in \S\ref{sec:vantage point}, \S\ref{sec:features}, and \S\ref{sec:inference models}.

Detecting operating systems is another aspect of identification which is traditionally used for IT assets due to their limited number of operating systems. Tools like Nmap \cite{nmap} and p0f \cite{p0f} use passive and/or active methods to collect network traffic patterns from networked assets and match them against fingerprints of known operating systems.  
IP TTL, TCP window size, and TCP options are some of the traffic features that are used in fingerprinting operating systems.

In this thesis, I develop models to infer the make and model of IoT devices, which is the finest level of identification (classification) in contrast to role detection or abstract clustering. Fine-grained identification helps us develop a better monitoring strategy, like developing one-class models for each IoT device type (will be discussed in Chapter~\ref{ch:smart infer}). Additionally, knowing the devices' make and model can help us utilize public vulnerability databases like \cite{cve} and discover vulnerabilities (risks) specific to each device type.

\subsection{IoT Asset Behavior Monitoring}
 
Some of the prior works focused on developing inference models to monitor network activities of IoT devices, looking for malicious behavioral patterns (a blacklisting approach traditionally used for IT networks).  
Work in \cite{Doshi2018} simulated behaviors of the Mirai botnet by generating DDoS traffic similar to what Mirai-infected devices generate, including TCP SYN, UDP, and HTTP {\myverb{GET}} flood. The authors used the collected data to train an inference model to learn DDoS patterns. Work in \cite{Hasan2019} used synthetic traffic collected from a virtual IoT environment \cite{Pahl2018} that includes eight types of network behaviors, of which seven are malicious (\eg scanning and DoS), and one type is the combination of all normal behaviors. 
Work in \cite{Eirini2019} developed a multi-stage inference model that first, a model identifies the device type, and then another inference model monitors individual packets to find anomalous behaviors within five classes of malicious activities like replay attacks and spoofing. Works in \cite{Brun2018, Enkhtur2021} also employed a blacklisting method to monitor the behavior of IoT assets.

In contrast to blacklisting, there is a separate body of works that use a whitelisting (allowlisting) approach, which develops models based on benign behaviors of IoT devices, so any deviation (activities that violate normal or benign boundaries) is considered abnormal. Works in \cite{Arunan2020a, Hamza2019, Meidan2018, Arunan2020b, Hamza2018, Guo2020, Nguyen2019} developed benign profiles of IoT devices and tuned their models to flag severe deviations from the profiles as anomaly alerts. Each of the mentioned works differentiates itself in aspects like traffic features and types of inference models that will be explained in more detail in \S\ref{sec:features} and \S\ref{sec:inference models}, respectively.

The blacklisting method is an established technique widely applied to IT networks, available in many firewalls and intrusion detection systems (IDS) \cite{suricata, zeek, snort}. However, the main limitation of blacklisting is that they cannot detect zero-day attacks whose patterns were not seen before. On the other hand, the whitelisting method can detect novel attacks as long as benign behaviors are modeled by limited boundaries, which is relatively feasible for IoT devices due to their narrow benign activities. We note that whitelisting is fairly challenging (almost infeasible) for IT assets as they can show a growing set of activity patterns.

In this thesis, I focus on the identification and monitoring objectives. In Chapter~\ref{ch:smart infer}, I develop inference models for both identification and monitoring purposes. In Chapters~\ref{ch:obscure infer} and \ref{ch:scalable infer}, my main focus becomes the identification of IoT devices; and in Chapter~\ref{ch:secure infer}, I study the resiliency of decision tree-based monitoring models. For all monitoring models developed in this thesis, I use the whitelisting method.

\section{Traffic Measurement Methods}
Network traffic can be measured in two ways: active and passive. Active measurement requires scanning the entire network or target devices by sending crafted packets and receiving their responses. On the other hand, passive measurement receives network traffic to/from devices by different means, like sniffing multicast/broadcast communications or configuring SPAN (Switched Port Analyzer) ports on network switches/routers to mirror a copy of the entire traffic to a server for analysis.
 
\subsection{Active Measurement}
Active measurement is carried out by sending specific query packets to network devices and analyzing the received responses for inference. Examples of active measurements include (but are not limited to) port scans, IP scans, and operating system fingerprinting. For port and/or IP scanning, certain packets (\eg TCP, UDP, SNMP, ARP, ICMP) are sent to target ranges of transport-layer port numbers or potential IP addresses (within a subnet). The returned responses indicate the status of ports (open/closed) or IP addresses (assigned/unassigned). Active measurement has been used as a well-known technique to fingerprint operating systems and web servers. Nmap \cite{nmap} uses well-crafted TCP, UDP, and ICMP probes that will result in unique responses depending on the target device's operating system. Httprint \cite{httprint} is another tool used for web server fingerprinting. Httprint infers web servers by looking into their HTTP server headers and the responses to different HTTP queries, including malformed ones.

In the context of IoT traffic inference, very few works utilize active measurement. Work in \cite{Arunan2018} used TCP port scan to identify IoT devices based on their open TCP ports. Works like \cite{Yang2019, Feng2018} developed internet-wide scanners to probe all public IPv4 addresses to collect plaintext protocol headers (like HTML, FTP, and Telnet) from connected devices. Work in \cite{Kumar2019}, probes IoT devices to collect information from DHCP, UPnP, mDNS, Telnet, and HTML banners.

\subsection{Passive Measurement}
Passive measurement is done by measuring devices' network traffic without necessarily disrupting them with probe packets. There are different ways of passive measurement: (1) Some works like \cite{Hamza2018, Hamza2018b, Hamza2019, Hamza2022, Arunan2020b, Arunan2020b, Arunan2020a} programmed SDN (software-defined-networking) switches by specific flow rules to periodically measure the statistics of network activities (packet and byte counters) across inserted flow rules. (2) Works in \cite{LCN2021, Saidi2020, Meidan2020} used flow exporting agents like IPFIX \cite{ipfixRFC5103} and NetFlow \cite{netflow} (installed on network switches/routers) to collect aggregate statistics (total packet/byte count, average payload size, and average inter-arrival time) of unique IP flows. (3) Traffic mirroring is another method \cite{Marchal19, Nguyen2019, Hasan2019, Kotak2021, Engelberg2021, Xiaobo2022, Divakaran2020} which takes a copy of all network packets. Although packet mirroring is the finest-grained passive measurement, it can be prohibitively expensive compared to SDN-based and flow export methods, and therefore, it might not be desirable for inference from high-rate network traffic. In Chapters~\ref{ch:obscure infer} and \ref{ch:scalable infer}, I elaborate on some of the challenges of IoT traffic inference at the scale of ISP networks and use the IPFIX tool as a cost-effective measurement method.

Note that active measurements come with traffic overheads (probing packets sent on the network). Such overheads might not affect the operation of IT assets; however, they can be detrimental (or even disruptive) to sensitive IoT devices with limited computing resources. Therefore, as this thesis focuses on IoT assets, passive traffic measurement is used throughout the entire thesis. That said, a recent work \cite{23nomsActivePAssive} outlined a preliminary effort to experiment with selective and dynamic scans driven by passive traffic inference to improve IoT asset characterization while overheads are managed.

\section{Visibility Levels}\label{sec:vantage point}

IoT traffic can be measured at different vantage points, either inside or outside the network. The measurement vantage point is an important factor in the quality of traffic inference as it dictates the visibility, granularity, and richness of traffic features. For instance, in a segmented enterprise network with a hierarchy of core, edge, and access switches, suppose the traffic is collected on a core switch, then local traffic exchanged within subnets may not be captured. In another scenario, some metadata (\eg IP and MAC address) of the measured traffic could be obfuscated by a network middlebox (\eg router). A typical obfuscation that makes inference challenging is the presence of a NAT-enabled (network address translation) appliance before the traffic measurement point. NAT is a technology (commonly in home networks and some enterprise networks) that maps local (private) IP addresses to public IP addresses, resulting in the identity of devices behind NAT getting obfuscated.

\subsection{Full Visibility}
The majority of existing works like \cite{Sharma2022,Miettinen2017,Arunan2018,Anat2020, Hamza2022, IoTInspector, Kumar2019,Guo2020,Nguyen2019,Mazhar2020,Divakaran2020,Arunan2020a,Arunan2020b} need {\textbf{full}} visibility into the traffic in order to extract traffic features necessary for their high-performing models. Some of these works like \cite{Sharma2022,IoTInspector,Kumar2019,Mazhar2020} extract features from local traffic (\eg ARP, DHCP, and SSDP), only available with full visibility into the entire traffic. Works in \cite{Miettinen2017, Anat2020, Nguyen2019,Arunan2020a,Arunan2020b} rely on devices' identity to group their network features over a time window which is not achievable with obfuscated traffic (partial visibility). Other works like \cite{Arunan2018,Hamza2022} need to observe services locally exposed by IoT devices that are not visible when traffic is measured outside the gateway.

\subsection{Partial Visibility}
Few works like \cite{Saidi2020, Meidan2020,Guo2018,Kotak2021,LCN2021} developed their prediction methods to infer from {\textbf{partially}} visible traffic. Works in \cite{Saidi2020, LCN2021, Meidan2020} used IPFIX and/or NetFlow data records for extracting flow-based telemetry. IPFIX and NetFlow can aggregate packets based on their five-tuple metadata and generate statistical measures of bi-directional flows. However, if measured post-NAT (\eg outside home networks), connected devices' identity (IP/MAC addresses) gets obfuscated. Hence, collating multiple flows of the same device to obtain richer traffic features becomes infeasible. We note that IPFIX and NetFlow data records are relatively coarse-grained as they report statistical measures of an aggregate of packets. Other works like \cite{Guo2018,Kotak2021} inspect individual obfuscated packets (\ie post-NAT) to extract headers and/or payload for inferencing purposes; however, they neither rely on device identities nor local traffic.

Both full and partial measurements have their own pros and cons. Full visibility gives us more amount of data with precise identity of devices. However, gaining full visibility in certain deployments may not be practical and cost-effective. Therefore, inferring based on partial data becomes inevitable. In Chapters~\ref{ch:smart infer} and \ref{ch:secure infer}, I will address some of the practical challenges of inference when we have full visibility; and in Chapters~\ref{ch:obscure infer} and \ref{ch:scalable infer}, I will develop techniques to infer from partially-visible traffic.

\section{Traffic Units}
\label{sec:features}

Network traffic can be measured in different units: network flows (\eg from programmable switches or IPFIX/NetFlow records from traditional switches) or raw network packets. These traffic units are analyzed to extract the required features used as inputs of inference models. In what follows, I explain some related works that use packet and/or flow traffic units to predict IoT behaviors on the network.

\subsection{Network Packets}
\label{subsec:packet}

Extracting packet-based features requires some form of inspection (shallow or deep) depending on where (in the TCP/IP protocol stack) the intended feature is located. For instance, extracting IP headers like IP addresses or Time To Live (TTL) is considered a relatively shallow inspection. In contrast, deeper inspection is needed for TLS cipher suites or DNS domain names. Deep packet inspection requires more processing time and resources. 
We note that packet-based features come in different types: some, like packet size, TTL, and TCP window size, have a numerical value and order, while others, like IP address, port number, and DNS domain name, are of nominal (categorical) values without any specific order \eg TCP ports 80 and 81 have no semantic order.

Work in \cite{IoTInspector} used the first three bytes of connected devices' MAC address, known as Organizational Unique Identifier (OUI), to detect the manufacturer of IoT devices. OUI is administrated by IEEE and is unique to each manufacturer; therefore, it is feasible to infer device manufacturers by looking up their MAC addresses. The authors of \cite{IoTInspector} also used plaintext headers in mDNS, SSDP, and UPnP for detecting device types. Works in \cite{Kumar2019, Mazhar2020} used headers of DHCP, mDNS, HTTP, and UPnP for detecting IoT device types. Work in \cite{Feng2018} used headers in HTTP, FTP, Telnet, and RTSP responses from IoT devices to craft a search query to find device labels through the Google search engine.

Relying entirely on transparent data from application headers comes with challenges: (a) Certain IoT device types may not communicate via those application protocols. Hence, corresponding features can be completely missing, and, (b) Encryption (\eg TLS) can potentially obfuscate information expected from those protocols.

\vspace{4mm}

To overcome the challenges mentioned above regarding application headers, work in \cite{Yang2019} incorporated data of headers from other layers like IP TTL, IP ToS (Type of Service), TCP maximum segment size, and TCP window size, in addition to transparent application headers. Work in \cite{Sharma2022} used WiFi 802.11 protocols headers to detect IoT devices on wireless networks. Authors of \cite{Guo2018} used domain names from DNS packets to detect and classify IoT devices. The same work was extended in \cite{Guo2020} to detect novel remote endpoints by whitelisting benign IoT servers. In \cite{Trimananda19}, authors developed a signature format based on request and response packets in TCP flows exchanged between an IoT device and a remote endpoint. Each signature includes information about all packets sent/received (chronologically ordered within a TCP flow) while indicating the sender (IoT device or remote server) and their sizes.

Authors of \cite{Xiaobo2022} developed a packet signature format using direction, TTL, payload size, TCP flags, TCP window size, and TCP options. They collated consecutive packets over a short window (one second long) and used the common parts of packet signatures to classify IoT devices. Work in \cite{Anat2020} used features like the number of DNS queries, number of unique DNS domain names, and size of HTTP User-Agent to distinguish IoT from non-IoT traffic.

Some works like \cite{Bezawada2018, Kotak2021} extracted features from encrypted packet payloads. Authors of \cite{Bezawada2018} used Shannon's Entropy of byte values. Work in \cite{Kotak2021} transformed TCP payloads to gray-scale images and developed an inference model using image recognition techniques to identify device types. Authors of \cite{Engelberg2021} used distributions of packet size and inter-arrival time to characterize IoT behaviors.

\vspace{-5mm}
\subsection{Network Flows}\label{subsec:flow}
In contrast to packet-based features that require inspection of individual packets, flow-based features can be computed on an aggregate of packets. Flow-based features may seem to carry less information than their packet-based counterparts. Instead, they are attractive because of their relatively lower computation costs. Also, due to aggregation, the number of flows becomes significantly less than the number of packets to process \cite{Madanapalli2018} by inference models. A network flow is a uni-directional or bi-directional network communication between two endpoints (\eg an IoT device and another IoT device or a cloud/local server). Existing works can be categorized into two groups based on how they form network flows: (1) Some like \cite{Hamza2018,Hamza2019,Arunan2020a,Arunan2020b, Divakaran2020,Hamza2022} employ programmable or software-defined networking techniques to insert flow rules into programmable (control plane or data plane) switches, measuring flow counters to construct time-series signals or compute custom-grained statistical features in real-time; and (2) Others like \cite{Muraleedharan2010, Meidan2020,Saidi2020,LCN2021} employ traditional network telemetry tools like IPFIX  and NetFlow available on switches/routers to export summarized IP flow records. These tools can be configured (of course, at extra computation costs) to measure more than just byte/packet counts, like the average time between packets or the number of packets containing non-empty payloads.

Authors of \cite{Hamza2018,Hamza2022} used Manufacturer Usage Description (MUD) profiles as fingerprints for detecting IoT devices and monitoring their behavior. A MUD profile of an IoT device contains a finite set of benign flows (access control entries specifying the intended network flows) for the given device that can be enforced at the network level, reducing its attack surface. Work in \cite{Hamza2019} trained clustering models on the counters of individual MUD flows (a model per flow) to detect volumetric attacks on IoT devices. In \cite{Arunan2020a,Arunan2020b}, counters of specific flows (\eg DNS, NTP, SSDP) computed at different time scales were used as features of multi-class classifiers to determine IoT device types and detect changes in their normal behavior. Work in \cite{Divakaran2020} used five-tuple flows (source/destination IP address, source/destination port number, and IP protocol number) with their corresponding average and standard deviation packet/byte count for detecting novel flows and behaviors.

In \cite{Meidan2020}, NetFlow features, including flow size/duration, TCP flags, and IP protocol, were used for classifying IoT devices. In \cite{Saidi2020}, authors extracted the IP address of cloud servers from NetFlow and IPFIX records, followed by reverse lookups via public tools like Censys \cite{Durumeric2015} and DNSDB \cite{dnsdb} to obtain domain names and create a profile for each IoT device type. Authors of \cite{Muraleedharan2010} computed distributions of some IPFIX features, like packet count and average packet size based on benign data. They showed how benign and attack data can be distinctly determined by configurable thresholds -- this work focused on non-IoT devices.

\subsection{Combination of Packets and Flows}

Prior work discussed in \S\ref{subsec:packet} and \S\ref{subsec:flow} used either network packets or network flows to compute features for traffic inference. Some other works attempted a combination of packet-based and flow-based features to predict the class of network traffic and/or detect anomalous activities. The traffic units used by works in \cite{Gu2020,Thangavelu2019,Engelberg2021,Marchal19,arunan19} are packets, and authors developed techniques to construct flow-level data from the aggregate of packets measured within a time window, obtaining their packet-based or flow-based features. In \cite{Gu2020}, authors used a similar technique and argued that the variability (measured by Shannon entropy) of protocols, IP addresses, and port numbers is different in benign versus malicious traffic. Work in \cite{Thangavelu2019} considered combined features like the number of DNS error responses, the number of HTTP response codes, SSDP packet count, and MQTT flow duration across sliding time windows.

Work in \cite{Marchal19} developed techniques to focus on autonomous IoT traffic \ie periodic activities generated automatically independent of user interactions, and computed features like the number of periodic flows, frequency of port changes, and flow duration. Work in \cite{arunan19} developed a two-stage scheme for classifying IoT traffic: starting to incorporate packet-based features like transport-layer port numbers, DNS domain names, and cipher suites (from TLS packets), followed by focusing on flow-based features like sleep time, NTP inter-arrival time, and flow size. In one of my papers \cite{Arman2020}, I employed a combination of TCP SYN signature (obtained from TCP options), DNS domain names, and packet/byte count of ten specific flows for identifying IoT device types.

I note that prior work selected traffic features of their inference objective merely based on the prediction accuracy. However, other factors like their computational costs, how frequently they emerge in network traffic, or how prevalent they are in traffic of heterogeneous IoT classes remain unexplored. In this thesis, I use both packet-based and flow-based features to leverage their unique advantages in an efficient manner. In Chapter~\ref{ch:smart infer}, specifically, I demonstrate how combining different packet-based and flow-based traffic features into a single inference model can be detrimental to prediction performance and computing costs. Hence, my approach is to develop separate specialized models into a multi-view inference scheme. In other chapters of this thesis, I will use flow-based features as they are more conducive (\eg cost-effectiveness at scale) to the context of those chapters.

\section{Inference Models}\label{sec:inference models}

Once a set of traffic features is determined, one needs to employ algorithms to learn patterns in those traffic features resulting in an inference model (for classifying IoT assets and/or monitoring their cyber health). A variety of strategies and algorithms have been used in the literature on IoT traffic inferencing. We categorize them into two main types:

(1) {\textbf{Stochastic}} models that learn the distribution data in traffic features. Machine Learning (ML) models are a subset of stochastic models and are widely used for IoT traffic inference \cite{Thangavelu2019,Miettinen2017, Bezawada2018,Marchal19,Doshi2018,Kotak2021,Nguyen2019,Meidan2020,Sharma2022,Gu2020, Arman2020}. Stochastic models are often applied to features of (typically) random values but with certain distributions, like packet count, packet inter-arrival time, and packet size.

(2) {\textbf{Deterministic}} (also known as heuristic) models \cite{Saidi2020,Divakaran2020,IoTInspector,Arunan2018,Guo2018,Hamza2018,Hamza2022} are often used for features with limited observable values \eg IP address of cloud servers, domain names, and port number of network services. The intuition behind deterministic modeling is that it is fairly practical (though not guaranteed to be optimal), and required features do not often change frequently (remain persistent to some extent) -- therefore, their exact values can be used as a fingerprint for inference.


\vspace{-3mm}
\subsection{Stochastic Inference Models}
\vspace{-3mm}
Some works like \cite{Muraleedharan2010, Engelberg2021} used Chi-square, Cosine, and Jensen–Shannon distance metrics to measure the distance between new (unseen) distributions and an expected distribution. If the measured distance is less than a configurable threshold, then it is inferred that the two distributions match. In \cite{Muraleedharan2010}, authors developed an inference model for non-IoT traffic that learns the distribution of some IPFIX features seen in benign traffic. During the testing phase, if the observed feature distribution significantly deviates from the expected distribution, it is considered an anomaly. In \cite{Engelberg2021}, authors used packet size and inter-arrival time distributions for detecting IoT device types. 

ML models are a subset of stochastic models that automatically learn distribution for several features. Their learning process can be tuned using hyperparameters specific to their learning algorithms. Decision tree \cite{Doshi2018}, Random Forest \cite{Arunan2020a}, Support Vector Machine (SVM) \cite{Meidan2020}, Neural Networks \cite{Kotak2021}, and Naive Bayes \cite{arunan19} are some examples of the ML models used in the literature.

ML models can be trained and used differently for inference purposes. In what follows, I explain two types of ML models used for IoT traffic inference.

\vspace{-2mm}
(1) {\textbf{Supervised models:}} For this model type, labeled datasets are needed for training. Depending on the inference objective, classes could be different \eg in works that aim to identify IoT device types \cite{arunan19, Kotak2021, Arunan2020a}, each class corresponds to a device type, whereas in works like \cite{Doshi2018} with anomaly detection objective, the classes are defined as benign and malicious. Some works like \cite{Eirini2019,Hasan2019} further expanded the malicious class into specific attack sub-classes like DoS and port scan. The works mentioned above used multi-class supervised models which train a single ML model on a labeled dataset, including traffic features of all target classes. Alternatively, one can train a separate model for each class using a dataset containing traffic features of the corresponding class and all other classes. This method is known as one-vs-rest, used by works in \cite{Meidan2020,Sharma2022,Miettinen2017}. However, the accuracy of one-vs-rest can be impacted by imbalanced datasets.

\vspace{-2mm}
(2) {\textbf{Unsupervised models:}} These models do not require the training dataset to be labeled; instead, they create a number of clusters that instances with similar traffic features fall into the same cluster. Unsupervised models are used for different purposes like capturing benign behavior of each IoT device type \cite{Arunan2020b}, distinguishing IoT from non-IoT traffic \cite{Nader2021} or labeling an unlabeled traffic dataset \cite{Kumar2019}. In \cite{Thangavelu2019}, unsupervised models are used as a fallback mechanism to correct the output of a supervised model when the latter is not confident about its output.

\vspace{-2mm}
Due to the dynamics of network traffic, ML models trained on packet/flow features are often imperfect, causing some degree of misprediction. Various techniques can be used to boost ML models' prediction quality. In Chapter~\ref{ch:smart infer}, I employ several ML models (multi-view), each with a different perspective on the network traffic, and a consensus mechanism draws the final prediction. In Chapter~\ref{ch:obscure infer}, I employ the classification score reported by ML models to discard low-confident predictions.

\subsection{Concept Drifts}
Though IoT devices display finite and distinct network behaviors, those identifiable patterns are subject to change for various reasons. For instance, how users interact with their devices can directly impact device behaviors. Also, IoT device manufacturers may issue software/firmware updates leading to behavioral changes. Inference models are generated based on a limited amount of training data; hence, their accuracy may degrade when the distribution of data changes. This challenge is known as concept drift which researchers have studied \cite{Lu2019, Bifet2007, Bifet2009, Gama2004, Oza2005, Xu2017, Liu2016, Oza2001, Bifet2010} in contexts like handwriting, electricity market, and synthetic data. However, to the best of my knowledge, no existing work has explored this challenge in the context of IoT traffic inference.

In the field of machine learning (not specifically IoT traffic inference), there exist works that aim to manage concept drift. Some of the existing works \cite{Bifet2007, Bifet2009, Gama2004, Oza2005, Ying2005, Gama2009} assume that the true labels are available during the testing phase so they can evaluate the model's performance and if they observe a considerable amount of error rates in the model's predictions, they flag concept drifts. Another body of works like \cite{Dasu2006, Qahtan2015, Gu2016, Shao2014, Leeuwen2008, Song2007} use the underlying data distribution to measure the distance between the training and testing data, so if the distance is greater than a threshold, they consider it as a concept drift. Although this method is more realistic than relying on the testing labels, it can become computationally expensive when features (\ie dimensions) grow \cite{Lu2019}.

Authors in \cite{Reis2018a, Reis2018b} explained that obtaining true labels could be delayed or even infeasible in some cases. Therefore, they used statistical tests to measure the similarity between classification score distributions without requiring true labels during the testing phase. A classification score is a number in the range of $[0,1]$ that is returned by classification models for each prediction -- higher values indicate the model is more confident in its predicted label. The authors showed that matching the distribution of scores yields acceptable results in detecting concept drifts, while it is computationally more desirable than computing the distance in the distribution of multi-dimensional feature space.

Once concept drifts are detected, there are techniques to address their impact on the quality of inference. Works like \cite{Bach2008, Bifet2007, Xu2017, Han2015} leveraged the labels in the testing dataset to train a new model, replacing the original one. Authors of \cite{Oza2001, Oza2005, Gomes2017} used an ensemble model that includes several models trained on past data, and once the drift occurs, they train a new model purely using the fresh data and add it to the existing ensemble. Contextualized modeling is another method for handling concept drifts \cite{Ying2005, Gama2009, Reis2018a, Reis2018b, Nasc2018} that identifies a number of recurring contexts and generates separate models per individual context. During the testing phase, the best model is selected (using a selection strategy) from the previously trained models based on the characteristics of the testing data. Works in \cite{Ying2005, Gama2009} assumed that true labels for the testing instances are available to help them select the model that has the best performance for the given test data. In \cite{Nasc2018}, authors used temporal information to select the model that is trained specifically for a certain month of the year or a specific time of the day. Works in \cite{Reis2018a, Reis2018b} select the best model by comparing the classification score distributions of the models on their training dataset versus a given testing dataset -- the model with the closest score distribution (between its training dataset and the testing dataset) is selected as the best model.

In Chapter~\ref{ch:scalable infer}, I study the challenge of concept drift in a practical use case where an ISP aims to detect IoT devices in its subscribed home networks. I experiment with two inference techniques, namely contextualized data modeling (one classifier per individual training homes) and global modeling (one classifier trained on combined data from all training homes), and analyze their performance under spatial and temporal concept drifts. For contextualized modeling, I develop and quantify the efficacy of strategies (idealistic and pragmatic) to select the best model.

\subsection{Resiliency of Machine Learning Models}

Although ML models have shown promising results in IoT traffic inference, their integrity is vulnerable to a type of attack known as {\textbf{adversarial machine learning attacks (adversarial attacks)}}. This type of data-driven attack was first studied in machine vision (image recognition), where the objective is to create an adversarial copy of a benign image indistinguishable from human eyes. However, the image classifier gets subverted and mispredicts the adversarial copy. In network security, there could be a similar malicious objective whereby the adversary changes the attack traffic in a specific manner (with prior knowledge) that can go undetected by an ML model monitoring the network traffic. 

The process of finding how to manipulate the input data (images in computer vision and network packets in the context of this thesis) to change the output of the model is called adversarial learning. Adversarial learning in network traffic inference differs from computer vision: (1) for image recognition, the similarity of adversarial instances to benign instances (expected to look identical to human eyes) is the primary constraint of the problem. However, this constraint can be relaxed in cyber security and traffic inference problems; (2) there is no requirement for the impact (intensity) of intended attacks in computer vision problems, \ie any adversarial instances are accepted as long as they subvert the model and look identical to benign images; (3) practicality of adversarial attacks (\ie executing them in a real environment) is of concerns in cyber security since every adversarial instance (a set of numerical values) may not be necessarily realized as network packets, which imposes a new set of constraints that do not exist in the computer vision context.

Adversarial attacks on differentiable models like neural networks have been widely studied in image recognition and cyber security research problems \cite{Caminero2019,Abusnaina2019,Ferdowsi2019,Ibitoye2019,Goodfellow2020}, while non-differential models like decision tree-based models remain relatively under-explored. In fact, adversarial learning techniques (computing gradients of the model loss function) for differentiable models cannot be readily applied to non-differentiable models; hence, new algorithms and methods are needed for the latter.

\vspace{-3mm}
Some prior works studied adversarial attacks on decision tree-based models \cite{Chen2019, Kantchelian2016, Zhang2020}. Authors in \cite{Chen2019} showed the vulnerability of Gradient Boosting and Random Forest models against adversarial attacks using methods developed by works in \cite{Cheng2018, Kantchelian2016}. They also developed a robust technique to avoid adversarial attacks by incorporating adversarial instances into the training phase, shaping the ensemble model, so it will still give high accuracy for attack inputs. Work in \cite{Zhang2020} defined its adversarial attack in a way to find minimal changes to a single feature of a benign instance that would result in misclassification irrespective of the model confidence. Work in \cite{Kantchelian2016} developed mixed-integer linear programming techniques to solve an optimization problem for adversarial attacks on tree ensembles. All existing works applied their methods to public datasets like \cite{MNIST} and \cite{fashion_mnist}, which are widely used in computer vision research studies.

\vspace{-3mm}
Authors of \cite{Caminero2019} employ reinforcement learning to generate adversarial instances against their multi-class inference model (four classes of cyber attacks plus a class of benign traffic). In that work, the adversary iteratively adds random noises to samples of an attack dataset and presents them to the inference model until they get classified as benign. However, randomly generated numerical instances do not necessarily represent cyber attacks that can be truly launched. Authors of \cite{Abusnaina2019} studied the impact of adversarial attacks on models that detect malware-infected IoT applications. Their neural network-based model infers from a graph-like data structure of features computed from the binary file of applications. Work in~\cite{Ferdowsi2019} trained a set of binary-class (malicious versus benign) inference models, each per unit of IoT devices. The authors used GAN (Generative Adversarial Networks), a well-known adversarial learning technique for neural networks (originally designed for image classification). Their primary objective was to improve the resilience of their model during training by including adversarial instances in the training set.

Another work~\cite{Ibitoye2019} focused on neural networks for detecting network-based attacks. The inference model is a multi-class classifier with four classes of cyber attacks and one class of benign traffic. Their objective was to manipulate numerical instances of either attack or benign for which the model makes incorrect predictions. The authors directly borrowed techniques from image recognition and did not demonstrate the feasibility of subverting the model on a real network. Importantly, they do not attempt to refine their vulnerable inference model.

In Chapter~\ref{ch:secure infer}, I address some of the shortcomings in existing works on adversarial attacks. First, I focus on decision tree-based models as they are relatively popular (\cite{Bezawada2018, arunan19, Eirini2019, Hasan2019, Doshi2018}) in IoT traffic inference and have not been studied as comprehensively as other ML models. Second, I introduce a new set of constraints relevant to the cyber security context (nonexistent in computer vision and image recognition) that ensure the generated adversarial attacks are feasible to launch in IP (Internet Protocol) networks. Finally, I develop a technique that patches (refines) a vulnerable decision tree-based model without re-training the model.

\vspace{-6mm}

\subsection{Deterministic Inference Models}
\vspace{-5mm}
While some traffic features, such as packet and byte count, display properties like following distribution probabilities in their value ranges, some features have nominal (categorical) values. For instance, IP addresses, domain names, and service port numbers that a given IoT device communicates with are considered nominal features. Some existing IoT traffic analysis works used deterministic (heuristic) models to infer from nominal features. Works in \cite{Hamza2022,Hamza2018,Hamza2018b} developed deterministic models to detect IoT device types and abnormal network flows using MUD profiles. The model creates a run-time profile for each IoT device of which the type is yet to be detected. Run-time profile is a tree data structure that includes observed network flows. Then, the model can select the closest type by comparing a given run-time profile against previously learned profiles of the known device types.

In \cite{Arunan2018}, authors developed a tree structure based on open TCP ports on IoT devices. The type of a given IoT device on the network is determined by progressively checking its open ports on the pre-populated tree.
The process continues until a leaf node is reached. Each leaf either contains a known device type or is blank which means that the given combination of open/closed TCP ports is unknown (unseen).

Some works like \cite{IoTInspector, Kumar2019} applied signature matching and regular expressions to the content of network packets for detecting IoT device types. In \cite{IoTInspector}, authors used NetDisco \cite{NetDisco} tool that scans the local network by mDNS, SSDP, and UPnP queries and looks the received responses up in a database of signatures inferring the type of corresponding IoT devices. In \cite{Kumar2019}, authors developed regular expressions for different device types based on the class ID field in DHCP packets, device type field in UPnP packets, title field in HTML packets, and name field in mDNS packets.

Works in \cite{Guo2018, Guo2020, Saidi2020} developed models based on IP addresses and domain names of cloud servers communicating with IoT devices. In \cite{Guo2018, Saidi2020}, the authors extracted a list of server IP addresses and domain names for each IoT device type during training. For testing, if a certain number of servers in a list was observed, they concluded the presence of the device type on the network. In \cite{Guo2020}, the deterministic model detected DDoS attacks by measuring excessive communications with new servers that were not captured in the training phase.

This thesis employs both deterministic (Chapter~\ref{ch:obscure infer}) and stochastic (Chapters~\ref{ch:smart infer}, \ref{ch:scalable infer}, and \ref{ch:secure infer}) models. In Chapter~\ref{ch:obscure infer}, I demonstrate that deterministic models can help mitigate false positives generated by stochastic models.

\begin{table}[t!]
	\caption{Summary of the existing literature on IoT traffic inference.}
	\centering
	\renewcommand{\arraystretch}{1.3}
	\begin{adjustbox}{max width=1\textwidth}  	
		\begin{tabular}{|l|l|l|l|l|l|}
			\hline
			{\textbf{work}} & {\textbf{objective}} & {\textbf{measurement}} & {\textbf{visibility}} & {\textbf{traffic unit}} & {\textbf{inference model}} \\ \hline
			\cite{Miettinen2017,Bezawada2018,Meidan2017,Sharma2022, Anat2020} & identification & passive & full & packet & stochastic \\ \hline
			\cite{Trimananda19,Mazhar2020} & identification & passive & full & packet & deterministic \\ \hline
			\cite{Hamza2018b} & identification & passive & full & flow & deterministic \\ \hline
			\cite{Thangavelu2019,Marchal19,arunan19,Nizar2019} & identification & passive & full & combined & stochastic \\ \hline
			\cite{Yang2019} & identification & active & full & packet & stochastic \\ \hline
			
			\cite{Arunan2018,IoTInspector} & identification & active & full & packet & deterministic \\ \hline
			\cite{Feng2018} & identification & active & partial & packet & deterministic \\ \hline
			
			\cite{Kotak2021,Xiaobo2022} & identification & passive & partial & packet & stochastic \\ \hline
			\cite{Guo2018} & identification & passive & partial & packet & deterministic \\ \hline
			\cite{Meidan2020} & identification & passive & partial & flow & stochastic \\ \hline
			\cite{Saidi2020} & identification & passive & partial & flow & deterministic \\ \hline
			\cite{Engelberg2021} & identification & passive  & partial & combined & stochastic \\ \hline
			
			\cite{Doshi2018,Hasan2019,Eirini2019,Enkhtur2021} & monitoring & passive & full & packet & stochastic \\ \hline
			\cite{Brun2018,Meidan2018} & monitoring & passive & full & flow & stochastic \\ \hline
			\cite{Nguyen2019} & ident.+monitor. & passive & full & packet & stochastic \\ \hline
			\cite{Guo2020} & ident.+monitor. & passive & full & packet & deterministic \\ \hline
			\cite{Arunan2020a,Arunan2020b} & ident.+monitor. & passive & full & flow & stochastic \\ \hline
			
			\cite{Hamza2018,Hamza2022} & ident.+monitor. & passive & full & flow & deterministic \\ \hline
			\cite{Hamza2019,Divakaran2020} & ident.+monitor. & passive  & full & flow & stochastic+deterministic \\ \hline
			\cite{Kumar2019} & ident.+risk assess & active & full & packet & stochastic+deterministic \\ \hline
			
			\cite{Loi2017,Mandalari2021,Lyu2017}  & risk assess & active  & full & packet & deterministic \\ \hline
			\cite{Ren2019}  & risk assess & passive  & full & packet & stochastic+deterministic \\ \hline
			\cite{Anand2021} & risk assess & passive  & full & packet & deterministic \\ \hline
			
		\end{tabular}
	\end{adjustbox}
	\label{tab:relatedworks}
\end{table}

\section{Research Gaps in IoT Traffic Inference}

Table~\ref{tab:relatedworks} summarizes existing relevant works on IoT traffic inference. Some have their objective as identification, monitoring, risk assessment, or a combination of identification and monitoring/risk assessment. Next, I conclude this chapter by pointing out some of the research gaps in the literature, which this thesis will address in the following chapters.

\begin{figure}[t!]
	\begin{center}
		\includegraphics[width=0.99\textwidth]{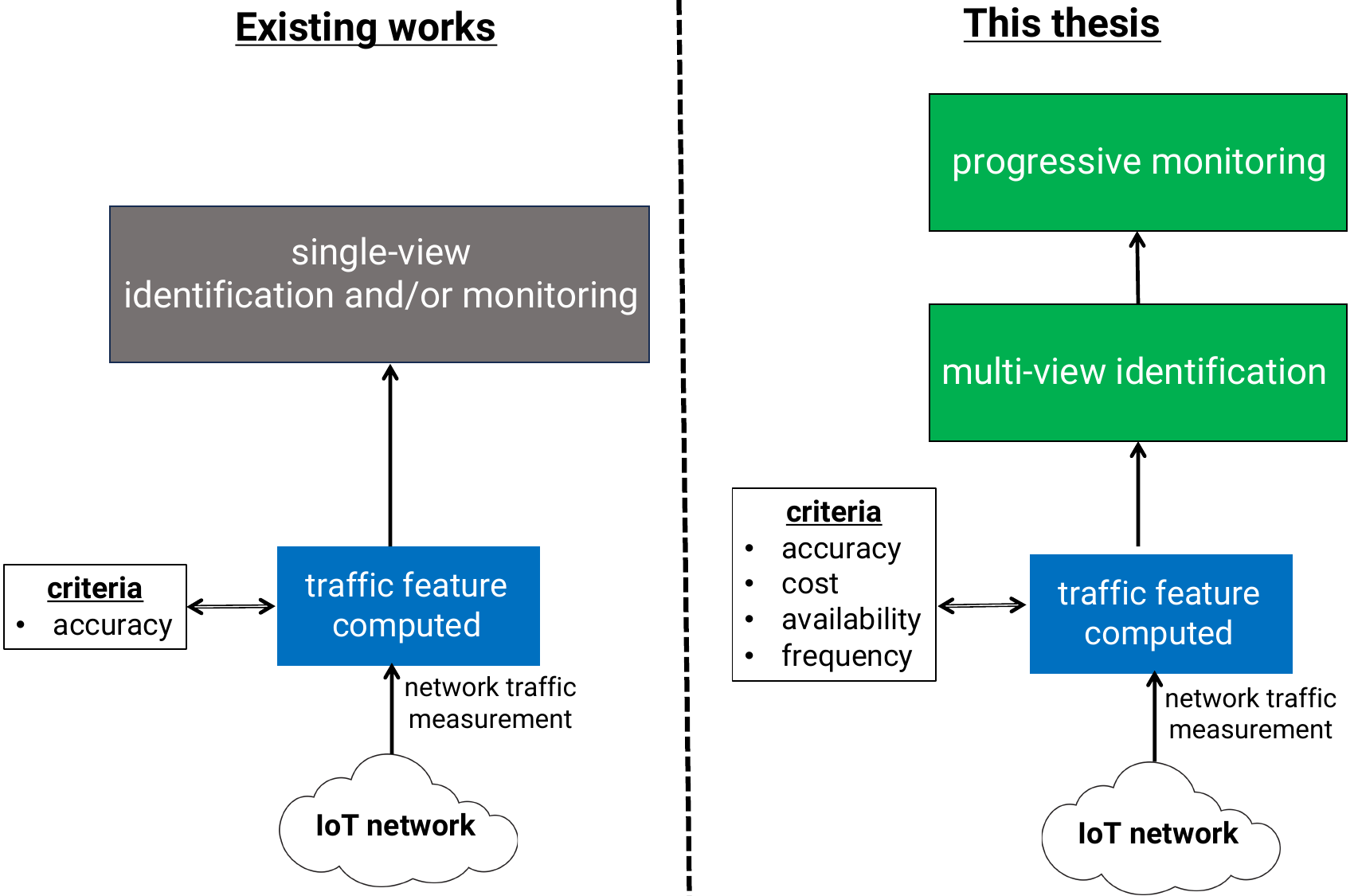}
		\caption{Differentiating this thesis from existing works in terms of feature selection criteria, and inference tasks and techniques.}
		\label{fig:gap1}
	\end{center}
\end{figure}

(1) The majority of existing works either focused on identification or monitoring -- even those listed in Table~\ref{tab:relatedworks} with both objectives did not clearly distinguish the two tasks and used an inference model for both of them. This is also shown in Fig.~\ref{fig:gap1} which compares the methods used for identification/monitoring in the existing works with the techniques I develop in Chapter~\ref{ch:smart infer}. Also, as explained in \S\ref{sec:features}, one can use a variety of traffic features for traffic inference. However, existing works only compared traffic features based on their prediction power by metrics like accuracy (mentioned as criteria in Fig.~\ref{fig:gap1}). However, features differ in other aspects, like computation cost, frequency, and availability. Hence, without considering different characteristics of traffic features, existing inference models are developed with a mixture of various features which can negatively impact the performance. In \S\ref{ch:smart infer}, I develop a generic inference system with multi-view identification and progressive monitoring stages distinctly defined and performed. Then, I thoroughly analyze the performance of specialized models and demonstrate how they can outperform a single model with a combined set (superset) of features. The progressive monitoring model can improve explainability, which is considered a challenge of machine learning models \cite{kok2022expai}).

\begin{figure}[t!]
	\begin{center}
		\includegraphics[width=0.99\textwidth]{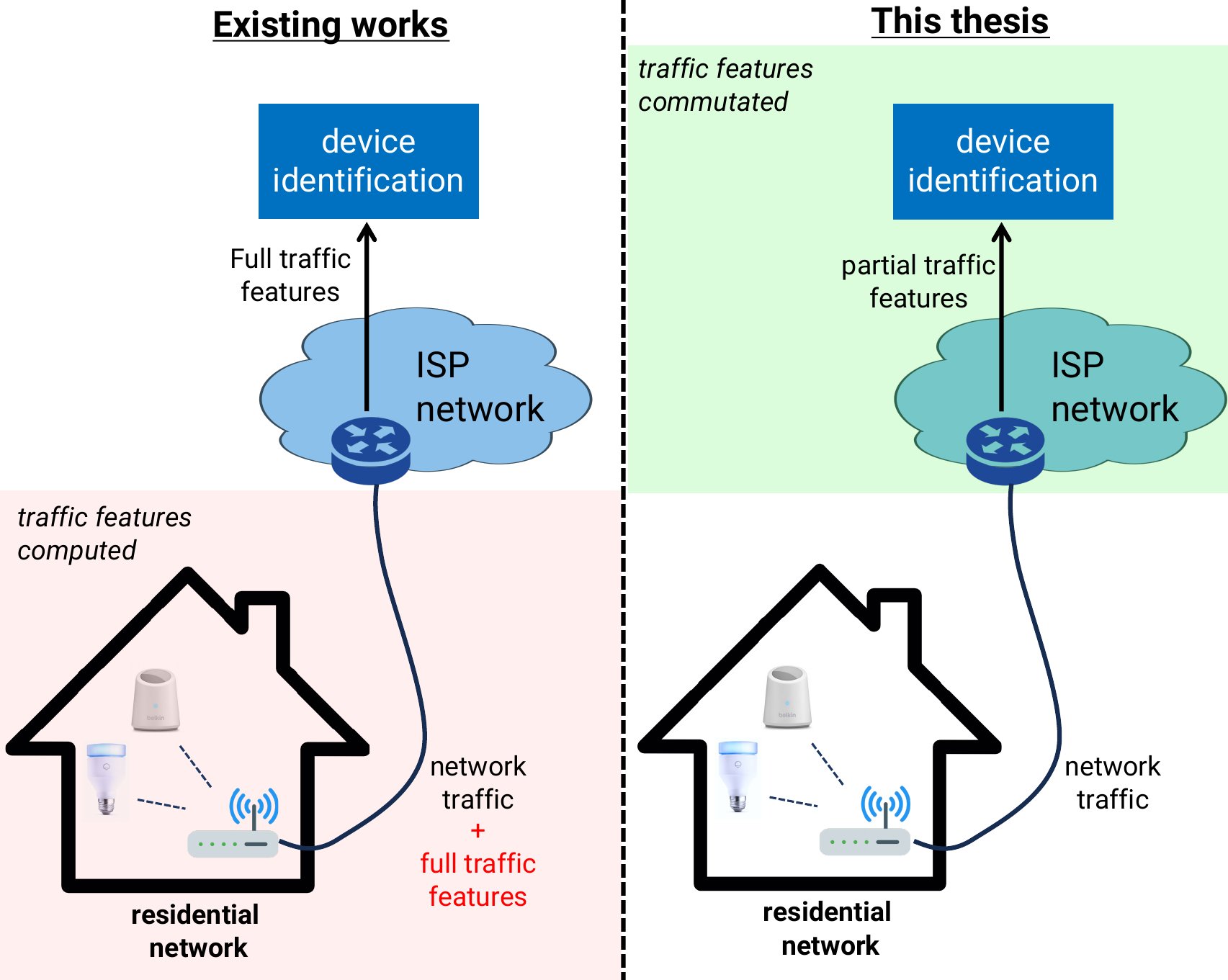}
		\caption{Differentiating this thesis from existing works in terms of traffic visibility.}
		\label{fig:gap2}
	\end{center}
\end{figure}

\begin{figure}[t!]
	\begin{center}
		\includegraphics[width=0.99\textwidth]{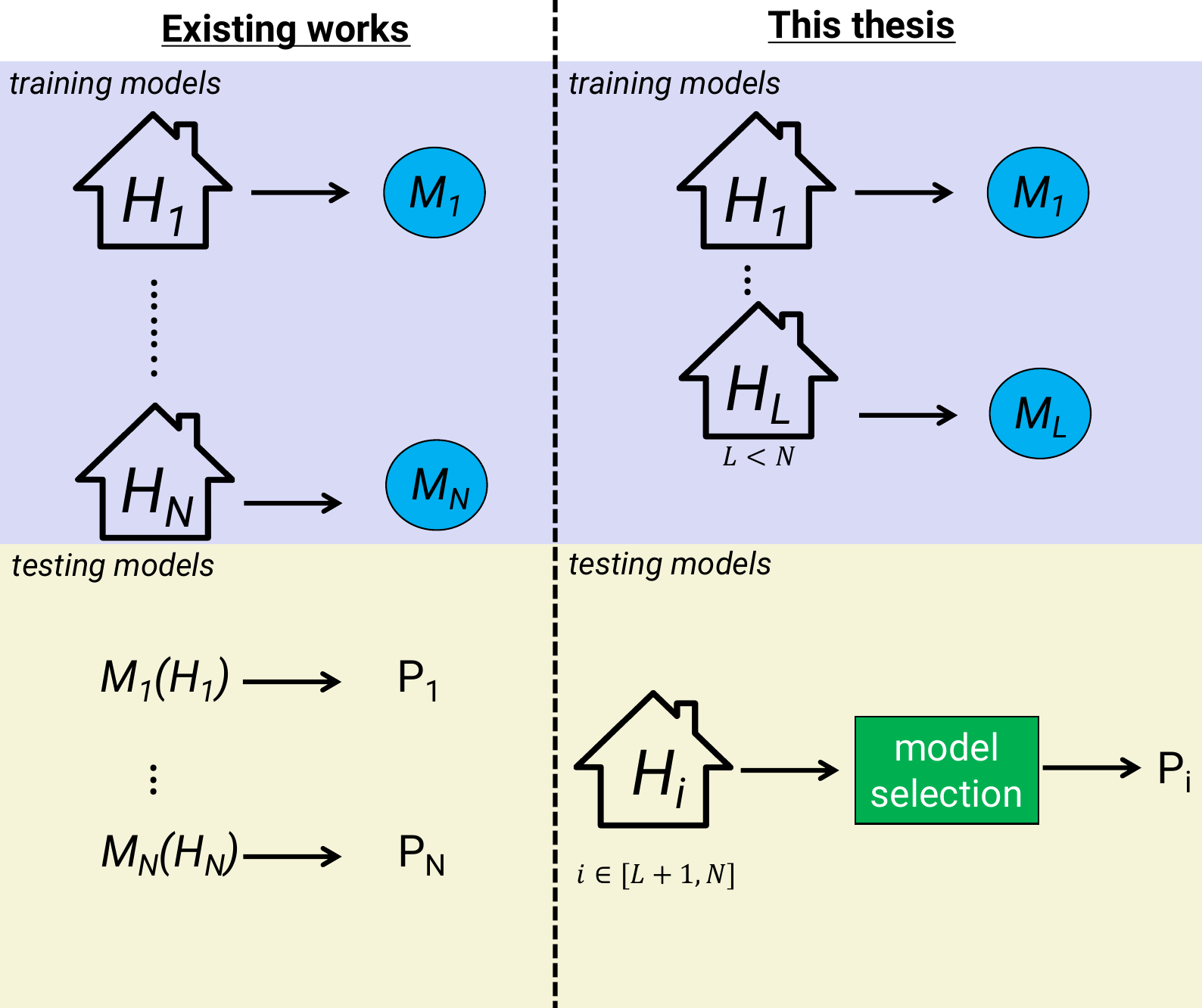}
		\caption{Differentiating this thesis from existing works in terms of concept drift.}
		\label{fig:gap3}
	\end{center}
\end{figure}

\begin{figure}[t!]
	\begin{center}
		\includegraphics[width=0.99\textwidth]{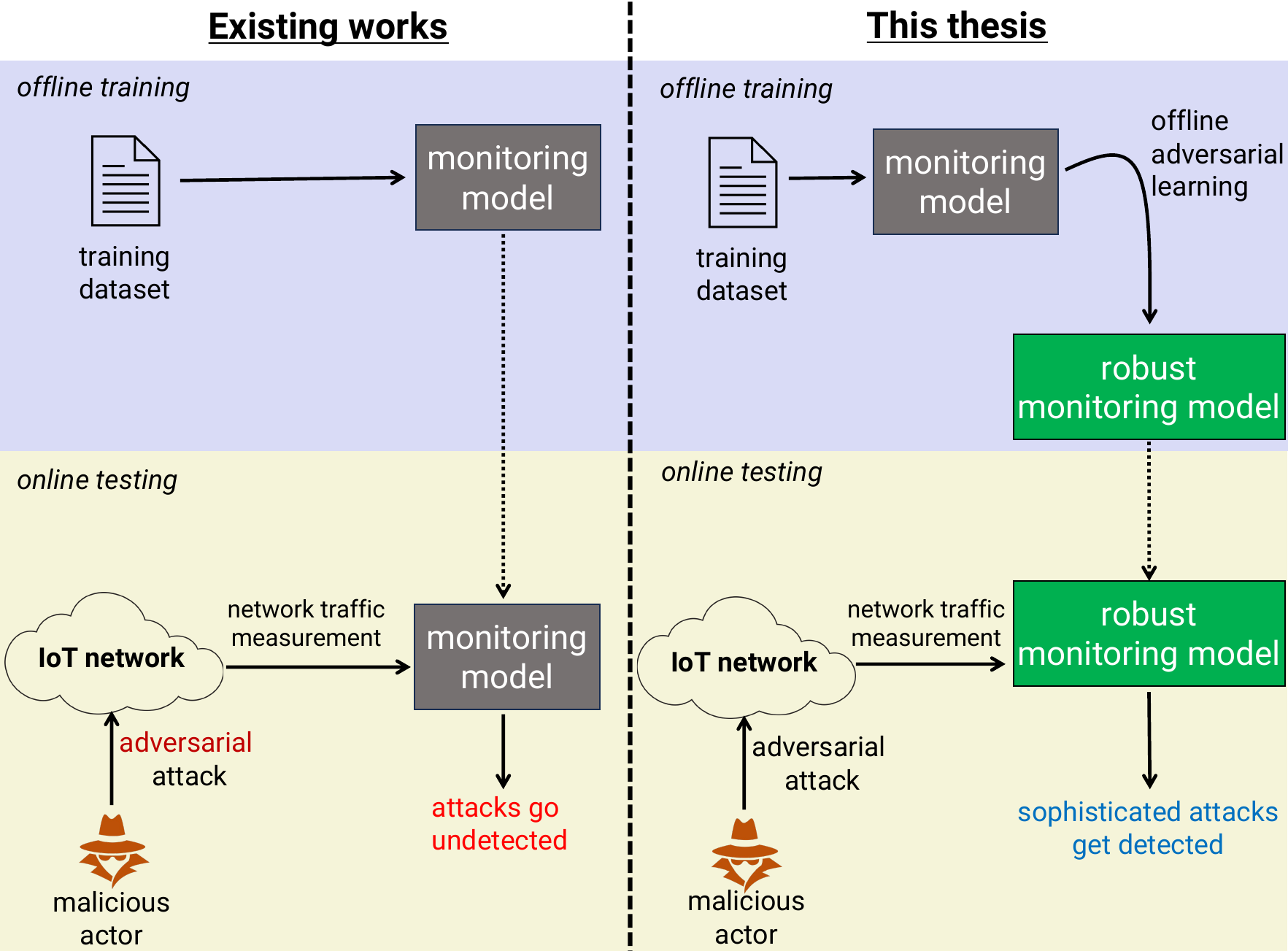}
		\caption{Differentiating this thesis from existing works in terms of resiliency against adversarial attacks.}
		\label{fig:gap4}
	\end{center}
\end{figure}

\vspace{10mm}
(2) Most existing works require full visibility into IoT traffic (measurement is done close to connected devices) to achieve their inference objectives. This is depicted in Fig.~\ref{fig:gap2} where full traffic features are computed at the home network. Existing models and methods that infer from partial traffic have shortcomings that make them practically challenging in some instances due to a lack of scalability. In Chapter~\ref{ch:obscure infer}, I consider an ISP that aims to detect vulnerable IoT devices in the home networks of its subscribers at scale. This is particularly challenging due to the heterogeneity of devices and network settings in home networks \cite{trendSmartHomeRisk}. To address this, I develop a scalable inference model that detects IoT devices using obfuscated and coarse-grained flow data (noted as partial traffic feature in Fig.~\ref{fig:gap2}). I also demonstrate that deterministic and stochastic models can be combined for this problem to reduce false positives significantly.

\vspace{10mm}
(3) Existing works have shown the impact of temporal concept drift on the performance of their inference models. Existing works, as shown in Fig.~\ref{fig:gap3}, train a separate model $M_i$ for each given context $H_i$ (home network in this case) and then test it based on testing data collected from the same context to obtain a performance metric $P_i$. This can be prohibitively challenging as the number of smart homes is growing significantly, with about 300 million smart households in 2022 \cite{smartHome2022}. Still, to my knowledge, no prior work has studied concept drift in IoT traffic or developed methods to handle its challenges in machine learning-based traffic inference. In Chapter~\ref{ch:scalable infer}, I focus on detecting vulnerable IoT devices in home networks from an ISP perspective (originally defined in Chapter~\ref{ch:obscure infer}) to highlight the existence of concept drift and quantify its impact on behaviors of IoT devices. As depicted in Fig.~\ref{fig:gap3}, given $N$ home networks, I train $L$ models ($L<N$) and develop some methods to select the closest model for any unseen home $H_i$ which $i\in[L+1,N]$. I develop two inference techniques to manage this practical challenge and explain how they can be dynamically merged to yield better performance. I also demonstrate how the concept drift can be managed when there is no access to true labels in the testing phase.

(4) To my knowledge, no prior work has systematically and practically studied the impact of adversarial attacks on IoT traffic inference models. Few works studied decision tree-based models compared to neural networks, even in computer vision (\ie the original context of adversarial attacks). Therefore, as illustrated in Fig.~\ref{fig:gap4}, existing IoT monitoring models are vulnerable to adversarial attacks which can lead to inefficacy in detecting attacks. In Chapter~\ref{ch:secure infer}, I develop a framework to quantify the resiliency of decision tree-based models against adversarial attacks in theory and practice. Then, I develop a technique to patch vulnerable decision tree-based models that can make them robust against data-driven adversarial attacks.

%% file: tex/Chapter3-Smart-Infer/main.tex
\chapter[Efficient and Progressive IoT Traffic Inference]{Efficient and Progressive \\IoT Traffic Inference}
\label{ch:smart infer}


Machine learning-based techniques have proven to be effective in IoT network behavioral inference. Existing works developed data-driven models based on features from network packets and/or flows, but mainly in a static and ad-hoc manner, without adequately quantifying their gains versus costs. In the first technical chapter of my thesis, I develop a generic architecture for two distinct inference tasks, namely IoT network behavior classification and monitoring, that flexibly accounts for various traffic features, modeling algorithms, and inference strategies. I argue that quantitative metrics are required to systematically compare and efficiently select various traffic features for IoT traffic inference. 

This chapter makes three contributions:
\textbf{(1)} For IoT behavior classification, I identify four metrics, namely cost, accuracy, availability, and frequency, that allow us to characterize and quantify the efficacy of seven sets of packet-based and flow-based traffic features, each resulting in a specialized model. By experimenting with traffic traces of 25 IoT devices collected from our testbed, I demonstrate that specialized-view models can be superior to a single combined-view model trained on a plurality of features by accuracy and cost. I also develop an optimization problem that selects the best set of specialized models for a multi-view classification;
\textbf{(2)} For monitoring the expected IoT behaviors, I develop a progressive system consisting of one-class clustering models (per IoT class) at three levels of granularity. I develop an outlier detection technique on top of the convex hull algorithm to form custom-shape boundaries for the one-class models. I show how progression helps with computing costs and the explainability of detecting anomalies; and, \textbf{(3)} I evaluate the efficacy of the optimally-selected classifiers versus the superset of specialized classifiers by applying them to our IoT traffic traces. I demonstrate how the optimal set can reduce the processing cost by a factor of six with insignificant impacts on the classification accuracy. Also, I apply the monitoring models to a public IoT dataset of benign and attack traces and show they yield an average true positive rate of 94\% and a false positive rate of 5\%. Finally, I publicly release the data (training and testing instances of classification and monitoring tasks) and code for convex hull-based one-class models.

\section{Introduction}
\label{ch3:sec:intro}
New IoT devices are emerging each week in the market and are increasingly being deployed in a variety of smart environments like homes, buildings, enterprises, and cities. Many of these networked assets often lack in-built security measures and are ridden with security holes, as amply demonstrated by prior research work \cite{Sivanathan2017,Loi2017}. These assets pose grave risks to the security and privacy of personal and organizational data gleaned from sensors \cite{Sivaraman2018}, but also provides a launching pad for attacks to systems within \cite{Sivaraman2016} and outside \cite{Lyu2017} the organization. Examples of such risks are the attacks by a University’s vending machines on its own campus network \cite{vending_attack}, and IoT botnets DDoS-attacking DynDNS \cite{mirai}. In fact, 57\% of IoT devices can become targets of medium to high impact cyber-attacks \cite{Palo2020}. A survey \cite{forescout2016} revealed that about two-thirds of enterprises do not have adequate visibility over IoT devices connected to their network, leading to at least five security incidents per year. 

Obtaining visibility into IoT network activities is essential \cite{Palo2020} to securing these networks. IoT devices often perform a limited set of tasks. They come with power and computing constraints, and their network activity is relatively less but more distinct than traditional IT devices \cite{arunan19}. This creates an opportunity to develop models \cite{arunan19,Trimananda19,Miettinen2017,Thangavelu2019,Bezawada2018,Saidi2020,Yang2019,Feng2018,Marchal19,Hamza2022,Hasan2019,Doshi2018,Hamza2019,Arunan2020b,Nguyen2019} that learn IoT benign behavior relatively well.
About 85\% of organizations \cite{cyberreport} are planning to utilize artificial intelligence or machine learning models to develop monitoring systems for their assets.

Automatic IoT traffic inference is often achieved in three distinct phases: (1) \textit{discovery} where the identity (\eg MAC and/or IP address) of connected devices is determined (via passive analysis or importing an asset inventory list from active scans), (2) \textit{classification} or \textit{identification} that maps device identities to their types/roles (\eg {\myverb{device D $\longmapsto$} Camera}), and (3) \textit{monitoring} where their network behavior is continuously measured and checked against expected patterns, ensuring their cyber health or flagging any anomalous behaviors. Device discovery has been long studied in computer networks (especially using active scans for IT asset management), and there are tools like Nmap \cite{nmap} for this purpose. Given the overheads and disruption risks introduced by active scanners \cite{Anand2021}, non-invasive or passive\footnote{Analytical techniques are applied to network traffic from a SPAN (Switched Port Analyzer) port.} inference methods sound more appropriate for IoT asset management. 
Some works like \cite{arunan19,Trimananda19,Miettinen2017,Thangavelu2019,Bezawada2018,Saidi2020,Yang2019,Feng2018,Marchal19} focused on traffic classification while others like \cite{Hamza2022,Hasan2019,Doshi2018,Hamza2019,Arunan2020b,Nguyen2019} developed methods for traffic monitoring. A recent work \cite{23nomsActivePAssive} attempted to dynamically leverage the combined capabilities
offered by these two active and passive approaches to characterize IoT assets.

Machine learning algorithms have been widely applied to packet/flow-based traffic features for capturing recurring patterns of IoT devices. Most existing works used a single-combined-view approach  (training a single model on a plurality of traffic features). Single-combined-view models have some drawbacks, like dealing with missing features \cite{Emmanuel2021}, demanding more data \cite{Arman2020}, and overfitting \cite{Zhao2017}. On the other hand, a multi-view approach employs a collection of models (``views''), each specializing in certain aspects of data. For example, in image recognition, views can be images captured with varying angles, texture data, or color information of an object \cite{Zhao2017}. In the context of IoT traffic inference, the single-combined-view approach is dominantly used for developing inference models \cite{Miettinen2017,Nguyen2019,Doshi2018,Hamza2019,Arunan2020b,Marchal19,Yang2019,Bezawada2018,Thangavelu2019}.

Prior research work developed IoT traffic inference models based on various traffic features, some are packet-based, and others are flow-based. The existing literature tends to incorporate traffic features solely based on their prediction accuracy for modeling purposes. I argue that more factors, like how prevalent is a feature in the network traffic across device classes, how frequently certain features emerge to measure, or how expensive it would be to compute features, should also be considered for developing inference models for IoT devices. 
For IoT behavioral monitoring specifically, models are applied in a static manner -- all required network traffic telemetries are collected and checked in one shot for anomalies. Static approaches do not scale cost-effectively. I instead consider a dynamic approach that progresses in multiple stages, ranging from coarse-grained to fine-grained. Progressive monitoring assists IoT network operators in finding the cause of anomalies (explainable inference), particularly at scale with a large number of connected devices.

\newpage
My specific contributions are as follows:

\begin{itemize}
    \item The \textbf{first} contribution (\S\ref{ch3:sec:identification}) focuses on IoT traffic classification. I begin by identifying four metrics that help us systematically compare prediction gains versus measurement costs of specific traffic (packet/flow) features, each resulting in a specialized model. By experimenting with traffic traces of 25 IoT devices collected from our testbed, I demonstrate how specialized models outperform a single-combined-view model (trained on a plurality of features) in prediction accuracy and overall computing costs. I also develop a method to select an optimal set of specialized models for multi-view classification.

    \vspace{1mm}
    \item The \textbf{second} contribution (\S\ref{ch3:sec:monitor}) focuses on IoT behavioral monitoring. I develop a progressive architecture of one-class models at three levels of granularity, improving computing costs as well as the explainability of detecting anomalous traffic patterns. I employ the convex hull algorithm and develop an outlier detection technique on top of it for tightening the benign boundaries in the one-class models.
    
    \vspace{1mm}
    \item  Finally, the \textbf{third} contribution (\S\ref{ch3:sec:eval}) begins by evaluating  the efficacy of the optimally-selected classifiers versus the superset of specialized classifiers by applying them to our IoT traffic traces. The results show how the optimal set of models can reduce the processing cost by a factor of six with insignificant impacts on the prediction accuracy. I also apply the monitoring models to a public IoT dataset of benign and attack traces and show they yield an average true positive rate of 94\% and a false positive rate of 5\%. The data (training and testing instances of classification and monitoring tasks) and enhancement code for the convex hull algorithm are publicly released \cite{smartinfer_data} to the research community.

\end{itemize}

\begin{figure}[!t]
	\centering
	\includegraphics[width=0.85\linewidth]{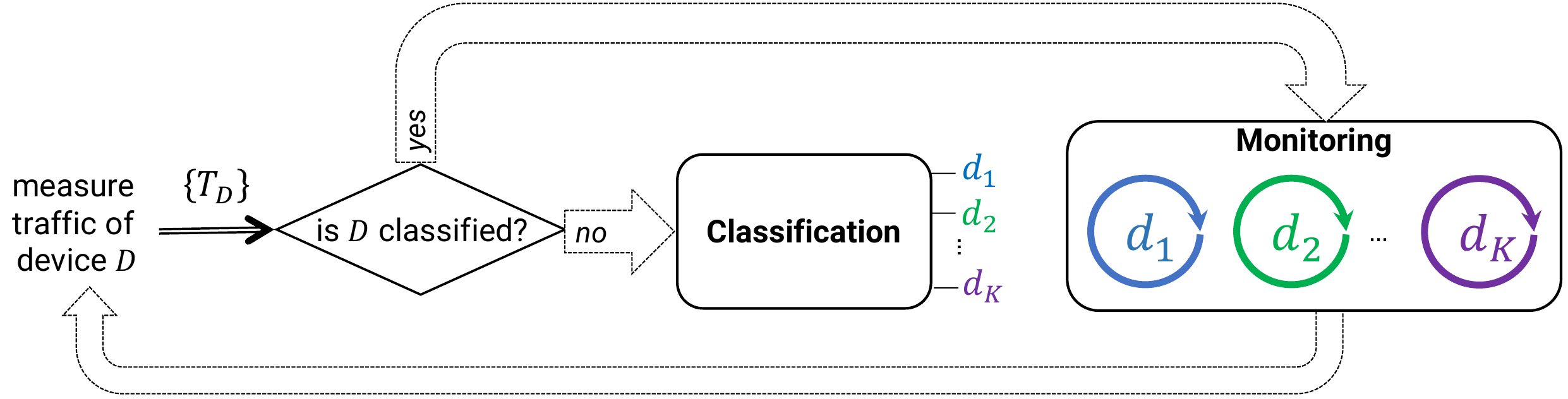}
	\caption{A high-level overview of my system architecture for IoT traffic inference.}
	\label{ch3:fig:sys_arch}
\end{figure}

	Fig.~\ref{ch3:fig:sys_arch} illustrates the inference architecture that I develop in this chapter. The measured network traffic ($\{T_D\}$) of a given IoT device $D$ is presented to the classification model if the type (class) of $D$ is unknown. Once the type of device $D$ is determined, $\{T_D\}$ is given to the one-class monitoring model (specific to the type of $D$) to periodically monitor its network behavior.

\vspace{-9mm}
\section{Multi-View Classification of IoT Traffic}\label{ch3:sec:identification}
\vspace{-3mm}
Discovering the IoT devices connected to a network is the first step for traffic inference. Asset discovery can be made actively (a response from connected devices to probing packets \cite{nmap}) or passively (the first packet sent by each connected device on the network). For this chapter, let us assume a list of identities (\eg  MAC/IP addresses) for the connected devices in the network is available. However, this assumption may not be valid for certain deployments where devices' identities are obfuscated by a middle-box (\eg a NAT-enabled router). In the next chapter, I will study some of the challenges of classifying IoT devices using opaque traffic data and develop some techniques to address them.

\begin{table*}[t!]
	\centering
	\caption{Description of symbols used in this chapter.}
	\vspace{-2mm}
	\renewcommand{\arraystretch}{1.1}
	\begin{adjustbox}{width=\textwidth}
	\begin{tabular}{|l|l|}
		\hline
		{\textbf{symbol}} & {\textbf{description}} \\ \hline
		$d$ & known IoT device type \\ \hline
		$K$ & number of known IoT device types \\ \hline
		$D$ & a given IoT device \\ \hline
		$\{T_D\}$ & network traffic of $D$ \\ \hline
		$\{F_i\}$ & traffic features needed for $M_i$ \\ \hline
		$M$ & multi-class classifier \\ \hline
		$N$ & number of multi-class classifiers \\ \hline
		$X_N$ & a binary vector with size $N$ that shows each model is selected or not \\ \hline
		$A$ & accuracy threshold \\ \hline
		$F_{N{\times}K}$ & a matrix with $N$ rows and $K$ columns that shows if model $i$ is feasible for IoT device type $j$ \\ \hline
		$E$ & number of epochs in each resolution period \\ \hline
		$d_{i}^e$ & predicated device type $d_i$ for epoch $e$ \\ \hline
		$s_{i}^e$ & predicated confidence score $s_i$ for epoch $e$ \\ \hline
		$C_i$ & total number of times device type $d_i$ has been predicted over the resolution period \\ \hline
		$S_i$ & sum of confidence scores for device type $d_i$ over the resolution period \\ \hline
		$\{F_{fg}\}$ & set of fine-grained traffic features for the monitoring phase \\ \hline
		$\{F_{mg}\}$ & set of medium-grained traffic features for the monitoring phase \\ \hline
		$\{F_{cg}\}$ & set of coarse-grained traffic features for the monitoring phase \\ \hline
		$m_{x,y}^{d_i}$ & one-class monitoring model number $y$ for device type $d_i$ with granularity level of $x$ \\ \hline
		$D_{P{\times}P}$ & a matrix including the mutual distance of $P$ points within a given cluster \\ \hline
		$MD$ & average mutual distance of points within a cluster \\ \hline
		$CD$ & average distance of points from the cluster's centroid \\ \hline
		$\alpha$ & sparsity coefficient \\ \hline
		$\beta$ & outlier coefficient \\ \hline

	\end{tabular}
\end{adjustbox}
%
\end{table*}

\begin{figure}[!t]
	\centering
	\includegraphics[width=0.80\linewidth]{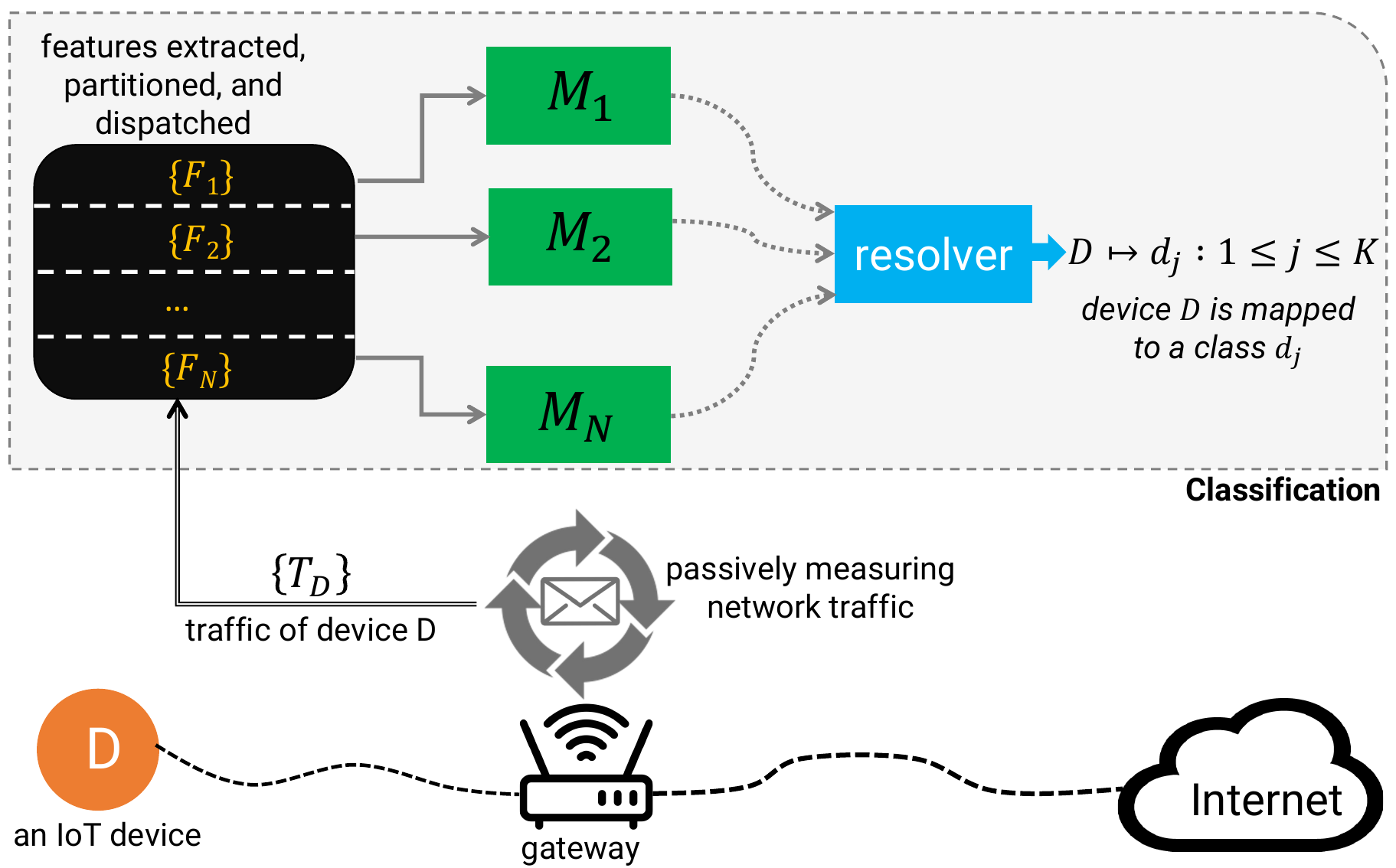}
	\caption{Architecture of multi-view classification of IoT behaviors.}
	\label{ch3:fig:classification_arch}
\end{figure}

As we saw earlier in Fig.~\ref{ch3:fig:sys_arch}, the inference begins with classification (determining the type of individual networked devices), followed by monitoring their network behavior which will be discussed in \S\ref{ch3:sec:monitor}.
Fig.~\ref{ch3:fig:classification_arch} illustrates a zoomed-in version of the classification module. It is the system architecture I develop to infer the mapping between devices and their class. The input of the classification task is $\{T_D\}$, which is the network traffic of an IoT device $D$ that we aim to predict its type. This architecture is designed to be generic, catering to a variety of traffic features and behavioral modeling algorithms. Though I use representative features and models in this chapter for evaluation, one may choose a subset of them or extend them with additional features. 

\vspace{4mm}

For classification, I consider a multi-view approach.  
This means that a collection of models ($M_i: 1\leq i\leq N$), each specializing in a certain aspect (view) of network traffic. Examples of views can be (but are not limited to) domain names, TCP SYN/SYN-ACK signatures, or packet/byte count of flows at different resolutions (\eg 3-tuple, 4-tuple, or 5-tuple). Multi-view learning comes with some benefits compared to a single-combined-view approach: (a) Having multiple distinct models helps with reducing the number of features needed to train each model and decreases the chance of overfitting \cite{Yaser2012}; and (b) In network traffic, certain features may not always be available or appear less frequently for some devices. For example, one IoT device may never use HTTP (unavailable) for network communications, or a
device may generate TCP packets every minute (frequent) while it sends UDP packets once an hour (less frequent). Thus, combining features with diverse availability across heterogeneous classes of IoT assets (single-combined-view) can cause features to be missing in some of the traffic instances measured from the network, which is not desirable for machine learning algorithms \cite{Emmanuel2021}.

Given predictions provided by $N$ models (a multi-view approach), we need a mechanism to consolidate results of (or perhaps, resolve conflicts among) models -- I call it a ``resolver'' module.
In this chapter, I use multi-class models to minimize conflict resolution, but one may use one-class models \cite{Arunan2020b} at this stage too.   
Different strategies, like majority voting, weighted voting, and average weighted voting, can be employed to resolve predictions obtained from $N$ models. Later in this section, I will evaluate the impact and efficacy of several resolution strategies with various lengths of resolution periods (\ie the time needed to resolve device types) in \S\ref{ch3:sec:eval}.
The classification phase ultimately maps each networked device $D$ to one of $K$ known classes $d_j$. Needless to say that the classification architecture does not require or make any assumptions for traffic features, the type of models, or resolver techniques. 
For instance, one may choose a Random Forest model for TCP SYN signatures, a neural network model for domain names, and a Naïve Bayes model for TLS cipher suites. Additionally, views are extensible to incorporate other necessary or desired features.

I note that classical supervised learning algorithms (this thesis included) aim to train a model in the closed-set world, where training and test samples share the same class space (\ie $K$ IoT devices classes). Open-set learning is a more challenging and realistic problem, where test samples from the IoT classes are unseen during training. Automatic recognition of open sets \cite{Reis2018c} is the sub-task of detecting test samples that do not come from the training, which is beyond the scope of this thesis.

Let us begin by identifying metrics to characterize traffic features leading to specialized views. These metrics can help us better understand the difference and quantify the efficacy of various traffic features. Following that, I compare specialized IoT traffic classifiers against a single combined-view one. Finally, I develop an optimization problem for selecting the best set of views.

\subsection{Characterization Metrics and Traffic Features}\label{ch3:sec:MetricsFeatures}

In order to assess various traffic features, we need quantitative metrics. Existing works mainly used the notion of accuracy to characterize and compare different types of features. Though it is an important factor, the accuracy metric is insufficient for a comprehensive assessment. In what follows, I identify four metrics that enable us to compare traffic features.

\vspace{-4mm}
\begin{description}[font=\mdseries\bfseries]
\item[Cost:] Measuring network traffic and extracting the required features incur processing and computing costs. Packet-based features demand packet inspection (by hardware or software). 
For a packet-based feature, I define cost as the average amount of CPU time required to extract it per packet. Flow-based features, on the other hand, can be obtained from programmable or software-defined networking (SDN) switches \cite{Arman2020,Arunan2020a,Hamza2019} without a need for packet inspection. Programmable/SDN switches (when programmed appropriately \cite{iTelescope2019}) can provide inference engines with flow-level telemetries (packet/byte counters) at configurable granularities. 
Note that programmable switches come with a certain capacity of TCAM (Ternary Content Addressable Memory) which could be a limiting factor (cost) for flow-based features. Therefore, I use an approximation for the required TCAM size in bytes for flow-based features.

\vspace{1mm}
\item[Accuracy:] Traffic features differ in terms of the amount of information they carry. For example, some DNS domain names can be unique to a device type, whereas NTP packets typically have a size of 90 bytes (48 bytes of payload plus 42 bytes of UDP, IP, and Ethernet headers) according to the NTP standard specifications \cite{rfc1305} as long as the optional authenticator field is not used. The amount of information can affect the quality of the predictions, which can be measured using various metrics like accuracy, precision, recall, and F1-score. This chapter uses the average of correct predictions across device classes for the accuracy metric.

\vspace{1mm}
\item [Availability:] The heterogeneity of IoT devices makes it challenging to presume what network protocols they use. Therefore, certain traffic features may not be available for some IoT device types (classes). For example, the August doorbell cameras do not send any HTTP traffic, and Samsung smart cameras never send a User-Agent header in their HTTP requests. For a given traffic feature, I define availability as the fraction of total $K$ device classes that we can find that feature from their network traffic.

\vspace{1mm}
\item [Frequency:] Although some traffic features are available for certain device classes, they may not be present all the time (appear infrequently). For instance, JA3 signature is extracted from TLS handshake, which occurs once per TLS session \cite{Arman2020} whereas, Ethernet packet and byte counts are always present for measurement as long as the device is active on the network. Relying on less frequent features can increase the response time of the inference model -- due to waiting time before they emerge in the traffic. For a given feature, I define frequency as the fraction of all time epochs across $K$ classes it is present in training instances. Note that epoch is a configurable parameter. In my experimental evaluations, I use 1-min epochs for classification and 5-min epochs for monitoring.
\end{description}

\vspace{-3mm}
To make our discussion more tangible, I next focus on a representative set of network telemetries, including four packet-based and three flow-based feature sets (for classifying IoT behaviors) to showcase how characterization metrics work. I also look at how specialized-view models (each trained on a set of features) perform versus a single-combined-view model (trained on a combination of all features).

Largely inspired by prior works \cite{Arunan2020a,Arunan2020b,Saidi2020,Arman2020}, I consider four packet-based traffic features as follows.

\vspace{-5mm}
\begin{description}[font=\mdseries\bfseries]
\item [Domain Name:] The name of cloud servers that IoT devices communicate with can be used to identify device types \cite{arunan19, Saidi2020}. Domain names (\eg ``{\myverb{chat.hpeprint.com}}'' for HP printers or ``{\myverb{broker.lifx.co}}'' for LiFX lightbulbs) are extracted from DNS requests sent by IoT devices, and they are less likely (compared with IP addresses) to change over time \cite{Guo2020}.

\vspace{1mm}
\item [SYN signature:] TCP standard \cite{rfc9293} defines an optional field at the end of TCP packets called {\myverb{TCP Options}}. TCP options include a list of fields, some of which may have a value assigned to them. For example, the Maximum Segment Size option must have a value, while the SACK Permitted option (selective acknowledgment) has no value. My recent work \cite{Arman2020} showed that IoT devices often display unique and consistent SYN signatures, revealing their type.

\vspace{1mm}
\item [HTTP User-Agent:] HTTP packets may contain a User-Agent header that allows clients to introduce themselves to the servers they connect. The User-Agent header can highlight the browser and/or operating system so the server can customize its response to clients accordingly. Work in \cite{IoTInspector} employed the User-Agent header to distinguish IoT device types.

\vspace{1mm}
\item [JA3 signature:] TLS connections are initiated by a {\myverb{Client Hello}} handshake packet. This packet contains information about the client's capabilities and preferences for the TLS connection, including the TLS version, cipher suites, a list of extensions, elliptic curves, and elliptic curve formats. Work in \cite{ja3} developed a signature format based on these values to detect malware activities in TLS traffic. For the first time (to my knowledge), this chapter uses JA3 signatures to determine the class of IoT devices.
\end{description}

In what follows, I explain three sets of flow-based features (some employed by prior works \cite{Arunan2020b, Arunan2020a}) that capture distinct patterns of IoT behaviors on the network. Note that ``$\downarrow$'' and ``$\uparrow$'' signs, respectively, refer to the incoming direction of traffic ``to'' the networked IoT device and the outgoing direction of traffic ``from'' the networked IoT device.

\begin{description}[font=\mdseries\bfseries]
\item [Datalink:] This set captures statistical measures of total network activities per each device unit (discovered on the network), including four specific features: $\downarrow$total packet count, $\downarrow$total byte count, $\uparrow$total packet count, $\uparrow$total byte count. I obtain these coarse-grained features from statistics of two flow entries (one matching source MAC address and one matching destination MAC address of packets against a given device MAC address), covering all traffic of layer-2 and above, including (but not limited to) ARP, IPv4, IPv6, UDP, and TCP.

\vspace{1mm}
\item [Transport:] This set increases the granularity of statistics by extending its focus to transport-layer protocols. For this set, four flow entries are needed that match: (a) source/destination MAC address and TCP transport protocol\footnote{IP protocol number = 6.} (two flows) and (b) source/destination MAC address and UDP transport protocol\footnote{IP protocol number = 17.} (two flows). I obtain a total of eight medium-grained features: $\downarrow$TCP packet and byte count, $\uparrow$TCP packet and byte count, $\downarrow$UDP packet and byte count, $\uparrow$UDP packet and byte count, per each device unit (a given device MAC address).

\vspace{1mm}
\item [Application:] This set focuses on fine-grained features specific to network activities of three application-layer protocols, namely TCP/80 (a proxy for HTTP), TCP/443 (a proxy for HTTPS), and UDP/53 (a proxy for DNS). Similar to what was discussed above for the other two sets, each protocol demands two flow entries (matching TCP/UDP based on IP protocol, a corresponding port number, source/destination MAC address), and provides four features meaning a total of 12 application-layer features.
\end{description}

It is important to note that all the flow-based features discussed above can be obtained concurrently from a programmable switch using multiple flow tables, each table corresponding to a specific flow granularity (one for datalink, one for transport, and one for application).

\subsection{Quantifying the Efficacy of Specialized-View Classifiers}\label{ch3:sec:QntfyFeatures}
\vspace{-5mm}

I now quantify the four characteristic metrics of features identified above using the traffic traces of our IoT testbed. I collected PCAP traces (consisting of about 100 million packets exchanged locally and remotely) on our testbed serving 25 consumer IoT devices (a unit per each device type) during February and March 2020. I use a script written in Golang to analyze the collected PCAP files and discretize them into 1-minute epochs (configurable). The script runs on a machine using an 8-core Intel Core i9 CPU, 32GB of memory, and storage of 1TB.


\vspace{-3mm}
My Golang script measures the average CPU time needed for extracting packet-based features of each packet. For flow-based features, the cost is an estimated amount of TCAM required to monitor the activities of each device unit connected to the network. For a given set of flow-based features, the amount of TCAM is proportional to the number of flow rules inserted into the programmable switch. Furthermore, the number of matching fields is another factor determining the amount of required TCAM. 
Let us approximate the total byte size of TCAM required by each set of flow-based features to monitor the network activities of a single IoT device. The size of each matching field comes from the standard size of the respective field. I note that a MAC address is six bytes, an IP protocol field is one byte, and a transport-layer port number field takes two bytes of memory.
Therefore, the TCAM cost for datalink, transport, and application flow features are estimated as 12 bytes (two flows of 6 bytes each), 28 bytes (four flows of 1 plus 6 bytes), and 54 bytes (six flows of 2 plus 1 plus 6 bytes), respectively. Note that increasing the granularity (from datalink to application) requires more matching fields and hence more TCAM memory. I note the cost is measured in different units for packet-based versus flow-based features. One may attempt to unify the cost measures by converting CPU times and TCAM sizes to dollar costs. However, this is beyond the scope of this chapter. I will use a simpler cost proxy for optimization in \S\ref{ch3:subsec:optimization} in order to unify packet and flow-based costs.

\textbf{Classification Dataset:} During each 1-min epoch, I extract the traffic features discussed above for individual active devices. I use the MAC address to identify connected devices and associate traffic features with them. I maintain a list of measured values for packet-based features and compute packet/byte counters for flow-based features during 1-min epochs. Finally, I export the traffic features (list of packet values and flow counters) of a device per epoch to a CSV file row. 

{\vspace{10mm}}
I split the entire processed data into three CSV files: (1) {\myverb{TRAIN}} dataset, containing instances from the month of February that I use to quantify the cost, availability, and frequency metrics, and also to train  machine learning models (specialized ones and the single-combined one); (2) {\myverb{TEST1}} dataset, containing instances from the first ten days of March that are used to measure the accuracy metric for the rest of this section; and, (3) {\myverb{TEST2}} dataset, containing instances from March 11 to 15 that are used to evaluate the performance of the multi-view classification in \S\ref{ch3:sec:eval}. I publicly release these three CSV files to the research community \cite{smartinfer_data}.

Given four sets of packet-based features and three sets of flow-based features, I train a Random Forest model (proven to be effective in learning patterns from IoT traffic \cite{Miettinen2017, Safi2022, arunan19, Arunan2020a}) per each feature set, meaning a total of seven specialized models.
Also, for packet-based features, I employ the bag of words technique, which is a well-known method to handle categorical data in machine learning applications \cite{arunan19, Kumar2019, Holland2020}. 
To obtain the baseline measure of accuracy for specialized models, I apply them to {\myverb{TEST1}} dataset. For each model, I compute three measures of accuracy: average accuracy per IoT class ($\mu$), the standard deviation of accuracy across IoT classes ($\sigma$), and the minimum accuracy across IoT classes ($min$).

\begin{table*}[t!]
	\centering
	\caption{Characteristics of feature sets I quantified by applying specialized classifiers to the data.}
	\vspace{-2mm}
	\renewcommand{\arraystretch}{1.1}
	\begin{adjustbox}{width=\textwidth}
	\begin{tabular}{|lc|cccc|ccc|}
	
		\cline{3-9}
		\multicolumn{2}{l|}{}& \multicolumn{4}{c|}{\textbf{Packet-based 
		Views}} & \multicolumn{3}{c|}{\textbf{Flow-based Views}}\\ 
		\multicolumn{2}{c|}{} & Domain name & SYN sign. & User-Agent & JA3 sign. & Datalink & Transport & Application \\ \hline
		\multirow{3}{*}{\textbf{Accuracy}} & $\mu$ & \low $88$\% & \low $85.9$\% & \high $100$\% & \high $98$\% & \midL $94.4$\% & \midH $97.2$\% & \midH $97.0$\% \\
		& $\sigma$ & \low $29$\% & \low $33$\% & \high $0$ & \midH$5$\% & \midL $15$\% & \midL $9$\% & \midH $7$\% \\ 
		& $min$ & \low $3$\% & \low $0$\% & \high $100$\% & \midH $75$\% & \midL $27$\% & \midL $55$\% & \midH 69\% \\ \hline
		\multicolumn{2}{|l|}{\textbf{Cost}} & \high $127$ ns & \midH $220$ ns & \midL $234$ ns & \low $285$ ns & \midH $12$ bytes & \midL $28$ bytes & \low $54$ bytes \\ 
		\multicolumn{2}{|l|}{\textbf{Availability}} & \midH $96$\% & \midH $96$\% & \low $36$\% & \midL $88$\% & \high 100\% & \midH 96\% & \midH 96\% \\ 
		\multicolumn{2}{|l|}{\textbf{Frequency}} & \low $28$\% & \midL $42$\% & \midL $34$\% & \low $15$\% & \high $100$\% & \high $97$\% & \midH $62$\% \\ \hline
		
	    \hline
	\end{tabular}
\end{adjustbox}\label{ch3:tab:telemetries}
\end{table*}

\textbf{Performance of Specialized-View Classifiers:} Table~\ref{ch3:tab:telemetries} summarizes the four characteristic metrics of the traffic features when I applied their corresponding specialized classifiers to our data. Each row is color-coded in which yellow cells highlight weaker performance, while greener cells show stronger/better performance. Note that the color interpretation varies across metrics, \eg higher average accuracy, lower standard deviation of accuracy, lower cost, and higher availability are considered as better results and hence are highlighted with greener color.

Starting with the domain name model (under packet-based views), we see an average accuracy of $88$\% with relatively large variations across classes that is mainly due to certain domain names shared across different device types.  Manual investigations of the {\myverb{TRAIN}} dataset revealed that Belkin camera, Belkin motion sensor, and Belkin switch communicate with the same domain of their manufacturer (Belkin: ``{\myverb{belkin.com}}'') and CDN service provider (CloudFront: ``{\myverb{cloudfront.net}}''). I also found that devices like Google Home and Google Chromecast perform reverse DNS lookups for eight specific IP addresses, 
which result in DNS queries with a domain name of format ``{\myverb{<IPaddress>.in-addr.arpa}}''. Finally, we observe that queried domain names are available in the traffic of almost every ($96\%$) IoT device and seem to be relatively predictive of IoT classes, but DNS packets are only seen in their traffic just less than $30$\% of the time (not highly frequent).

The specialized model trained on SYN signatures gives an accuracy slightly lower than the domain name model. I manually verified that this is again mainly attributed to the shared SYN signatures in the {\myverb{TRAIN}} dataset. I found ten IoT device types shared their SYN signature with another device type. Similar to what we saw above in domain names, Belkin devices display identical SYN signatures in their network traffic. For example, the Belkin switch emits three unique SYN signatures: two were found in SYN-ACK packets sent locally in response to Samsung SmartThings, Belkin camera, and Belkin motion sensor, and one in SYN packets sent to access cloud-based services remotely.
From these signatures, the first two (local) are shared with the Belkin camera and the Belkin motion sensor, and the third one (remote) is shared with the TP-Link camera, leaving no unique SYN signature for the Belkin switch. After manually investigating the traffic traces of the TP-Link camera, I found no common cloud servers with those of the Belkin switch. I think the shared SYN signature (remote) is probably attributed to using similar firmware or software libraries in the TP-Link camera and the Belkin switch.

\vspace{-2mm}
The two models trained on User-Agent headers and JA3 signatures yield fairly high accuracy (\ie 98-100\%) for those classes in which these features are available ($36$\% for User-Agent and $88$\% for JA3). Interestingly, no common User-Agent or JA3 signatures were found (by manual verification) across all IoT classes. The reason for an imperfect accuracy ($98$\%) of the JA3 model is that the Philips Hue lightbulb comes with two different JA3 signatures in the {\myverb{TEST1}} dataset, of which one was not seen in the {\myverb{TRAIN}} dataset. The two signatures are used when the lightbulb communicates with two different IP addresses in the Google cloud. I manually verified that the contacted server in the {\myverb{TRAIN}} dataset was located in the Netherlands, and the one that was newly seen in the {\myverb{TEST1}} dataset was located in the US.

\vspace{-2mm}
Moving to flow-based views, we see those three specialized models perform relatively well in terms of accuracy but cannot beat User-Agent and JA3 models. As expected, flow-based features are more available and frequent compared to their packet-based counterparts. The accuracy (particularly by measures of $\sigma$ and min) increases steadily in granularity. Fine-grained application models outperform coarse-grained datalink models. Although a slight decrease ($0.2$\%) in the average accuracy ($\mu$) is seen from transport models to application models, the standard deviation and min measures highlight notable improvements.

Table~\ref{ch3:tab:telemetries} highlights the fact that none of the specialized models (and their corresponding features) is perfect by all four metrics. User-Agent and JA3 models seem highly accurate but not desirable by the metrics of availability and frequency. Flow-based models perform moderately better. However, all of them showed a relatively low accuracy in predicting the class of the Nest smoke sensor (the minimum value). However, the domain name model perfectly classified (100\% accuracy) the DNS packets of the Nest smoke sensor. This highlights the fact that a reliable traffic classification for a growing range of device types demands collective specialization -- several models, each measuring and analyzing a unique aspect of the network traffic. Now, the question is ``\textit{how can/should one best employ these specialized features/models for IoT traffic inference?}". In what follows, I train a single combined model on all features discussed above and evaluate its performance against individual specialized models.

\textbf{Specialized Classifiers versus Single-Combined-View Classifier:} 
With the specialized models having their performance evaluated individually and compared with each other, let us look at how they compete with a single-combined model. To have a fair comparison, I train a Random Forest model on all the seven features: domain names, SYN signatures, User-Agents, JA3 signatures, and $\downarrow\uparrow$ packet/byte counts for Datalink, Transport (TCP, UDP), Application (HTTP, HTTPS, and DNS) traffic flows, for the single-combined-view classifier. 
Like I did earlier in this section, I use the {\myverb{TRAIN}} and {\myverb{TEST1}} datasets for training and testing of the combined-view model.

As the combined-view model includes all features shown in Table~\ref{ch3:tab:telemetries}, its cost becomes the sum of the cost of each feature set, meaning about $866$ns CPU time and $94$ bytes TCAM memory. This cost is far higher than any of the specialized models, as reported in Table~\ref{ch3:tab:telemetries}. 
Moving to accuracy, the combined-view model gives an average accuracy (across all 25 IoT classes) of about $96$\% with a standard deviation of $9$\%. The lowest class accuracy ($min$) of the combined-view model is $55$\% for the Nest smoke sensor. 
Comparing these measures with those reported in Table~\ref{ch3:tab:telemetries}, it can be clearly seen that three specialized-view models, namely User-Agent, JA3 signature, and Application models, outperform the combined-view model, even though their features are a subset of those used to train the combined-view model. This could be attributed to many instances with missing-value features due to the variability of features' availability and frequency. Missing values are known to be detrimental to the performance of machine learning models \cite{Kamrul2021, Wang2021}.

\subsection{Optimal View Selection}\label{ch3:subsec:optimization}

With some specialized-view models outperforming the single-combined-view model, I now aim to improve inference quality and efficiency by selecting the optimal set of specialized models (realizing a multi-view inference). The optimal set aims to minimize the total computation cost while it meets a certain level of accuracy, which is explained in detail in what follows.
Needless to say that with sufficient computing and processing powers (CPU and TCAM) and relaxing the cost constraint, one can employ all the specialized models I have discussed so far. However, two considerations are of utmost importance in real practice (particularly at scale): (1) processing costs are always capped; therefore, only a subset of specialized models can be employed to meet constraints, and (2) improving the prediction power (accuracy of classification) realized by a multi-view approach must be balanced against other factors and additional costs. For example, IoT device types that generate HTTP User-Agent headers can be perfectly classified by the corresponding model. However, those types are a subset of devices that generate JA3 signatures, which can also be classified perfectly by the JA3 model. In this scenario, no accuracy gain will be realized if we select the User-Agent model along with the JA3 model.

To formulate an optimization problem, one may attempt to minimize the total cost and/or maximize the prediction accuracy. I note that for my choice of packet-based and flow-based features, processing costs (CPU and TCAM) are in different units and hence demand some forms of variable adaptation or unification. In this chapter, I choose a simple objective: minimizing the number of specialized models considered. Note that this can be approximately considered a total-cost-minimization problem. 
Let $X = \left[x_1,x_2,\ldots,x_N \right]$ denote a vector of binary values, where $x_i=1$ indicates the specialized-view model $M_i$, where $i\in[1,N]$, is selected for inference, and $x_i=0$ otherwise. Therefore, the objective function is given by:
\vspace{-12mm}

\begin{equation}
\begin{split}
& \min \sum_{i=1}^{N} x_i\\
\end{split}
\end{equation}

\vspace{-10mm}

In terms of constraints, every IoT class is expected to realize a prediction accuracy of more than a configurable threshold. That way, a minimum prediction quality is secured across individual classes by incorporating necessary specialized models. The threshold value can be set specifically on a per-class basis. In this chapter, I choose a global threshold (denoted by $A$) across all classes. Therefore, each specialized model is considered either feasible (producing the desired accuracy) or infeasible (not delivering the desired accuracy) per each known class, determined from prior validation/testing on labeled data. Let $F=[f_{i,j}]_{N{\times}K}$ denote the feasibility matrix of $N$ models across $K$ IoT device classes. 

\newcommand{\twopartdef}[4]
{
	\left\{
		\begin{array}{ll}
			#1 & \mbox{if } #2 \\
			#3 & \mbox{otherwise}
		\end{array}
	\right.
}

\begin{equation}
\label{ch3:eq:feasibility}
f_{i,j} = \twopartdef
{1}      {\mbox{model~} M_i \mbox{~is feasible~(accuracy~}{\ge}~A\mbox{)~for class~}d_j}
{0}      {}
\end{equation}
where, $i\in[1,N]$ and $j\in[1,K]$. The constraints, therefore, can be stated by:


\begin{equation}
\begin{split}
 \forall j\in[1,K],~\sum_{i=1}^{N} x_i.f_{i,j}~\ge 1\\
\end{split}
\end{equation}

In other words, this constraint essentially checks all elements in the vector of $X\times F$ are non-zero. 
In \S\ref{ch3:sec:eval}, I show how this optimization problem can be mapped to a convex problem and solved efficiently using off-the-shelf solvers such as CVXPY tool \cite{cvxpy}, initialized by data from {\myverb{TEST1}} dataset. Having our problem defined as a convex optimization problem is attractive as convex problems offer polynomial-time solutions as opposed to non-convex ones that are NP-Hard in general. I will show that the optimally-selected multi-view will incur a sixth of the total cost of all specialized models with almost the same level of prediction accuracy.

\subsection{Resolving Multi-View Predictions}\label{ch3:subsec:resolve}

With the multi-view approach, specialized models work independently and in parallel. Each model makes its prediction whenever the necessary features become available.  
For example, in an epoch, a connected device may generate a DNS query but no TCP SYN packet is sent. 
Given models are often imperfect (may make incorrect and/or less confident classification), a robust strategy is we collect predictions (from various specialized models for a given device $D$) over a period of time (say, six hours) and resolve discrepancies among models and across predictions for a final prediction the class for $D$.

\begin{figure}[!t]
	\centering
	\includegraphics[width=0.82\linewidth]{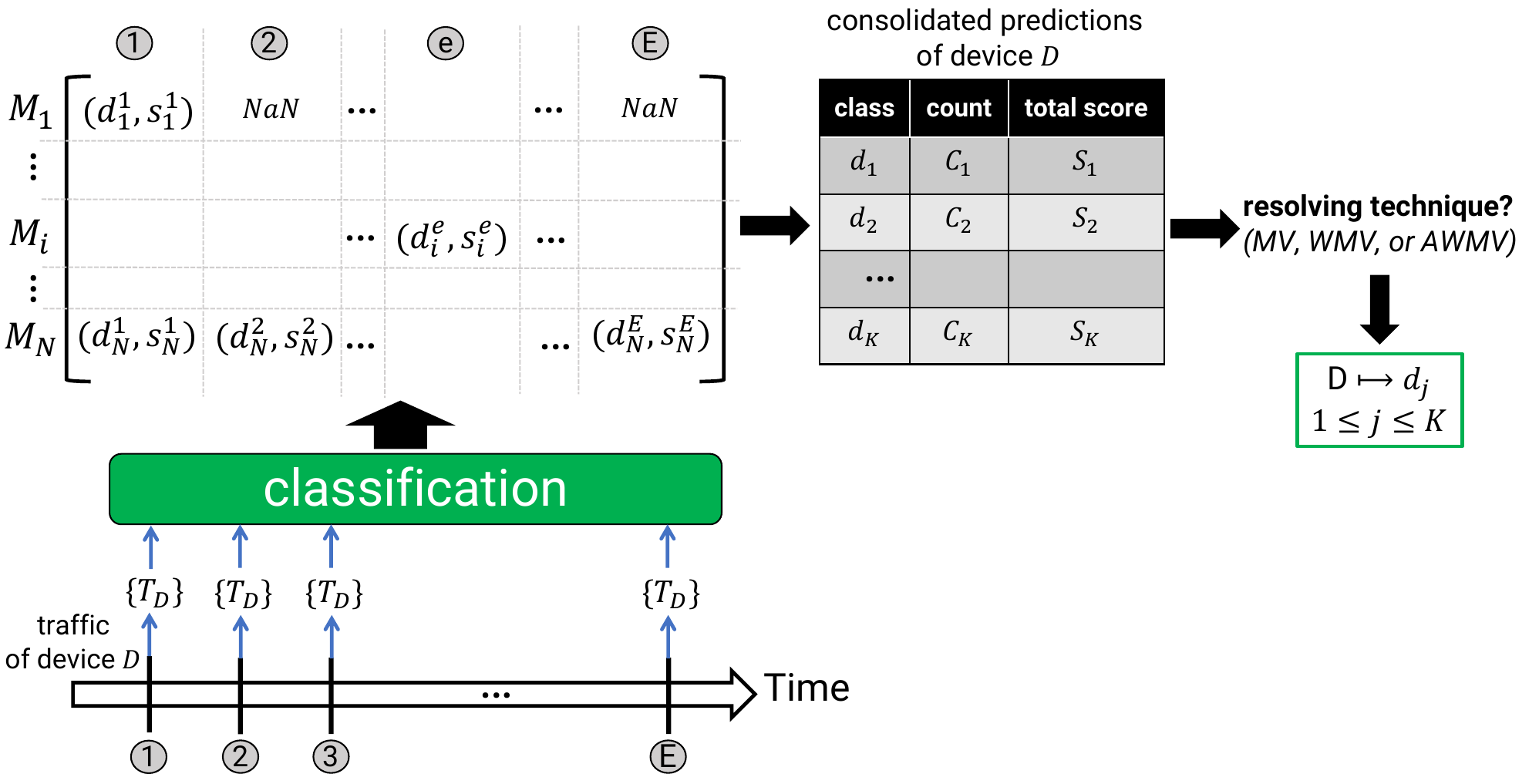}
	\caption[Resolution method for a multi-view traffic classification]{Resolution method for a multi-view traffic classification of IoT device $D$.}\label{ch3:fig:resolution}
\end{figure}

Fig.~\ref{ch3:fig:resolution} illustrates how multi-view predictions are resolved for a given device $D$. The device traffic $T_D$ is measured every epoch time and presented to $N$ specialized models to make their inference (the green classification box).
In total, device $D$ can receive a maximum of $N\times E$ raw predictions within $E$ epochs, represented by the matrix on the top left of Fig.~\ref{ch3:fig:resolution}. Note a cell with value ``{\myverb{NaN}}'' highlights a model (in rows) that is unable to predict at the corresponding epoch (in columns) due to missing features. The prediction $(d_i^e,s_i^e)$ is a two-tuple, consisting of the predicted class $d_i^e$ by model $M_i$ at epoch $e$ with a corresponding confidence score $s_i^e$.

At the end of epoch $E$, raw predictions of device $D$ are consolidated into a chart with $K$ rows and three columns. The first column represents unique classes $d_j$ ($j\in[1,K]$) various models may classify the device. The second column shows the count ($C_j$) of raw predictions for each class $d_j$.
The third column is the total sum of scores in raw predictions for each class $d_j$, computed by:

\vspace{-6mm}

\newcommand{\onepartdef}[2]
{
		\begin{array}{ll}
			#1 & \mbox{if } #2 \\
		\end{array}
}

\begin{equation}
S_j = \onepartdef{\displaystyle \sum_{e=1}^E\sum_{i=1}^N s_{i}^e}{d_{i}^e=d_j}
\end{equation}


I acknowledge that one may want to take a more conservative approach, considering raw predictions with scores higher than their desired level \cite{LCN2021}. In Chapters~\ref{ch:obscure infer} and \ref{ch:scalable infer}, I use classification score to filter high-confidence predictions; however, in this chapter, I do not filter predictions.

In ideal situations, we would expect to see a single row of non-zero counts in the consolidated predictions chart, meaning $D$ is consistently mapped to one class by all specialized models. However, finding $D$ mapped to two or more classes (\eg $D\longmapsto d_1$ and $D\longmapsto d_2$) is a reasonably likely outcome in practice. Therefore, resolving those conflicting predictions becomes important. In this chapter, I consider three resolution strategies discussed next.

\begin{description}[font=\mdseries\bfseries]
\item[Majority Voting (MV):] This method considers the most popular label as the final result. In other words, regardless of the classification score, the class with the largest count (in the second column of the consolidated chart) is the winner. For example, assume at the end of the resolution period, I obtain $D\longmapsto d_1$ hundred times, $D\longmapsto d_2$ ninety times, and $D\longmapsto d_3$ eighty times from the models. In this case, MV selects $d_1$ as the final prediction.
\vspace{1mm}
\item[Weighted Majority Voting (WMV):] This method primarily promotes confident predictions. In other words, regardless of counts, the class with the largest total score (in the third column of the consolidated chart) is the winner.
Suppose in the example above, the total score for $D\longmapsto d_1$ is 75, for $D\longmapsto d_2$ is 80, and for $D\longmapsto d_3$ is 75. In this case, WMV infers $d_2$ as the device type of $D$.
\vspace{1mm}
\item[Average Weighted Majority Voting (AWMV):] This method considers popularity and confidence by taking the ratio of the total score to the count. In other words, the class with the largest average score (the third column divided by the second column of the consolidated chart) is the winner.
In the examples above, the average weight of $D\longmapsto d_1$ is 0.75, and  $D\longmapsto d_2$ is 0.89, and $D\longmapsto d_3$ is 0.93. In this case, AWMV selects $d_3$ as the final label.

\end{description}

In \S\ref{ch3:sec:eval}, I will show how these resolving methods affect the  accuracy of the classification phase.

\newpage

\section[Progressive Monitoring of IoT Network Behaviors]{Progressive Monitoring of \\IoT Network Behaviors}
\label{ch3:sec:monitor}

We saw earlier in Fig.~\ref{ch3:fig:sys_arch} that the monitoring phase commences once the class of device $D$ is determined. This phase is a continuous process for validating the behavior of classified IoT devices against their benign profile. 
With the device class determined, we can employ models trained on expected benign behaviors specific to that class (one-class models) for monitoring. 
At this stage, my objective is to infer from device traffic continuously and close to real-time; therefore, I incorporate traffic features with the highest measure of availability and frequency.
That said, one may choose to employ some additional models that make inferences at slower timescales (catering to less frequent features).
We recall from Table~\ref{ch3:tab:telemetries} that pack-based features are relatively less available and less frequent compared to their flow-based counterparts. Hence, focusing only on those packet-based features can come at the risk of missing real-time short-lived anomalies in the behaviors of connected devices. Flow-based features, instead, are available for real-time monitoring. Therefore, I choose to develop the monitoring inference models based on the flow-based features. 

\begin{figure}[!t]
	\centering
	\includegraphics[width=0.85\linewidth]{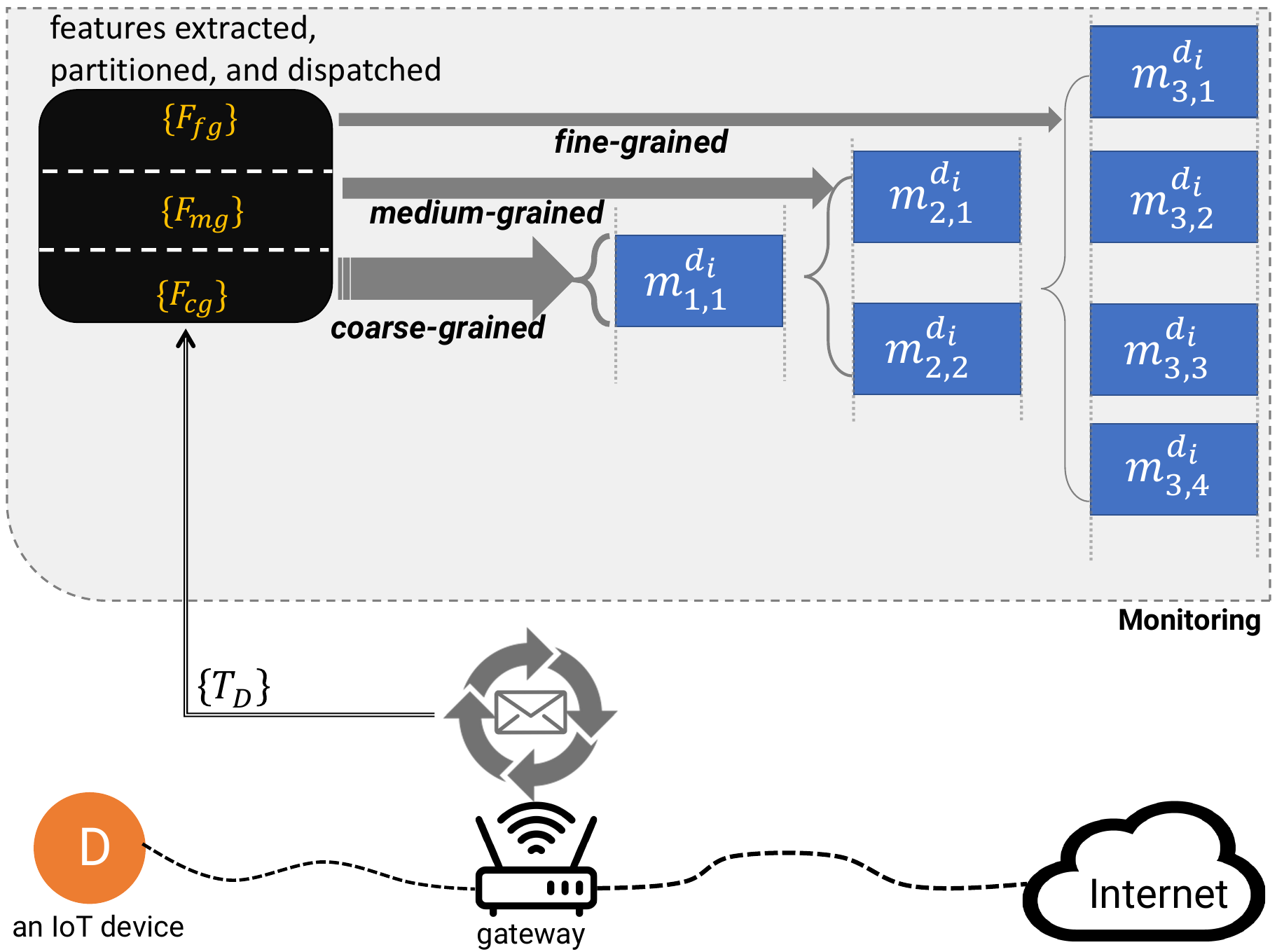}
	\caption{Architecture of progressively monitoring IoT behaviors with three levels of granularity.}
	\label{ch3:fig:monitor_arch}
\end{figure}

Fig.~\ref{ch3:fig:monitor_arch} illustrates the architecture of the monitoring stage of my IoT traffic inference. Given IoT device $D$  mapped to class $d_i$, a number of device-specific one-class models denoted by $m^{d_i}$ are trained and employed to ensure $D$ behaves according to its expected patterns; if not, anomalies are flagged by one or more of those models.

It can be seen in Fig.~\ref{ch3:fig:monitor_arch} that the architecture advocates an approach whereby monitoring progresses in stages, each with models of specific granularity (from coarse-grained to medium-grained to fine-grained) to manage processing costs at scale. Given IoT class $d_i$, let $m^{d_i}_{x,y}$ denote model number $y$ at granularity level of $x$, where $1{\leq}x{\leq}3$ in this chapter. Progression can be triggered in different ways. For example, one may choose to increase the granularity (more expensive monitoring) whenever coarser models flag anomalies for a specific device $D$ (while other devices continue to be monitored by coarse-grained as long as they conform to their expected patterns). Another may want to frequently check the traffic of certain devices against all or just fine-grained models depending upon their available computing resources.

It is important to note that, similar to the classification architecture (\S\ref{ch3:sec:identification}), the monitoring architecture is flexible and generic in terms of its features, models, and progression. 
For example, at the finest level, it allows for the choice of application flows (\eg HTTPS and DNS versus SSH, HTTPS, and DNS).
Even in an environment where a device type communicates over a non-standard protocol (\eg TCP/6000), a one-class model specific to that application flow (service) can be trained and added to this layer. In this chapter, I consider and experiment with a three-layer architecture, but one can freely choose more or fewer layers at different granularities.

At each granularity, a set of one-class models (trained on a corresponding set of features) will be applied. Coarse-grained models are trained on expected behaviors at an aggregate level, while behaviors across certain protocols (transport and/or application layers) are analyzed as granularity progresses. In this chapter, I use models across the three layers as follows.

\begin{description}[font=\mdseries\bfseries]

\item[Coarse-grained:] For this layer, I use a model $m^{d_i}_{1,1}$ that is trained on an aggregate measure of traffic activities, namely the four features of datalink flows (\ie {$\downarrow$}{$\uparrow$} total packet/byte counts of a given MAC address $D$) discussed in \S\ref{ch3:sec:MetricsFeatures}.
\vspace{1mm}
\item[Medium-grained:] For this layer, I use two models: $m^{d_i}_{2,1}$ trained on the feature of TCP flows (\ie {$\downarrow$}{$\uparrow$} TCP packet/byte counts), and $m^{d_i}_{2,2}$ trained on the feature of UDP flows (\ie {$\downarrow$}{$\uparrow$} UDP packet/byte counts).
\vspace{1mm}
\item[Fine-grained:] For this layer, I use four models: $m^{d_i}_{3,1}$ trained on {$\downarrow$}{$\uparrow$} HTTP packet/byte counts, $m^{d_i}_{3,2}$ trained on {$\downarrow$}{$\uparrow$} HTTPS packet/byte counts, $m^{d_i}_{3,3}$ trained on {$\downarrow$}{$\uparrow$}  DNS packet/byte counts, $m^{d_i}_{3,4}$ trained on  {$\downarrow$}{$\uparrow$} NTP packet/byte counts.

\end{description}
\vspace{10mm}

\begin{figure}[t!]
	\begin{center}
		\mbox{
			\subfigure[Default convex hulls.]{
				{\includegraphics[width=0.40\textwidth]{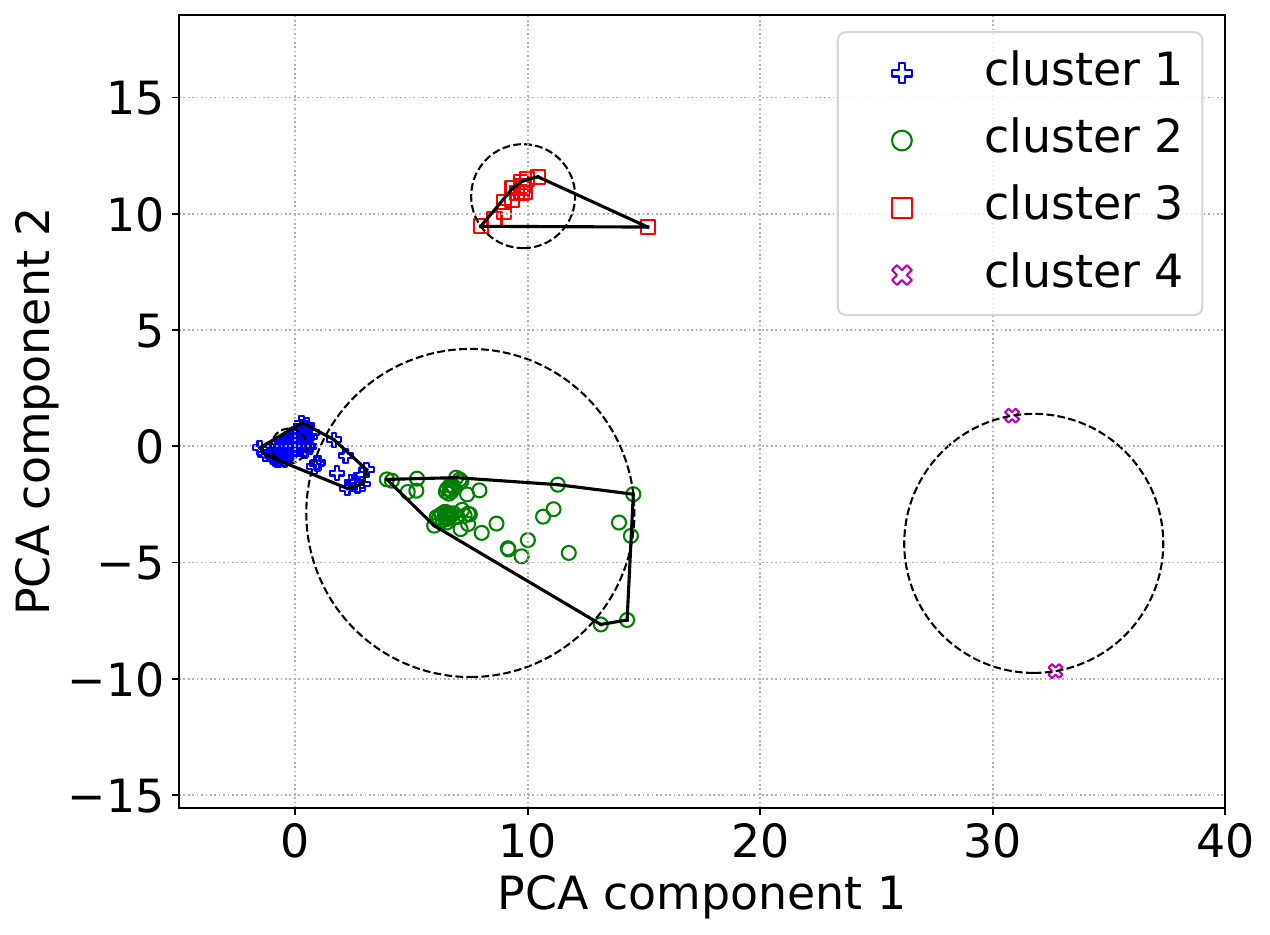}}\label{ch3:fig:default convex}\quad
			}
		}
		\hspace{-4mm}
		\mbox{
			\subfigure[My refined convex hulls.]{
				{\includegraphics[width=0.40\textwidth]{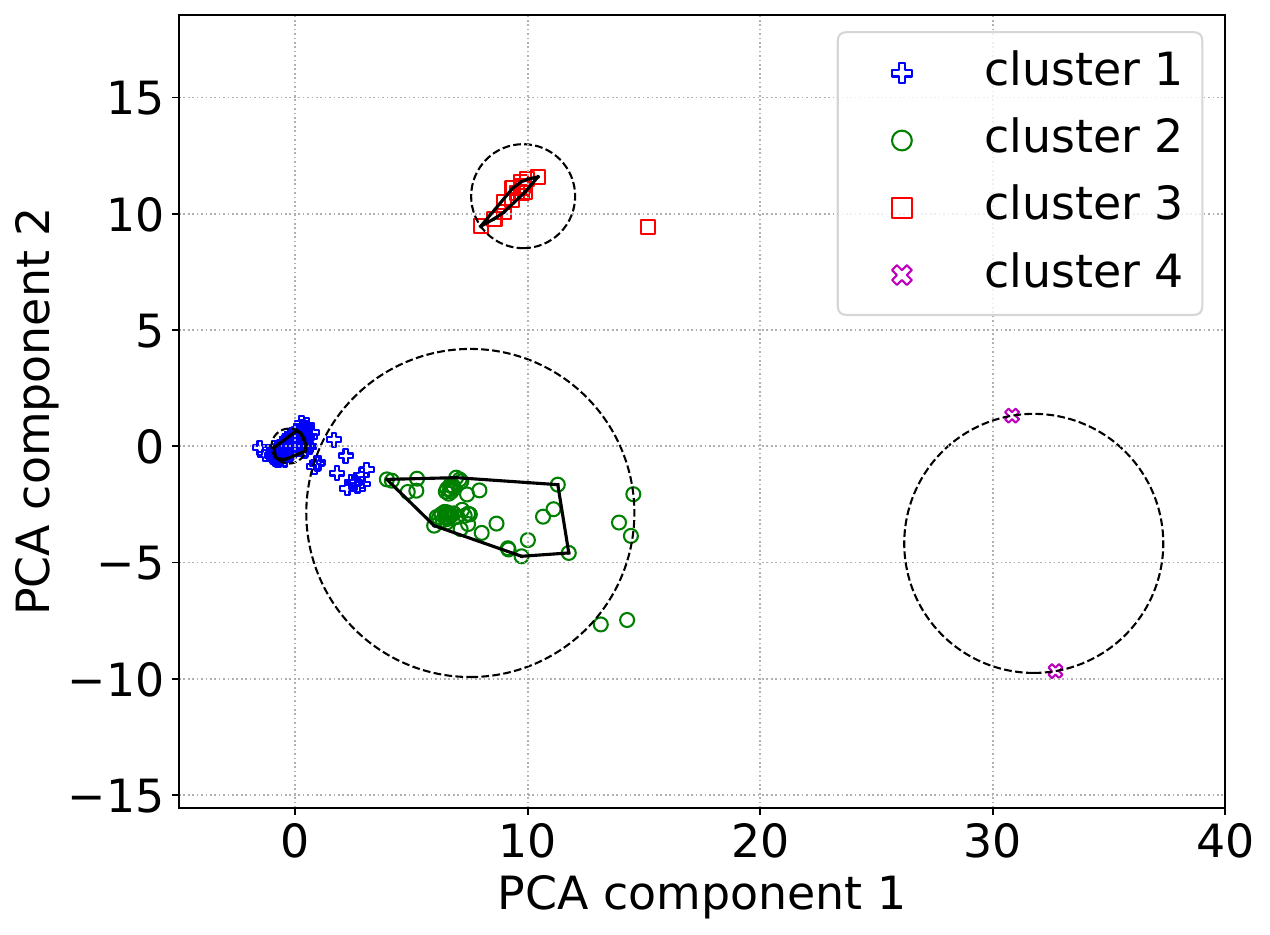}}\label{ch3:fig:tuned convex}\quad
			}
		}
		\caption[Convex hulls versus spherical boundaries for training instances of Belkin switch]{Convex hulls (solid lines) versus spherical boundaries (dashed lines) for training instances of Belkin switch: (a) default convex hulls contain all data instances of corresponding clusters, and (b) my refined version of convex hulls exclude noises from benign boundaries.}
		\label{ch3:fig:convex vs spherical}
		\vspace{-6mm}
	\end{center}
\end{figure}

Note that some of these models may not be applicable to some IoT device classes. For example, Belkin devices, the LiFX lightbulb, the Nest smoke sensor, the Netatmo weather station, and the TP-Link camera do not use HTTPS traffic (based on the dataset I analyzed for this chapter). Therefore, no HTTPS model is trained for those classes.

One-class models are often trained on data purely from an objective class (\eg benign instances \cite{Arunan2020b,Hamza2019}). However, some algorithms like  Isolation Forest \cite{Ahmed2020} require some levels of impurity in training data (\eg benign instances dominate, but some malicious instances are present). 
I argue that including malicious instances in the training dataset may limit the model's learning to only ``known'' malicious behaviors \cite{Robin2010}. Hence, models cannot react reliably to growing attacks with unbounded behaviors (zero-day attacks).
A better strategy, particularly for IoT devices with bounded normal behaviors, is to train the models on only benign data, so any instance outside the benign boundary would be considered abnormal.


Clustering algorithms have proven to be effective in modeling expected patterns. However, some like $K$-Means inherently lack the notion of boundary. In other words, a data instance, no matter where it appears in the space and how far it is from expected instances, will be assigned (with a probability) to its nearest cluster. Therefore, to employ these models for anomaly detection applications, we need to augment them with custom boundaries. Work in \cite{Arunan2020b} used spherical boundaries around the centroid of each cluster by covering 97.5\% of the closest points to the centroid. Spherical boundaries are easy to form. However, there are many cases where the area they cover is unnecessarily bigger than the actual cluster, which makes the boundaries relatively loose. I will show in \S\ref{ch3:sec:eval} that loose boundaries can increase the chance of false negatives and reduce anomaly detection rates.

In this chapter, I use the concept of convex hull \cite{convex_hull} for developing custom-shaped clusters. Given a set of points $P$, the convex hull is the smallest boundary that contains all points in $P$. Note that the complexity of checking whether an instance falls inside convex hull boundaries is higher than spherical boundaries -- the time complexity is $\mathcal{O}(P)$ versus $\mathcal{O}(1)$. There are existing tools like SciPy library \cite{scipyqhull} that generate convex hull for a given set of points. In this chapter, I employ PCA (Principle Component Analysis) to reduce the dimensions of the feature space into a 2D space. This dimension reduction helps with the training/testing time and also enables us to visualize the results. Note that one can also generate convex hulls on a multi-dimensional feature space with less than nine dimensions \cite{scipyqhull}.

With convex hulls, a practical challenge is that training instances may still contain unintended noises that can lead to undesirable boundaries.
Collecting purely benign traffic traces of networked assets is almost infeasible as incidents such as device misconfiguration, network disruption, or even uncaught attacks can poison the benign dataset. Therefore, we need a data purification (refinement) method to avoid including noise instances in the benign boundaries. In this chapter, I first use $K$-Means on raw benign data (including noises) to get the initial set of clusters. Next, I apply two filters (explained below) to remove some odd clusters (with characteristics too dissimilar to their cohort) and then prune the remaining clusters by tightening their boundary.

Fig.~\ref{ch3:fig:convex vs spherical} illustrates the clusters (formed by $K$-Means) and their boundaries. In Fig.~\ref{ch3:fig:default convex}, the boundaries computed by the default convex hull are compared with the spherical boundaries. The convex hull algorithm requires at least three data points to form a boundary. Cluster 4, therefore, is excluded by default due to having only two data instances. It can be seen in Fig.~\ref{ch3:fig:default convex} that default convex hulls include all instances which can skew the benign boundaries (\eg cluster 3 and, to some extent, cluster 2) toward seemingly noisy instances. Such performance is enhanced in Fig.~\ref{ch3:fig:tuned convex}, highlighting the refined version of the convex hull I developed for this chapter.

\subsection{Refined Convex Hull}\label{ch3:sec:refined}

Noises are inevitable in the training dataset which is intended to be purely benign. Traffic noises are often a few short-lived instances that are caused due to intermittent outages or variations in the network services. Therefore, the common observation is that noisy instances are usually scattered and appear far from most benign instances. Noise can lead to two detrimental impacts on clusters:

(1) Some clusters emerge that do not share the characteristics of other benign clusters. These clusters are often {\textit{sparse}}, as opposed to most benign clusters that are formed with a significantly larger number of instances ({\textit{dense}}). An indicator for distinguishing sparse from dense clusters is the mutual distance of the instances. In sparse clusters, instances are far from each other; hence, we expect to see a more prominent measure of mutual distance than the dense clusters.

(2) Although some clusters are dense, they may include noisy instances that can skew/loosen their benign boundary. I refer to these noise instances as {\textit{outliers}}, which are often far from the centroid of their corresponding cluster compared to other instances in the same cluster.

 In what follows, I explain the methods I develop for detecting sparse clusters and outlier instances in dense clusters:

{\textbf{Detecting Sparse Clusters:}} For this, I use the distribution of mutual distance (denoted as $MD$) across all clusters. Having a total of $C$ clusters, for a given cluster $c \in \left[1, C\right]$ with $P$ instances in it, I compute a mutual distance matrix $\left[D_{a,b}\right]_{P\times P}$  in which $D_{a,b}$ refers to the distance of instances $a$ and $b$. From the matrix $D$, I measure $MD_c$ (\ie the average mutual distance across $P$ instances in the cluster $c$). Computing $MD_c$ for all $C$ clusters gives a distribution from which I compute an average ($\mu_{MD}$) and standard deviation ($\sigma_{MD}$) across all $C$ clusters.

A cluster $c$ is sparse if \mbox{$MD_c > \mu_{MD} + \alpha . \sigma_{MD}$}, where $\alpha$ is a configurable parameter to specify how much deviation from the average is acceptable. 
Increasing $\alpha$ will allow for the inclusion of sparse clusters, and decreasing it will aim for highly-dense clusters.

{\textbf{Detecting Outliers:}} I focus on distances of points from the cluster centroid to detect outliers inside a relatively dense cluster. Let $CD_a$ denote the distance of instance $a$ (for all $a \in \left[1, P\right]$) from the centroid. 
The distribution of $CD_a$ is represented by $\mu_{CD}$ and $\sigma_{CD}$, referring to the average and standard deviation of distances, respectively.

Now, for a given cluster, instance $a$ becomes an outlier if \mbox{$CD_a > \mu_{CD} + \beta . \sigma_{CD}$}, where $\beta$ is a configurable parameter like $\alpha$ explained above. 

I also publicly release my Python source code for generating the convex hull of one-class models \cite{smartinfer_data}. In \S\ref{ch3:sec:eval}, I will evaluate the refined convex hull model over a range of $\alpha$ and $\beta$ values and demonstrate how they can affect the performance of anomaly detection.

\section{Evaluation of Efficient IoT Traffic Inference}
\label{ch3:sec:eval}

In this section, I first evaluate the multi-view classification method using three resolving techniques discussed in \S\ref{ch3:subsec:resolve}. Then, I evaluate the progressive one-class models using a public dataset.

\begin{figure}[!t]
	\centering
	\includegraphics[scale=0.55]{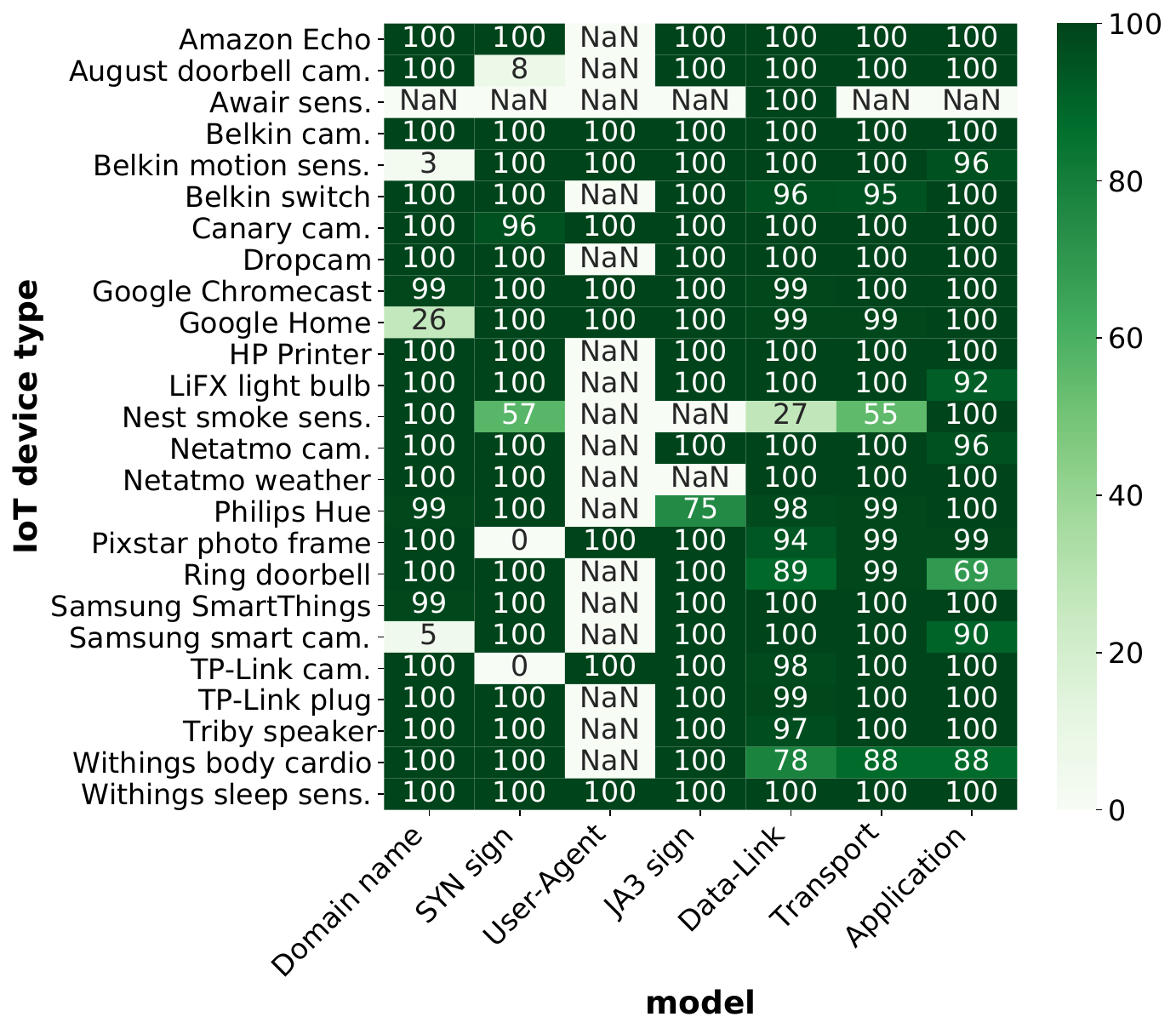}
	\caption{Testing accuracy of each specialized model for IoT device types with {\myverb{TEST1}} dataset.}
	\label{ch3:fig:accuracy heatmap}
\end{figure}

\subsection{Classification Phase Results}

Before evaluating the multi-view approach, I need to solve the optimization problem formulated in \S\ref{ch3:subsec:optimization} to select our optimal set of specialized models. I implement the optimization problem in two different Python libraries, namely Pyomo \cite{pyomo} and CVXPY \cite{cvxpy}. CVXPY assures that our optimization problem is convex; otherwise, it would not be able to solve it. 
The feasibility matrix $F$ of our optimization is initialized according to Eq.~ \ref{ch3:eq:feasibility} using the threshold parameter $A$ and the accuracy values obtained from the {\myverb{TEST1}} dataset (aggregate results were reported in Table~\ref{ch3:tab:telemetries}).
I choose the accuracy threshold parameter equal to a fairly high value $99$\%. The optimal set of models returned by both Pyomo and CVXPY consists of the domain name and datalink models, as they meet the accuracy threshold of $99$\% for each IoT class in {\myverb{TEST1}} dataset --  such a performance was expected from the optimization problem. I note that the actual accuracy can vary for a new unseen dataset. 
I will shortly perform experiments with the optimization problem varying the thresholds and specialized models available to us.

Let us better understand how the optimal set of specialized models is computed.  
Fig.~\ref{ch3:fig:accuracy heatmap} shows the performance (prediction accuracy) of the seven models we discussed in \S\ref{ch3:sec:MetricsFeatures} across 25 IoT device classes in the {\myverb{TEST1}} dataset. Cells with a ``{\myverb{NaN}}'' value highlight device classes $d$ that miss traffic features necessary for model $m$, in the {\myverb{TRAIN}} dataset -- hence, $m$ is unable detect the class $d$. That said, Amazon Echo is an exception with the User-Agent model. In fact, Amazon Echo has a unique User-Agent header in {\myverb{TRAIN}} dataset, but it does not exchange any User-Agent header in the {\myverb{TEST1}} dataset. The SYN signature model mispredicts all testing instances of Pixstar photo frame as Google Home and those of TP-Link camera as Belkin switch. This is because of their shared SYN signatures. 
For the Awair sensor, all datasets (\ie {\myverb{TRAIN}}, {\myverb{TEST1}}, and {\myverb{TEST2}}) only include EAPol (Extensible Authentication Protocol over LAN) traffic which does not use TCP/UDP protocols. No traffic except Ethernet is seen for this sensor. Therefore, only the datalink model provides predictions for this IoT class.

\begin{table}[t]
	\centering
	\caption{Performance of multi-view classification at different resolution periods: all models versus the optimal set of models.}
	\renewcommand{\arraystretch}{1.1}
	\begin{adjustbox}{width=0.75\textwidth}
		\begin{tabular}{l|c|c|c|c|c|c} \hline
		
			\hline
			Resolver & \multicolumn{3}{c|}{All models} & \multicolumn{3}{c}{Optimal models} \\
            technique & 1h & 6h & 24h & 1h & 6h & 24h\\ \hline
			MV & $99$\% & $99$\% & $98$\% & $94$\% & $93$\% & $89$\% \\ 
			WMV & $100$\% & $100$\% & $100$\% & $99$\% & $99$\% & $98$\% \\ 
			AWMV & $95$\% & $91$\% & $85$\% & $93$\% & $89$\% & $83$\% \\ \hline
			
			\hline
		\end{tabular}
	\end{adjustbox}
	\label{ch3:tab:ident_accuracy}
\end{table}

			
%

According to Fig.~\ref{ch3:fig:accuracy heatmap}, the only model that is able to provide inference for all IoT classes (device types) is the datalink model. In other words, the datalink model is the only one that offers an availability metric of 100\% 
for the multi-view classification. 
At the same time, the datalink model falls short of expectations in accurately predicting traffic for several classes, including the Belkin switch, the Nest smoke sensor, the Philips Hue, the Pixstar photo frame, the Ring doorbell, the TP-Link camera, the Triby speaker, and the Withings body cardio. The prediction accuracy for those individual classes is less than the desirable threshold of $99$\%. Considering alternatives for these classes, we find that the domain name model gives an accuracy greater than $99$\% for all of them. Hence, having at least two models, namely the domain name model and the datalink model, would satisfy the optimization constraints.

\textbf{Experimenting with Optimization:} We saw in Fig.~\ref{ch3:fig:accuracy heatmap}, the seven specialized models I consider in this chapter perform fairly well on the majority of IoT classes. Such high performance from individual models may not highlight optimization's benefits to the multi-view inference. 
Note that the multi-view architecture and optimization readily allow for various specialized models (an extended version or a subset of my list of models) and IoT classes. 
In order to showcase the dynamics of my optimal model selection, let us experiment with a list of models when the domain name and SYN signature models are assumed to be excluded. I solve the optimization problem with the remaining five specialized models. This time the optimal set consists of JA3, datalink, and application models. Again, this selection is not overly unexpected.
Similar to what we saw earlier, the datalink model must be selected as it is the only model that can predict traffic for the Awair sensor. JA3 and application models are needed to be in the mix for the Withings body cardio and the Nest smoke sensor, respectively. Note that we expect an accuracy of at least $99$\% per class.

My optimization problem, with no exception, may become infeasible when it is applied to another dataset, set of models, or IoT classes. In that case, relaxing the accuracy threshold (less than 99\%) can help make the problem feasible. To showcase an infeasible scenario, let us limit the superset of models to User-Agent, datalink, transport, and application with the original accuracy threshold of $99$\%. In this case, no model can meet the threshold for two classes, namely the Ring doorbell and the Withings body cardio -- the problem becomes infeasible to solve. However, if we relax the accuracy threshold to a value like $85$\%, the datalink and application models satisfy the accuracy constraints for all devices and become our optimal set.

\textbf{Performance of Optimal Multi-View Classification:} 
I now evaluate the efficacy of the optimal multi-view classification by comparing its performance and cost against those of a classification scenario where all specialized models are employed. For this purpose, I use {\myverb{TEST2}} dataset.
Table~\ref{ch3:tab:ident_accuracy} summarizes the prediction performance (detection rate) of all models versus optimal models selected at three resolution periods: short (one-hour), medium (six-hour), and long (24-hour), with three resolver techniques discussed in \S\ref{ch3:subsec:resolve}. Given five days' worth of data available in {\myverb{TEST2}}, I can evaluate the performance across 120 short, 20 medium, and 5 long resolution periods. For the evaluation metric, I measure the detection rate, which is the fraction of total active IoT devices that are correctly classified for a given resolution period. Table~\ref{ch3:tab:ident_accuracy} shows the average detection rate computed across all resolution periods. Note that the detection rate metric slightly differs from the accuracy metric (per one-minute epoch) I initially used to evaluate the prediction of specialized models in \S\ref{ch3:sec:identification}. This new metric is computed at the end of each resolution period, where a device is mapped to a class after resolving several predictions.

The first observation from Table~\ref{ch3:tab:ident_accuracy} is that making classification inferences using all models is slightly more accurate than the optimal models. The lowest gap is $1$\% when the WMV resolver technique is used for the short and medium resolution periods. It can be as high as $9$\% for MV over the long resolution period. Another interesting observation is that the inference (using all models and optimal models) performs better when the resolution period is shorter, indicating that mispredictions accumulated over time can be detrimental to the ultimate inference. The impact of such accumulation is more pronounced in MV and AWMV techniques, whereas WMV yields a fairly robust detection rate across the three resolution periods. The advantages of the optimally-selected models become more apparent when computing and processing costs are considered. Table~\ref{ch3:tab:ident_cost} shows the CPU time and TCAM size required for all models versus optimal models. It can be seen that the cost of optimal models is significantly lower than all models by a factor of $6$ (by the measure of CPU) and $7$ (by the measure of TCAM size).

\begin{table}[t]
	\centering
	\caption{Cost of multi-view classification: all models versus the optimal set of models.}
	\vspace{-2mm}
	\renewcommand{\arraystretch}{1.1}
	\begin{adjustbox}{width=0.75\textwidth}
		\begin{tabular}{l|c|c} \hline
		
			\hline
		  Cost	& All models & Optimal models\\ \hline
			CPU time (ns) & 866 & 127 \\ 
			TCAM size (bytes) & 94 & 12 \\ \hline
			
			\hline
		\end{tabular}
	\end{adjustbox}
	\label{ch3:tab:ident_cost}
	\vspace{-1mm}
\end{table}

As discussed throughout this chapter, classification is the foundational step for the monitoring phase. Therefore, it is essential to correctly determine the class of all devices before proceeding to the monitoring phase. As we saw in Table~\ref{ch3:tab:ident_accuracy}, this assurance is given by using all models with the WMV technique; however, the optimal set may miss some devices in a few resolution periods. For example, out of the total of 120 short (1hr) resolution periods, the optimal scheme with WMV technique mispredicts the Google Home as the Google Chromecast in $9$ periods, the Pixstar photo frame as the Belkin switch in $7$ periods, and the Nest smoke sensor as the Google Home in $2$ resolution periods.

Although the chance of incorrect inference (after resolution) is relatively minimal, it can be avoided by adding another layer of consolidation and/or resolution on top of the current inference. For example, one can keep obtaining the classification outputs for several resolution periods (\eg six periods of each one hour) and then resolve conflicts (\ie different claimed for given MAC address) using similar techniques explained in \S\ref{ch3:subsec:resolve}. Another technique that could be effective in resolving conflicts is developing a consistency score \cite{Arunan2020b} that helps select the class that is consistently given to a device for a number of consecutive resolution periods. Such an additional resolution is beyond the scope of this chapter.
In the context of obtaining a perfect classification for all devices, we can compare the performance of all models versus optimal models in terms of response time. It is seen in Table~\ref{ch3:tab:ident_accuracy} that all models can achieve $100$\% detection rate in a single 1-hour resolution period; however, the optimal models need extra 1-hour periods for obtaining the same result. Even with some delays, the optimal approach is still attractive in terms of computing costs -- we often prefer to spend a few more hours to make a reliable (accurate and confident) inference but keep the computing cost to a minimum.


\subsection{Monitoring Phase Results}

In order to evaluate the performance of the monitoring phase, I first train seven one-class models per IoT class using their benign traffic data and then quantify their false positive rates (incorrectly predicting benign instances as anomalies) when applied to unseen benign data and their true positive rates (correctly detecting anomalies) when applied to attack (malicious) traffic data. For this purpose, I use public traffic traces \cite{ayyoob_data} that was collected from our own testbed in 2018 and includes benign and attack traffic traces of the same set of IoT classes (except for the Withings sleep sensor) we had in classification phase of this chapter. I use benign traffic traces collected in May, June, and October 2018 to train the one-class models. 

I found some devices were inactive during the capture period for several days. Hence, the benign dataset could not be split chronologically into training and testing portions, as it would have resulted in a heavily imbalanced training or testing dataset. To overcome this challenge, I shuffle the benign dataset to distribute data instances across all device classes evenly. Next, I take 70\% of the benign dataset and create {\myverb{BENIGN1}} instances used for training the one-class models, and the remaining 30\% of the data will create {\myverb{BENIGN2}} instances used for measuring false positive ratio. 
Note that portions of the public traces \cite{ayyoob_data} I obtain are a mixture of benign and attack traffic. The annotation files released by the same source \cite{ayyoob_data} helped me extract only attack traffic, constructing my {\myverb{ATTACK}} dataset, which I use for measuring the true positives.
The CSV files of {\myverb{BENIGN1}}, {\myverb{BENIGN2}}, and {\myverb{ATTACK}} instances are publicly released \cite{smartinfer_data}.

For the monitoring phase, I experimented with one-minute and five-minute epochs. I found that an epoch duration of five minutes gives better performance in terms of false/true positives. Hence, I use five-minute epochs for the rest of this subsection. As discussed in \S\ref{ch3:sec:monitor}, some of the intended models cannot be trained for certain IoT device classes due to the unavailability of the features required. I also noticed that some device types generate certain features infrequently. For instance, the Ring doorbell only generates NTP traffic in 20 epochs, in contrast to 823 epochs with DNS traffic. Models trained with insufficient epochs yield poor performance. Therefore, I empirically found that at least 100 epochs (about 8 hours' worth of traffic) are needed for a reliable one-class model.

\begin{figure}[!t]
	\centering
	\includegraphics[scale=0.45]{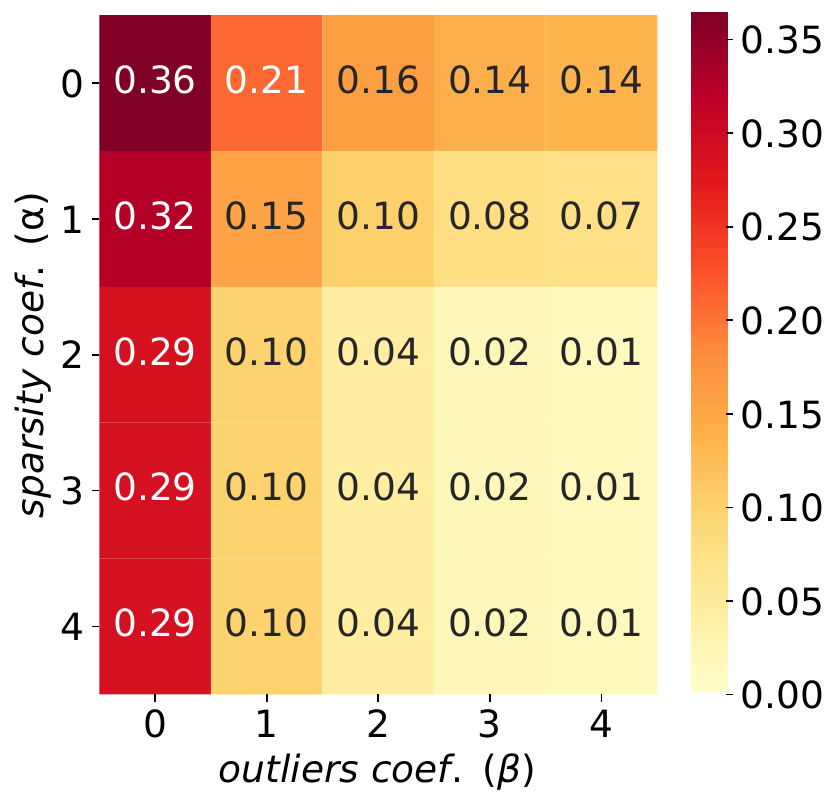}
	\caption{Average false positive (per IoT class) of the coarse-grained models applied to the benign testing dataset with varying combinations of coefficients $\alpha$ and $\beta$.}	\label{ch3:fig:coef_heatmap}
\end{figure}

Let us begin by evaluating the one-class models with varying coefficients $\alpha$ (sparsity) and $\beta$ (outliers) to analyze their impact on the performance of models.  
For this evaluation, I only focus on the coarse-grained model (introduced in \S\ref{ch3:sec:monitor}) as Ethernet traffic features are available for all device classes.
That said, one can perform the same evaluation on other one-class models and fine-tune parameters $\alpha$ and $\beta$ for them separately. 
I use the best combination of $\alpha$ and $\beta$ obtained for the Ethernet one-class models as default settings for other models (medium-grained and fine-grained).

Fig.~\ref{ch3:fig:coef_heatmap} illustrates a heat map indicating the average false positive ratio of one-class models (of all IoT classes) when applied to the benign testing dataset for 25 combinations of values for coefficients $\alpha$ and $\beta$. Recall smaller $\alpha$ values result in more clusters being considered sparse (noisy) and removed, meaning a reduced number of benign clusters (conservative approach). The coefficient $\beta$ has a similar effect in restricting benign boundaries (of those clusters considered as dense). Smaller $\beta$ results in clusters with tighter boundaries, increasing the chance of some truly benign instances falling outside the benign boundaries and contributing to false positives.

\begin{figure}[!t]
	\centering
	\includegraphics[scale=0.4]{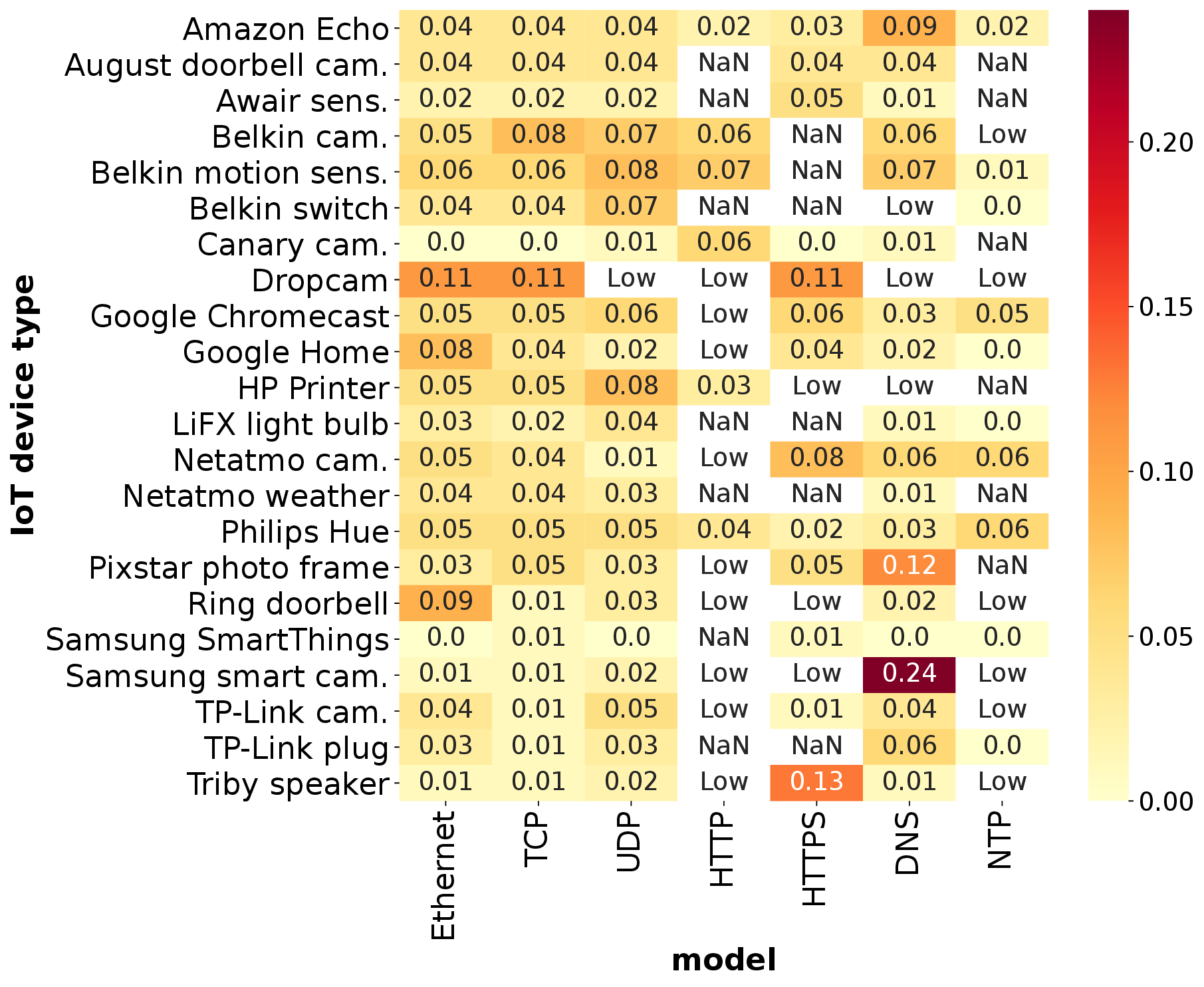}
	\caption{False positives of each model per IoT class when applied to the benign testing dataset.}
	\label{ch3:fig:fp heatmap}
\end{figure}
\vspace{4mm}

Although larger values for $\alpha$ and $\beta$ can improve false positives with the benign dataset, they may loosen boundaries and cause false negatives with the attack dataset. Therefore, balancing these two cases, I choose $\alpha=2$ and $\beta=2$ (the middle cell in Fig.~\ref{ch3:fig:coef_heatmap}), which gives a relatively low average false positive ratio equal to $0.04$.

With parameters $\alpha$ and $\beta$ fine-tuned empirically, let us quantify the performance of individual one-class models per each IoT class.
Fig.~\ref{ch3:fig:fp heatmap} shows the false positive ratio of each model across IoT classes on the benign testing dataset. Cells denoted by ``NaN'' and ``Low'', respectively, indicate no training instances and a low number ($<100$) of training instances. On average, each model offers a reasonably acceptable false positive ratio of less than $0.05$ across all device classes. Among eight models, the DNS model performs the worst for packets of the Samsung smart camera with about $0.24$ false positive ratio -- this could be due to the marginally small number ($108$) of DNS training instances for this class which is just slightly higher than the minimum requirement of 100 instances discussed above. Two IoT classes, namely the Nest smoke sensor and the Withings body cardio, were only active over $22$ and $15$ epochs, respectively. Therefore, they are excluded from the evaluation due to insufficient training instances.

I now evaluate the true positive rate of my models by applying them to the {\myverb{ATTACK}} dataset. Note that the public traffic traces \cite{ayyoob_data} I used to create {\myverb{ATTACK}} dataset do not include any attacks via HTTP, HTTPS, DNS, or NTP. However, there are a few instances of attacks via SSDP and SNMP, but I did not train one-class models associated with those specific application-layer protocols. Still, Ethernet and/or UDP models can flag behavioral deviations due to SSDP/SNMP-based attacks. That said, one may train SSDP and SNMP one-class models and utilize the progressive monitoring architecture to detect such attacks. 
\vspace{4mm}

Table~\ref{ch3:tab:attack detection} shows the true positives for each attack instance on their respective IoT device type. Some attacks in the dataset were launched on specific device types, leading to blank cells in the table. Each attack was launched at three different rates of 1, 10, and 100 packets per second (pps). For each attack vector, I report the true positives of those models whose features could be affected by the attack. For example, the ping-of-death attack uses ICMP to send ping requests, it only appears in Ethernet traffic without affecting features of TCP/UDP and above. In most cases, the Ethernet model yields the same true positive rate as TCP/UDP models. For 10 attacks, TCP or UDP models show a higher true positive rate than the Ethernet model. For example, the TCP model can detect all SYN and SYN reflection attacks (the two middle columns in Table~\ref{ch3:tab:attack detection}) on the Google Chromecast while the Ethernet model can, at best, detect three-quarters of high rates attacks (100pps). Ethernet, TCP, and UDP models have an average true positive rate of $0.85$, $0.96$, and $1.00$ (across all device types and attacks available in the dataset). This highlights the fact that the performance of detecting anomalies is better when the model is finer-grained (and more expensive computationally).

\begin{table*}[t!]
	\centering
	\caption[True positive rate of the models when applied to attack traces]{True positive rate of the models when applied to attack traces. ``E'', ``T'', and ``U'' denote Ethernet, TCP, and UDP models, respectively.}
	\renewcommand{\arraystretch}{1.1}
	\begin{adjustbox}{width=\textwidth}
		\begin{tabular}{l|ccc|ccc|ccc|cccc}
			\hline
			
			\hline
			& \multicolumn{3}{c|}{\textbf{Ping of Death}} & \multicolumn{3}{c|}{\textbf{SYN attack}} & \multicolumn{3}{c|}{\textbf{SYN reflection}} & \multicolumn{3}{c}{\textbf{UDP DoS}} \\ 
			& 1pps & 10pps &100ps & 1pps & 10pps &100ps & 1pps & 10pps &100ps & 1pps & 10pps &100ps\\ \hline
			\vtop{\hbox{\strut {Belkin}}\hbox{\strut {switch}}}
			& [E]:$1.00$
			& [E]:$1.00$
			& [E]:$1.00$ 
			& \vtop{\hbox{\strut {[E]:$1.00$}}\hbox{\strut {[T]:$1.00$}}}
			& \vtop{\hbox{\strut {[E]:$1.00$}}\hbox{\strut {[T]:$1.00$}}} 
			& \vtop{\hbox{\strut {[E]:$1.00$}}\hbox{\strut {[T]:$1.00$}}} 
			& \vtop{\hbox{\strut {[E]:$1.00$}}\hbox{\strut {[T]:$1.00$}}}
			& \vtop{\hbox{\strut {[E]:$1.00$}}\hbox{\strut {[T]:$1.00$}}} 
			& \vtop{\hbox{\strut {[E]:$1.00$}}\hbox{\strut {[T]:$1.00$}}} 
			& --
			& --
			& --  \\ \hline
			\vtop{\hbox{\strut {Belkin}}\hbox{\strut {motion sens.}}}
			& [E]:$0.50$
			& [E]:$1.00$
			& [E]:$1.00$ 
			& \vtop{\hbox{\strut {[E]:$0.00$}}\hbox{\strut {[T]:$0.17$}}}
			& \vtop{\hbox{\strut {[E]:$1.00$}}\hbox{\strut {[T]:$1.00$}}}
			& \vtop{\hbox{\strut {[E]:$1.00$}}\hbox{\strut {[T]:$1.00$}}} 
			& \vtop{\hbox{\strut {[E]:$0.67$}}\hbox{\strut {[T]:$0.50$}}}
			& \vtop{\hbox{\strut {[E]:$1.00$}}\hbox{\strut {[T]:$1.00$}}} 
			& \vtop{\hbox{\strut {[E]:$1.00$}}\hbox{\strut {[T]:$1.00$}}} 
			& \vtop{\hbox{\strut {[E]:$0.50$}}\hbox{\strut {[U]:$1.00$}}}
			& \vtop{\hbox{\strut {[E]:$1.00$}}\hbox{\strut {[U]:$1.00$}}} 
			& \vtop{\hbox{\strut {[E]:$1.00$}}\hbox{\strut {[U]:$1.00$}}}  \\ \hline 
			\vtop{\hbox{\strut {Samsung}}\hbox{\strut {smart cam.}}}
			& [E]:$1.00$
			& [E]:$1.00$
			& [E]:$1.00$ 
			& \vtop{\hbox{\strut {[E]:$1.00$}}\hbox{\strut {[T]:$1.00$}}}
			& \vtop{\hbox{\strut {[E]:$1.00$}}\hbox{\strut {[T]:$1.00$}}} 
			& \vtop{\hbox{\strut {[E]:$1.00$}}\hbox{\strut {[T]:$1.00$}}} 
			& \vtop{\hbox{\strut {[E]:$1.00$}}\hbox{\strut {[T]:$1.00$}}}
			& \vtop{\hbox{\strut {[E]:$1.00$}}\hbox{\strut {[T]:$1.00$}}} 
			& \vtop{\hbox{\strut {[E]:$1.00$}}\hbox{\strut {[T]:$1.00$}}} 
			& \vtop{\hbox{\strut {[E]:$1.00$}}\hbox{\strut {[U]:$1.00$}}}
			& \vtop{\hbox{\strut {[E]:$1.00$}}\hbox{\strut {[U]:$1.00$}}} 
			& \vtop{\hbox{\strut {[E]:$1.00$}}\hbox{\strut {[U]:$1.00$}}}  \\ \hline 
			\vtop{\hbox{\strut {Netatmo}}\hbox{\strut {cam.}}}
			& --
			& --
			& --
			& \vtop{\hbox{\strut {[E]:$0.50$}}\hbox{\strut {[T]:$1.00$}}}
			& \vtop{\hbox{\strut {[E]:$1.00$}}\hbox{\strut {[T]:$1.00$}}}
			& \vtop{\hbox{\strut {[E]:$1.00$}}\hbox{\strut {[T]:$1.00$}}} 
			& \vtop{\hbox{\strut {[E]:$0.34$}}\hbox{\strut {[T]:$1.00$}}}
			& \vtop{\hbox{\strut {[E]:$1.00$}}\hbox{\strut {[T]:$0.84$}}} 
			& \vtop{\hbox{\strut {[E]:$1.00$}}\hbox{\strut {[T]:$1.00$}}} 
			& --
			& --
			& --  \\ \hline
			\vtop{\hbox{\strut {Google}}\hbox{\strut {Chromecast}}}
			& --
			& --
			& --
			& \vtop{\hbox{\strut {[E]:$0.25$}}\hbox{\strut {[T]:$1.00$}}}
			& \vtop{\hbox{\strut {[E]:$0.25$}}\hbox{\strut {[T]:$1.00$}}}
			& \vtop{\hbox{\strut {[E]:$0.75$}}\hbox{\strut {[T]:$1.00$}}} 
			& \vtop{\hbox{\strut {[E]:$0.00$}}\hbox{\strut {[T]:$1.00$}}}
			& \vtop{\hbox{\strut {[E]:$0.25$}}\hbox{\strut {[T]:$1.00$}}} 
			& \vtop{\hbox{\strut {[E]:$0.50$}}\hbox{\strut {[T]:$1.00$}}} 
			& --
			& --
			& --  \\ \hline
			\vtop{\hbox{\strut {LiFX}}\hbox{\strut {lightbulb}}}
			& [E]:$1.00$
			& [E]:$0.50$
			& [E]:$1.00$ 
			& --
			& --
			& --
			& --
			& --
			& --
			& \vtop{\hbox{\strut {[E]:$1.00$}}\hbox{\strut {[U]:$1.00$}}}
			& \vtop{\hbox{\strut {[E]:$1.00$}}\hbox{\strut {[U]:$1.00$}}} 
			& \vtop{\hbox{\strut {[E]:$1.00$}}\hbox{\strut {[U]:$1.00$}}}  \\ \hline 
			\vtop{\hbox{\strut {Amazon}}\hbox{\strut {Echo}}}
			& --
			& --
			& --
			& --
			& --
			& --
			& --
			& --
			& --
			& \vtop{\hbox{\strut {[E]:$1.00$}}\hbox{\strut {[U]:$1.00$}}}
			& \vtop{\hbox{\strut {[E]:$1.00$}}\hbox{\strut {[U]:$1.00$}}} 
			& \vtop{\hbox{\strut {[E]:$1.00$}}\hbox{\strut {[U]:$1.00$}}}  \\ \hline 
			
			\hline
		\end{tabular}
	\end{adjustbox}\label{ch3:tab:attack detection}
\end{table*}


To compare the impact of convex hull boundaries versus spherical boundaries on the performance of models in detecting attacks, I generated benign clusters with spherical boundaries that contain the 97.5th percentile of benign instances. I found that spherical boundaries cause the models to fall short in detecting some of the network attack instances. For example, the TCP model with spherical boundaries gives a true positive of $0.71$ for SYN and SYN reflection attacks on the Google Chromecast, while this measure is $1.00$ for the TCP model with convex hull boundaries. In fact, the spherical boundaries lead to lower true positives for all four types of attacks, with the Ethernet, TCP, and UDP models having average true positives of $0.73$, $0.80$, and $0.96$.

\section{Conclusion}\label{ch3:sec:con}
In this chapter, I developed a generic traffic inference architecture consisting of a multi-view classification stage followed by progressive monitoring phases. I developed a systematic approach for the multi-view classification to quantify the prediction power versus computing costs of the specialized-view models, each trained on specific features (packet or flow). I demonstrated that specialized models could be more accurate and cost-effective than a single combined-view model. I also formulated an optimization problem to select the best views for the multi-view classification that minimizes the computing cost subject to prediction accuracy requirements. For progressive monitoring, I employed one-class models and developed configurable convex hulls for them, tightening benign behavioral boundaries. I evaluated the efficacy of the models (classification and monitoring) by applying them to real traffic traces collected from our IoT testbed. I found that the optimal set of specialized classifier models can reduce the processing cost by a factor of six with insignificant impacts on the prediction accuracy. Also, the monitoring models yielded an average true positive rate of 94\% and a false positive rate of 5\%. Finally, I publicly released the data (training and testing instances of classification and monitoring tasks) and enhancement code for the convex hull algorithm to the research community.

For this chapter, I assumed that we have full visibility over network traffic so we can select from a range of available traffic features for multi-view classification and progressive monitoring. However, there might be some situations whereby full visibility becomes infeasible due to measurement costs and challenges. In the next chapter, I will explain a scenario in which we need to rely on partial visibility and coarse-grained data for our traffic inference. I will also demonstrate how heuristic-based (deterministic) models can help us correct mispredictions made by machine learning models.

%% file: tex/Chapter4-Obscure-Infer/main.tex
\chapter{Inference from Opaque Flows}
\label{ch:obscure infer}


%
%
%
%


In the previous chapter, I showed that the cost of computing traffic features for IoT traffic inference could be optimally balanced against prediction quality when complete or close-to-complete visibility into the network traffic can be obtained, and fine-grained traffic features can be computed. However, obtaining high-quality data in certain situations becomes impractical. Therefore, we are left with partially-visible traffic and/or coarse-grained traffic features to infer from. In this chapter, I elaborate on this challenge from an ISP viewpoint aiming to detect known IoT devices (say, a list of perhaps vulnerable/risky ones) in their subscriber home networks. Offering such a proactive service could benefit both the ISP and subscribers.

This chapter makes three contributions: (1) I analyze more than nine million IPFIX flow records of 26 IoT devices collected from a residential testbed over three months and identify 28 flow features pertinent to their network activity that characterize the network behavior of  IoT devices -- the IPFIX records dataset is released as open data to the public; (2) I train a multi-class classifier on stochastic attributes of IPFIX flows to infer the presence of certain IoT device types in a home network with an average accuracy of 96\%. On top of the machine learning model, I develop a \textit{Trust} metric to track network activity of detected devices over time; and (3) Finally, I develop deterministic models of specific and shared cloud services consumed by IoTs, yielding an average accuracy of 92\%. I demonstrate a combination of stochastic and deterministic models mitigates false positives in 75\% of incidents at the expense of an average 7\% reduction in true positives.

\section{Introduction}\label{ch4:sec:intro}
In the previous chapter, I argued that selecting the best combination of traffic features for identifying IoT devices is a nontrivial task that requires a thorough analysis of the features and optimization. This is true when a transparent and comprehensive copy of network traffic can be measured with minimum or no abstraction (obfuscation), leading to a rich set of traffic features that capture different aspects of device characteristics. However, there might be cases where capturing network traffic close to devices or computing packet-based features (especially those that require deep packet inspection) become practically infeasible due to high traffic volume or distributed networks. Without relying on high-quality data, traffic inference tasks can become even more challenging.
	
An example of a situation where gaining high-quality data becomes practically challenging, is residential networks. Home networks are becoming increasingly complex, yet neither ISPs nor subscribers have much visibility into connected devices and their network-level behavior. Technologies like NAT present only an opaque view of home network to the global Internet \cite{Grover2013,Meidan2020}. This makes it surprisingly challenging for ISPs to even detect and discover connected devices, as the traffic transmitted by every active device in a home would have the same IP and MAC address of the home gateway.

	IoT devices in homes are able to connect (via the ISP network) with their intended servers on the Internet. Stable and fast Internet connections can indulge botnets and malware in launching coordinated volumetric attacks using millions of infected IoTs \cite{Mirai20}. This can affect the performance of ISPs' core network and potentially make them accountable for carrying traffic of illegal campaigns \cite{Himal2019}. On the other side, residential subscribers gradually express willingness in paying to ISPs \cite{SaaS2020} for improved security of their Internet-connected assets. Therefore, ISPs seem to be relatively incentivized to play a role in providing network monitoring and security services to households \cite{Nthala2018}.

		One may argue that IoT manufacturers and/or cloud providers can also offer these services. It is worth to note that the maturity of IoT manufacturers and/or cloud providers varies widely, so the entire consumer IoT market is particularly vulnerable to low-end, low-quality manufacturing and high-risk data handling practices \cite{Harkin2022}. Additionally, unlike ISPs, IoT manufacturers and cloud providers often do not establish strong and long-term business relationships with end-users. Lastly, manufacturers and cloud providers can, at best, obtain visibility into software components embedded in their devices. However, ISPs are best poised to monitor and analyze the network behavior of connected devices and hence become capable of detecting deviations from the norm. Therefore, from both business and technical viewpoints, ISPs are a better option than IoT manufacturers and cloud providers. For users, engaging with one entity (their ISP) becomes easier than with multiple entities (manufacturers/cloud providers) to receive security services.

	User privacy is an important factor for ISPs. Firstly, the inference of connected devices in residential networks must be offered as a subscription service \cite{Habibi2017} to manage privacy concerns arising from the amount of personal information collected by ISPs. Users will provide their consent (and they may opt out in the future) to activate this service by accepting the terms and conditions provided by their ISP. Moreover, the ISP may choose to narrow its focus on a limited number of well-known vulnerable devices (\eg obtained from public resources like CVE or NVD). Therefore, detecting only high-risk IoT devices would benefit both users and the ISP. Secondly, the method developed in this chapter (in contrast to some of the prior works \cite{Guo2020, Miettinen2017, Thangavelu2019, Marchal19}) only utilizes flow-based metadata (\eg flow volume and protocol) without requiring packet payloads that contain more sensitive and private information.

	
	Network behaviors of connected IoT devices highly depend on their expected functionality (\eg security camera, power switch, or motion sensor) and the implementation of functionalities by their manufacturer and/or developer. Therefore, one can describe device behaviors by two categories of features: (a) \textbf{stochastic} features such as distribution of packet count/inter-arrival time or total flow volume/duration; and (b) \textbf{deterministic} features such as specific services or domain names that a device communicates with them.
	Stochastic features display patterns in their distribution, making them conducive to machine learning techniques for modeling purposes. Hence, both industry and academia widely use them for enhancing network visibility in a variety of domains and use-cases \cite{cyberreport}. On the other hand, deterministic features \cite{arunan19,Arunan2018,Saidi2020,Hamza2022} can be distinctly predictive and may not necessarily need a machine-learning algorithm to generate a model for inferring connected devices. In fact, machine learning algorithms cannot be directly applied to some deterministic features like domain names as there is no finite superset for them.
	

	There exist prior works \cite{Meidan2020,Guo2020, Miettinen2017,Thangavelu2019,Marchal19, Saidi2020} on systems and methods that ISPs can employ to detect IoT devices in home networks; however, as mentioned in Chapter \ref{chapter:LiteratureReview}, they have some limitations that make them impractical. This chapter employs IPFIX data to detect consumer IoT devices in households. IPFIX is a well-known protocol for collecting flow-based metadata. IPFIX has been used by industry for many years, and hence employing its features at the edge of ISP networks requires no change to home networks. More importantly, IPFIX records do not have any payload or other sensitive (\eg user-specific) information, so that privacy concerns are minimized.
	
	In this chapter, I make the following contributions: \textbf{(1)} I analyze (\S\ref{ch4:sec:IPFIX}) near three million IPFIX records of 26 IoT devices in a testbed. Then, I extract 28 flow-level stochastic features from the IPFIX records. The dataset is publicly released \cite{ipfixdataset}, consisting of more than nine million IPFIX records; \textbf{(2)} I use these features (\S\ref{ch4:sec:stochastic}) to train and tune a multi-class classifier model using a decision tree-based machine learning algorithm to infer the type of IoT devices from IPFIX records. I also deduce thresholds specific to each device class for reducing the misclassification rate and develop a trust metric per class, enabling us to monitor the behavioral health of detected devices; \textbf{(3)} Finally, I develop two types of deterministic models (\S\ref{ch4:sec:dt}), each with differentiated capabilities in predicting connected IoT devices by analyzing the cloud service information (embedded in outgoing IPFIX records) that IoT devices consume. I evaluate the efficacy of the deterministic models independently. I also show how deterministic models can mitigate false positives when combined with the stochastic (machine learning-based) model.

	\section[Inferring Connected IoT Devices from IPFIX Records]{Inferring Connected IoT Devices\\from IPFIX Records}\label{ch4:sec:IPFIX}
	Identifying the composition of IoT devices in households can be done either by collecting network telemetry data (packet-based and/or flow-based) from inside (pre-NAT) or outside home networks (post-NAT). The pre-NAT telemetry, (used in the previous chapter of this thesis) indeed, reveals more information about the connected devices, particularly their unique identifiers (\eg MAC and/or IP addresses), enabling network operators to combine various telemetry records and map them to their unique device identifier.
	However, collecting pre-NAT telemetry is a nontrivial exercise without making changes to legacy home gateways -- prohibitively expensive for ISPs to deploy at scale for millions of households. On the other hand, the post-NAT approach provides a limited amount of data, primarily because NAT hides the entire internal network and identity of active devices -- this makes it infeasible to directly associate flow records with their respective end-devices. That said, post-NAT inferencing and monitoring would be practically more attractive for ISPs, particularly for the ease of deployment at scale. 
	
	IPFIX \cite{ipfixRFC5103} is a standard protocol for transmitting flow-level metadata from network switches and routers for network monitoring and traffic analysis. IPFIX uses a 5-tuple key (source and destination IP address, source and destination transport-layer port number, and transport-layer protocol) to uniquely identify and aggregate packets belonging to the same flow. 
	For connection-oriented flows, IPFIX uses session termination signals such as {\myverb{TCP FIN}} to detect the end of flows.
	For connection-less flows (\eg {\myverb{UDP}}), on the other hand, IPFIX uses two parameters, namely \textit{idle-timeout} and \textit{active-timeout}. If no packet is exchanged over a flow for the idle-timeout or if a flow is active for more than the active-timeout, the flow is terminated and the corresponding IPFIX record is exported. 
	
	I choose IPFIX telemetry to infer connected IoT devices in home networks for the following reasons: (i) the industry has increasingly adopted IPFIX as a multi-vendor universal lightweight protocol for performance and security monitoring, and hence it is perceived less challenging for ISPs to enable it on their network; (ii) IPFIX telemetry can be employed at (the edge of) ISP networks without the need to make any changes to subscriber home networks (gateways), and hence scalable; and (iii) IPFIX records only contain metadata of network communications without revealing any payload or other sensitive  information, and hence relatively low risks to user privacy. Nevertheless, relying on IPFIX comes with its own challenges: (a) collecting IPFIX telemetry from the (edge of) ISP networks will provide partial visibility into the activity of home networks as the identity (MAC/IP)  of connected devices will be hidden by NAT routers, and (b) IPFIX records provide aggregate information about flows, yielding coarse-grained data compared to fine-grained packet traces.

	\begin{figure}[t!]
		\centering
		\includegraphics[width=0.90\linewidth,height=0.70\linewidth]{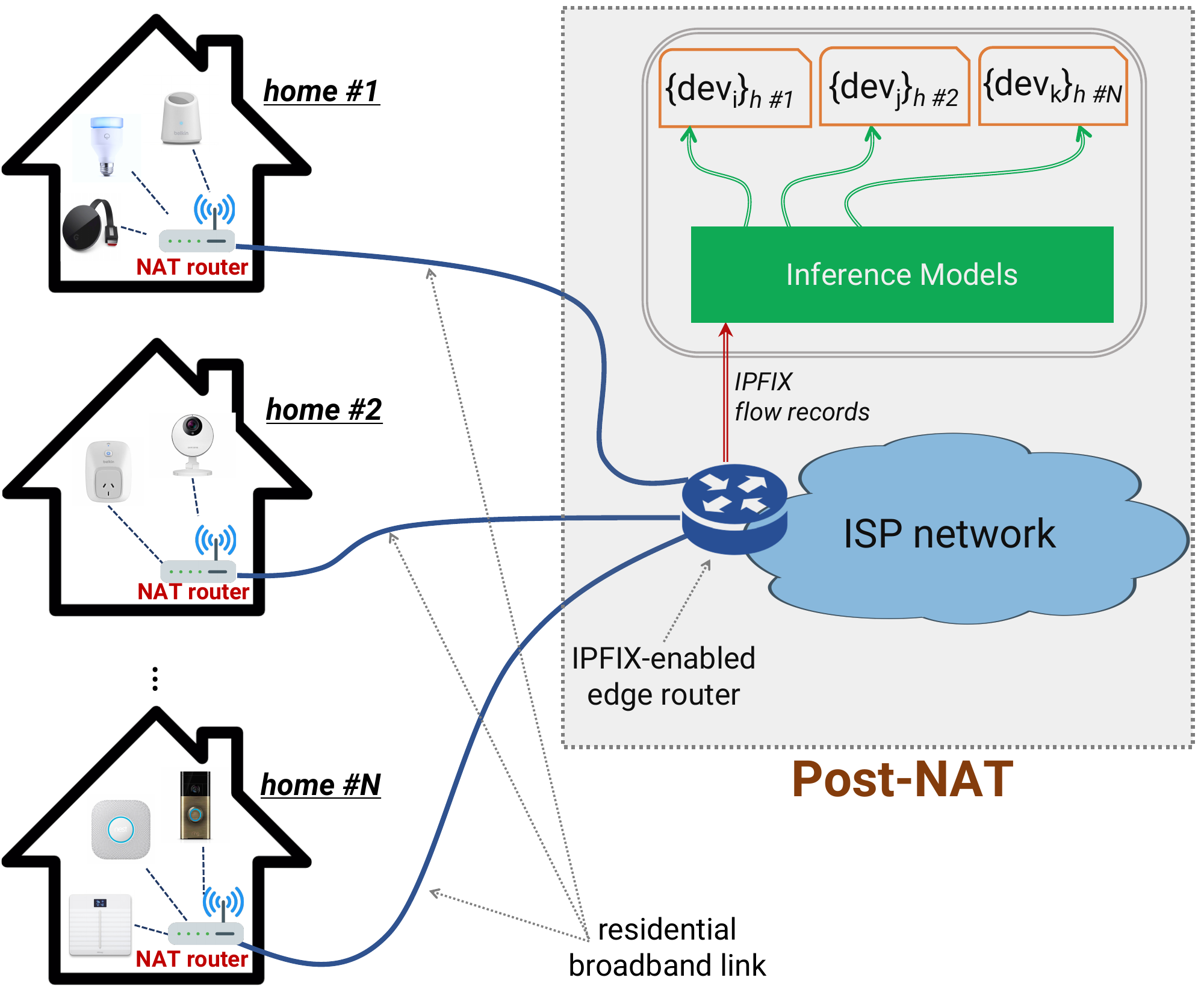}
		\caption{System architecture of post-NAT inference from IPFIX records.}
		\label{ch4:fig:arch}
	\end{figure}
	
	Fig.~\ref{ch4:fig:arch} illustrates the architecture of my inference system. Internet traffic of home networks is exchanged with the ISP network, where IPFIX-enabled edge routers are configured to export the respective IPFIX records. The inference can be made from a single IPFIX record (identifying the corresponding device that generated the flow) or from a collection of IPFIX records during a fixed epoch time (say, daily). Details will be discussed in \S\ref{ch4:sec:stochastic} and \S\ref{ch4:sec:dt} where I explain the stochastic and deterministic models. The ISP will progressively construct a set of IoT devices for each home upon discovering a new device type. 
	It can be seen in Fig.~\ref{ch4:fig:arch} that how post-NAT inferencing (the scope of this chapter) can be employed by the ISP, capturing only remote flows via which an IoT device inside a home network communicates with a remote server on the Internet. I note a  pre-NAT approach (practically challenging for ISPs, and beyond the scope of this chapter), in contrast, may provide richer visibility into all communications, including both local and remote flows. Obtaining visibility into local traffic of home networks, however, can raise privacy concerns for users.

	In practice, traffic inference models (particularly for determining the type of connected IoT devices) may encounter challenges in response to changes in (a) the composition of IoT devices (users may add or remove one or more devices) or (b) the network behavior of an existing device can evolve over time and/or space domains due to firmware upgrades,  configuration updates, dynamic user behaviors. 
	
	Let us start with the first scenario of change. The inference task is unaffected if a known type of device is added to or removed from the network. However, connecting an entirely new type (unknown to the model) will go undetected. I note that traditional supervised learning algorithms (this chapter included) aim to train a model in the closed-set world, where training and test samples share the same label space. Open set learning is a more challenging and realistic problem, where there exist test samples from the classes that are unseen during training. Automatic recognition of open set \cite{Reis2018c} is the sub-task of detecting test samples that do not come from the training, which is beyond the scope of this chapter. Therefore, in this chapter, it is not unexpected if some devices in a given home network remain undetected (false negative), but it is important for the ISP that their model does not detect nonexistent devices (false positive). This is crucial since inferring the composition of devices in a home network may lead to subsequent actions like notifying users about a vulnerable device on their network. Hence, reducing false positives is one of the primary objectives of this chapter.

	Moving to the second scenario which is assumed to be relatively easier to detect by a Trust metric (will be discussed in (\S\ref{ch4:sec:trust} and \S\ref{ch4:sec:eval}). Again, re-training the model by incorporating behavioral changes \cite{Arunan2020b} and/or variations \cite{Hamza2018} is beyond the scope of this chapter. In the next chapter, I will specifically focus on the challenge of concept drift (behavior changes of IoT devices in temporal and spatial domains) and develop multiple strategies to minimize its impact on the efficacy of the traffic inference model.

	\subsection{IPFIX Dataset and Features}\label{ch4:sec:ipfix}
	
		The residential testbed in Japan\footnote{Cybersecurity lab of KDDI Research in Saitama, Japan.}  comprises 29 consumer IoT devices ranging from smart cameras, speakers, door phones, and TV to lightbulbs, sensors, and vacuum cleaners. PCAP (Packet Capture) traces were collected from the testbed (pre-NAT) during January, March, and April 2020. Due to technical issues in the lab, traffic collection was stopped during February 2020.
	It is important to note that the data was collected in a pre-NAT setup to obtain ground-truth label (device MAC/IP address) of traffic data; however, I do not take any advantage of those identities in my inference process. I exclude the local flows from the raw data collected pre-NAT, and only keep remote flows for developing my inference models in this chapter. In other words, the processed data truly represent a post-NAT telemetry scenario.
	
	The IoT network traces constitute: (i) traffic generated due to human users interacting with the devices at least twice a week in January and March -- for example, asking questions from smart speakers, checking air quality from smart sensors, switching the color of smart lightbulbs, and watching a live stream from cameras; as well as (ii) traffic generated by the devices autonomously -- for example, {\myverb{DNS/NTP}} activities that are unaffected by human interactions.
	During April, due to the COVID outbreak, devices were operated completely autonomously. For April 12-13 and 21-22, there is no traffic trace of the devices due to a power shutdown and the server outage, respectively.
	Note that PCAP files were configured to record up to a size limit of 9.6 MB (meaning time slots with a duration of 20-30 minutes). Hence extracting IPFIX records from individual PCAPs may lead to some flows that cross the boundary of subsequent slots and break into shorter incomplete flows. We, therefore, stitch PCAPs on a monthly basis, and then extract IPFIX records from the monthly PCAPs. YAF was used (Yet Another Flowmeter)\footnote{YAF was created as a reference implementation of the IPFIX protocol, providing a platform for experimentation and rapid deployment of new flow meter capabilities.} \cite{yaf} to generate IPFIX binary files from the PCAP files, followed by applying Super Mediator tool \cite{supermed} to generate IPFIX JSON (JavaScript Object Notation) files from the binaries. YAF comes with an option called {\myverb{flow-stats}} that allows exporting a richer set of flow-level information. This option was used to embed all the possible features into individual IPFIX records. YAF tool defines default values of idle-timeout and active-timeout parameters (for {\myverb{UDP}} flows) equal to 5 and 30 minutes, respectively.

		\begin{figure}[t!]
		\centering
		\includegraphics[width=0.95\linewidth]{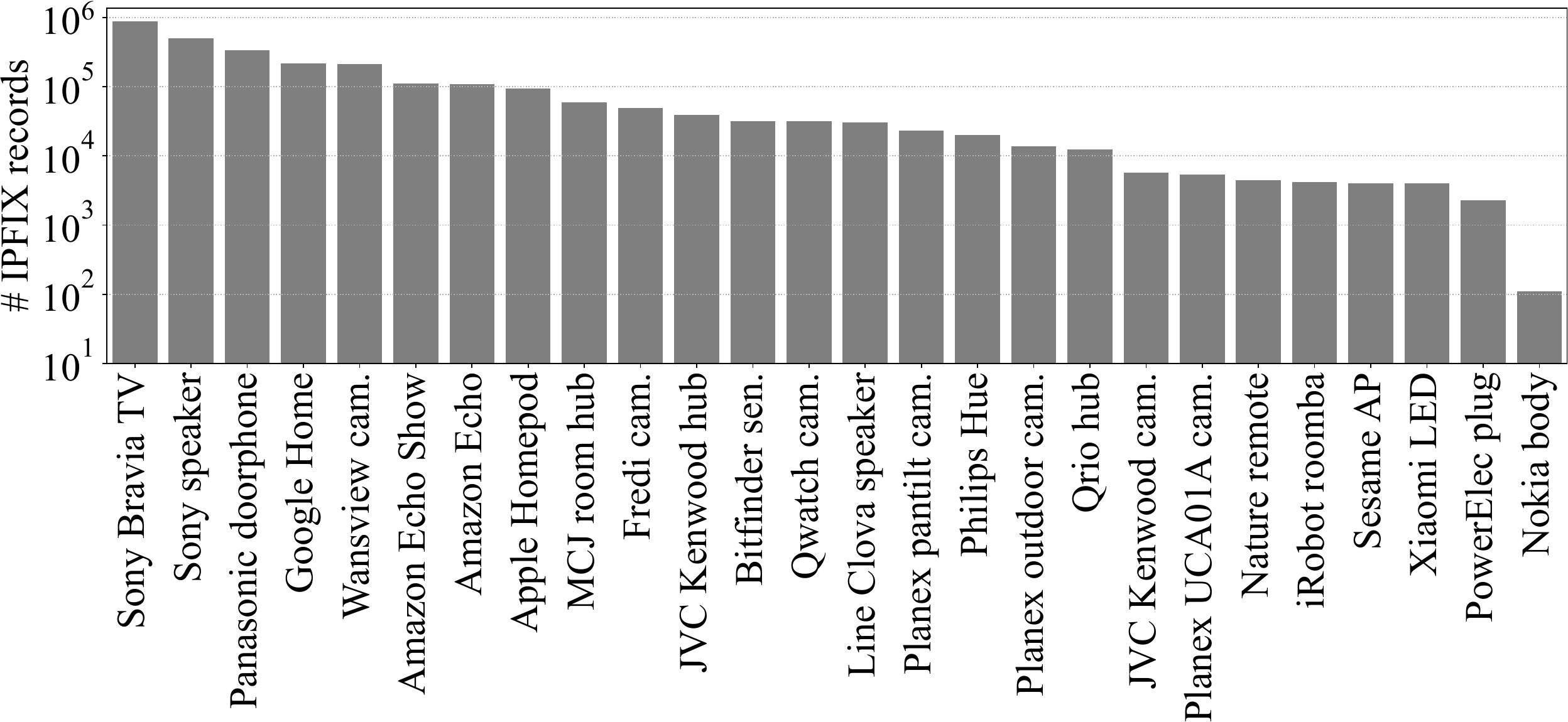}
		\vspace{-4mm}
		\caption{Distribution of remote IPFIX records across 26 IoT devices.}
		\label{ch4:fig:records count}
	\end{figure}	
	\vspace{-4mm}
	The IPFIX dataset \cite{ipfixdataset} I used for this chapter contains more than nine million IPFIX records in which about $70\%$ of them are local flows that do not leave the NAT gateway (cannot be captured by the ISP edge routers). I use IPv4 address element of IPFIX records to filter local flows, including: (i) unicast flows with both source and destination from private address space (\ie {\myverb{10/8}}, {\myverb{172.16/12}}, {\myverb{192.168/16}}) reserved by the Internet Assigned Numbers Authority (IANA) \cite{rfc1918}, (ii) multicast flows (\ie with address in the range of  {\myverb{224/4}}), (iii) link-local flows (\ie with address in the range of {\myverb{169.254/16}}), and (iv) broadcast flows. Three devices, namely Sony camera, Planex one-shot camera, and LinkJapan eSensor are found with only local flows in the dataset, hence excluded from our study in this chapter. After excluding the local flows, there are about three million remote IPFIX records. For the rest of this chapter, I will only focus on remote flows (nearly three million IPFIX records). Note that IPFIX records are bi-directional, but we can determine the initiator of each communication flow by checking the source IP address. In the dataset, $94\%$ of the flows are initiated by IoT devices (inside the home network), and the remaining $6\%$ (traversing NAT) are initiated by remote cloud servers. I will explain in \S\ref{ch4:sec:dt} how NAT traversing flows add noise to inputs of my deterministic model.

	\vspace{-3mm}
	Fig.~\ref{ch4:fig:records count} illustrates the number of IPFIX records (remote flows) per device collected during January, March, and April 2020. Highly active devices like Sony Bravia TV, Sony speaker, Panasonic doorphone, Google Home, and Wansview camera generated over 100K flows. Devices like MCJ room hub, Fredi camera, JVC Kenwood hub, Philips Hue, and Qrio hub are moderately active with 10K-100K flows. Lastly, we observe relatively less-active devices like Nokia body that generated only 111 flows during these three months.
	
		\begin{figure}[t!]
		\centering
		\includegraphics[width=0.88\linewidth,height=0.58\linewidth]{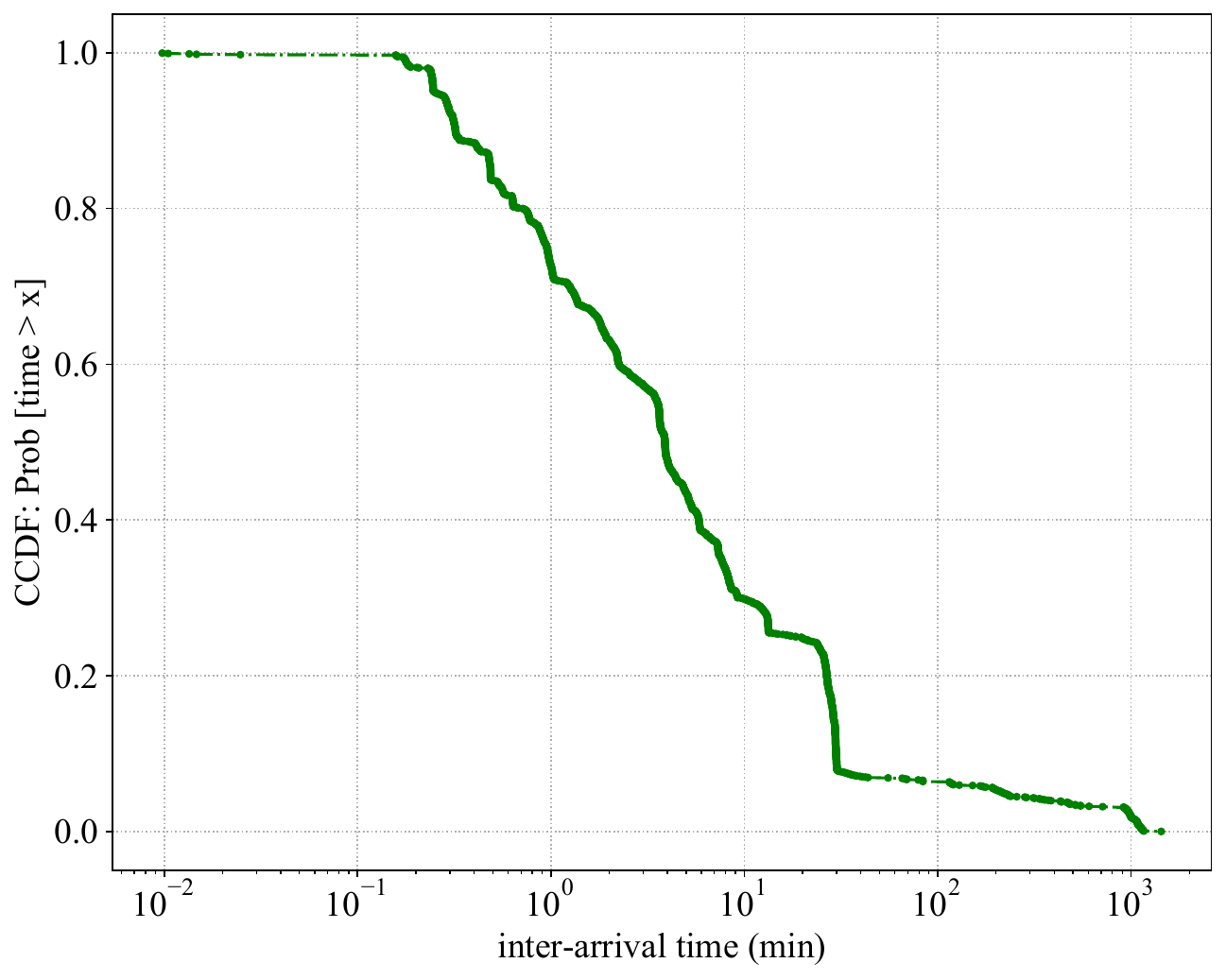}
		\caption{CCDF of (daily) average inter-arrival time of IPFIX records per IoT device in the dataset.}
		\label{ch4:fig:IAT}
	\end{figure}
	
	Considering the frequency of network activities per device type, I plot in Fig.~\ref{ch4:fig:IAT}, CCDF (Complementary Cumulative Distribution Function) of average flow inter-arrival time, on a daily basis per device during the three months. The average inter-arrival time across all device types is about $50$ minutes, while 14 device types generate a flow every $10$ minutes on average. We observe that the longest daily average of flow inter-arrival time belongs to Nokia body with more than 14 hours because this device only becomes active whenever the user interacts with it.

	Different IoT devices display distinct behaviors on the network. Researchers have shown how network activity (both in terms of total traffic volume and temporal patterns) of a smart camera or TV would differ from that of a motion sensor or air quality monitor \cite{arunan19}. In another case, \cite{Hamza2018} showed how various manufacturers tend to embed a (relatively unique) combination of standard and propriety network services (\eg  {\myverb{TCP/443}} for {\myverb{TLS}}, or {\myverb{UDP/56700}} for cloud control messages) in the firmware of their devices. In other words, the behavior of IoT devices can be profiled using a combination of stochastic and deterministic features. Generally, stochastic features are suited for machine learning models as their patterns can be characterized by certain distributions. In contrast, deterministic features may not always be a good fit for machine learning algorithms, especially when their dimension is high. For example, in the case of network services, we require over 65K binary features (total number of possible port numbers \cite{IANAservice}). For domain names, on the other hand, one cannot even define a bounded set of features. Deterministic features, therefore, need to be modeled differently.
	
	In this chapter, I use cloud services consumed by IoT devices for my deterministic model. Each cloud service (\eg {\myverb{UDP/10086}}, {\myverb{TCP/8443}}) is a pair of transport-layer protocol and port number on the server-side. One may choose to use other deterministic features like IP address and/or domain name of cloud servers. However, cloud IP addresses can be relatively dynamic \cite{Saidi2020}, hence unreliable over time.
	Also, domain names are not readily available in IPFIX records, though can be retrieved by way of reactive lookups. Cloud services, instead, can be directly deduced from IPFIX records and are more likely to be persistent during the lifetime of IoT devices. In fact, network services that IoT devices offer and/or consume along with local/Internet endpoints they communicate with, constitute a formal description, called Manufacturer Usage Descriptions \cite{rfcMUD}, that can be used to track IoT devices \cite{Hamza2018} or limit their attack surface\cite{Hamza2019}.

	\section{Stochastic Model of IoT Network Behavior}\label{ch4:sec:stochastic}
	
	\begin{table}[t!]
		\caption{``Unidirectional'' activity features extracted from IPFIX records.}
		\renewcommand{\arraystretch}{1.2}
		\begin{adjustbox}{max width=0.95\textwidth}
			\begin{tabular}{|l|l|}
				\hline
				\textbf{Feature} & \textbf{Description}\\ \hline 
				\myverb{packetTotalCount}& \# packets\\ \hline 
				\myverb{octetTotalCount} & \# bytes \\ \hline 
				\myverb{smallPacketCount} & \# packets containing less than 60 bytes payload\\ \hline
				\myverb{largePacketCount} & \# packets containing at least 220 bytes payload\\ \hline
				\myverb{nonEmptyPacketCount} & \# packets containing non-empty payload\\ \hline
				\myverb{dataByteCount} & total size of payload\\ \hline 
				\myverb{averageInterarrivalTime} & average time (ms) between  packets\\ \hline 
				\myverb{firstNonEmptyPacketSize} & payload size of the first non-empty packet\\ \hline 
				\myverb{maxPacketSize} & the largest payload size \\ \hline 
				\myverb{standardDeviationPayloadLength} & standard-deviation of payload size for up to the first 10 non-empty packets\\ \hline
				\myverb{standardDeviationInterarrivalTime} & standard-deviation of inter-arrival time for up to the first 10 packets\\ \hline
			\end{tabular}
		\end{adjustbox}
		\label{ch4:tab:features}
	\end{table}

	\subsection{Baseline model}

	My stochastic model (generated by a machine learning algorithm) characterizes the activity behavior of connected IoT devices on the network from flow-level statistical features provided by IPFIX records. I use this as a baseline and also representative of stochastic models used in prior works like \cite{Meidan2020}. Table~\ref{ch4:tab:features} summarizes a list of all potential elements (indicative of flows' network activity) that will be used as a part of the features in \S\ref{ch4:sec:eval}. It is important to note that each IPFIX record represents a bi-directional flow \cite{ipfixRFC5103}, hence there are two sets of these features -- each specific to a direction (outbound and inbound). For example, {\myverb{smallPacketCount}} (third row in Table~\ref{ch4:tab:features}) shows the number of small packets that an IoT device (the initiator of flow) transmitted to an Internet-based server (remote endpoint), and accordingly, {\myverb{reverseSmallPacketCount}} indicates the number of small packets that the remote endpoint sent to the IoT device on the same flow. 
	
	In addition to the activity features listed in Table~\ref{ch4:tab:features}, I capture six binary identity features of each flow record (Table~\ref{ch4:tab:ports}), highlighting whether their network service is one of four popular services ({\myverb{HTTP}}, {\myverb{TLS}}, {\myverb{DNS}}, {\myverb{NTP}}) or \textit{other} {\myverb{TCP/UDP}} services -- specific network services will be analyzed by a separate deterministic model in \S\ref{ch4:sec:dt}. 
	Considering popular services in the dataset, {\myverb{TLS}} dominates by contributing to $35$\% of flows, followed by {\myverb{DNS}} ($20$\% of flows), {\myverb{HTTP}} ($6$\% of flows), and {\myverb{NTP}} ($2$\% of flows). Also, in terms of other services, {\myverb{UDP}} outweighs {\myverb{TCP}} by contributing to $30$\% of flows versus $3$\%.
	Note that the remaining $4$\% of flows use network control protocols like {\myverb{ICMP}} and {\myverb{IGMP}} that are not associated with either of the dominant transport-layer protocols namely {\myverb{TCP}} and {\myverb{UDP}}.
	
	{\vspace{5mm}}
	I found some device types display distinct behaviors by certain features in their IPFIX records. For example, IPFIX records with {\myverb{reverseSmallPacketCount}} larger than 500 would belong to Sony speaker with a probability of $95$\%; also, $97$\% of the IPFIX records which their {\myverb{reverseDataByteCount}} falls between $300$ to $400$KB, belong to Sony Bravia TV. That said, we observe some overlap of features across device types. This is mainly because each IPFIX record contains an ``aggregate'' set of information on the activity and identity of a single network traffic flow. For example, I found 16 devices share the {\myverb{averageInterarrivalTime}} feature, having a value less than $10$ seconds on average; five devices have this feature valued between $10$ and $20$ seconds; and, five other devices have a value greater than $20$ seconds in their IPFIX records.
	
	\begin{table}[t!]
		\centering
		\caption{``Binary'' identity features of popular services extracted from IPFIX records.}
		\renewcommand{\arraystretch}{1.2}
		\begin{adjustbox}{max width=0.65\textwidth}
			\begin{tabular}{|l|l|l|}
				\hline
				\textbf{Network service} & \textbf{Protocol} & \textbf{Transport-layer port numbers}\\ \hline
				\myverb{HTTP} & \myverb{TCP} &80, 8080, 8008, or 8888\\ \hline
				\myverb{TLS} &\myverb{TCP} & 443, 1443, 8443, or 55443\\ \hline
				\myverb{DNS} &\myverb{UDP} & 53, or 5353\\ \hline
				\myverb{NTP} &\myverb{UDP} & 123\\ \hline
				others &\myverb{TCP} & \textit{any} \\ \hline
				others & \myverb{UDP} & \textit{any}\\ \hline
			\end{tabular}
		\end{adjustbox}
		\label{ch4:tab:ports}
	\end{table}
	
	In another example, there are three categories of devices in the dataset by only considering the {\myverb{reversePacketTotalCount}} feature: (a) Planex UCA01A camera with a high count (more than 200) of packets, (b) Xiaomi LED and iRobot Roomba with medium packet counts (between 100 and 200), and (c) other 23 device types with low packet counts (less than 70). Also, in terms of network services, devices like Fredi camera, JVC Kenwood camera, Sesame Access Point (AP), Xiaomi LED, and all three Planex cameras never use {\myverb{TLS}}. Instead, all IPFIX records of a device like Qrio hub are {\myverb{TLS}}.

	These examples highlight that determining the device type from IPFIX records would certainly require a collection of stochastic features. We, therefore, employ machine learning techniques that have been widely adopted by researchers for network traffic inferencing \cite{Mahdavinejad2018} to learn the stochastic attributes of IPFIX records for inferring their device type.
	
	I choose Random Forest, a powerful machine learning model that builds a decision tree ensemble, allowing for the automatic creation of a sequential combination of feature rules (similar to what we did in the examples above) to predict device classes. In order to maximize the prediction performance, I extract every possible feature from the IPFIX records and let the Random Forest algorithm construct the optimal decision trees.

	\subsection{Confident Classification}
	\label{ch4:sec:ct}
	As mentioned earlier, analyzing flows in a post-NAT scenario will reveal partial information about actual IoT devices connected to the home network; therefore, a machine learning-based model may misclassify some IPFIX records. Note that the primary objective of my inferencing is to determine whether an IoT device is present and active in a home network, or not. Therefore, the quality of each prediction is of top priority. This objective necessitates a method for discarding inconfident predictions, made primarily due to a lack of distinct features in the subject IPFIX flow. 

	Note that one may attempt to aggregate individual IPFIX records over a time window in order to increase the amount of data passed into the inference model. However, this is not applicable in post-NAT deployment as there is no association between individual IPFIX records and devices. 
	
	
	Machine learning models often produce a measure of confidence for their prediction. The confidence-level is a number between 0 and 1. Higher confidence values indicate more reliable predictions. I, therefore, use the confidence-level of the Random Forest model to decide whether a prediction is ``accepted'' or ``discarded'' -- certain thresholds are determined and applied to the confidence-level of the model's prediction. I note that thresholds can be applied globally \cite{arunan19} across all classes, or specifically per individual classes. Given the diversity of network behaviors as well as richness of training data across various IoT types, I choose to apply class-specific thresholds that are obtained from the training phase. 
	
	$ C_L$ denotes the model's confidence for correctly predicting instances of class $L$, and the corresponding threshold is denoted by $\lambda$, which is calculated for each class separately. To set the thresholds, I use the distribution (average: $\mu_{C_L}$ and standard deviation: $\sigma_{C_L}$) of confidence per each class during the training phase as the baseline. 
	In this chapter, I consider two ways of thresholding per each class: (a) greater than the average confidence: $\lambda_1=[{\mu_{C_L}},1]$, and (b) within $\sigma_{C_L}$ from the average confidence:  $\lambda_2=[{\mu_{C_L}-\sigma_{C_L}},{\mu_{C_L}+\sigma_{C_L}}]$. Note that $\lambda_1$ is only concerned about its lower-bound $\mu_{C_L}$ without imposing an upper-bound. Instead, $\lambda_2$ aims to conform to the distribution obtained from training by imposing both lower and upper bounds to the model's confidence. Note that one can choose other thresholds for this purpose. In \S\ref{ch4:sec:eval}, I will show how each of these thresholds would improve the accuracy of the model by discarding the majority of mispredictions.

	\subsection{Behavioral Changes of IoT Devices}
	\label{ch4:sec:trust}
	In this chapter, my primary objective is to identify IoT devices connected to home networks using IPFIX flow records collected post-NAT. ISPs may wish to obtain additional insights into the network behavior of classified devices in each home. For example, they may want to track the activity pattern of discovered devices for purposes like ensuring their cyber health or triggering the model re-training process due to some behavioral changes. Those follow-up actions (health assurance and/or model re-training) are beyond the scope of this chapter; however, in the next chapter, I develop and compare various techniques to have an accurate inference under network behavioral changes of IoT devices. I note that tracking the behavior of IoT devices post-NAT can be relatively challenging as our unit of data (IPFIX flow record) does not carry the identity (e.g., MAC/IP address) of their respective end-device. In the absence of a solid meta-data for associating the IPFIX records with devices, I again resort to the output of the inference model.
	
	For reasons discussed above, a behavioral change can only be determined based on a set of observed IPFIX records per home. As our unit of data is a single IPFIX record, it is needed to accumulate a handful of them in order to determine a behavior change. Therefore, I choose  a configurable epoch (\eg a day) to construct a collection of model predictions and look for changes in the time domain.
	One may choose to employ the widely-used \textit{Jaccard index} to compare two sets of categorical values (predicted devices across successive epochs). This index quantifies the ratio of the size of the intersection of two sets to the size of their union. Let us start by formally defining the baseline Trust metric ($T_{base}$) as follows:
	\vspace{15mm}
	
	\begin{equation}
		\mathnormal{T_{base}}  = \frac{|P_{i-1} \cap P_{i}|}{|P_{i-1} \cup P_{i}|}
	\end{equation}
	
	where $P_{i-1}$ and $P_{i}$ are the sets of devices classified for epochs $i-1$ and $i$, respectively, and $|.|$ operator returns the size of the given set.
	
	$T_{base}$ is relatively simple and keeps track of predicted devices per household over consecutive epochs, making it a stateful measure. However, there are four caveats where $T_{base}$ falls short of expectations: (1) $T_{base}$ is computed across all devices of a home network for a given period without giving any Trust indication for individual devices; (2) if a device is removed from the network (turned off) by the user following  days of being active (and hence correctly predicted), $T_{base}$ will immediately drop for the next day, which is not highly desirable; (3) suppose an active device generates less (or more) records than its normal pattern (captured behavior during the training phase). In that case, $T_{base}$ will not detect such changes as long as the device is inferred by at least one IPFIX record; and (4) when the number of flows for a particular device remains consistent but flow features like the number of packets/bytes changes, $T_{base}$ would be unable to detect those dynamics.
	
	With the shortcomings of the baseline metric discussed above, I now aim for a more comprehensive metric that focuses on capturing the state of predictions within an epoch without directly incorporating temporal dynamics. With respect to the first problem, the Trust metric needs to be device-centric instead of home-centric. For the second problem, I define Trust in such a way as to be stateless, meaning it is measured for each epoch purely based on the predictions of that epoch without looking at the past. In order to address the third and fourth problems, I incorporate the number of records being predicted as the target device I aim to measure Trust for, as any severe deviation in the number of records and/or flow features will be reflected in the model's output. In \S\ref{ch4:sec:eval}, I evaluate $T_{base}$ against a more desirable metric that I will develop next and demonstrate how Trust metric can reveal more information about changes in the behavior of individual devices.
	
	\vspace{3mm}
	
	Considering the points highlighted above, I define Trust for individual devices over each epoch. For a device, the Trust metric is computed based on the number of predictions for that type during a given epoch (\eg a day) while taking into account class-specific patterns obtained from the training phase. For a device class $L$ in the model, I measure the expected (average) number of IPFIX records (denoted by $N_{e,L}$) and the expected rate of discarded IPFIX records (denoted by $D_{e,L}$) during training. Note that some predictions are discarded due to a lack of confidence reported by the model. A raw measure of trust is computed by:

	\begin{equation}
		\mathnormal{T_{L}}  = \frac{N_{o,L}}{N_{e,L} \times D_{e,L}}
	\end{equation}

	where, $N_{o,L}$ is the number of ``observed'' flows predicted as class $L$ -- obviously, those predictions receive an accepted level of confidence from the model.
	The intuition behind this raw measure of trust is to check the number of expected flows from a device with the number of flows that are indeed classified as that device type during the epoch. Trust values around one imply that corresponding devices are behaving as expected. However, the measure of raw trust can become larger than one when a device is over-active (\eg under attack), or its flows are mispredicted as another device class. Either of these cases (\ie having the raw trust value more than one) needs to be interpreted as misbehavior. Therefore it is needed to normalize the raw trust so that it decays in case of deviation from normal behavior. I define normalized trust (will be referred to as trust in the rest of this chapter) by:
	\begin{equation}
		\mathnormal{T_{norm,L}} = \exp\left({-\frac{(T_L - 1)^2}{2\sigma_{T_L}^2}}\right)
	\end{equation}
	
	\vspace{3mm}
	where $\sigma_{T_L}$ is the standard-deviation of $T_L$, computed during the training phase. Use of $\sigma_{T_L}$ allows for customizing the change rate of trust across various device types -- for a complex device like smart TV that displays a wide range of network activities the trust metric changes slower, while for a simple device like smart sensor it is will change relatively faster.
	
	It is important to note that trust is measured for individual IoT device types, which is fine-grained and can help highlight changes in the behavior of specific device types. One may take another step by combining trust measures of different device types of a given home network to detect changes in a cluster/group of device types. This augmentation is beyond the scope of this thesis.

	\subsection{Evaluating Efficacy of Stochastic Modeling}
	\label{ch4:sec:eval}
	For performance evaluation, I split the data into two groups: training (1,664,994 IPFIX flows recorded in January and first half of March), and testing (1,416,479 IPFIX recorded in the second half of March and full April). I use Scikit-Learn Python library to train the Random Forest classifier.

	\subsubsection{Stochastic Model Training and Tuning}
	I tune the parameters of the Random Forest model to maximize its performance for the chosen features (Table \ref{ch4:tab:features} and \ref{ch4:tab:ports}). I tune three important parameters, namely the number of decision trees, the maximum number of attributes to consider at each branch split, and also the maximum depth of each decision tree.
	For each combination of parameters, I quantify the model accuracy by 10-fold cross-validation, whereby the training dataset is randomly split into training (90\% of total instances) and validation (the remaining 10\% of total instances) sets, and the accuracy is averaged over 10 runs to produce a single performance metric.
	 
\begin{figure}[t!]
		\centering
		\includegraphics[width=0.99\linewidth]{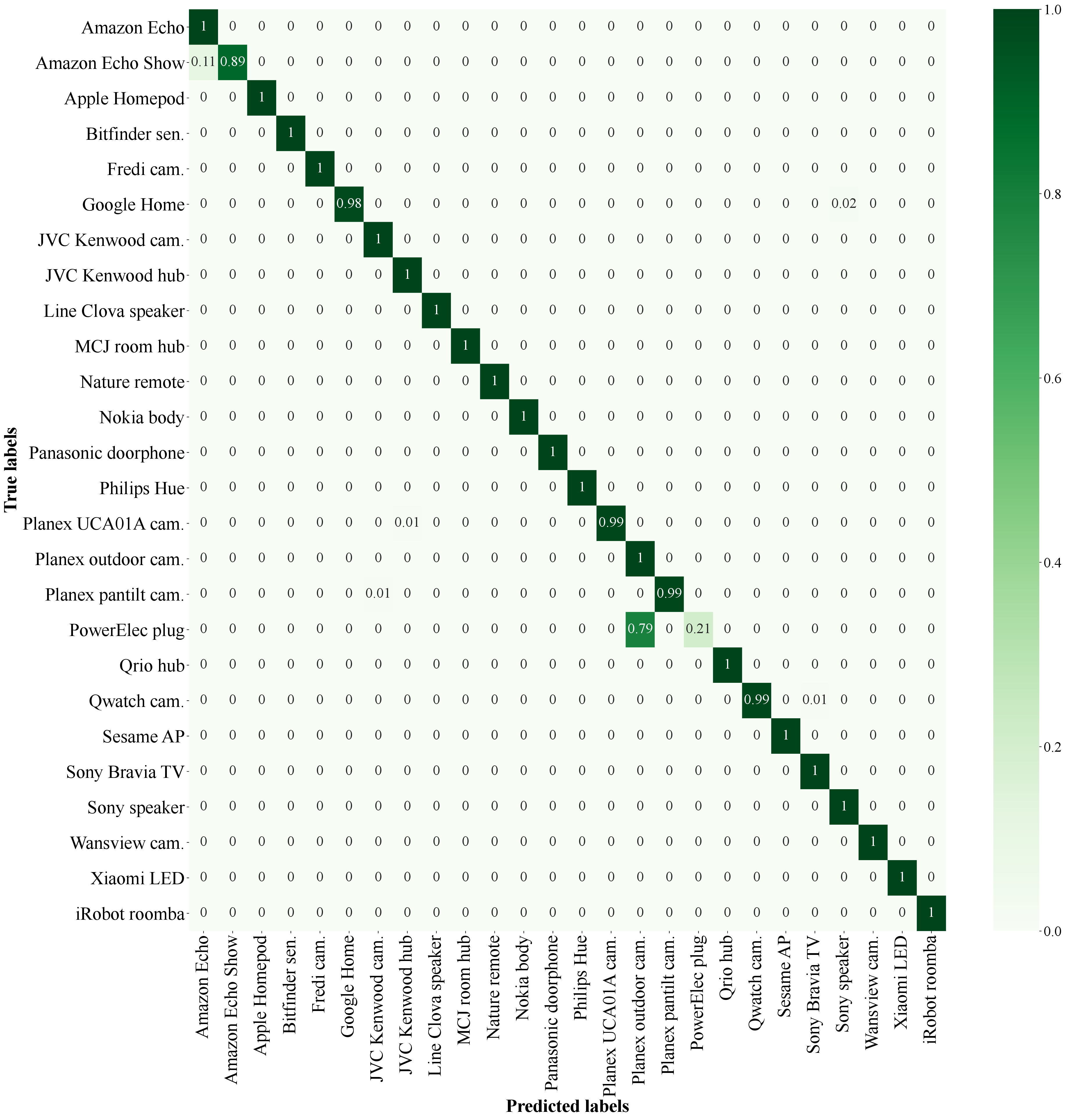}
		\caption{Confusion matrix of accepted predictions -- Random Forest model followed by applying class-specific confidence thresholds ($\lambda_1$).}
		\label{ch4:fig:final cm}
	\end{figure}

	 In this chapter, accuracy is my main metric to quantify the performance of inference because the models developed in this chapter will classify individual IPFIX records of a single home network. Hence, other measures like $F_1$-score are not applicable. However, in the next chapter, as there will be multiple home networks, I will use both accuracy and $F_1$-score to measure the inference performance. I compute the model's accuracy by averaging the rate of correct predictions (true-positives) across individual classes of the model. In other words, the average of all values across the main diagonal of the model's confusion matrix, shown in Fig.~\ref{ch4:fig:final cm}. 
	I found the best tuning parameters which yield the highest prediction accuracy ($88$\%) to be 100 decision trees, maximum of five features for finding the best split, and maximum depth of 20 for each decision tree.

		\begin{figure}[t!]
		\centering
		\includegraphics[width=\linewidth]{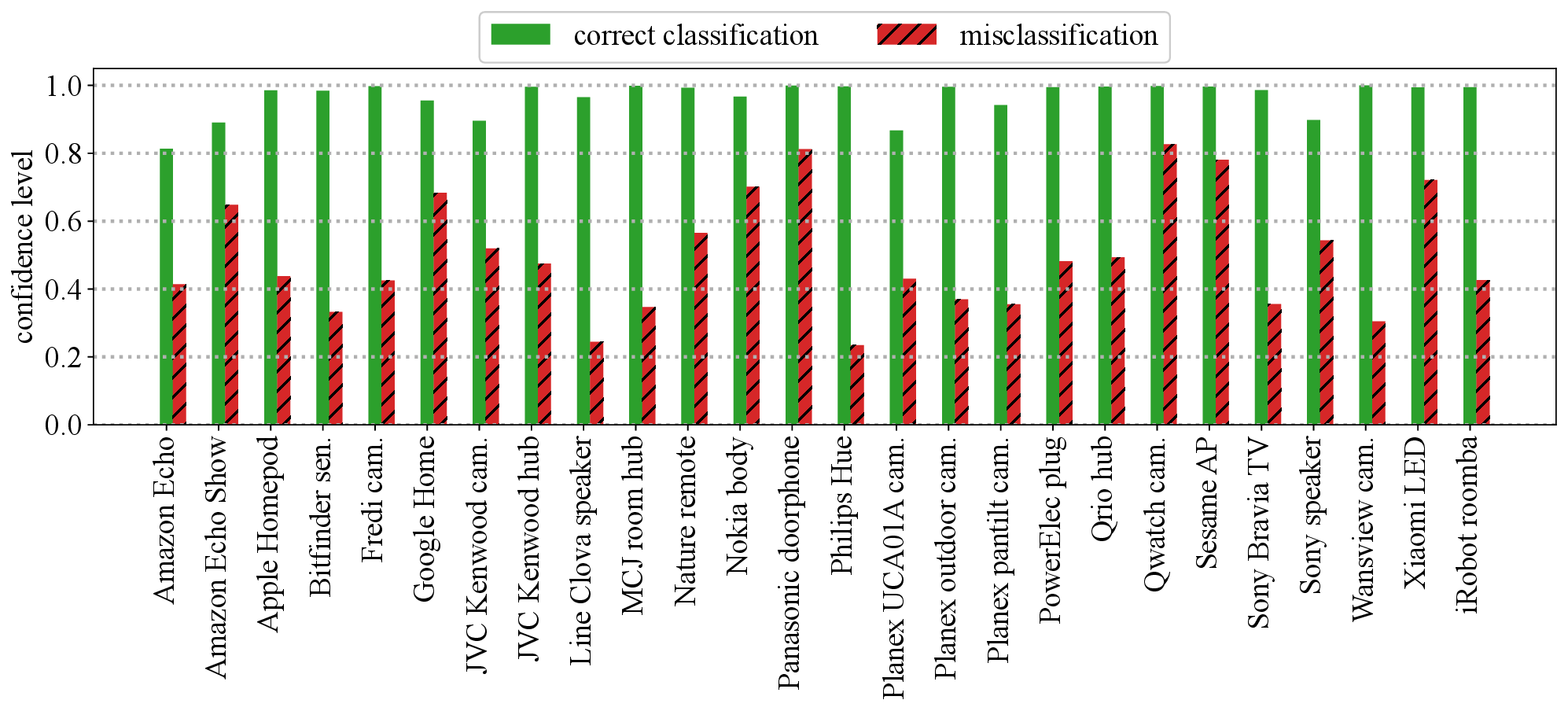}
		\caption{Average confidence-level of correctly and incorrectly classified instances per device type for the training dataset.}
		\label{ch4:fig:avr conf}
	\end{figure}
	
	I next quantify the confidence of the model (on the training dataset) for correct classifications as well as incorrect classifications across various classes of device type. As shown in Fig.~\ref{ch4:fig:avr conf}, green bars indicate the average confidence for correctly predicted flows and red bars indicate the average confidence for mispredicted flows. 
	A clear takeaway from this analysis is that the model is more confident when a flow is correctly classified and is relatively less confident when a flow is misclassified.
	
	\begin{table}[t!]
		\centering
		\small
		\caption{Impact of $\lambda_1$ and $\lambda_2$ thresholds on training instances.}
		\renewcommand{\arraystretch}{1.2}
		\begin{adjustbox}{width=0.65\textwidth}
			\begin{tabular}{|l|c|c|c|c|}
				\hline
				& \multicolumn{2}{c|}{\cellcolor{green!25}\textbf{Accepted}} & \multicolumn{2}{c|}{\cellcolor{red!25}\textbf{Discarded}} \\ \hline
				& \textbf{Correct} & \textbf{Incorrect} & \textbf{Correct} & \textbf{Incorrect} \\ \hline
				$\lambda_1$ & 1,325,787 & 924 & 237,724 & 100,559 \\ \hline
				$\lambda_2$ & 1,408,600& 28,214 & 154,911& 73,269 \\ \hline
			\end{tabular}
		\end{adjustbox}
		\label{ch4:tab:lambda}
	\end{table}
	
	The second observation is that the two threshold methods ($\lambda_1$ and $\lambda_2$ defined in \S\ref{ch4:sec:ct}) affect the inference differently, as shown in Table~\ref{ch4:tab:lambda}: $\lambda_1$ is found to be more conservative by discarding $20\%$ of predictions, compared to $14\%$ discard rate resulted from applying $\lambda_2$; $\lambda_1$ is proven to discard more of incorrect predictions compared to $\lambda_2$ ($99\%$ versus $72\%$); considering the accepted predictions, $\lambda_1$ gives a slightly better accuracy than $\lambda_2$ ($99.91\%$ compared to $99.02\%$). Given the primary objective being the quality (purity) of predictions, I use the first method of thresholding ($\lambda_1$) for the testing phase of the evaluation.
	
	\subsubsection{Stochastic Model Testing}
	Having tuned the parameters and obtained the class-specific confidence thresholds, I now evaluate the efficacy of the stochastic model by applying it to the testing dataset.
	Fig.~\ref{ch4:fig:final cm} shows the confusion matrix of the inferencing (the model combined with $\lambda_1$ thresholds).  The average accuracy across all classes is $96\%$, with 20 classes receiving more than $99\%$ accuracy -- the rate of correct predictions across accepted predictions. I found that applying the thresholds significantly improves the quality of model prediction. Note that the average accuracy of the model alone (without $\lambda_1$ thresholding) is $82\%$. Also, only $1\%$ of the accepted instances are incorrectly predicted.
	However this quality of prediction is achieved at the cost of discarding a part of the correctly predicted flows \ie $26$\% of correct predictions when the model is less confident than expected thresholds. Discard rate of correct predictions, on average, is about $17$\% per class.

	Though the stochastic model gives an overall acceptable performance, I find flow records of the Power Elec plug are not well predicted. I found that the model mispredicts most of Power Elec plug instances (relatively confidently) as Planex outdoor camera. Further investigations revealed that almost all of those misclassified flows correspond to {\myverb{UDP/10001}} service which is the second top service in IPFIX records of Planex outdoor camera (the mispredicted class). Analyzing the features of {\myverb{UDP/10001}} records from these two classes showed that the inter-arrival time and the number of outgoing packets in the flows of Power Elec plug significantly changed from the training to the testing phase, becoming similar to those of Planex outdoor camera flows, and hence resulted in the misprediction. Note that the ISP may choose to re-train the model \cite{Arunan2020b} after verifying certain legitimate changes, but re-training is beyond the scope of this chapter.

	\begin{figure}[!t]
		\centering
		\includegraphics[width=0.92\linewidth, height=0.80\linewidth]{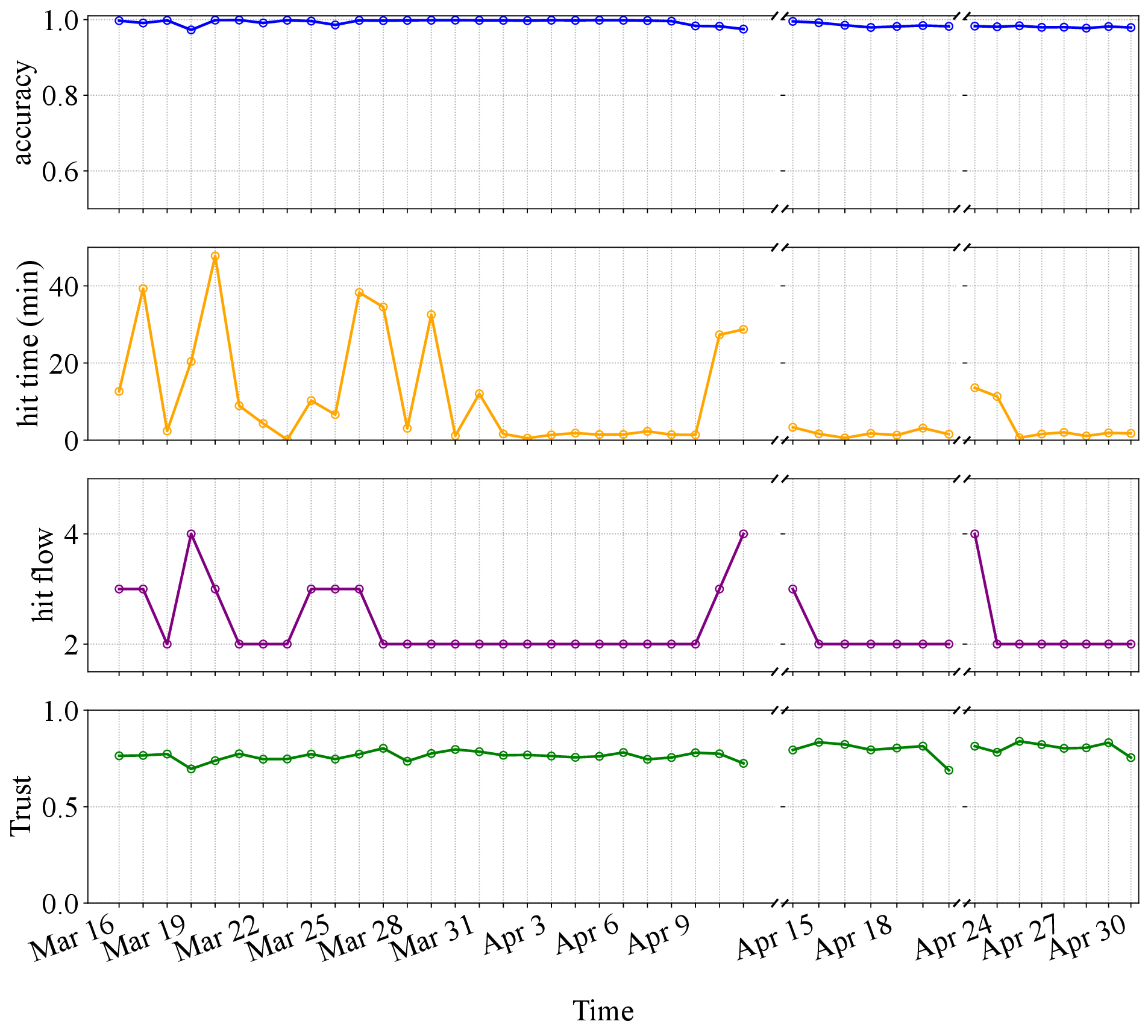}
		\caption{Time-trace of average accuracy, trust, hit-time, and hit-flow of the inferencing across all classes on the testing dataset.}
		\label{ch4:fig:time trace}
	\end{figure}
	
	\vspace{-4mm}
	Another takeaway from the evaluation is that the performance of machine learning-based inference model varies across different network services. The scheme yields a better accuracy in {\myverb{TCP}} flows compared to that of  {\myverb{UDP}} flows. Focusing on {\myverb{TCP}}, we see the highest true-positive rate in both {\myverb{TLS}} and ``other'' {\myverb{TCP}} flows with $99$\%, followed by {\myverb{HTTP}} flows with $94$\%. For {\myverb{UDP}} flows, ``other'' {\myverb{UDP}} services have $100$\%, followed by {\myverb{DNS}} and {\myverb{NTP}} with $98$\% and $72$\% true-positives, respectively. It is important to note that {\myverb{NTP}} flows receive the highest discard rate ($99$\% of the {\myverb{NTP}} flows are predicted with fairly low confidence). This is mainly because the vast majority of the {\myverb{NTP}} flows have the exact size of 76 bytes (including payload, {\myverb{UDP}} and {\myverb{IP}} headers) across all IoT classes in the dataset. I note that most of these IoT devices use the minimum packet structure defined by the {\myverb{NTP}} standard without any extension field, resulting in identical byte count feature for this specific type of flows.

	\subsubsection{Operational Insights}
	I track the efficacy of the inference model in operation on a daily basis, as shown in Fig.~\ref{ch4:fig:time trace}. In addition to overall accuracy (top plot), which remains fairly high during the testing period, I capture two metrics: (a) \textit{hit-time}, the time taken for the model to detect a device since its first-seen on the network, and (b) \textit{hit-flow}, which indicates the number of flows generated by a device before the model correctly detects it. It is seen from the second top plot in Fig.~\ref{ch4:fig:time trace} that the average hit-time across all classes varies between $9$ seconds and $48$ minutes, highlighting the responsiveness of the inference scheme. Also, I observe that the average hit-flow mostly equals $2$ (with a maximum of $4$) --
	this suggests that discarding about a third of correctly classified flows does not incur an excessive delay in the inference.
	
	Moving to the trust metric shown in the bottom plot of Fig.~\ref{ch4:fig:time trace}, we see consistently high values during the testing period on average (highlighting the overall health of devices). Only on 20\nth  April, we observe a slight drop in the trust due to the early shutdown of the capturing server in preparation for a planned power outage on 21\nst and 22\nd April -- this resulted in just 3 hours worth of data captured on 20\nth April.
	
		\begin{figure}[!t]
		\centering
		\includegraphics[width=0.92\linewidth]{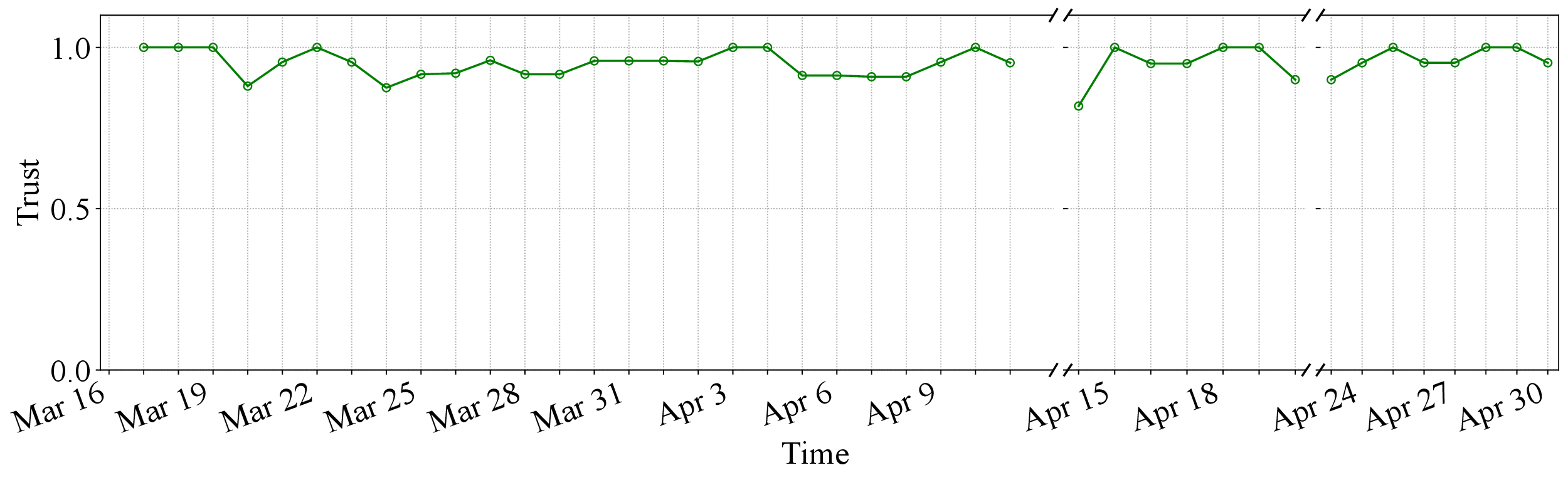}
		\caption{Time-trace of baseline trust ($T_{base}$) during the testing period.}
		\label{ch4:fig:base_trust}
	\end{figure}
	
	Fig.~\ref{ch4:fig:base_trust}  shows the temporal dynamics of baseline Trust ($T_{base}$). Based on its definition, $T_{base}$ starts getting value from the second day of the testing period (17th March). It can be seen that $T_{base}$ displays relatively high and consistent values across the entire testing period, with a daily average of $95\%$, indicating most devices are consistently present and active in the network. In what follows, we will see certain changes in the behavior of some devices that will go undetected by $T_{base}$  (mainly because of its aggregate nature). In contrast,  the metric $T_{norm,L}$ (by incorporating fine-grained contexts) is able to highlight them.

	\begin{figure}[!t]
		\centering
		\includegraphics[width=0.92\linewidth, height=0.80\linewidth]{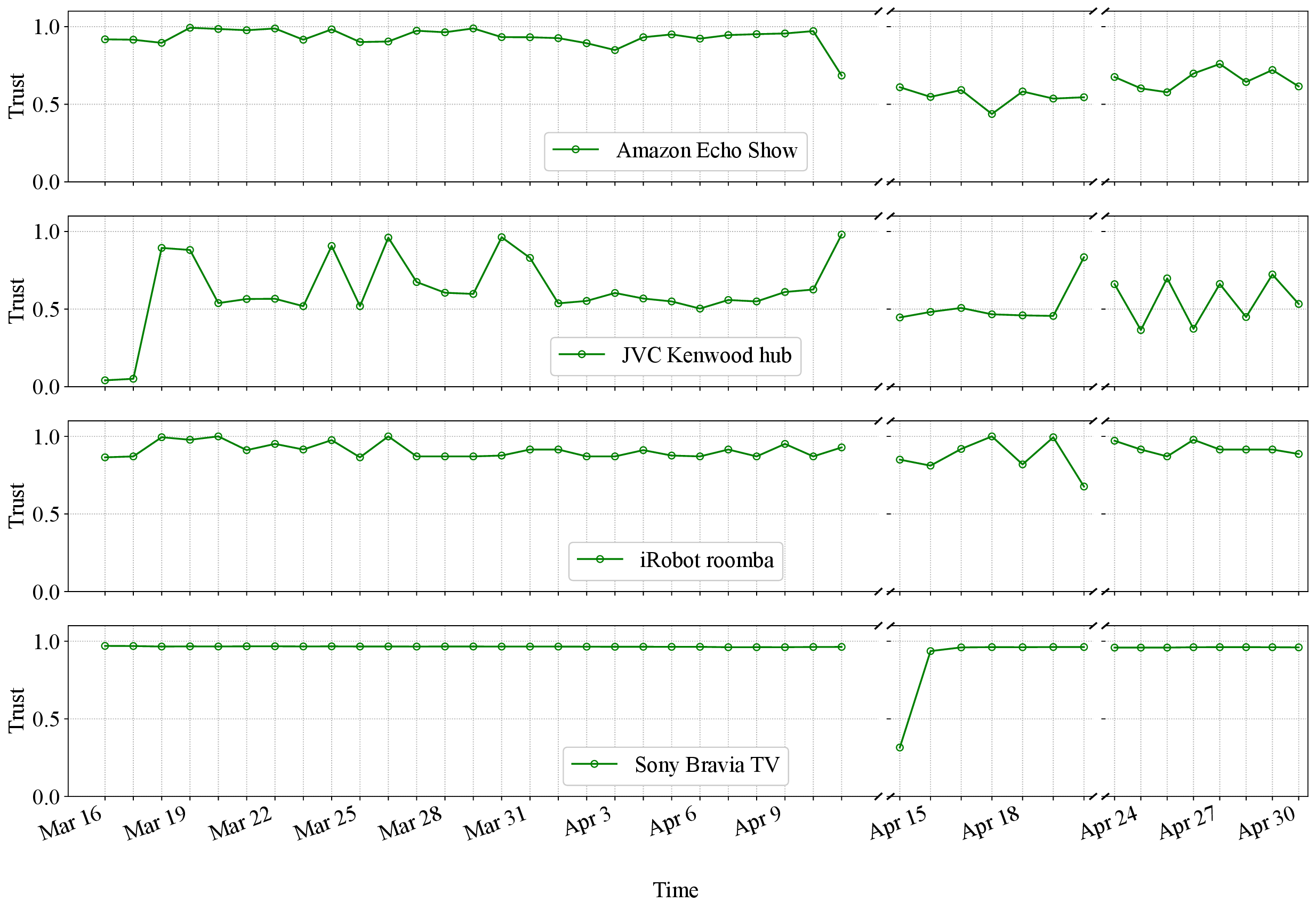}
		\caption{Daily measure of Trust for four representative IoT devices on the testing dataset.}
		\label{ch4:fig:trust trace}
	\end{figure}
	
	\vspace{-4mm}
	\textbf{Dynamics of Trust Metric:}
	To better understand the temporal dynamics of the Trust metric, I plot in Fig.~\ref{ch4:fig:trust trace} the daily Trust for a few representative classes which display noticeable changes in their behavior.
	
	\vspace{-4mm}
	We can categorize the behavioral changes by trust metric in two groups: (i) permanent declines which are probably caused by a legitimate firmware upgrade, leading to change in traffic features and/or the number of generated flows, and (ii) temporary declines that can be caused by an intermittent network condition or device usage/configuration for a short period of time, leading to generation of significantly different number of flows without necessarily affecting the traffic features. 
	Fig.~\ref{ch4:fig:trust trace} illustrates the daily trust of four representative devices from the testbed: Amazon Echo Show (change in traffic features), and JVC Kenwood hub (change in the number of flows) represent a firmware upgrade scenario, while iRobot roomba and Sony Bravia TV represent temporary behavioral/usage changes affecting their flow count.

	Amazon Echo Show has an average trust of $0.94$ before April 11 on which it starts to decay and remains around $0.61$ until the end of April. By further analysis, I found that the daily flow count was almost expected, but some significant changes were observed in certain traffic features ({\myverb{reverseOctetTotalCount}} and {\myverb{reverseDataByteCount}} dropped by about $80$\%) compared to the training phase. These changes caused the number of accepted flows to drop by $40$\%, compared to the expected value from the training phase.
	
	JVC Kenwood hub starts with a very low trust value on 16\nth and 17\nth March, followed by a fluctuating trust measure averaged at $0.66$ from 18\nth March to 11\nth April. Next, we observe a permanent drop, starting from 14\nth April with an average trust of $0.54$. JVC Kenwood hub generated 983 and 1019 flows during the first two days respectively (16\nth and 17\nth March) -- in each day, about 870 flows received accepted prediction. Moving to the second half of April, the hub generated on average 218 flows per day, and around 168 of flows (daily) are predicted with an acceptable confidence. These values are far from the expected behavior of JVC Kenwood hub based on the training dataset where the expected number of flows is about 415 per day.
	
	The iRobot Roomba displays a consistently high trust value over the testing period (about $0.91$ on average) except for 20\nth April (with trust value of $0.67$). This device has an average of $50$ flows per day over the testing period, except on 20\nth April, it only had $7$ flows. The shutdown of the capture server in preparation for a planned power outage on the following days caused this drop. As the expected flow count for iRobot Roomba is about $43$ per day, the significant deviation on 20\nth April resulted in a low trust value for this device.
	
	Sony Bravia TV shows a persistent trust measure throughout the testing phase except for April 14, when it generated a very high number of flows. On average, it generated $5,648$ flows per day during those days with a healthy trust metric. However, the flow count rose to $98,013$ on April 14 -- of those flows, $\approx$$44,000$ (about $3.8$ times the expected baseline) received an accepted prediction.
	
	\vspace{-4mm}
	
	\section{Deterministic Model of IoT Cloud Services}\label{ch4:sec:dt}
	\vspace{-4mm}
	In the previous section, I trained a multi-class ML model on stochastic features of IPFIX records. Although ML models are popular in the literature, they come with some open challenges: (a) They do not provide perfect accuracy. The best model would still result in some false positives \cite{Garadi2020}, depending on the complexity of patterns across classes and the amount of training data available. For our problem, a false positive arises when the model detects an IoT device that is not present on the network. As explained in \S\ref{ch4:sec:IPFIX}, false positive is detrimental for device inference at scale; (b) Activity behaviors of IoT devices are susceptible to firmware upgrade \cite{Arunan2020b}, which can significantly impact the stochastic features described in Table~\ref{ch4:tab:features}. In that case, the stochastic model may require re-training with a new dataset to learn certain changes in benign behaviors of the upgraded device; (c) Stochastic features can vary \cite{arunan19} depending on user activities and interactions with their devices at home. For example, considering a smart speaker, asking about the weather, or listening to a music track would result in different amounts of traffic generated on the network. Therefore, capturing a superset of all IoT behaviors at scale may not be practical; and (d) machine learning models are vulnerable to adversarial attacks (I will study this challenge in Chapter~\ref{ch:secure infer}) which makes them less attractive in critical infrastructures.

	On the other hand, IoT cloud services seem promising to overcome certain limitations of the stochastic model. Cloud services that certain IoT device types may share can be detected deterministically, reducing false positives. As cloud services are maintained on the servers (of manufacturers and/or application developers) and not on the device side, they are less likely to change by firmware upgrades or user activities. I denote a cloud service by a two-tuple information of {\myverb{IP protocol/port number}} (\eg {\myverb{TCP/80}}, or {\myverb{UDP/53}}) readily available in IPFIX records, that is not impacted by NAT operations. In fact, services that are exposed by IoT devices may also be helpful in detecting them; however, as they are overwritten by NAT routers and are not accessible post-NAT, hence beyond the scope of this chapter.

	\subsection{Fingerprint of IoT Cloud Services}\label{ch4:sec:cloud services}

	Let us begin by considering two groups of IPFIX records, namely ``outgoing'' flows initiated by IoT devices inside the home network to cloud servers on  the Internet and ``incoming'' flows initiated by remote endpoints (on the Internet) to household IoT devices. Cloud services, therefore, are captured from the source transport-layer port of incoming records and the destination transport-layer port of outgoing records.
	
	\begin{figure}[t!]
		\centering
		\includegraphics[width=\linewidth]{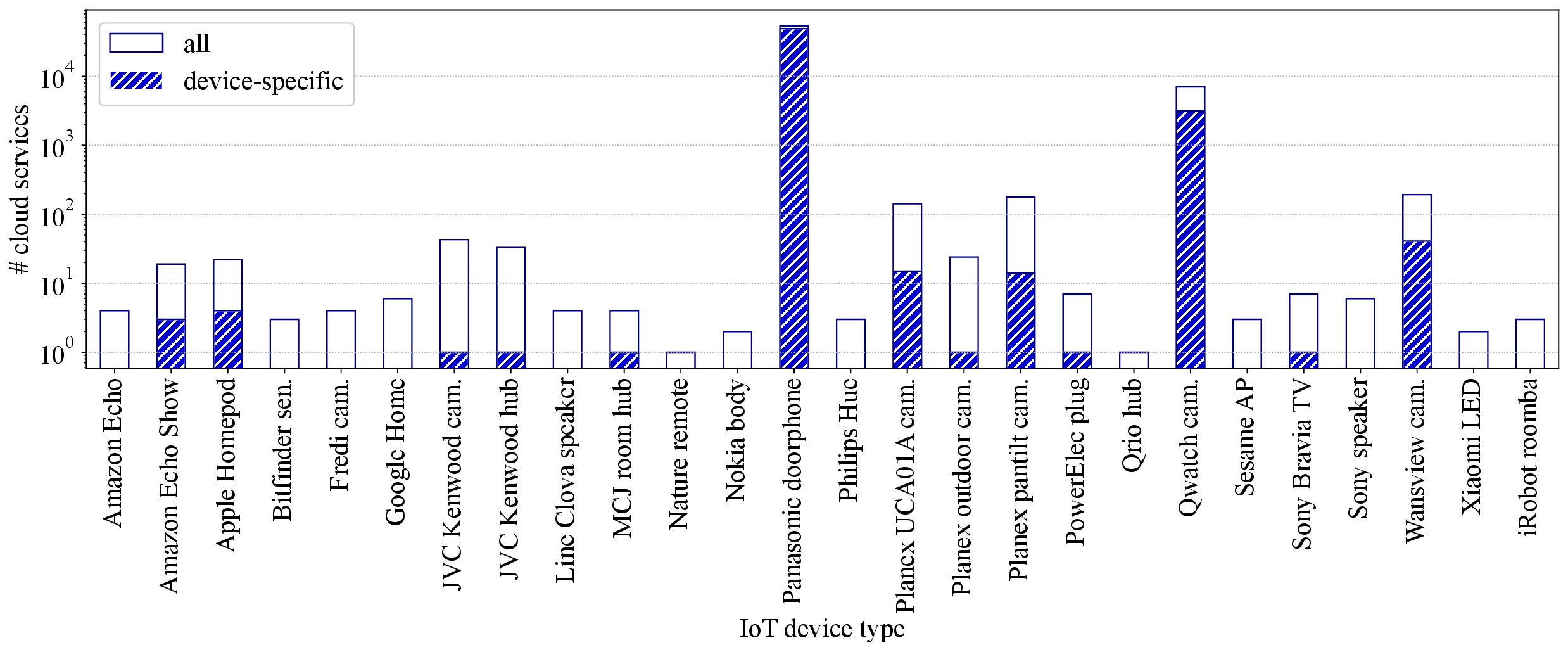}
		\caption[Count of cloud services per IoT device type in the training dataset]{Count of cloud services per IoT device type in \underline{the training dataset}.}
		\label{ch4:fig:service count unfiltered}
	\end{figure}
	
	Fig.~\ref{ch4:fig:service count unfiltered} shows the distribution of cloud services across various IoT device types in the training dataset (\ie January and first half of March). I find a total of 56,561 services under three groups: (1) \textbf{\textit{generic}} services like {\myverb{TCP/80}} and {\myverb{UDP/53}} that are found in traffic records of more than half of the IoT devices; (2) \textbf{\textit{mixed}} services that are shared across device types (only less than half of them) in the dataset -- for example, {\myverb{UDP/37038}} is only shared between JVC Kenwood camera and hub; and (3) \textbf{\textit{specific}} services that are only used by a single device type --  service {\myverb{UDP/41229}} is only seen in flow records of Amazon Echo Show. Note that some device types, such as Fredi camera, Google Home, and Amazon Echo come with no specific service (only generic and/or mixed services). In contrast, device types like Qwatch camera and Panasonic doorphone display thousands of specific services.
	Observing such an undesirable behavior \ie very wide distribution of cloud services for some devices like Qwatch camera and Panasonic doorphone led us to further analyze their corresponding flow records.

	A deeper investigation revealed that NAT traversing flows are the main cause for wide ranges of cloud services in the behavior of those IoT device types. Devices like smart cameras provide live video streams to their users. This feature requires the camera to be accessible behind NAT, which can be achieved either by having a relay server or directly traversing through NAT. In the former case, the server receives feed from the camera and passes it to users. In the latter case, user directly connects to the camera from the Internet, streaming the video feed -- direct connections initiated from outside home networks may not be feasible for all implementations of NAT gateways \cite{rfc7604}. If an IoT device behind NAT is accessible from outside, there will be communications flows destined to the device (\ie incoming flows), likely sourced from random transport-layer port numbers (cloud services). Therefore, I need to filter cloud services deduced from incoming flows and generate my deterministic model only based on cloud services obtained from outgoing flows (\ie initiated by IoT devices), avoiding random cloud services.
	
	\vspace{4mm}
	Filtering the incoming flows reduced the total number of cloud services to 5200. However, a device like Qwatch camera contributes to 4680 of cloud services. I further analyzed IPFIX records of Qwatch camera and noticed that this device uses Session Traversal Utilities for NAT (STUN) protocol \cite{stun}. STUN provides internal hosts with a set of tools to detect NAT type when communicating with remote endpoints. Upon completion of STUN tests, Qwatch camera performs NAT hairpinning (loopback) --  two hosts  behind the NAT (on the local network) communicate with each other via the external (public) IP address of the NAT gateway with port forwarding. I further exclude flows with public IP address of the home gateway to/from local IP addresses.

		\begin{figure}[t!]
	\centering
	\includegraphics[width=0.98\linewidth]{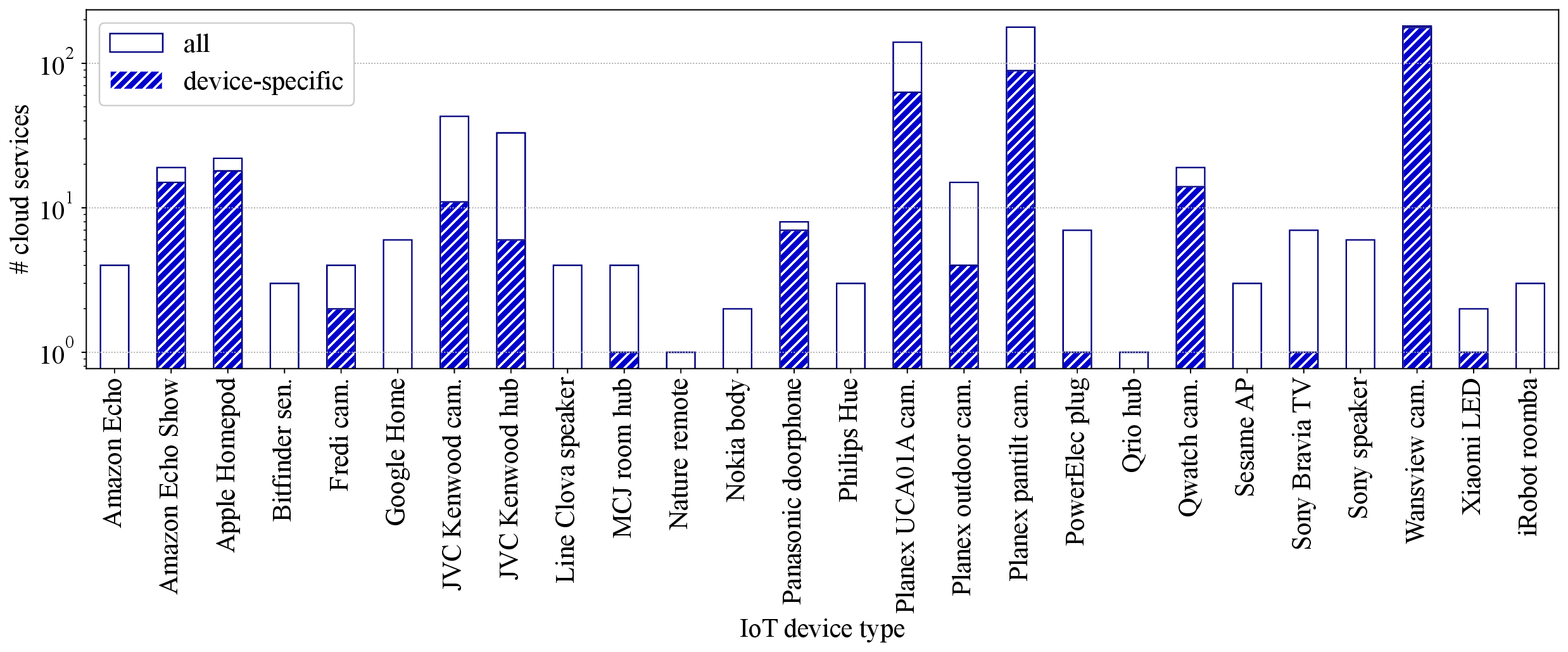}
	\caption[Count of cloud services after filtering NAT traversing flows]{Count of cloud services \underline{after filtering} NAT traversing flows.}
	\label{ch4:fig:service count}
\end{figure}
	
	\vspace{4mm}
	Fig.~\ref{ch4:fig:service count} shows the distribution of cloud services after removing incoming and hairpinning flows. It can be seen that the number of services (for certain device types) has significantly reduced by at least an order of magnitude (compared to Fig.~\ref{ch4:fig:service count unfiltered}) after filtering NAT traversing flows. For instance, the total count of cloud services for Qwatch camera and Panasonic doorphone has now become 19 and 8 services, respectively.
	
	

	\begin{table}[t!]
		\centering
		\small
		\caption[Distribution of cloud services across IoT device types after filtering NAT traversing flows]{Distribution of cloud services across IoT device types \underline{after filtering} NAT traversing flows.}
		\renewcommand{\arraystretch}{1.2}
		\begin{adjustbox}{width=0.65\textwidth}
			\begin{tabular}{|l|c|c|c|c|}
				\hline
				\multirow{2}{*}{} & \textbf{Generic} & \textbf{Mixed} & \textbf{Specific} & \textbf{Total} \\ \hline
				\# services & 4 & 110 & 410 & 524 \\ \hline
				\# flows & 914,157 & 138,077 & 233,474 & 1,285,708 \\ \hline
			\end{tabular}
		\end{adjustbox}
		\label{ch4:tab:service distribution}
	\end{table}

	Table~\ref{ch4:tab:service distribution} summarizes the number of generic, mixed, and specific services along with their corresponding flow count. We can see a total of 524 cloud services across device types in the dataset where about $78\%$ of them are specific to an IoT type. Four generic services, namely {\myverb{TCP/80}} ({\myverb{HTTP}}), {\myverb{TCP/443}} ({\myverb{TLS}}), {\myverb{UDP/53}} ({\myverb{DNS}}), and {\myverb{UDP/123}} ({\myverb{NTP}}) are shared between more than half of the device types. Note that generic services  contribute to only $0.7\%$ of discovered cloud services, but they are found in more than $70\%$ of all flows analyzed.
	
		\begin{figure*}[t!]
		\begin{center}
			\mbox{
				\subfigure[IoT device types with specific and shared services.]{
					{\includegraphics[width=0.45\textwidth]{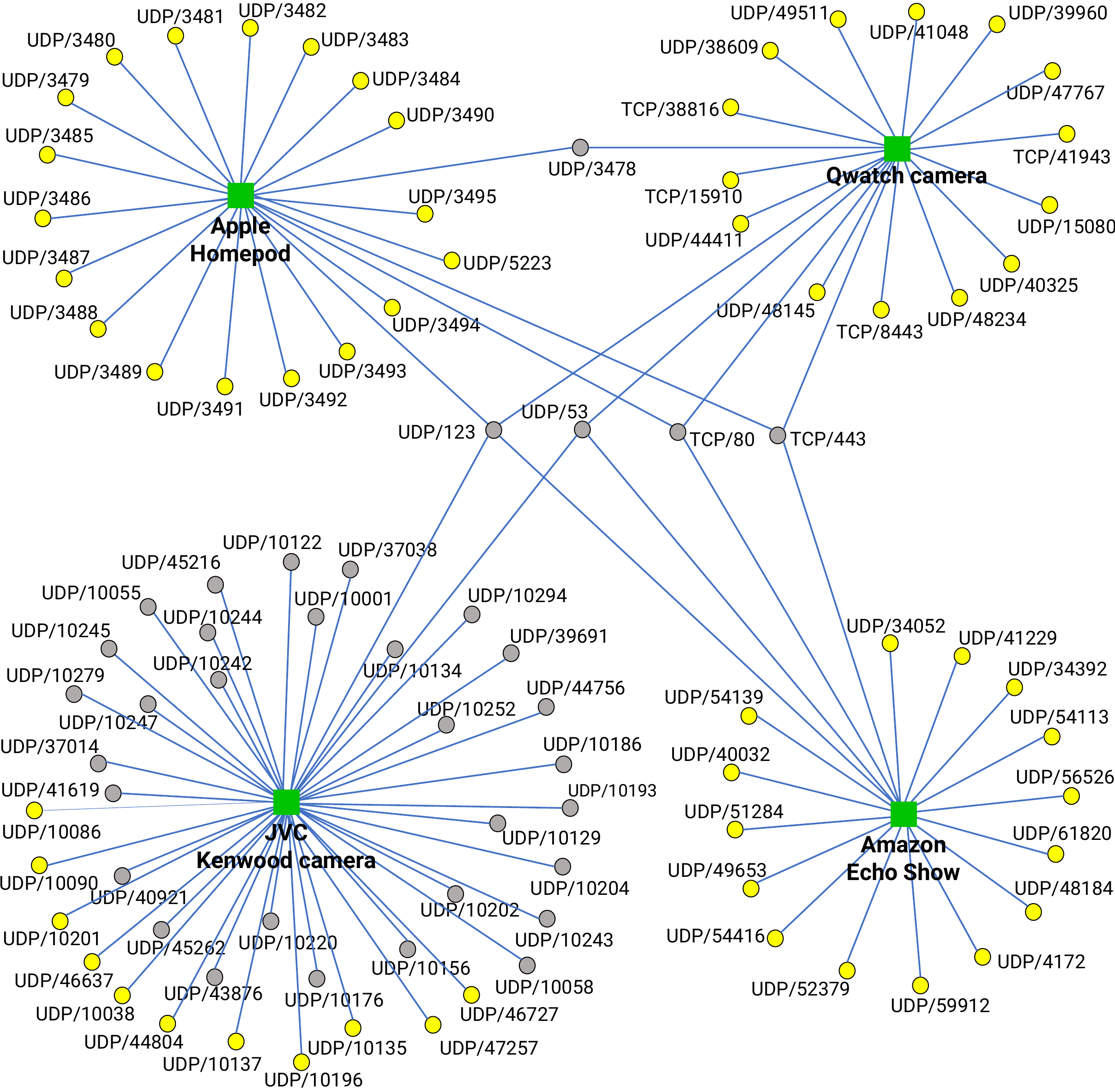}}\label{ch4:fig:unique network}\quad
				}
			}
			\hspace{-4mm}
			\mbox{
				\subfigure[IoT device types  with only shared (generic and/or mixed) services.]{
					{\includegraphics[width=0.45\textwidth]{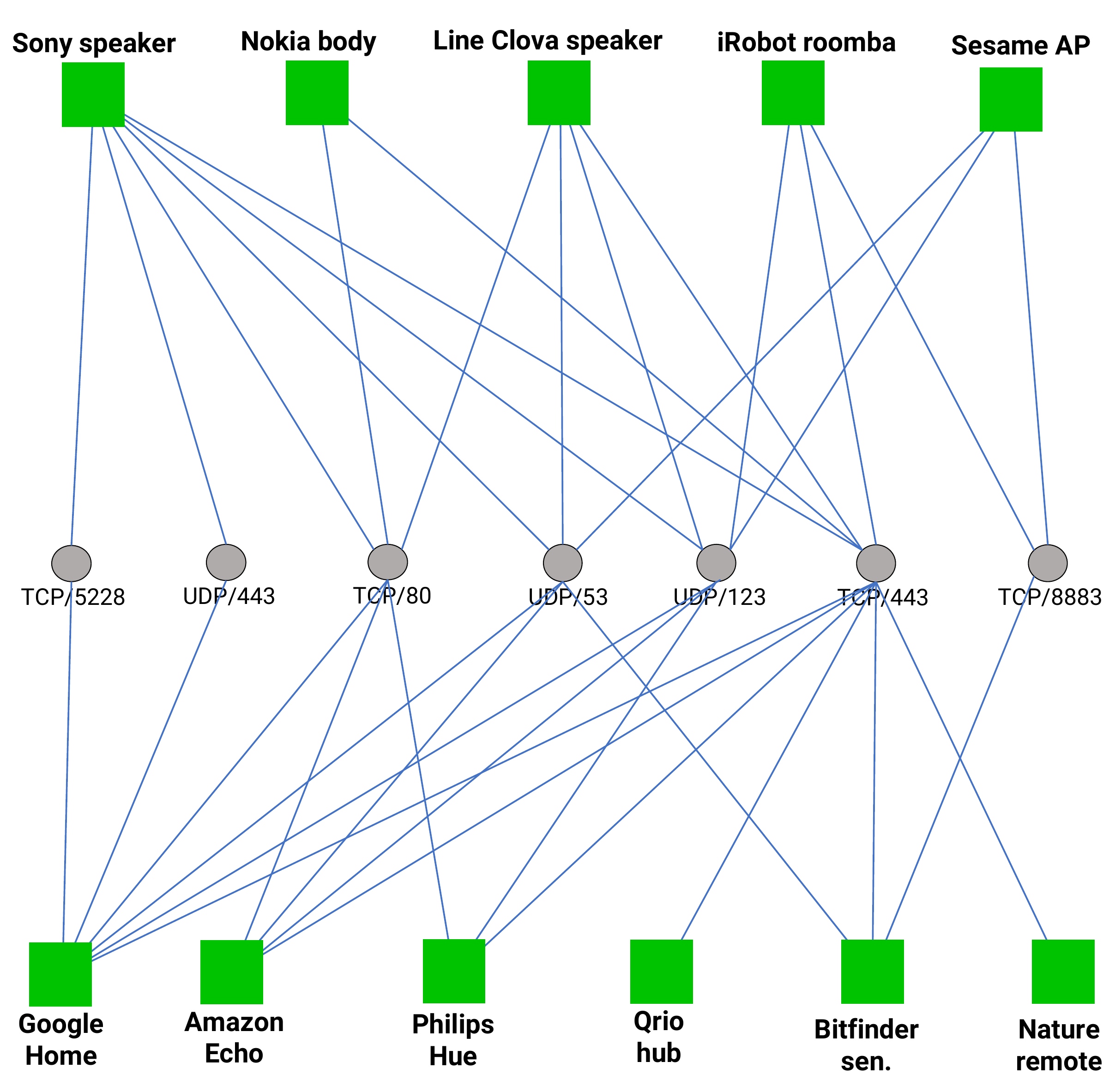}}\label{ch4:fig:shared network}\quad
				}
			}
			
			\caption[Fingerprint of IoT cloud services: (a) four representative device types with specific and shared services, and (b) 11 device types with only shared (generic and/or mixed) but no specific service]{Fingerprint of IoT cloud services: (a) four representative device types with specific and shared services, and (b) 11 device types with only shared (generic and/or mixed) but no specific service. Bright nodes highlight specific services and gray nodes highlight shared services.}
			\label{ch4:fig:service network}
		\end{center}
	\end{figure*}
	
	I visualize in Fig.~\ref{ch4:fig:service network} two representative fingerprints of IoT cloud services: (a) Four device types (Apple Homepod, Qwatch camera, JVC Kenwood camera, and Amazon Echo Show) with over ten specific services; and (b) device types in the dataset that have no specific cloud service. It can be seen in Fig.~\ref{ch4:fig:unique network} that even devices with a set of specific services (bright nodes) use a number of generic/mixed services (gray nodes) -- no device type in the dataset comes with a pure set of specific services.
	Fig.~\ref{ch4:fig:shared network} illustrates how device inference can be challenging with  shared cloud services across different device types. For example, we can see that Sony speaker and Google Home have the exact same set of services. The same happens to be the case for Line Clova speaker and Amazon Echo. In another example, we observe that the fingerprint of Nokia body (\ie {\myverb{\{TCP/80, TCP/443\}}}) is a subset of five other fingerprints. I will develop a weighted inference technique (next subsection) that associates a device-specific weight to each cloud service proportional to its occurrence frequency to tackle overlapped fingerprints.
	
	I note that the appearance and consistency of the fingerprint set over time can vary across device types in the training dataset. For example, the  fingerprint set of Panasonic doorphone was fully captured during the first two weeks of the training dataset and remained consistent until the end of the training period (about a month and a half). In another example, Amazon Echo uses three of its four shared services (it has no specific service) every day it is active. On the other hand, some devices grow their fingerprint set over time. For instance, Wansview camera uses new specific services on four days in March that were not seen during January. While new services emerged, parts of the fingerprint remained consistent throughout the training period -- for example, the camera communicated with its specific service {\myverb{UDP/60722}} every day it is active.
	
	\subsection{Inference Techniques}
	\label{ch4:sec:det inference methods}
	Detecting devices based on their cloud services can be done in different ways. In this section, I discuss a couple of these possible methods.
	
	\textbf{\underline{Specific Service}}: This inference is made on a per-flow basis. It needs to generate a map between individual cloud services and IoT device types from the training dataset for this method. For example, mappings {\myverb{(TCP/1111)}}$\rightarrow$\{$D_1$\} and {\myverb{(TCP/2222)}}$\rightarrow$\{$D_2,D_3$\} mean that the service {\myverb{TCP/1111}} exclusively belongs to the device type $D_1$ and  the service {\myverb{TCP/2222}} is shared between device types $D_2$ and $D_3$. Then I apply the generated map to each of the individual testing flows. If a given flow matches a one-to-one mapping rule (corresponding to a specific cloud service), then the unique type of the associated IoT device is inferred; otherwise if the cloud service is shared or unseen, no type is inferred.

	\textbf{\underline{Fingerprint Set}}: Inspired by \cite{Saidi2020}, this inference is made based on a collection of flows observed during a configurable epoch (say, daily) per household. Therefore, deterministic fingerprints (each is a set of cloud services obtained from the training dataset) for individual device types, similar to what is shown in Fig.~\ref{ch4:fig:service network}, are required as inputs to the inference. For testing during a given epoch, each cloud service (extracted from an IPFIX record) is checked against the fingerprint of all device types. If the service is found in a fingerprint set, I count a match for the corresponding device type (each service can be counted once per device type during an epoch). Note that a shared service (generic/mixed) will result in matches for more than one device type. At the end of the epoch, I compute the fraction of matched services for each device type. An IoT device is detected, if its computed fraction is above a configurable ``fingerprint threshold''. 

	The inference based on fingerprint sets can be employed in two ways: (a) ``\textit{unweighted}'', whereby the membership of cloud services to various fingerprint sets is measured as a binary value so all services have the same contribution in computing the fraction of matched services, and (b) ``\textit{weighted}'', whereby the service membership is measured by a weight (between 0 and 1), equal to the fraction of days that the service was seen in the training dataset over a total number of days on which the device was active. The final fingerprint match for each device is a weighted average of its observed services. That way, frequent services have more contribution in fingerprint matching. I consider the unweighted method as a baseline, representing deterministic models used in prior works like \cite{Saidi2020, Guo2020}.

	I note that the specific service inference method would be faster in detecting devices as it responds to each individual flow record independently (stateless). The fingerprint set inference, on the other hand, is stateful and needs to analyze a collection of records observed during the epoch before a set of devices are detected. The specific service method can only detect those IoT devices that consume at least one specific service, which is not the case for 11 IoT device types in the dataset, as highlighted by purely white bars in Fig.~\ref{ch4:fig:service count}.
	
	\subsection{Evaluating Efficacy of Deterministic Modeling}
	\label{ch4:sec:dt eval}
	I use the same training dataset (\ie traces from the month of January and first half of March) I used for the stochastic model to generate the deterministic models (\ie map of specific services to device types and a fingerprint set for each device). Similarly, the traces from the second half of March and April are used for testing. 
	
	Before evaluation, let us briefly look at the composition of cloud services in the testing set.
	Overall, about $19\%$ of testing flows have specific services that can be used to detect their corresponding IoT devices. A similar observation was made from the training dataset, reported by Table~\ref{ch4:tab:service distribution}.  Among all class types, three cameras, including Fredi camera, Panasonic doorphone, and Wansview camera, significantly dominate by having more than $94\%$ of their flows with services specific to their type. About $10\%$ of the total flows have services shared between two or three device types. 
	For example, more than three-quarters of flows for Qwatch camera are dedicated to the cloud service {\myverb{UDP/3478}} which is shared between Qwatch camera and Apple Homepod. That said, the specific service method (in what follows) will detect  both the camera and homepod with a perfect accuracy based on their individual specific services.
	In another example, {\myverb{TCP/8883}} is shared between Bitfinder sensor ($30\%$ of flows), Sesame AP ($98\%$ of flows), and iRobot roomba ($95\%$ of flows). I will show (in what follows) how the fingerprint set method overcomes the challenge of device types without specific services. Lastly, about $0.02\%$ of testing flows (\ie $268$ IPFIX records) contain $150$ new cloud services -- not seen in the training dataset; $97$ of them belong to Wansview camera. Such a dynamic in the  fingerprint of Wansview camera was observed in the training dataset (\S\ref{ch4:sec:cloud services}), but it does not affect my inference  -- we will soon see that this camera type is perfectly detected by its consistent use of specific cloud services like {\myverb{UDP/60722}}.
		\begin{figure}[!t]
		\centering
		\includegraphics[width=0.94\linewidth, height=0.80\linewidth]{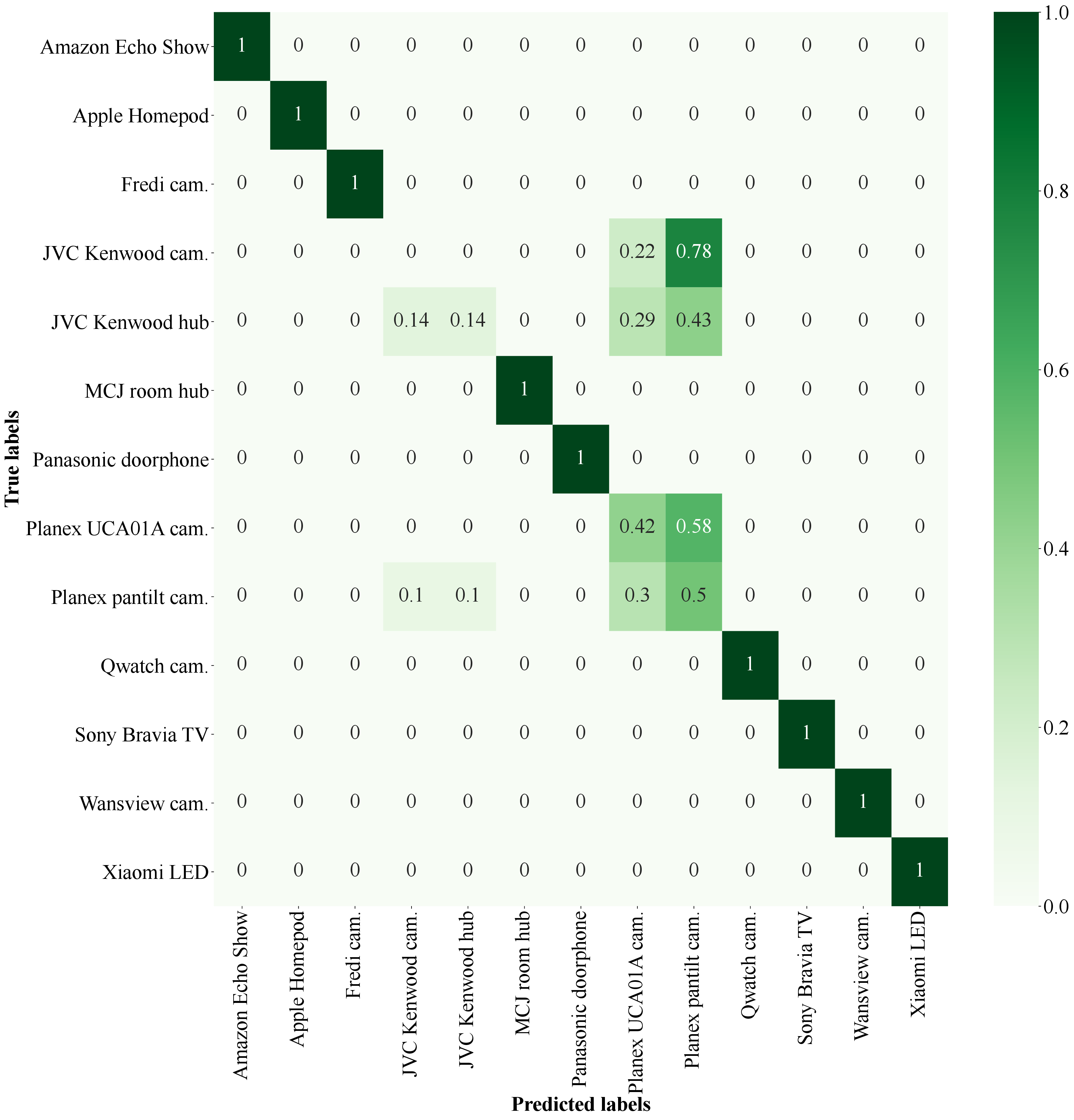}
		\caption{Confusion matrix of the deterministic model with the specific service inference method.}
		\label{ch4:fig:DET unique infer}
	\end{figure}
	
	\textbf{Evaluating the Specific Service Method}: As this method can only predict the device type of those flows that contain specific services, about $80\%$ of testing flows remain unclassified. Fig.~\ref{ch4:fig:DET unique infer} shows the confusion matrix of this method. It can be seen that the specific service method yields a perfect accuracy ($100\%$) for nine device classes and an overall average of  $77\%$ across all classes of devices that have specific services. Note that the Planex outdoor camera and PowerElec plug do not use any of their specific services (found in the training dataset) during the testing phase; therefore, these two classes are excluded from the confusion matrix. Also, we observe that devices like the JVC Kenwood camera and hub started using cloud services that were initially specific to Planex cameras (UCA01A and pantilt), and hence noticeable misclassification rates for them. For instance, during the training phase, JVC Kenwood camera was found to use 11 specific services, but none of them were used during the testing phase. Instead, it communicated with 2 and 7 services specific to Planex UCA01A and pantilt cameras, respectively -- this may be attributed to a firmware upgrade. This issue will be solved to some extent when I combine the two methods, namely specific service and fingerprint set.

	\textbf{Evaluating the Fingerprint Set Method}: This method is able to cover some of the shortcomings of the previous method, particularly in predicting device types that only use shared services. For this method to work, I need to configure a fingerprint threshold on the fraction of observed cloud services to determine whether a corresponding device type is present on a given home network. Note that the fingerprint set method operates on a collection of records observed during an epoch (chosen daily, ensuring a quality inference). I, therefore, demonstrate the efficacy of this method across epochs during the testing phase.
	
	\begin{figure}[!t]
		\centering
		\includegraphics[width=\linewidth]{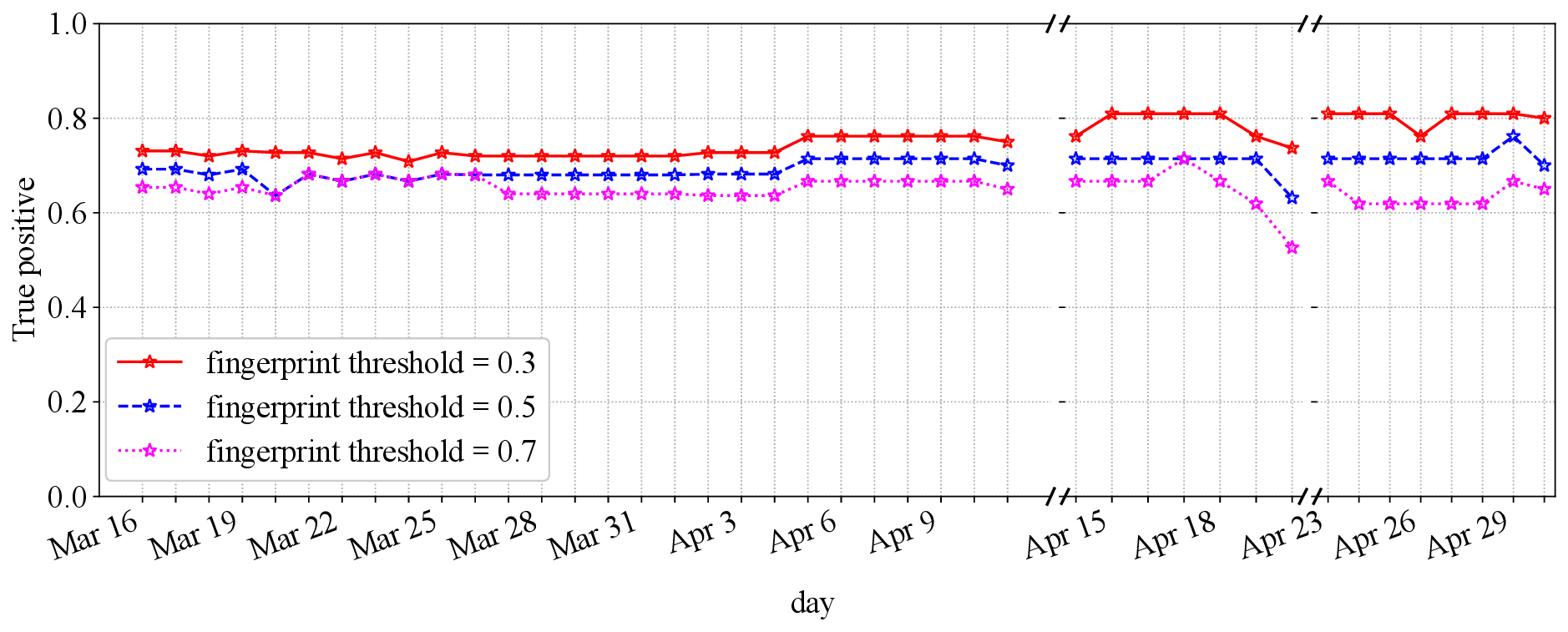}
		\caption[Daily trace of true positives from the deterministic model based on unweighted fingerprint set method]{Daily trace of true positives from the deterministic model based on \underline{unweighted} fingerprint set method.}
		\label{ch4:fig:DET obs threshold}
	\end{figure}

	Fig.~\ref{ch4:fig:DET obs threshold} shows a daily trace of true positives (\ie ratio of correctly detected active devices to all active devices) obtained from my deterministic model (with the fingerprint set method in unweighted mode) during the testing period.
	For the fingerprint threshold, I tested three values of 0.3, 0.5, and 0.7 that respectively detect $75\%$, $70\%$, and $65\%$ of active devices on average.
	Conservative operators may choose a higher value for the fingerprint threshold, providing them with a quality prediction at the cost of missing some devices types with weaker (shared) fingerprints. Setting the threshold value too low can be detrimental to the quality of inferencing, resulting in a high rate of false positives.

	\begin{figure}[!t]
		\centering
		\includegraphics[width=0.95\linewidth,height=0.63\linewidth]{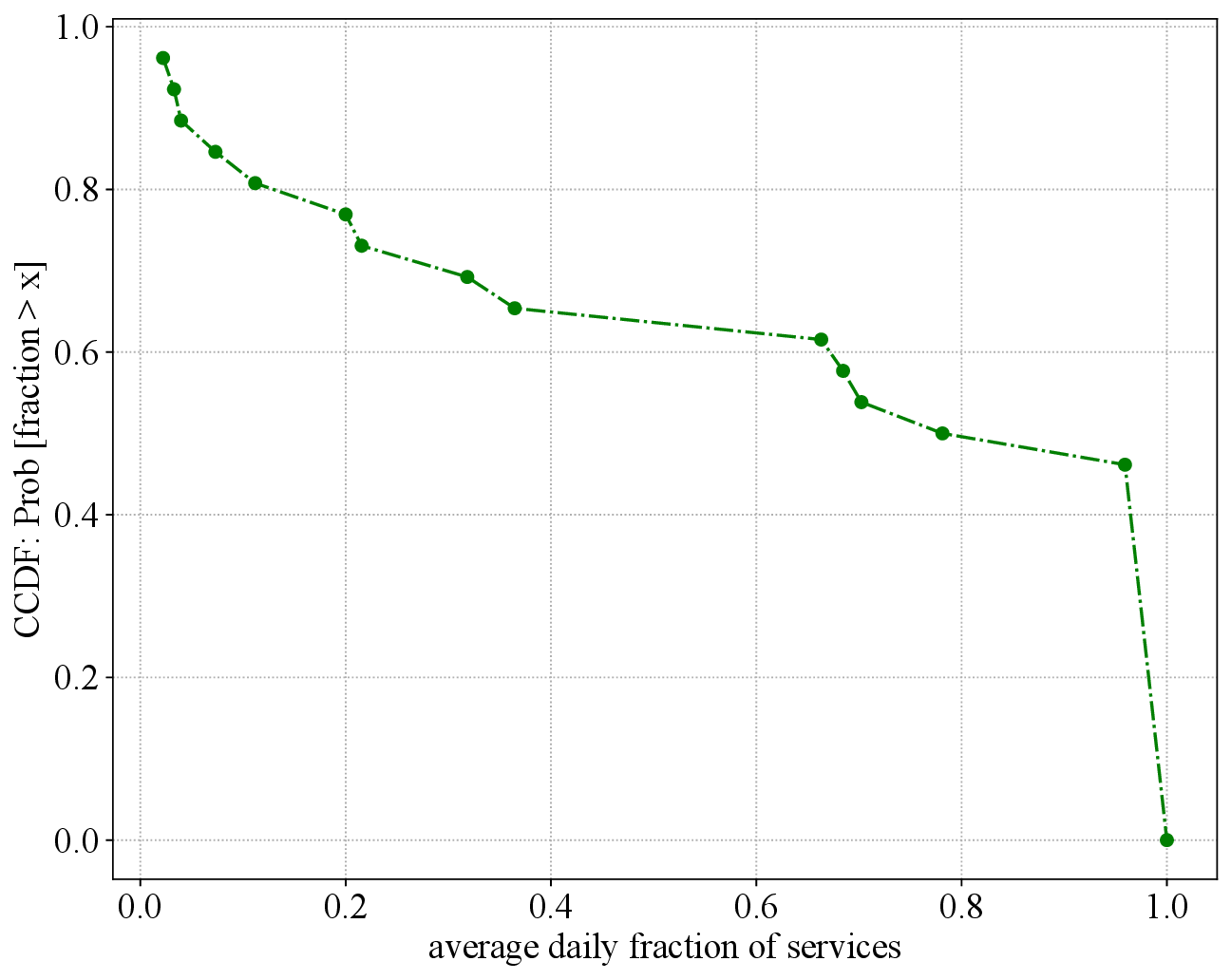}
		\caption{CCDF of average daily faction of services per IoT device during the testing period.}
		\label{ch4:fig:DET obs fraction CCDF}
	\end{figure}

	I found that the unweighted fingerprint set method can solve some of the shortcomings of the specific service method. For example, devices like the Bitfinder sensor, Sesame AP, and iRobot roomba which do not have any specific service, are correctly predicted on those days they are active.
	However, the detection rate at best reaches about $80\%$ during the entire testing period. To better explain the root cause of such a performance, I measured the daily observed fraction of cloud services for each device -- the fraction becomes equal to 1 if a subject device consumes all services from its fingerprint. I plot in Fig.~\ref{ch4:fig:DET obs fraction CCDF} the CCDF of the average daily fraction of services (per IoT device) during the testing period.
	I found that the measure of daily fraction for five (out of 26) devices, including the JVC Kenwood camera, JVC Kenwood hub, Planex UCA01A camera, Planex pantilt camera, and Wansview camera never exceeded 0.2, and hence are consistently missed by the model in the unweighted mode. Interestingly, twelve devices receive an average fraction equal to 1 -- they always get correctly predicted.

\begin{figure}[!t]
	\centering
	\includegraphics[width=\linewidth]{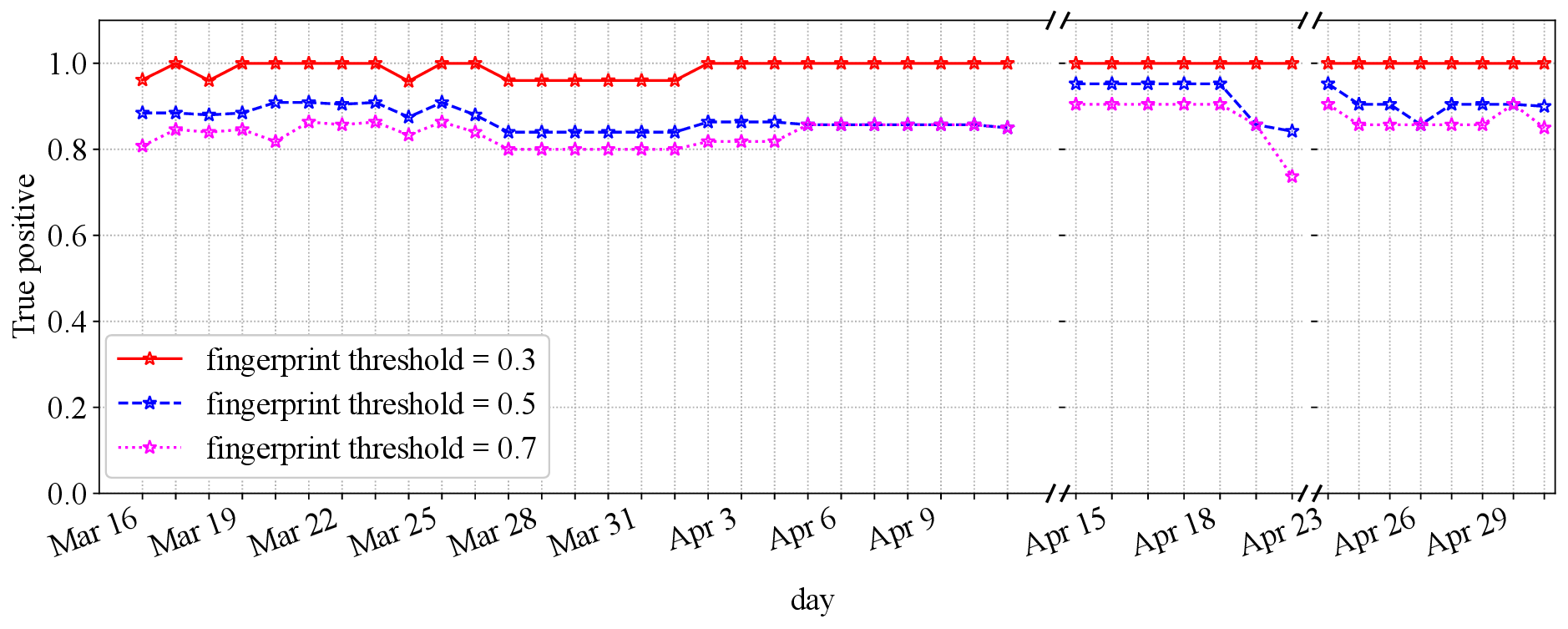}
	\caption[Daily trace of true positives from the deterministic model based on weighted fingerprint set method.]{Daily trace of true positives from the deterministic model based on \underline{weighted} fingerprint set method.}
	\label{ch4:fig:DET weighted obs threshold}
	\end{figure}

	The unweighted mode turns out to be discriminating certain device types that have shared cloud services in their fingerprint since this method misses an important factor of the usage frequency. 
	For example, a device type like Apple Homepod tends to use the generic service {\myverb{TCP/80}} every day, but the specific service {\myverb{UDP/3488}} only once or twice a week. 
	Assigning a corresponding frequency weight to each fingerprint service can help the model better infer connected devices. 
	Fig.~\ref{ch4:fig:DET weighted obs threshold} shows the daily trace of true positives obtained from the weighted method, given the same three fingerprint thresholds. It can be seen that giving more weight to frequent services can indeed improve the rate of true positives -- on average, the detection rate is increased by $24$\%, $19$\%, and $20$\% respectively for threshold values of 0.3, 0.5, and 0.7. Specifically, devices like the JVC Kenwood camera and hub, with a relatively poor score ($<0.2$) of the fingerprint match from the unweighted method, now receive about $0.64$ and $0.79$ match from the weighted method, and hence correctly predicted. Needless to say, JVC Kenwood camera consumed none of its specific cloud services, and JVC Kenwood hub contacted one of its specific services during the testing period.
	
	\begin{figure}[!t]
		\centering
		\includegraphics[width=\linewidth]{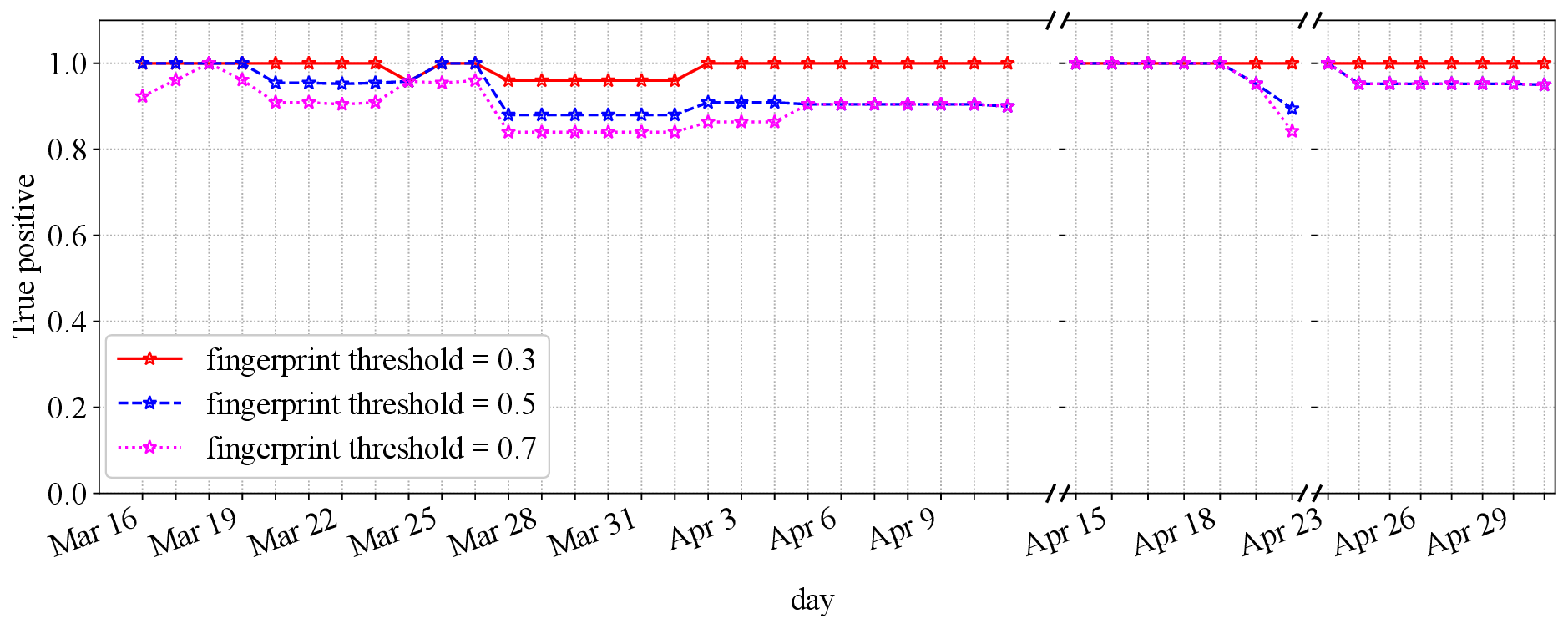}
		\caption[Daily trace of true positives for the deterministic model based on the combination of specific service and weighted fingerprint set methods]{Daily trace of true positives for the deterministic model based on the combination of \underline{specific service} and  \underline{weighted fingerprint set} methods.}
		\label{ch4:fig:DET combined obs threshold}
	\end{figure}

	\textbf{Specific Service and Fingerprint Set Methods Combined}: We have seen that both methods display certain limitations for some device types or in some circumstances. The specific service method offers a narrow coverage by excluding shared cloud services. On the other hand, the (weighted) fingerprint set method can fall short in predicting a device, which does not consume a sufficient number of its expected cloud services during the inference epoch.
	To benefit from their respective differentiated advantages, I combine these two methods. 
	Note that each operates differently (\ie the specific service method infers on a per IPFIX record basis, while the fingerprint set method infers from a collection of IPFIX records over an epoch). Therefore, we can combine their inferences at the end of each epoch by applying the union operation to the lists of predicted devices obtained from individual methods. That way, the chance of missing a connected IoT device will be minimized. 
	I show in Fig.~\ref{ch4:fig:DET combined obs threshold} the daily trace of true positives from the deterministic model when the two methods (specific service and weighted fingerprint set) are combined. It can be seen that the combined method further improves (compared to what was obtained in Fig.~\ref{ch4:fig:DET weighted obs threshold}) the quality of inference, yielding an average true positive of $99$\%, $94$\%, and $92$\% for the three thresholds of 0.3, 0.5, and 0.7, respectively. The combined model helps detect devices like Planex pantilt camera which had a relatively low score ($\approx$$0.39$ on average)  of the fingerprint match with the weighted method only.
	Importantly, the combined method reduces the average rate of false positives to relatively low values (\ie $10$\%, $9$\% and $4$\%) for the chosen thresholds (0.3, 0.5, and 0.7, respectively).
	
	
	\begin{table}[!t]
	\caption{Potential advantages of the deterministic model, augmenting the stochastic model.}
	\label{ch4:tab:ML DET}
	\renewcommand{\arraystretch}{1.2}
	\begin{adjustbox}{width=0.95\textwidth}
		\begin{tabular}{|l|cc|cc|}
			\hline
			\multicolumn{1}{|c|}{\multirow{2}{*}{{\ul \textbf{}}}} & \multicolumn{2}{l|}{\cellcolor{green!25}\textbf{Stochastic accepted}}                               & \multicolumn{2}{l|}{\cellcolor{red!25}\textbf{Stochastic discarded}}          \\ \cline{2-5} 
			\multicolumn{1}{|c|}{}                                 & \multicolumn{1}{c|}{\textit{correct}} & \multicolumn{1}{l|}{\textit{incorrect}} & \multicolumn{1}{c|}{\textit{correct}} & \textit{incorrect} \\ \hline
			Total flows                                            & \multicolumn{1}{c|}{870K}             & 8K                                      & \multicolumn{1}{c|}{370K}             & 170K               \\ \hline\hline
			{\cellcolor{cyan!25}specific-service flows {[}outgoing{]}}                  & \multicolumn{1}{c|}{223K}             & 3                                       & \multicolumn{1}{c|}{10K}              & 645                \\ \hline
			{\cellcolor{yellow!25}unseen-service flows {[}outgoing{]}}                    & \multicolumn{1}{c|}{155}              & 1                                       & \multicolumn{1}{c|}{74}               & 38                 \\ \hline
		\end{tabular}
	\end{adjustbox}	
\end{table}
	
	\vspace{-2mm}
	\subsection{Combining Stochastic and Deterministic Models}
	\label{ch4:sec:ml+dt}
	\vspace{-3mm}

	We saw that the stochastic model gives a reasonable accuracy during the testing period ($96$\% on average per class). However, it discards about $26$\% of the correctly classified (testing) flows due to a lack of confidence. Given its differentiated capabilities, the deterministic model has the potential to augment the stochastic model and improve the quality of inference. Table~\ref{ch4:tab:ML DET} highlights, at a high level, the potential advantage of the deterministic model that can help the stochastic model. The first row summarizes how the stochastic model correctly/incorrectly classifies all flows (incoming and outgoing); some are accepted while others get discarded. The second row (highlighted in blue) shows a potential benefit of the deterministic model: (i) 223K (a quarter of) correctly classified and accepted flows can be verified as they use cloud services specific to certain types of IoT devices; (ii) 10K ($\approx$3\%) of correctly classified but discarded flows can be ``revived''. Note that (as discussed in \S\ref{ch4:sec:cloud services}) the deterministic model infers from outgoing flows ($\approx$$85$\% of all flows), those that originate from the home network. The third row  (highlighted in yellow) shows some challenges that the deterministic model encountered with unseen services, but these types of flows were fairly negligible (less than $0.02$\%) in our dataset.
	
\begin{figure*}[!t]
		\centering
		\includegraphics[width=0.87\linewidth,]{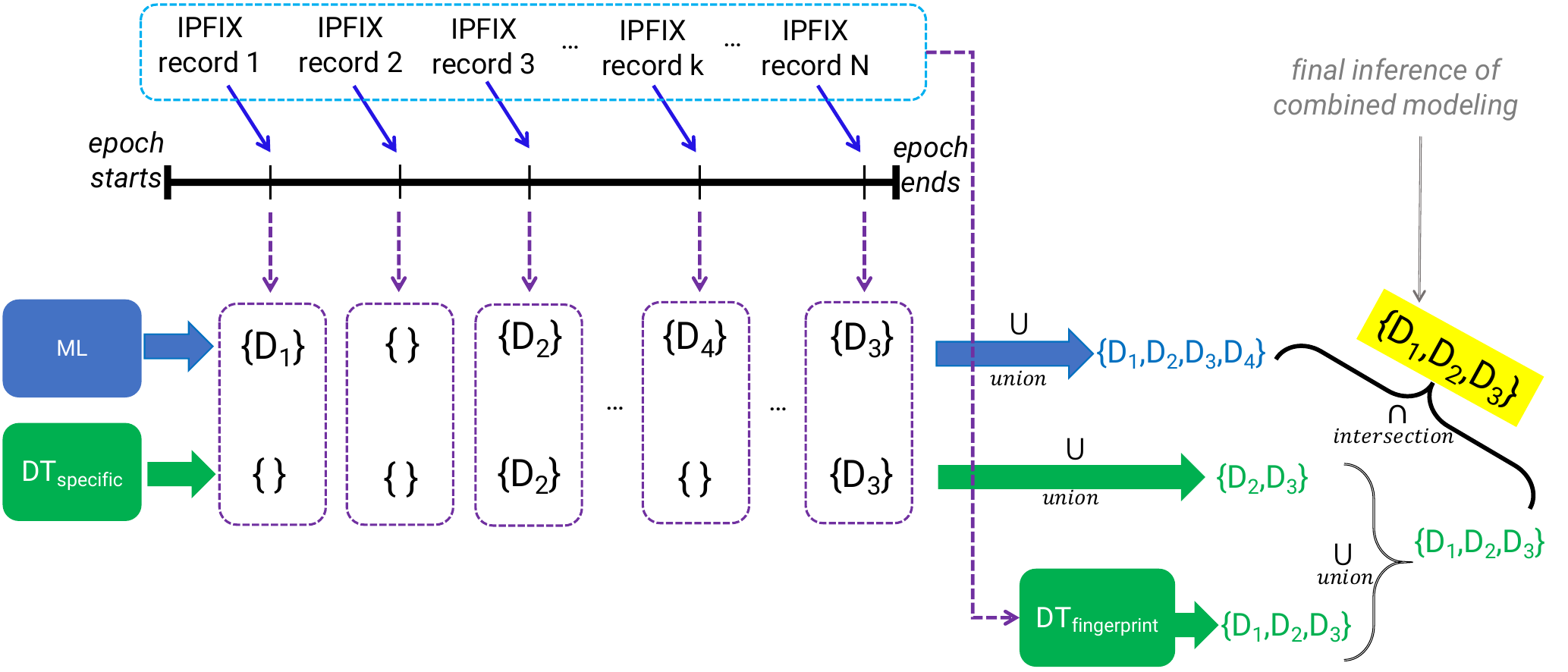}
		\caption{High-level architecture of a combined inference by stochastic and deterministic models.}
		\vspace{-2mm}
		\label{ch4:fig:ML DET combined}
\end{figure*}

	\vspace{-2mm}
	Fig.~\ref{ch4:fig:ML DET combined} shows the architecture of my combined inference from stochastic (\ie $ML$) and deterministic (\ie $DT$) models. I employ both specific service ($DT_{specific}$) and weighted fingerprint set ($DT_{fingerprint}$) methods for my deterministic inference. During a given epoch, the $ML$ and $DT_{specific}$ models work in parallel on a per-flow basis. Due to low confidence and/or an unseen/shared cloud service, some flows may result in no inference (\ie yielding an empty set) by either or both of the models. At the end of the epoch, the predicted set of devices is obtained by applying the union operation to their individual predictions during the epoch. 
	Additionally, the third model ($DT_{fingerprint}$) gives its prediction on all flows observed during the epoch.
	Next, I obtain a superset of detected devices by computing the union of outputs from the two deterministic models. Lastly, I intersect the two predicted sets yielded by the stochastic and deterministic models, obtaining the final inference of connected IoT devices.
	
	
	\begin{figure}[!t]
		\centering
		\includegraphics[width=\linewidth,]{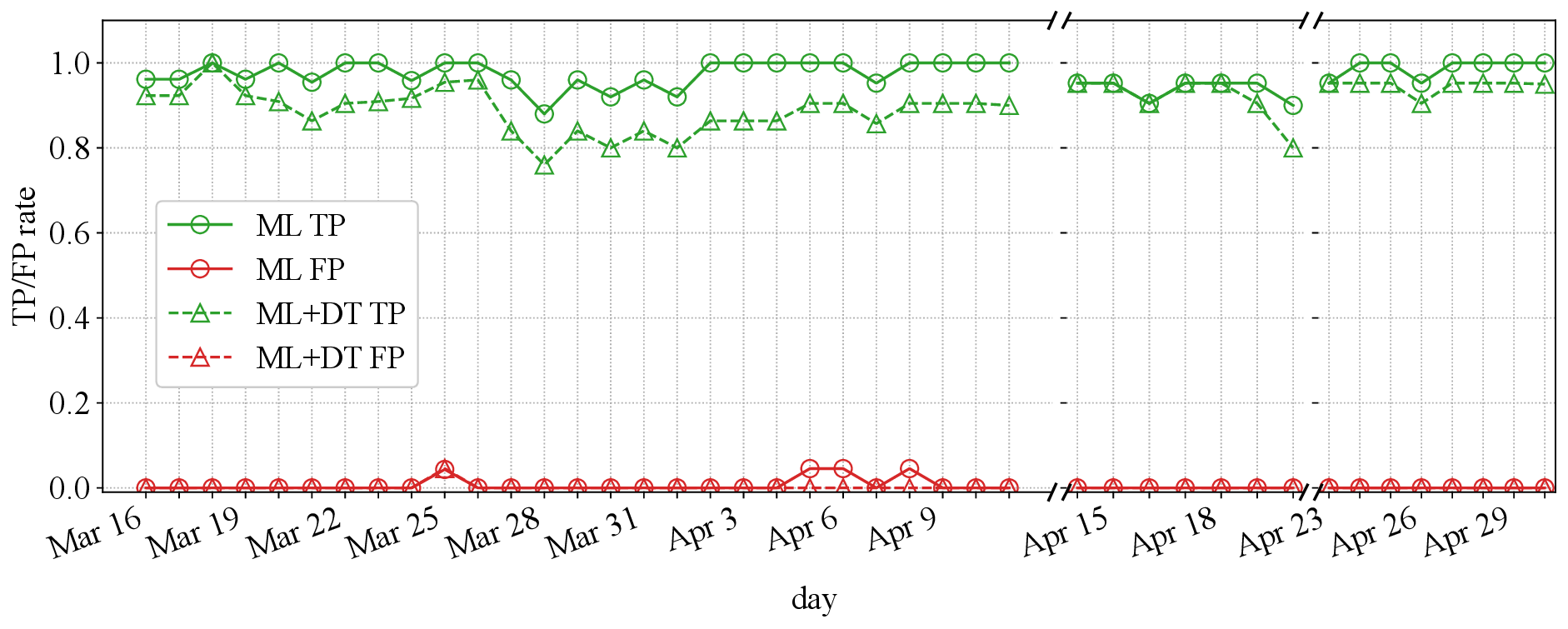}
		\caption{Prediction performance of stochastic modeling versus combined stochastic and deterministic modeling.}
		\label{ch4:fig:ML DET eval}
	\end{figure}
	
	\vspace{-2mm}
	Fig.~\ref{ch4:fig:ML DET eval} illustrates how the combined inference impacts the performance of prediction by two metrics, namely rates of true positive (TP) and false positive (FP). 
	TP indicates the correctly detected fraction of all devices that were active on the network, and FP indicates the incorrectly detected fraction of all active devices on the network.
	For this experiment, a comparison between two scenarios of stochastic inference ($ML$) versus the combined inference ($ML+DT$) is conducted.
	Also, I set the fingerprint threshold to 0.7 (a conservative setting) for the $DT_{fingerprint}$ model. We can see that the ML model is able to detect more devices on its own (with an average TP of $97$\%) compared to the combined scenario, given the conservative but confident nature of my deterministic model, a slightly lower TP ($90$\% on average) is achieved. Instead, the conservative deterministic modeling helped us mitigate false positives generated by the ML model during three days (April 5, 6, and 8) of evaluation.

	\begin{table}[!t]
		\centering
		\caption{Comparing performance of stochastic and deterministic baselines with that of their combination.}
		\label{ch4:tab:ML DET eval}
		\renewcommand{\arraystretch}{1.2}
		\begin{adjustbox}{width=0.75\textwidth}
			\begin{tabular}{|c|c|c|c|}
				\hline
				& \textbf{ML} (proxy of \cite{Meidan2020}) & \textbf{DT} (proxy of \cite{Saidi2020}) & \textbf{ML+DT} (this work) \\ \hline
				TP & $97$\% & $92$\% & $90$\% {\color{red} [$-7\%$]} \\ \hline
				FP & $0.4$\% & $4$\% & $0.1$\% {\color[rgb]{0.13, 0.54, 0.13} [$+75\%$]}\\ \hline
			\end{tabular}
		\end{adjustbox}
	\end{table}
	
	\vspace{-2mm}
	Table~\ref{ch4:tab:ML DET eval} shows true and false positive rates for ML only (representing prior work \cite{Meidan2020}), DT only (representing prior work \cite{Saidi2020}), and ML+DT combined models (this work). ML outperforms DT in terms of true positives by about $5$\%. Also, the false positive rate of ML is ten times better than that of DT ($0.4$\% compared to $4$\%). Higher false positives from the DT model are attributed to shared services explained in  \S\ref{ch4:sec:dt}. Even though DT is less accurate than ML when the two models are combined (ML+DT), false positives significantly decrease to an attractive  level of $0.1$\% -- this measure is 75\% better than that of ML only (highlighted in the green brackets). However, this improvement comes at the cost of losing about 7\% of true positives from ML (highlighted in the red brackets). 
	
\vspace{-2mm}
The combined inference may miss some devices (one out of ten), but the chance of incorrectly detecting a device not present on the network is fairly low. As discussed throughout this chapter, reducing false positives is crucial as they can incur unwanted management costs (taking actions on a vulnerable device that does not exist in the user premise). Thus, losing 7\% of true positives seems to be a reasonable trade-off for gaining 75\% fewer false positives.

\section{Conclusion}
Residential networks continue to become richer and more vulnerable with the widespread adoption of consumer IoT devices. ISPs recognize the need to address concerns associated with IoT security by obtaining visibility into these connected devices and their behavior. This chapter discussed how an ISP could leverage IPFIX telemetry to achieve this task post-NAT, at scale, without making changes to home networks. I analyzed about nine million IPFIX records (released as open data) from 26 IoT devices. I extracted 28 flow-level stochastic features for training a machine learning model to detect device types from IPFIX records with an average accuracy of 96\%. In addition to the stochastic model, I developed deterministic models based on cloud services captured from outgoing IPFIX records, yielding an average accuracy of 92\%. I demonstrated that combining stochastic and deterministic models can help reduce false-positive incidents by 75\% for an average cost of 7\% in true positives.
	
This chapter limited its focus to a single residential network, emulated using a testbed. It showed some of the challenges and possible solutions for an ISP aiming to detect IoT devices from their opaque flow data. However, in a real scenario, ISPs serve and manage thousands of residential networks where collecting labeled data and developing a separate inference model for each home is impractical. Therefore, there is a need for a scalable inference system to overcome this challenge which is the scope of the next chapter.

%% file: tex/Chapter5-Scalable-Infer/main.tex
\chapter{Scalable Network Flow Inference under Concept Drifts}\label{ch:scalable infer}


In the previous chapter, I explained that Internet service providers are best poised to play key roles in mitigating risks by automatically inferring active IoT devices per household and notifying users of vulnerable ones. I developed techniques for this purpose, focusing on a single home network. However, developing a scalable inference method that can perform robustly across thousands of home networks is a nontrivial task. This chapter focuses on the challenges of developing and applying data-driven inference models when labeled data of device behaviors is limited and the distribution of data changes across time and space domains (concept drifts). Contributions of this chapter are fourfold: (1) I analyze over six million network traffic flows of 24 types of consumer IoT devices from 12 real homes during six weeks to highlight the challenge of temporal and spatial concept drifts in network behavior of IoT devices; (2) I analyze the performance of two inference strategies, namely ``\textit{global inference}'' (a model trained on a combined set of all labeled data from training homes) and ``\textit{contextualized inference}'' (several models each trained on the labeled data from a training home) in the presence of concept drifts; (3) To manage concept drifts, I develop a method that dynamically applies the ``best'' model (from a set) to network traffic of unseen homes during the testing phase, yielding better performance in a fifth of scenarios when the labels are available for the testing data (ideal but unrealistic settings); and (4) I develop a method to automatically select the best model without needing labels of unseen data (a realistic inference) and show that it can achieve 94\% of the ideal model's accuracy.

\vspace{-2mm}
\section{Introduction}
\vspace{-2mm}
IoT devices are becoming popular in households. It is estimated that in 2021 there were about 258 million smart homes around the globe \cite{earthweb}. It is also anticipated that the smart home market will keep growing at least by the next two years \cite{secsales}. This means more IoT devices will be deployed in home networks by near future \cite{weforum} supplied by the overwhelming market of IoT device manufacturing that includes more than 1200 companies delivering software/hardware platforms for IoT devices \cite{iotbusiness}.

As explained in the previous chapter, IoT devices bring a number of vulnerabilities that make them attractive to cyber criminals. Weak (or default) passwords and open insecure service ports are considered the top two exploited vulnerabilities of IoT devices \cite{minim2020}. Now with the massive number of vulnerable IoT devices in households, they are becoming a risk challenge for ISPs \cite{bitdefender2020}. The first step in every risk assessment and security analysis is obtaining visibility by discovering IoT devices connected to the network and determining their type (\eg camera, TV, speaker) \cite{Palo2020} using  characteristics known a priori. In Chapter~\ref{ch:obscure infer}, I developed an inference technique by combining the outputs of stochastic and deterministic models for detecting IoT device types using obfuscated and coarse-grained IPFIX records.

{\vspace{5mm}}
However, network behavioral patterns of IoT devices may change \cite{Arunan2020b,Arunan2020a,TMA2021revistIoTclass} by the time and context of their use across various networks. The behavioral change is more pronounced when IoT traffic inference is the objective of an ISP tasked to serve and manage tens of thousands of home networks, each with a unique composition of assets and users, all distributed across sizable geography (city, state, or country). Several existing research works studied different methods that ISPs can leverage to detect IoT devices in residential networks \cite{LCN2021,Marchal19,Meidan2020,Miettinen2017,Thangavelu2019, ArmanIoTJ2022}.
Prior works (including the previous chapter) tend to train a global model with machine learning algorithms and fine-tune it by traffic data collected from a testbed (representing a single context). Due to the limitation of data or evaluation scenarios, they did not encounter context variations and could not highlight and/or address their impacts.
\vspace{7mm}

The challenge of data distribution change in data driven modeling is known as concept drift which has been studied in various contexts like image recognition \cite{Reis2018a}, insect species detection \cite{Reis2018b}, air quality detection \cite{Nasc2018}, and synthetic data \cite{Gomes2010, Ying2005}. Some of the existing methods for addressing concept drift \cite{Bifet2007, Bifet2009, Gama2004, Oza2005, Ying2005, Gama2009} require true labels during the testing phase to measure the error rate and re-train the model if the error rate increases significantly. This assumption is practically challenging for our context as obtaining labeled data for thousands of home networks in not a trivial task for an ISP. Therefore, a practical solution must be capable of addressing concept drift with a limited amount of labeled data from a few home networks for the training purpose.

Context-aware or contextualized modeling \cite{Reis2018a, Reis2018b,Nasc2018} is an alternative to global modeling where (1) a set of training contexts are identified; (2) models are trained on a per-context basis; and, (3) a given testing sample is predicted by the ``closest'' model selected from all available trained models. The definition of context is dependent on the application which in our case each home network can be considered as a context. Selecting the closest model is the key process where we can leverage classification scores of the models without requiring the true labels in the testing phase \cite{Reis2018a, Reis2018b} which makes this method more appealing for our problem.  Note that each contextualized model is trained by data collected from a single context which has narrow and relatively tight knowledge. In contrast, the global model captures data from multiple contexts which makes its knowledge broader and loose. Limiting a model to learn a single context may increase the chance of over-fitting, while exposing a model to a diversified data set may not necessarily result in better performance, especially when the data is noisy. Though contextualized modeling has been studied in other domains \cite{Reis2018a, Reis2018b,Nasc2018}, to the best of my knowledge, no relevant study is found in the area of IoT traffic inference.

\vspace{6mm}

In this chapter, I compare global and contextualized modeling for classifying IoT devices in home networks. Specifically, I aim to answer the following question: ``Given a labeled dataset (training) from $N$ homes
and an unlabeled dataset (testing) from other $M$ homes, which of the global versus contextualized modeling does yield better performance in classifying devices during the testing phase?'' My \textbf{first} contribution highlights the presence of concept drifts over time and space domains in IoT network traffic behavior by analyzing more than 6 million flow records (\S\ref{ch5:sec:difts}). For my \textbf{second} contribution, I develop global and contextualized models (aiming to manage concept drifts in the space domain) and compare their performance (\S\ref{ch5:sec:strategies}). For the \textbf{third} contribution, I demonstrate that a dynamic inference can be applied to a combination of global and contextualized models to address concept drifts in the time domain (\S\ref{ch5:sec:dynamic}). In the \textbf{fourth} contribution, I develop a selection strategy to select the closest model by using score distributions of contextualized models without using the true labels of the testing dataset.

\vspace{-3mm}
\section[Concept Drifts in Network Behaviors of IoT Devices]{Concept Drifts in Network Behaviors of \\IoT Devices}\label{ch5:sec:difts}
A prerequisite step for an ISP that aims at automatically detecting IoT devices in home networks is to collect labeled data on network traffic for an intended set of devices, generating one or more inference models. The ISP can encourage a \textit{limited} number of its subscribers to voluntarily (or by other incentives like discounting monthly bills) help with the labeling network traffic of their home. For example, software applications like ``IoT Inspector'' \cite{IoTInspector} can be installed on a computer (or home gateway) in each home and let users label the devices. Note that the data collection and labeling methods are beyond the scope of this chapter. Nevertheless, the more homes the ISP can collect labeled data from, its inference method becomes richer. However, collecting labeled data is expensive in both time and cost for the ISP, so this can only be done for a limited number of home networks. 

\begin{figure}[!t]
\centering
\includegraphics[width=0.93\linewidth]{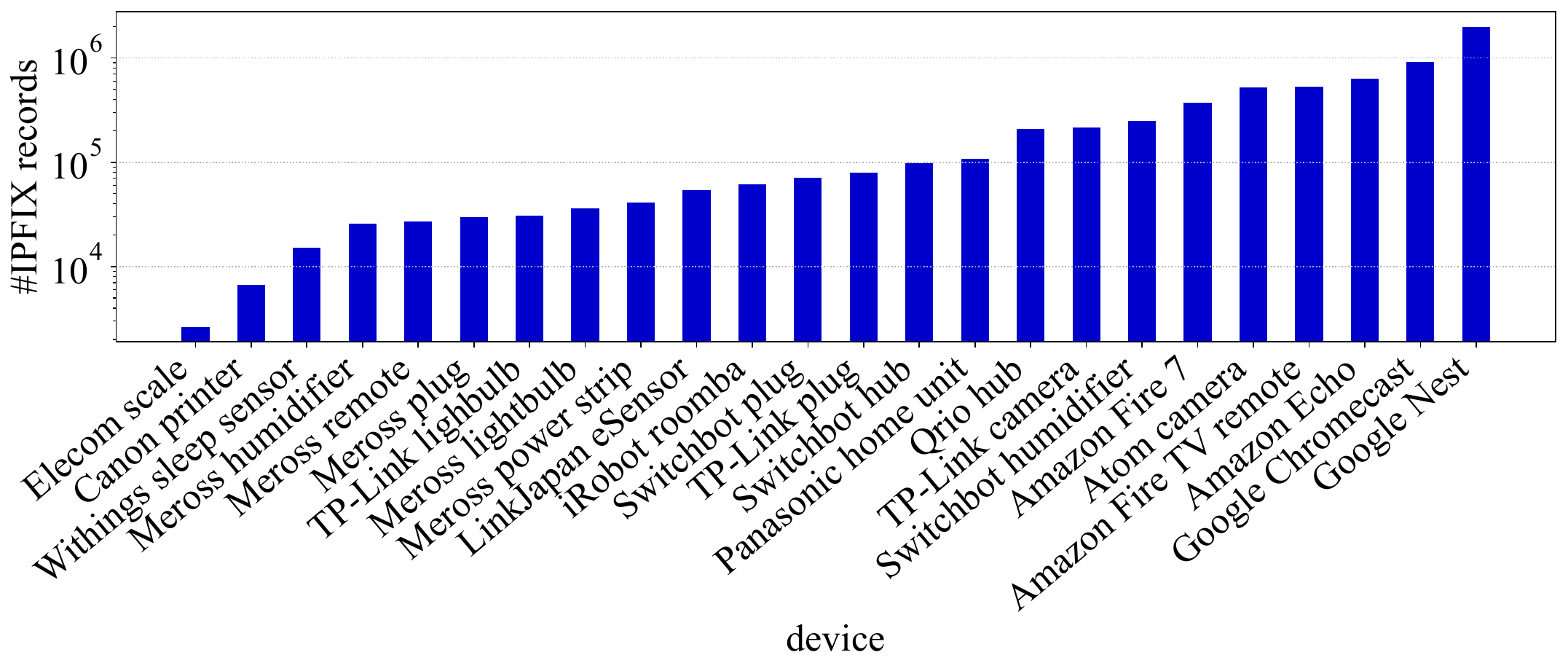}
\vspace{-3mm}
\caption{Number of IPFIX records per device type across 12 homes.}
\label{ch5:fig:records count}
\vspace{-4mm}
\end{figure}

Another approach for collecting the labeled data is that the ISP constructs a testbed in their lab (my previous work in \cite{LCN2021,ArmanIoTJ2022}), consisting of a handful of IoT devices (those vulnerable devices that they aim to detect), and collects labeled data from the testbed. The testbed approach is relatively easier, but the data quality may be synthetically higher than the data collected from real homes. Additionally, collecting traffic from multiple homes provides the ISP with multiple contexts of traffic data. In contrast, the testbed data only includes a single context that may be inadequate for thousands of new homes. In this chapter, I use data collected from real home networks. 

Concept drift is a known challenge of inference models that occurs when the distribution of data changes in the testing set from what it was in the training set. This challenge can be perceived differently in the time domain (when the model was trained on data collected some time ago) versus the space domain (when the model was trained on data of different homes with slight variations in context).

In the scope of this chapter (inferring IoT devices in home networks), we encounter both sources of concept drifts. The probability and intensity of concept drifts vary by how the ISP trains its inference models (composition of homes) and how often they get re-trained. Frequent re-training can make the models more tolerant to concept drifts in the time domain; however, the re-training process has its own practical challenges. Another point to consider is that the ISP may need to carefully select the training context (composition of homes) to cater to diversity, becoming relatively resilient to concept drifts in the space domain. 



\subsection{Data Collection}\label{ch5:sec:data}

For this chapter, the network traffic was collected\footnote{Traffic collection was outsourced to a third-party specialist contractor.} from 12 real homes. The households were selected from volunteer employees (who provided us with written consent) of the specialist contractor in Japan. The selected households have a diverse residential settings including single-person and family\footnote{No information about linking households to their corresponding dataset is available.}.
There is a set of 24 IoT device types (makes and models), including two cameras, four power switches, two humidifiers, an air sensor, two speakers, two media streamers, three hubs, two lightbulbs, a weighing scale, a tablet, a printer, a sleep sensor, a smart remote, and a vacuum cleaner. For each of these types, 12 units were procured, meaning a total of 288 IoT units.  Individual homes are given a unit of each device type. In other words, each of the 12 homes has its own set of 24 devices.  Data was collected from these homes for 47 days, of which I used the first 30 days for training and the remaining 17 days for testing.

To protect user privacy, precautionary actions, particularly for cameras (\eg ensuring the use of TLS-based encryption for their data transmission and placing them in less private locations of homes), have been taken. Also, the collected data excludes network traffic of other household devices (\eg personal computers, phones) by connecting a new WiFi access point (implemented with a Raspberry Pi) to the LAN interface of the existing home gateway that only serves the 24 experimental IoT devices (where other household devices are served by their existing gateway) and is the vantage point of data collection. Additionally, the process of traffic capture was configured by the MAC addresses (safelist) of the experimental IoT devices to avoid recording data of unintended devices.

For an ISP serving thousands of households, the choice of traffic data and the measurement location are nontrivial tasks mainly due to operational and scalability challenges. To address these challenges, similar to Chapter~\ref{ch:obscure infer}, I use post-NAT IPFIX records. That said, it is important to note that to evaluate the efficacy of my inference methods, raw packets (in PCAP format) were collected from inside home networks to obtain true device labels. The PCAP files were then transformed into post-NAT IPFIX records \cite{ipfixRFC5103} (emulating a real scenario) and use them for training and testing models.

The entire dataset consists of a total of 6,305,626 IPFIX records which is shown in Fig.~\ref{ch5:fig:records count} for each device type.
Each IPFIX record corresponds to a five-tuple flow which is distinguished with source/destination IP address, source/destination port number, and IP protocol number. IPFIX records are bidirectional which means they include activity features corresponding to both directions of a flow \eg incoming packet count and outgoing packet count. In this chapter, I extract the same set of activity (Table~\ref{ch4:tab:features}) and binary features (Table~\ref{ch4:tab:ports}) that I used in Chapter~\ref{ch:obscure infer}.

\begin{figure}[t!]
\centering
\includegraphics[width=0.95\linewidth]{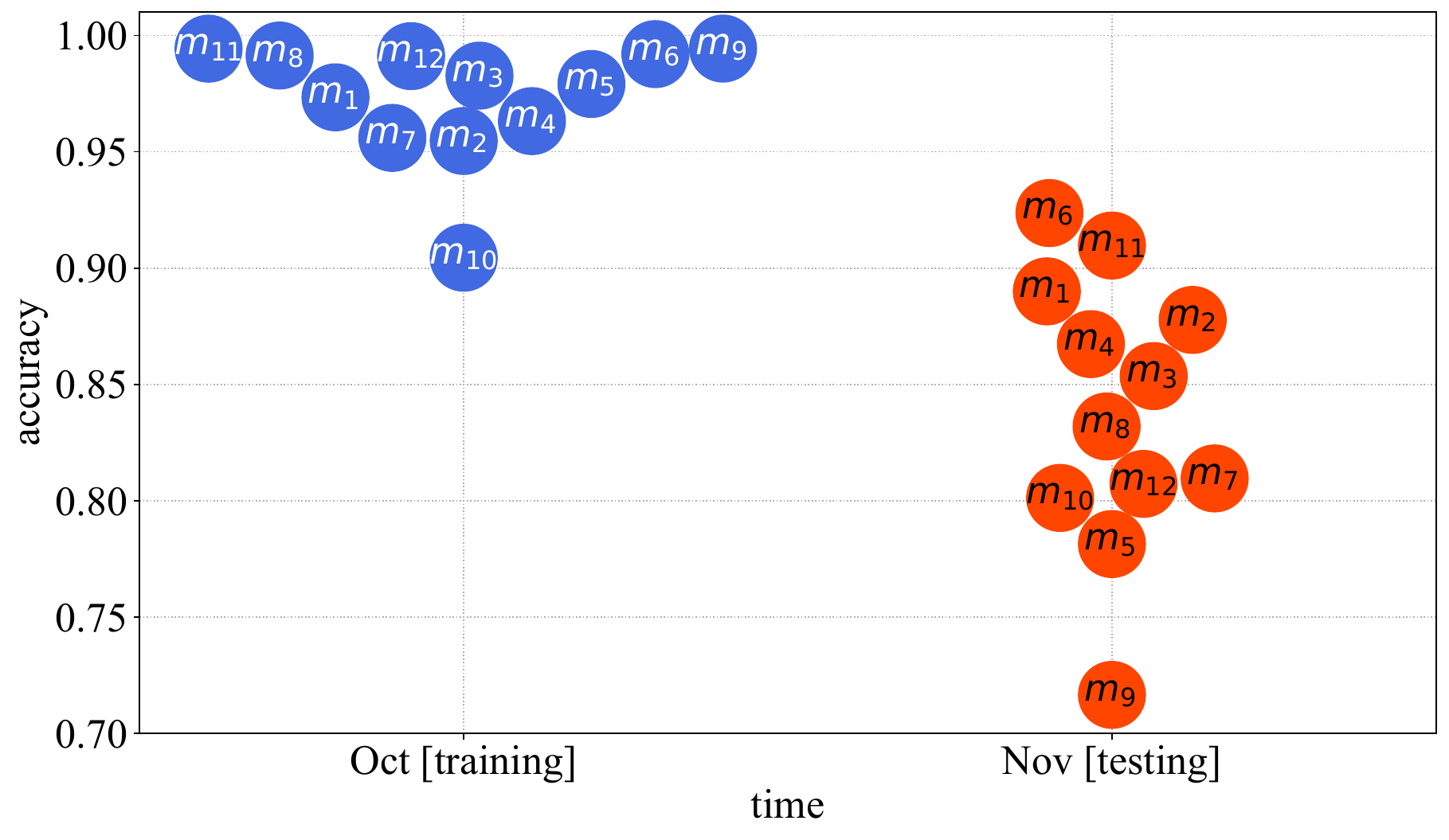}
\vspace{-3mm}
\caption{Temporal drifts in IoT network behaviors lead to performance decay across models (one per each home) from training to testing.}
\label{ch5:fig:temporal drift}
\end{figure}

In this chapter, I use multi-class Random Forest to develop the inference models as it has proven effective in network traffic inferencing \cite{zhao21}. Model inputs are IPFIX features, and outputs are a device type (from 24 classes) along with a classification score. I note that the inference is not bound to any specific algorithm, so one may choose to use other methods like neural networks for this purpose. As IPFIX records are coarse-grained, it is not unlikely for the model to be less confident in its prediction. To increase the prediction quality, similar to what I did in \S\ref{ch4:sec:ct}, I apply class-specific thresholds to the score given by the model, thereby accepting predictions accompanied by relatively high scores. I obtain class-specific thresholds during the training phase by taking the average score for correctly predicted training instances per class.

\subsection{Temporal Drifts in IoT Behaviors}\label{ch5:subsec:timedrift}

Let us begin with how the behavior of IoT devices changes over time. I train a Random Forest classifier for each home using its corresponding training dataset, obtaining 12 inference models \ie one model per home ($m_i: i\in[1,12]$). I use 10-fold cross validation technique to obtain average accuracy.

I apply these 12 models to their corresponding training and testing data to track their performance (average prediction accuracy across 24 classes). Fig.~\ref{ch5:fig:temporal drift} shows the accuracy (a value between 0 and 1) of models on average drops from training (shown in blue circles) to testing (red circles).
All models realize a lower accuracy in the testing phase than in the training phase. Temporal drifts  deteriorated the performance of models by about $13$\% on average. The smallest and largest gaps are seen for $m_6$ and $m_9$, where their testing accuracy drops by more than $6$\% and $27$\%, respectively.

\begin{figure}[t!]
\centering
\includegraphics[width=0.95\linewidth]{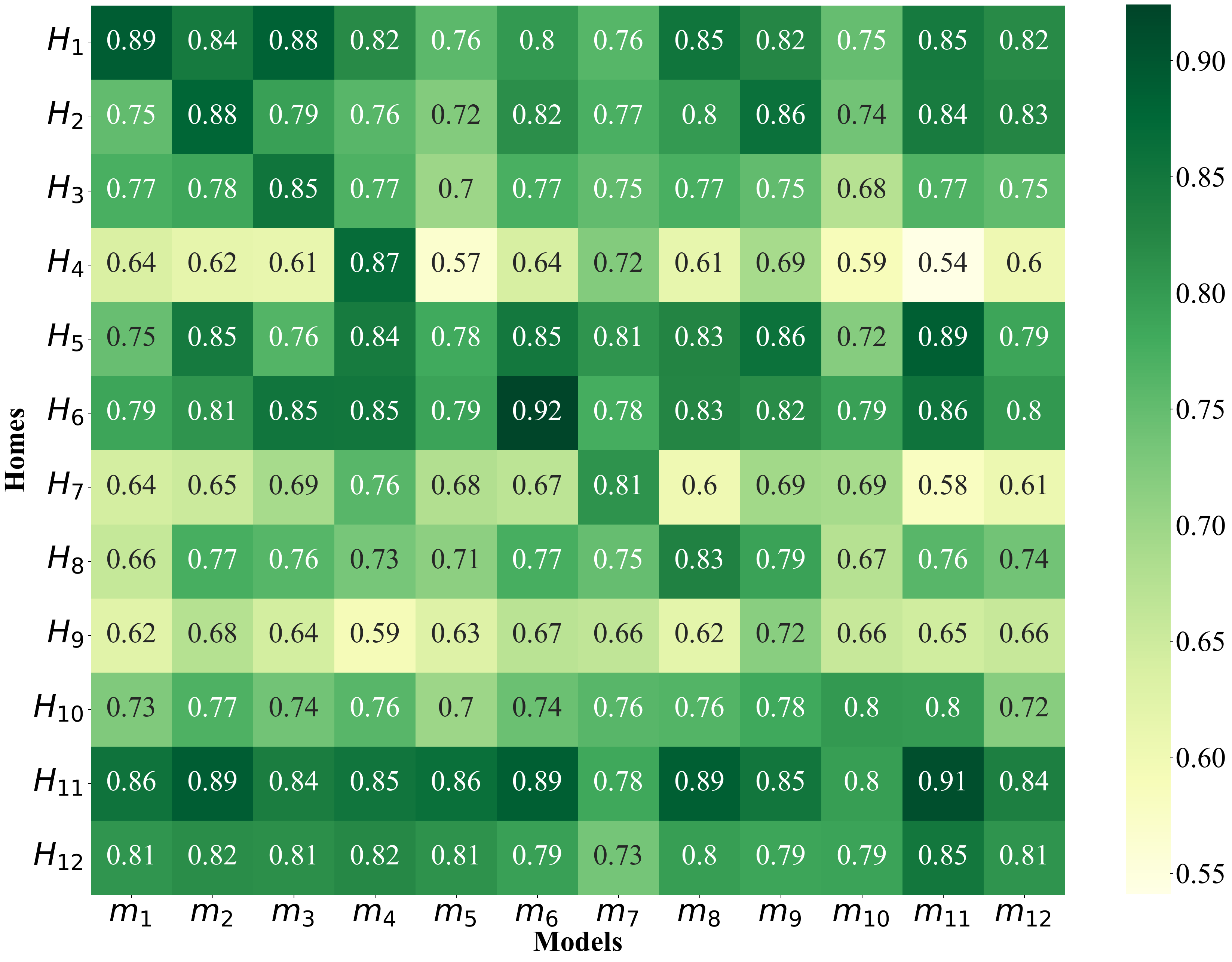}
\vspace{-3mm}
\caption{Spatial drifts in IoT network behaviors deteriorate model performance when tested against data of other contexts.}
\label{ch5:fig:spatial drift}
\end{figure}

\subsection{Spatial Drifts in IoT Behaviors}

The second form of drift is when models learn and infer across the spatial domain.
To analyze this scenario, I apply the models ($m_i: i\in[1,12]$) trained in \S\ref{ch5:subsec:timedrift} to the testing set of all homes ($H_i: i\in[1,12]$). Fig.~\ref{ch5:fig:spatial drift} shows the results of this experiment. Each cell is the average prediction accuracy of models (listed across columns) when tested against data of homes (listed across rows). 
\vspace{6mm}

Overall, models perform better when they apply to their corresponding context. We observe that diagonal cells are often among the highest in each row, with some exceptions. For example, in $H_{12}$, its own model $m_{12}$ is beaten by $m_{11}$, giving an accuracy of $0.81$ versus $0.85$. Such an unexpected pattern is more pronounced in $H5$, where the performance of $m_5$ is lower than that of eight other models.   

It can be seen that certain homes like $H_{11}$ and $H_{12}$ receive relatively consistent predictions (fairly green cells) from all models. Conversely, in homes like $H_4$ and $H_7$ inconsistent performance across models is evident (yellow versus green cells) -- spatial drifts deteriorate the performance more. Considering $H_4$, for example, the accuracy of its own model ($m_4$) is $0.87$, whereas all other models at best give an accuracy of $0.72$ -- a non-negligible gap of $0.15$. 

\textbf{Summary:} Concept drifts are unavoidable, particularly at the scale of an ISP with only a limited amount of labeled data available from a diverse set of homes. In the following sections, I will develop strategies and methods to manage the impact of concept drifts on the IoT traffic inference.

\section{IoT Traffic Inference Strategies}\label{ch5:sec:strategies}

\begin{figure}[t!]
	\centering
	\includegraphics[width=0.975\linewidth]{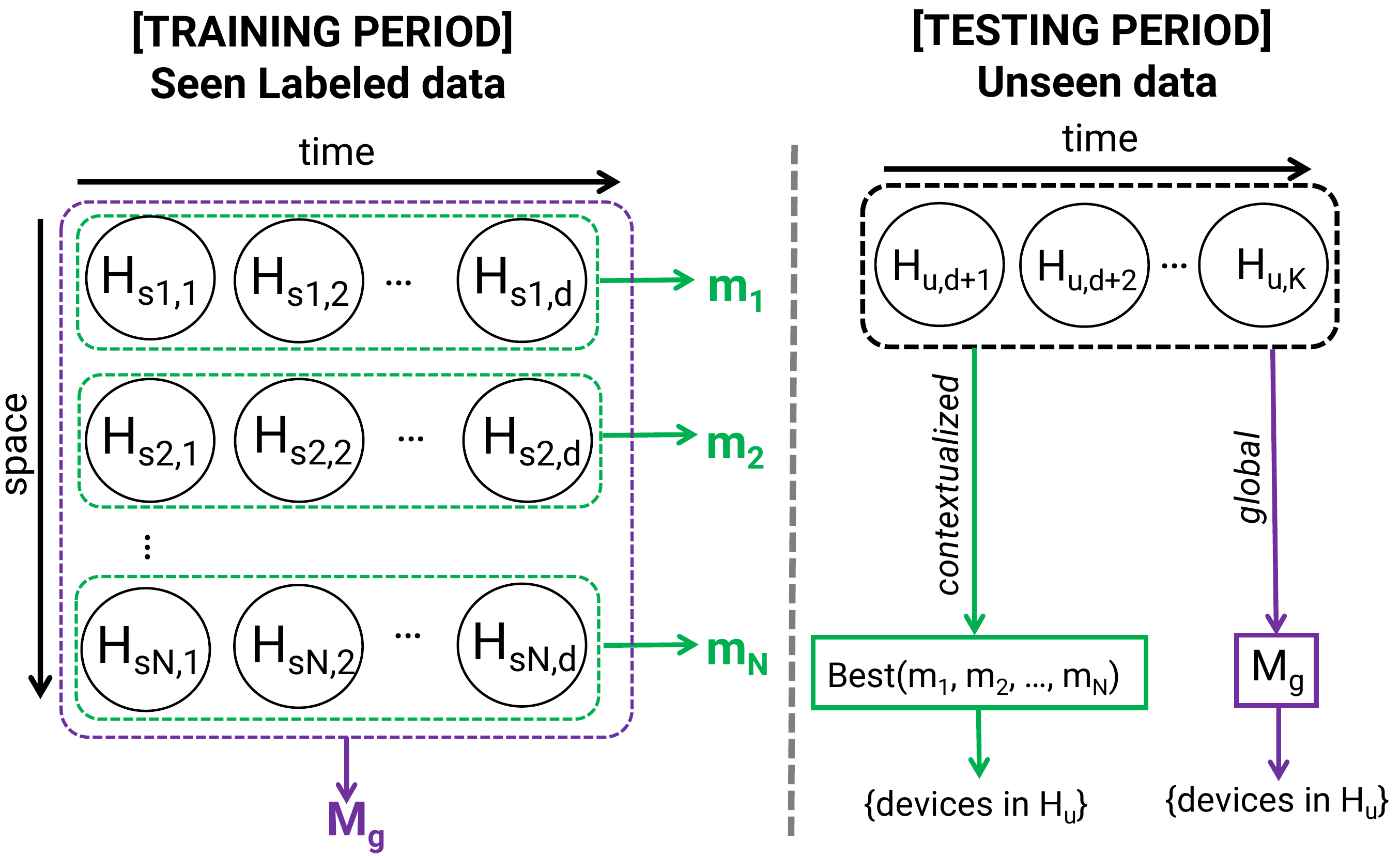}
	\caption{Global versus contextualized modeling for detecting IoT devices in home networks.}
	\label{ch5:fig:architecture}
\end{figure}
\vspace{6mm}

With high-level observations made in the previous section, let us assume labeled data is available from some seen contexts (say, $N$ homes). One may employ different strategies to train models and apply them to unseen contexts in practice. Fig.~\ref{ch5:fig:architecture} illustrates two representative strategies, highlighted by purple (global) and green (contextualized) colors. 
Suppose we have $d$ days worth of labeled data from $N$ homes, constructing our training set (shown in the left section of Fig.~\ref{ch5:fig:architecture}). Therefore, $H_{sN,d}$ denotes the unit of dataset collected from the seen home $N$ on day $d$. 

The most straightforward strategy (baseline) is that we combine all labeled data from $N$ homes and train a classifier (shown as purple $M_g$) based on the combined dataset. I call this global modeling throughout this chapter. Alternatively, one may go a step back and, instead of combing all data, trains a separate model per home (shown as green $m_1,..., m_N$). I call this method contextualized modeling. When it comes to the testing phase (detecting a set of IoT devices in an unseen home $H_u$), shown in the right section of Fig.~\ref{ch5:fig:architecture}, $M_g$ is readily applied to the daily data of $H_{u,K}$, giving prediction. For contextualized modeling, instead, an additional computation is required. It needs to ``select the best'' model from $N$ available models, before applying it to data of an unseen home.
In \S\ref{ch5:sec:model selection}, I develop a strategy for selecting the best model in the absence of labeled data in the testing period; but before that, the primary objective is to compare the efficacy of global versus contextualized modeling, assuming the best model can be selected (either automatically or ``given''). For now, I leverage ground truth labels (provided by an oracle\footnote{Note that this would be infeasible in real practice. I will develop a practical method in \S\ref{ch5:sec:model selection}.}) available for all contexts (seen and unseen) during the training and testing phases of my experiments. In other words, I choose the model that yields the highest accuracy (given ground truth) for unseen homes. 

\vspace{-3mm}
These two strategies come with key differences: (1) Suppose a labeled dataset becomes available from an unseen (or even a seen) home. Global modeling requires to re-train $M_g$ on past data combined with new data. Contextualized modeling, instead, trains an isolated model specific to a newly added/updated home, which is relatively faster and less expensive computationally; and (2) Although this chapter selects the best model by leveraging ground truth labels, a practical approach like what I will explain in \S\ref{ch5:sec:dynamic} may not always be able to select the best model from a set of available models (hence, affecting the inference performance).

\vspace{-4mm}
\textbf{Evaluating Inference Strategies:}
For the evaluation, I assume labeled data of five\footnote{Less than five seen homes result in an insufficient amount of data.} homes is available for training (seen), and data of the remaining seven homes is used for testing (unseen). To avoid creating bias, I ran the experiments ten times, and I randomly selected five training homes for each run. 
Let us call the first month (of the entire 47 days) training period. Note that I train models on data of seen homes during the training period. I train a global model ($M_g$) and five contextualized models ($m_i$) for the chosen seen homes, as shown in Fig.~\ref{ch5:fig:architecture}. 

It is important to note that contextualized modeling requires the best model assigned to an unseen home before the testing phase. Given an unseen home, the best model (one of five $m_i$'s in the evaluation) is selected and assigned based on the highest accuracy obtained by applying $m_i$'s to the labeled data of unseen homes (which is assumed to be available in this chapter).  
It is possible that the best model selected for an unseen home may not necessarily perform the best during the testing period. In fact, in nine runs (out of ten), I found at least one unseen home where the ideal model for its training period differs from that of the testing period due to temporal concept drifts discussed in \S\ref{ch5:subsec:timedrift}.

\begin{table}[t!]
	\caption{Performance of global versus contextualized modeling.}
	\renewcommand{\arraystretch}{1.6}
	\begin{adjustbox}{width=0.90\textwidth}
		\begin{tabular}{|c|c|c|c|c|}
			\hline
			Run & $M_g$ & $Best(\{m_i\})$ & $Best(M_g,\{m_i\})$ & $Best(M_g,\{m_i\})^d$\\ \hline
			\ciaoo{1} & $0.770$ &	$0.748$  & $0.769$ & $0.782$\\ \hline
			\ciaoo{2} & $0.827$ & $0.805$ & $0.834$ & $0.843$ \\ \hline
			\ciaoo{3} & $0.823$ & $0.803$ & $0.816$ &$ 0.834$ \\ \hline
			\ciaoo{4} & $0.825$ &	$0.791$ & $0.826$ & $0.840$ \\ \hline
			\ciaoo{5} & $0.771$ & $0.728$ & $0.768$ & $0.783$\\ \hline
			\ciaoo{6} & $0.822$ &	$0.795$ & $0.820$ & $0.819$ \\ \hline
			\ciaoo{7} & $0.844$ &	$0.782$ & $0.836$ & $0.837$ \\ \hline
			\ciaoo{8} & $0.840$ &	$0.787$ & $0.838$ & $0.836$ \\ \hline
			\ciaoo{9} & $0.841$ &	$0.788$ & $0.840$ & $0.831$  \\ \hline
			\ciaoo{10} & $0.826$ & $0.806$ & $0.820$ & $0.851$ \\ \hline \hline
			{\textbf{Avg.}} & {\textbf{0.818}}&	{\textbf{0.783}} & {\textbf{0.816}} & {\textbf{0.826}}\\ \hline
		\end{tabular}
	\end{adjustbox}
	\label{ch5:tab:runs}
\end{table}

Table~\ref{ch5:tab:runs} summarizes the prediction accuracy (averaged across testing homes) for 10 runs. 
The second and third columns show the performance of global modeling (\ie $M_g$) versus that of contextualized modeling (\ie $Best(\{m_i\})$), respectively. We observe that global modeling consistently outperforms contextualized modeling ``on average'' across all runs. However, this pattern is not necessarily present at individual home levels. Let us closely look at a representative run to draw detailed insights. Considering run \ciaoo{2}, homes $H_1$, $H_4$, $H_5$, $H_7$, and $H_9$ were randomly chosen as seen contexts (resulting in $m_i$'s) and therefore remaining homes $H_2$, $H_3$, $H_6$, $H_8$, $H_{11}$, $H_{10}$, and $H_{12}$ were considered unseen. For unseen home $H_2$, the global model gives an accuracy of $0.822$; however, the best selected contextualized model $m_9$ gives an accuracy of $0.860$. Similarly, in the same run, for unseen home $H_8$, the accuracy of the global model is $0.774$, but $m_9$ gives better accuracy of $0.790$.

Overall, in half of the ten runs, I found cases of unseen homes whereby the contextualized model outperforms the global model. Scaling my small-size experiment to the size of the operation of an ISP with thousands of homes, one would appreciate the whole argument that either global or contextualized modeling can be sub-optimal; hence, a better strategy is required.

\section[Combined and Dynamic Inference for Managing Concept Drifts]{Combined and Dynamic Inference for \\Managing Concept Drifts}\label{ch5:sec:dynamic}

In the previous section, we saw that global modeling might sound superior at a macro level but contextualized modeling is not always worse (if not better) at a micro level. 
Therefore, I choose a strategy whereby a combined capability of global and contextualized modeling is leveraged. In other words, I bring $M_g$ as an additional context to the mix of contextualized models $\{m_i: i\in[1..N]\}$ to select the best model, $Best(Mg,\{m_i\})$, for an unseen home.

Under the fourth column of Table~\ref{ch5:tab:runs} I report the accuracy of the two approaches combined.  
Unsurprisingly, the augmented contextualized modeling consistently outperforms its original form (third column of Table~\ref{ch5:tab:runs}) at a macro level (average accuracy across unseen homes).
Also, it can be seen that in some runs like \ciaoo{2} and \ciaoo{4}, the performance of the new strategy is slightly better than that of the global $M_g$. However, $M_g$ marginally wins the competition with the new strategy by the metric of aggregate accuracy. At a micro level (individual homes), I found that the two strategies yield almost the same prediction accuracy for about 60\% of testing homes across different runs. For the remaining 40\% of testing homes, the superiority of $M_g$ or the combined strategy is insignificant ($<1\%$). 
It is important to note that the best model (of the combine strategy) is selected based on observations during the training period. In other words, this selection of the best model, though it has improved by incorporating $M_g$ into the mix, is still ``static'', which makes it vulnerable to temporal concept drifts.

To overcome this challenge, I slightly refine my inference strategy and make it ``dynamic''. This means that the process of best model selection continues in the time domain (occurs, say, once a day). Such a change in my inference will certainly introduce additional computing costs (for periodic selection). Instead, the inference process becomes relatively more resilient to temporal concept drifts.  
Another practical challenge is that certain homes may not have sufficient instances (only a few IPFIX records due to the limited activity of the corresponding IoT devices) to make a meaningful selection for some days during the testing period. Therefore, I maintain a sliding window equal to the length of the testing period (a month). I consider the data of the sliding window for the dynamic selection of the best model. Note that we still have $M_g$ in the mix.


The fifth column in Table~\ref{ch5:tab:runs} shows the performance (average daily accuracy of all unseen homes) of the model selected dynamically.
Unsurprisingly, my dynamic combined approach consistently outperforms the baseline of contextualized modeling (third column). In six (out of ten) runs, the performance of the dynamic combined approach is better than that of both $M_g$ (second column) and static combined (third column) approaches. Although $M_g$ performs slightly better in the remaining four runs, its superiority is marginal. 
Averaging across all runs (the last row in Table~\ref{ch5:tab:runs}), the dynamic approach gives an aggregate accuracy of $0.826$, which is greater than all other approaches I evaluated.

\begin{figure}[t!]
	\centering
	\includegraphics[width=0.95\linewidth]{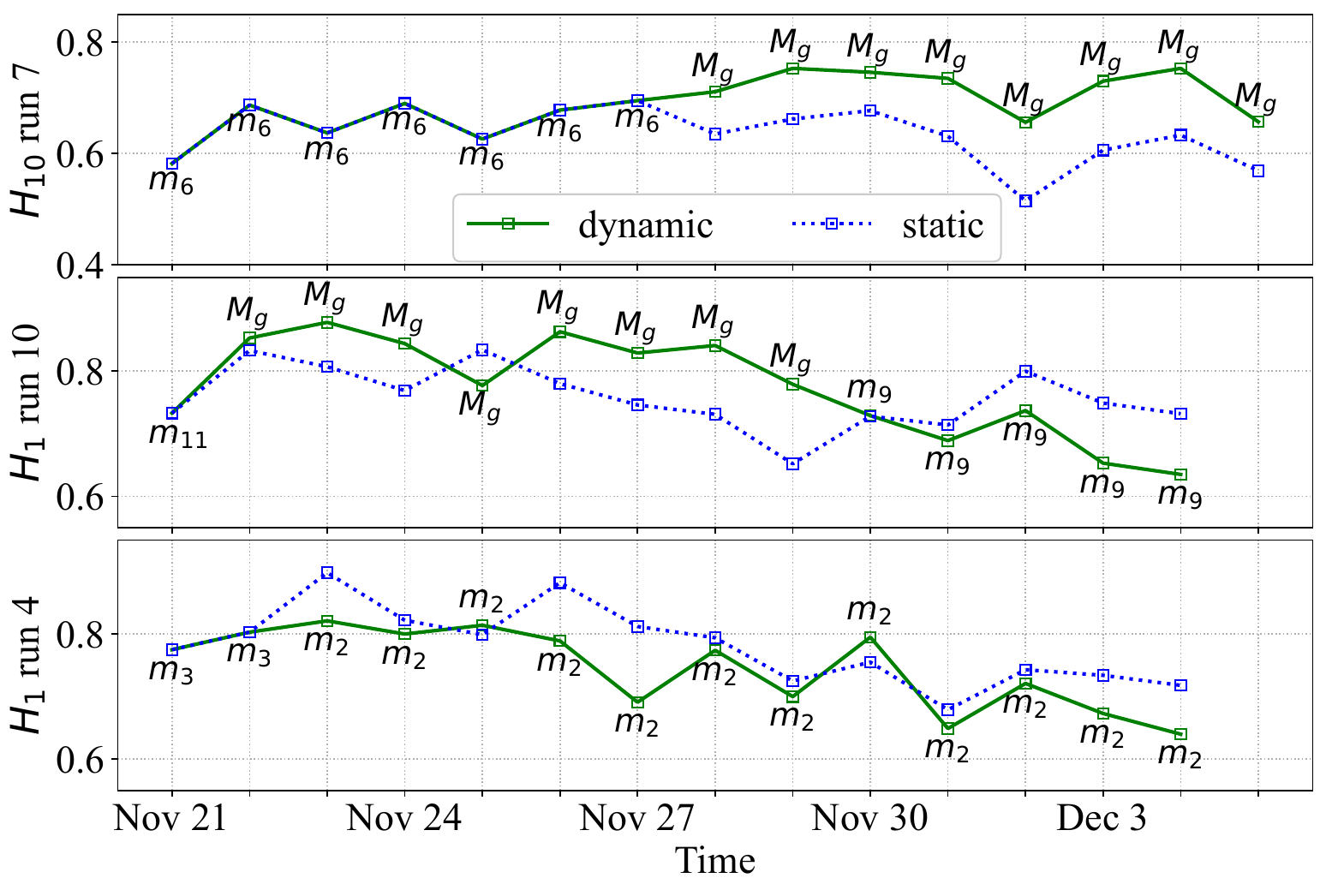}
	\caption{Accuracy of static versus dynamic models during testing period for three representative homes.}
	\label{ch5:fig:timetrace}
\end{figure}

Note that results in Table~\ref{ch5:tab:runs} are at an aggregate level. Fig.~\ref{ch5:fig:timetrace} helps us have a closer look at the performance of static versus dynamic approaches for three representative homes unseen to the models, each from a certain run, across the testing days. Blue dotted lines correspond to the prediction accuracy of the static combined approach, and green solid lines correspond to the accuracy of the dynamic combined approach. For the dynamic approach, the model selected on each day is annotated accordingly. Note that home $H_1$ was entirely inactive on 5\nth Dec, and hence no data-point for that day. 

{\vspace{8mm}}
Let us start from the top plot, corresponding to home $H_{10}$ in run \ciaoo{7}. For the first six days (between 21\nst and 27\nth Nov), $M_6$ is the selected model by both static and dynamic approaches. However, after that, the dynamic approach selects $M_g$, which gives higher accuracy compared to $M_6$. Moving to the second plot ($H_1$ in run \ciaoo{10}), we observe that the dynamic approach outperforms the static one on days between 21\nst Nov and 30\nth  Nov (except for 25\nth Nov). However, its superiority fades out from 1\nst Dec onward. Lastly, for home $H_1$ in run \ciaoo{4}, the static approach defeats the dynamic approach almost every day, except for 25\nth Nov and 30\nth. The key takeaway is that a definite winner cannot be easily concluded from these observations.

\begin{table}[t!]
	\caption{Average number of days across 10 runs that dynamic ($d$) approach gives higher ($>$), lower ($<$), or equal ($=$) accuracy compared to static ($s$) approach.}
	\renewcommand{\arraystretch}{1.4}
	\begin{adjustbox}{width=0.96\textwidth}
		\begin{tabular}{|c|c|c|c|c|c|c|c|c|c|c|c|c|}
			\hline
			&  $H_1$ &  $H_2$ & $H_3$  &$H_4$  &  $H_5$  &  $H_6$  & $H_7$  & $H_8$  & $H_9$  & $H_{10}$  & $H_{11}$  &  $H_{12}$   \\ \hline
			$d>s$&  5 &  1 &  0  &  0  &  0  & 0  & 4 &  4  &  0  &  2  & 0  &  1   \\ \hline
			$d<s$ & 3 &  2 &  1  &  0  &  0  &  1  &  1  &  1  &  0  &  1  &  0 &  1  \\ \hline
			$d=s$ & 6 &  12 &  14  &  14  &  9  &  14  &  10  &  10  &  15  &  12  &  16  &  14  \\ \hline		
		\end{tabular}
	\end{adjustbox}\label{ch5:tab:days}
\end{table}

Inspired by observations in Fig.~\ref{ch5:fig:timetrace}, I analyze the performance using a different lens. 
In each run, I count the number of days that the dynamic approach yields lower, higher, and equal accuracy compared to the static approach.
I then compute the average count across ten runs. Table~\ref{ch5:tab:days} shows the results of this evaluation. 
While most days, both dynamic and static techniques select the same model ($D=S$), the dynamic approach outperforms when the two differ. It can be seen that in 17 home days, the dynamic approach wins ($D>S$) versus 11 home days with the static approach winning ($D<S$), meaning more than $50$\% superiority for the dynamic approach.
I also quantified the accuracy delta between these two techniques. When the dynamic approach wins, it outperforms by $0.004$, on average. This metric is about $0.002$ for the static approach, meaning half of the dynamic approach.
With the dynamic approach, I found that for $20$\% of home days, one of $\{m_i\}$ (instead of $M_g$) is selected for inference. This means the dynamic combined method improves the baseline inference (by $M_g$) in a fifth of my experimented scenarios.

\section[Practical Method for Selecting the Best Contextualized Model]{Practical Method for Selecting the\\Best Contextualized Model}\label{ch5:sec:model selection}

So far, I have assumed that labels of unseen data are available, assisting us in selecting the best model (by measuring prediction accuracy). However, this assumption is unrealistic in practical settings, particularly at scale. Therefore, in this section, I develop and evaluate the efficacy of a method independent of unseen labels that selects the closest (best) model purely based on the seen data and labels. 

For this purpose, one approach is to compare the data distribution of a given unseen home against the seen (labeled) home datasets and select the model trained for the closest seen dataset. This method can be very complex and computationally expensive as our data is multi-dimensional, while statistical tests often work best for univariate distributions \cite{Scott1992}. Inspired by work in \cite{Reis2018b}, I use the distribution of classification scores (\eg prediction confidence),  a one-dimensional signal, obtained from the models as a proxy for selecting the best model. It is important to note that the model selection only applies to contextualized models where we have a set of models to choose from -- the single global model is free from a selection method. I will demonstrate in \S\ref{ch5:sec:metric eval} that this method may not pick the ideal model (from \S\ref{ch5:sec:dynamic}) in all cases, but the performance of the selected model is close to ideal (upper bound). I will also experiment with a baseline method that randomly selects a model and show that this method (matching distributions of classification scores) often performs better than the baseline.

\begin{figure}[t!]
	\centering
	\includegraphics[width=0.95\linewidth]{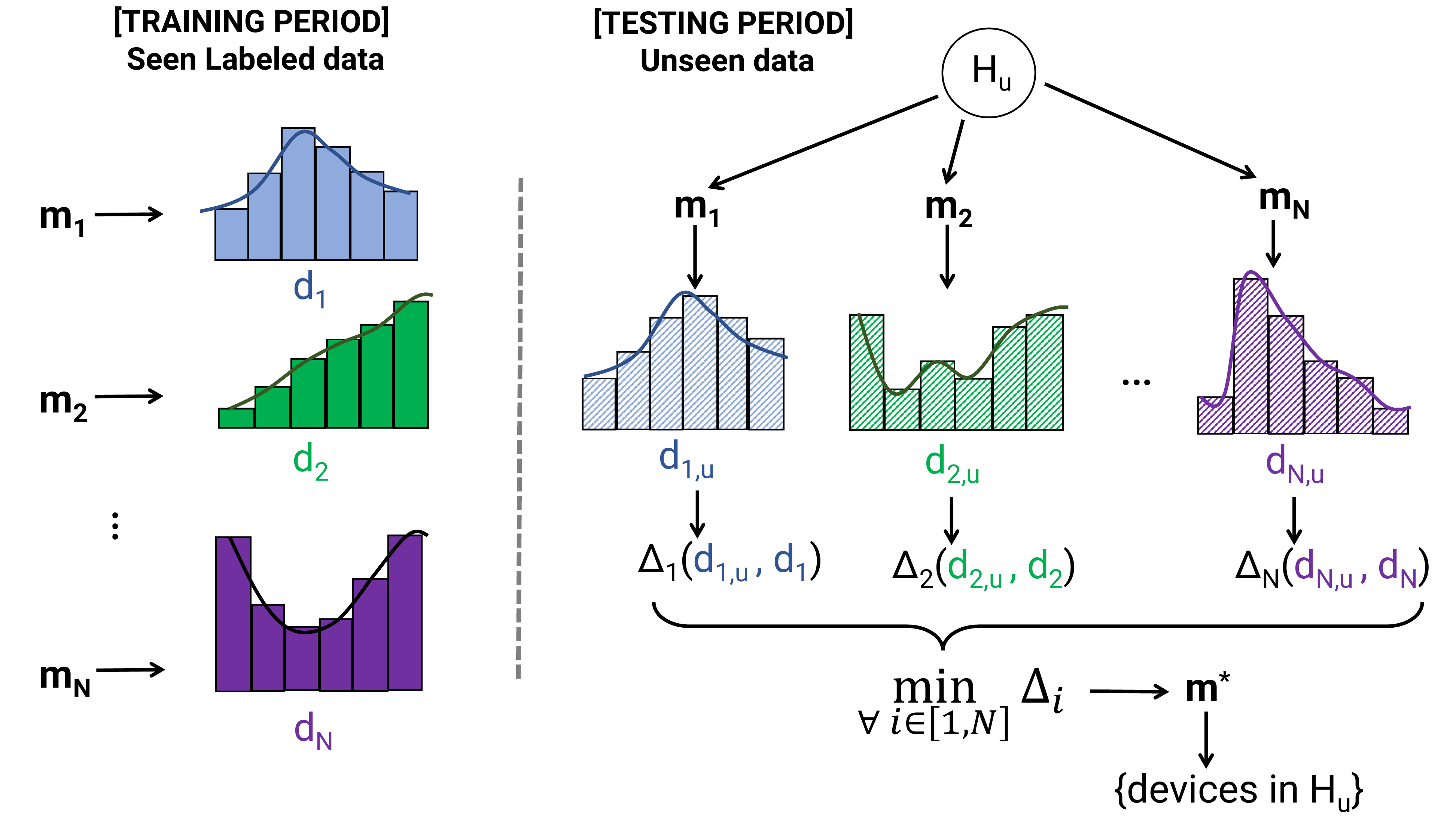}
	\caption{Selecting the best among contextualized models by measuring distances of score distributions.}
	\label{ch5:fig:modelSelection}
\end{figure}

{\vspace{5mm}}

Fig.~\ref{ch5:fig:modelSelection} illustrates the method of selecting the best model among contextualized models by measuring the distances of score distributions from individual $m_i$'s.  For each contextualized model $m_i$,  I obtain its score distribution by applying it to the seen labeled data of $H_i$ during the training period. 
To get unbiased score distributions, I use 10-fold cross validation to obtain 10 sets of score distributions (one set in each run). The final score distribution is a superset of all score values obtained from ten runs. Let $d_i$ denote the distribution of model $m_i$, shown on the left side of Fig.~\ref{ch5:fig:modelSelection}). 

Moving to the testing period with an unseen home $H_u$. I present its data to each of contextualized models ($m_i: 1{\leq}i{\leq}N$) trained on labeled seen data and compute a corresponding score distribution $d_{i,u}$, as shown at the top right of Fig.~\ref{ch5:fig:modelSelection}. I next measure the distance between each pair of $d_{i,u}$ and $d_i$, denoted by $\Delta(d_{i,u}, d_i)$, an approximate measure of distance between the underlying data distributions. Eventually, I select the best model $m^*$ giving the shortest distance $\min \Delta(d_{i,u}, d_i)$ and use it to determine the class of devices connected to home network $H_u$ from its IPFIX flow records. There are several statistical tests for measuring the distance (\ie the function $\Delta$) of two given distributions. I experiment with four well-known tests described in what follows.

\subsection{Distance Metrics}
Let us assume $F(x)$ and $G(x)$ are empirical Cumulative Distribution Functions (eCDF) computed from prediction scores when a classifier is applied to training ($d_i$) and testing instances ($d_{i, u}$), respectively.

\begin{itemize}
\item {\textbf{Kolmogorov-Smirnov (KS):}} KS is a non-parametric fitness test for continuous distributions that measures the greatest distance (supremum) between the two distributions \cite{Hodges1958}.
\begin{equation}
	KS(F, G) = \sup |F(x) - G(x)|
\end{equation}

\item {\textbf{Kantorovich–Rubinstein (KR):}} KR (also known as Wasserstein distance) is a non-parametric fitness test for two given distributions that measures the amount of change needed to transform a distribution to another \cite{Kantorovich1960}.
\begin{equation}
	KR(F, G) = \int_{-\infty}^{+\infty} |F(x) - G(x)|
\end{equation}
\vspace{-5mm}

\item {\textbf{Epps–Singleton (ES):}} ES is a parametric test, proved to be more accurate than KS in some cases \cite{Epps1986}. ES performs Fourier transform on the given distributions before measuring their distance which can be tuned using certain parameters. The inventors of ES suggested a set of parameters that performed well on different distributions and I use them in this chapter.

\item {\textbf{Jensen–Shannon (JS):}} This measure is a symmetric version \cite{lin1991} of Kullback–Leibler divergence that measures the average number of bits lost by approximating $F(x)$ using $G(x)$ \cite{Kullback1951}. Kullback–Leibler divergence can be measured for both continuous and discrete distributions. However, since the measurement for continuous distributions leads to a numerical approximation rather than an exact value \cite{Goldenberg2019}, I use the discrete version of it. For this purpose, I transform the score distributions to 10 discrete bins ($X$) with the size of $0.1$. Discrete Kullback–Leibler divergence  of $F(x)$ and $G(x)$ is defined by:

\begin{table*}[t!]
	\caption{Score distributions and distance measures for two representative unseen homes $H_7$ and $H_{10}$,  given two representative models $m_3$ and $m_8$.}
	\renewcommand{\arraystretch}{1.2}
	\centering
	\begin{adjustbox}{width=0.95\textwidth}
		\begin{tabular}{|>{\centering\arraybackslash}m{0.44cm}|>{\centering\arraybackslash}m{5.1cm}|>{\centering\arraybackslash}m{5.1cm}|}
			\arrayrulecolor{gray}\hline
			& \includegraphics[width=0.2\textwidth]{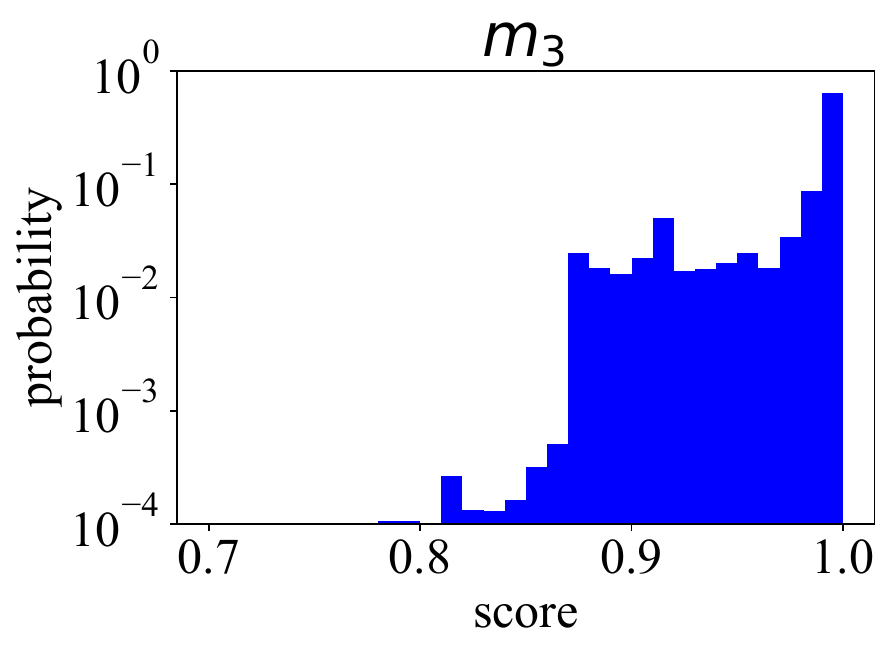} & \includegraphics[width=0.2\textwidth]{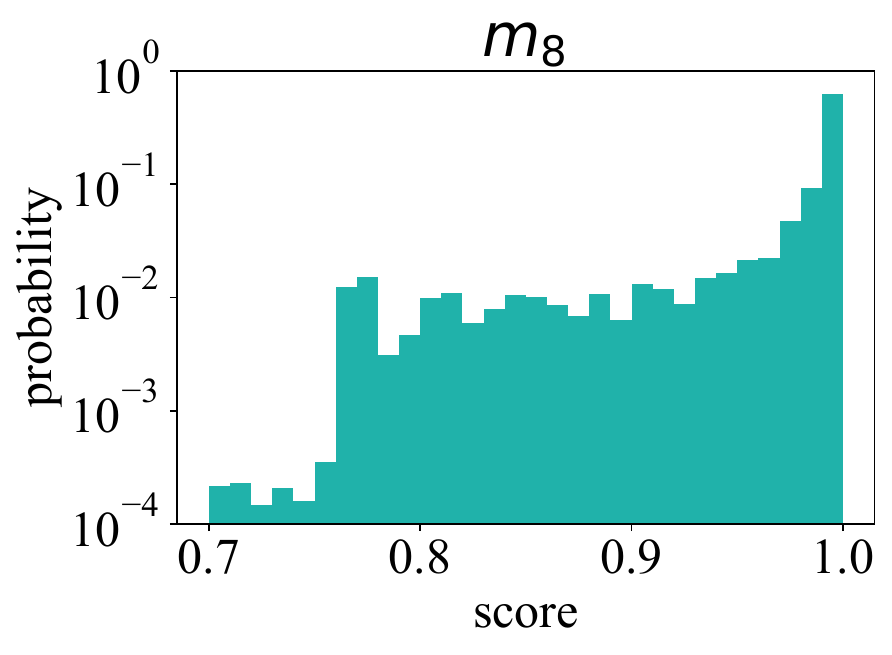} \\ \arrayrulecolor{gray}\hline
			$H_7$ & \makecell[b]{\fontsize{6}{48}\selectfont{KS=$0.04$, KR=$0.004$, ES=$0.57\times10^4$, JS=$0.05$} \\ \includegraphics[width=0.25\textwidth]{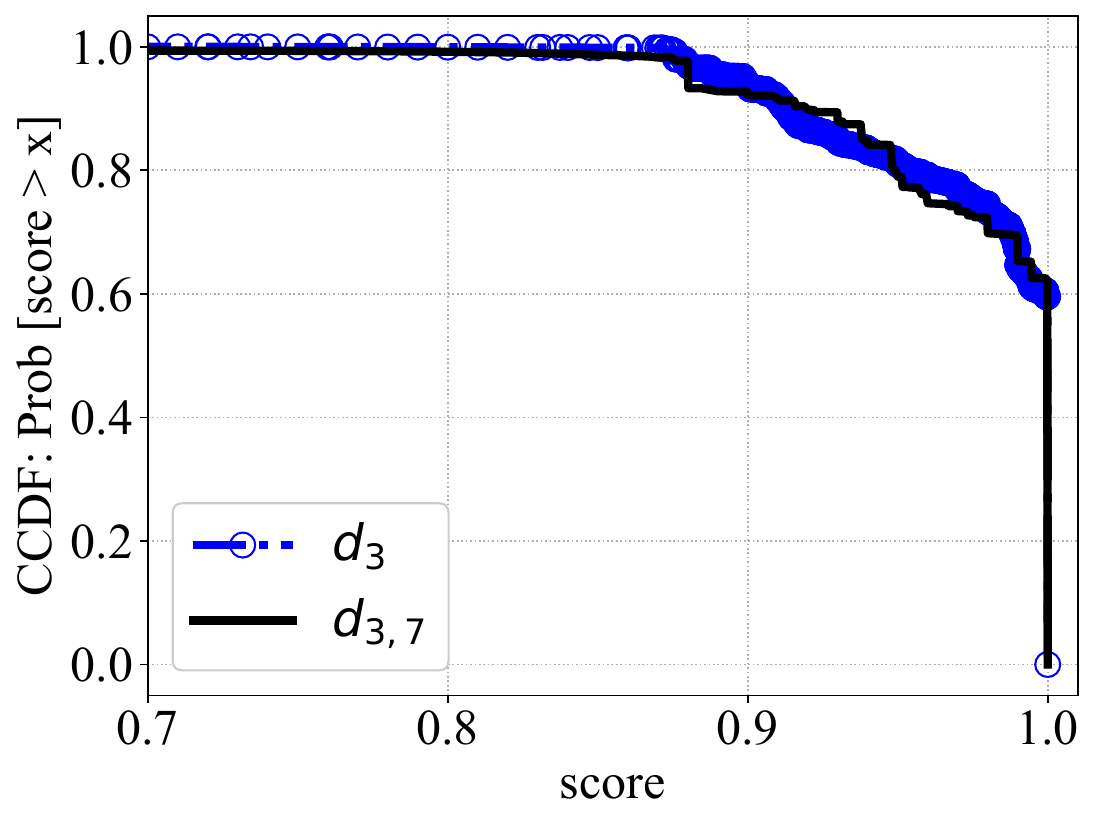}} & \makecell[b]{\fontsize{6}{48}\selectfont{KS=$0.13$, KR=$0.01$, ES=$0.50\times10^4$, JS=$0.12$} \\ \includegraphics[width=0.25\textwidth]{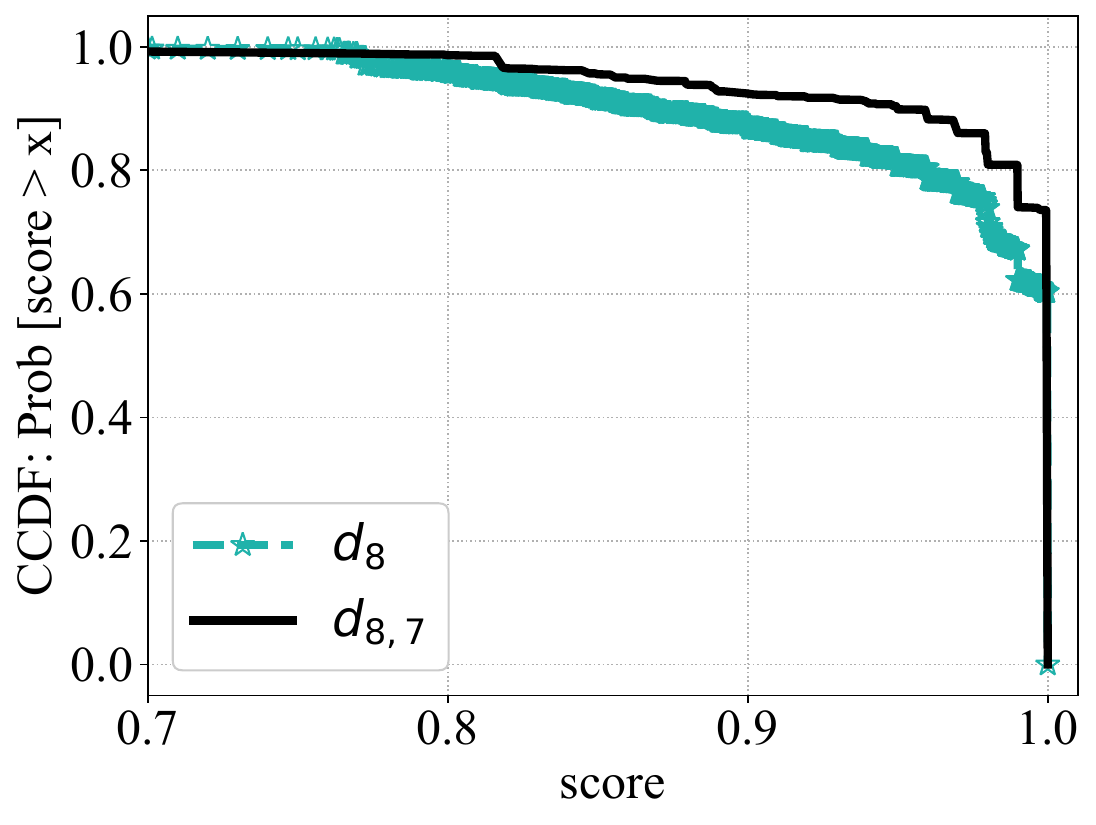}} \\ \hline
			$H_{10}$ & \makecell[b]{\fontsize{6}{48}\selectfont{KS=$0.40$, KR=$0.01$, ES=$0.57\times10^5$, JS=$0.25$} \\ \includegraphics[width=0.25\textwidth]{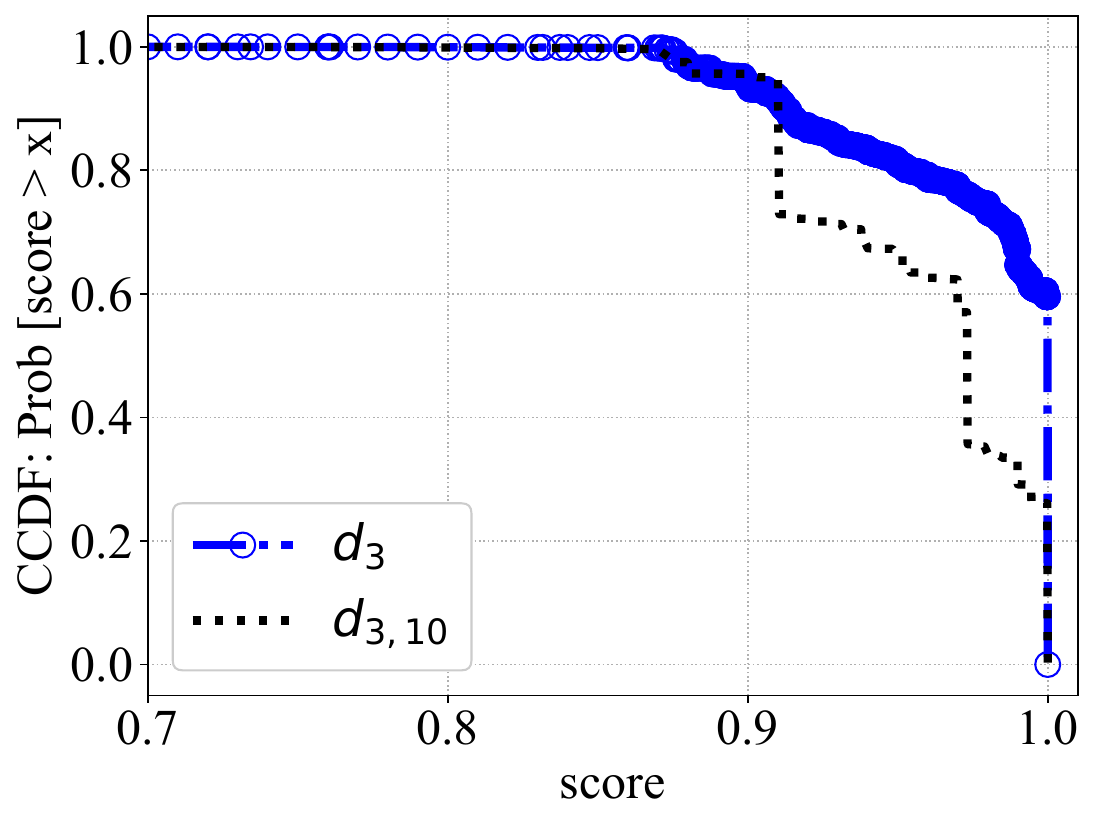}} & \makecell[b]{\fontsize{6}{48}\selectfont{KS=$0.05$, KR=$0.007$, ES=$0.21\times10^4$, JS=$0.12$} \\ \includegraphics[width=0.25\textwidth]{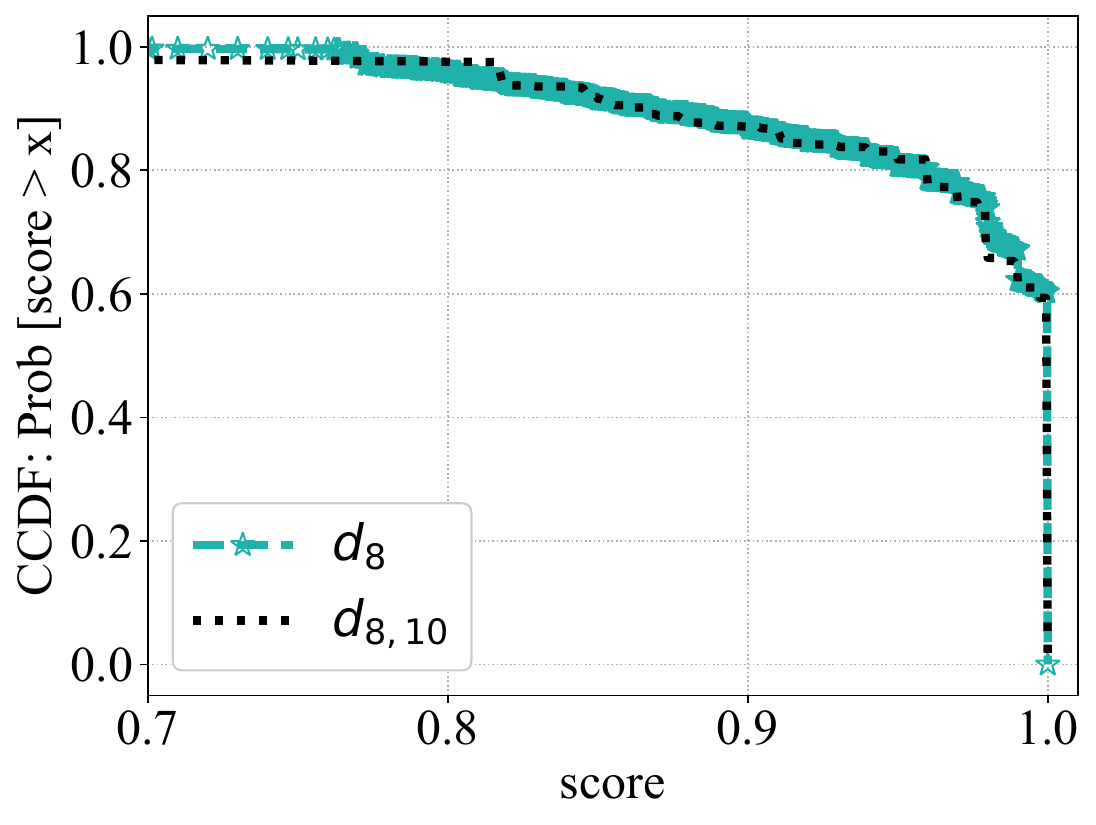}} \\ \hline
		\end{tabular}
	\end{adjustbox}
	\label{ch5:tab:metricsDistance}
\end{table*}

\begin{equation}
	D_{KL}(F, G) = \sum_{x \in X} F(x) \log\frac{F(x)}{G(x)}
\end{equation}

\vspace{-2mm}
I compute the JS distance as follows.
\begin{equation}
	JS(F, G) = \sqrt{\frac{D_{KL}(F, m) + D_{KL}(G, m)}{2}}
\end{equation}
\vspace{-2mm}
where, $m$ is the mean value of $F(x)$ and $G(x)$ combined. 
\end{itemize}

\vspace{-2mm}
To better understand how the score distributions work, let us consider a scenario where I apply two models $m_3$ and $m_8$ (trained on seen data of $H_3$ and $H_8$, respectively) to traffic data of two unseen homes $H_7$ and $H_{10}$. Table~\ref{ch5:tab:metricsDistance} summarizes and illustrates score distributions and distance measures for this representative scenario. The top row in this table shows the score distribution (histogram) of the two models from their respective training data. For a better illustration, I limit the scores (x-axis of the plots) to values greater than 0.7, as the majority of scores fall in this range.

In the second and third rows of this table, we have $H_7$ and $H_{10}$, our testing homes. CCDF plots visually highlight how closely the distribution of $H_7$ data (solid black lines) matches that of $m_3$ (dashed blue lines with circle markers). In other words, $d_{3,7}$ is closer to $d_3$ than $d_8$. We also see that $d_{8,10}$ (dotted black lines) matches $d_8$ (dashed green lines with star markers) better than $d_3$, meaning $m_8$ is the best model to infer from network traffic of home $H_{10}$. 

In addition, CCDF plots are accompanied by the four quantitative distance metrics (KS, KR, ES, and JS) I consider in this chapter. The smaller each metric value is, the closer the two distributions are to each other. For example, the KS metric measures the distance between $d_{3,7}$ and $d_3$ as $0.04$ where this distance is $0.13$ between $d_{8,7}$ and $d_8$, corroborating our visual observation from the CCDF plots in the second row. I also note that the KS distance shows that the data of home $H_7$ is ten times closer than that of $H_{10}$ to model $m3$ -- $KS=0.04$ versus $KS=0.40$. Although KS and KR distance metrics consistently support visual similarities in CCDF plots, ES and, to some extent, JS behave differently, particularly for the scenario of $d_8$ versus $d_{8,7}$. 
This is perhaps because minor similarities at a macroscopic level between $d_8$ and $d_{8,7}$ are more signified by ES and JS. In contrast, ES and JS better conform to KS and KR in the case of  $d_3$ versus $d_{3,10}$, where the two distributions are more distinct (dissimilar), highlighting relatively larger distances.

\subsection{Evaluating Efficacy of Practical Model Selection}
\label{ch5:sec:metric eval}

I now evaluate the performance of the distance metrics in selecting the best model for a given unseen home. I experiment with the selection strategy over the same ten runs, similar to what was discussed in \S\ref{ch5:sec:strategies}. To quantify how the selected model (by distance metrics) is close to the ideal model (the upper bound), I measure a ratio as $\frac{A_S}{A_I}$, where $A_S$ and $A_I$ are the average accuracy of the selected and the ideal model, respectively. A higher ratio indicates how my method performs close to ideal. Note that this ratio is expected to take a value of less than 1. However, it may go beyond 1 when the static inference strategy (\S\ref{ch5:sec:strategies}) is employed. It is because the static inference selects the best model purely based on data of the training period. When the selected model is applied to testing data, the upper bound $A_I$ is no longer applicable ($A_S$ can be greater than $A_I$). I indeed observed and explained similar patterns in Table~\ref{ch5:tab:runs}.

{\textbf{Baseline:}} To compare with my method, let us establish a baseline for selecting a model. I choose a random selection method whereby one of the six available models (five contextualized models $\{m_i\}$ plus a global model $M_g$) is picked randomly at the time of inference for a given unseen test home.

\begin{table*}[t!]
	\caption{Accuracy ratio of a distance-/random-based selected model to its corresponding ideal model.}
	\renewcommand{\arraystretch}{1.4}
	\begin{adjustbox}{width=0.99\textwidth}
		\begin{tabular}{|c|c|c|c|c|c|c|c|c|c|c|c|c|}
			\hline
			\multicolumn{2}{|r|}{Runs} & \ciaoo{1} & \ciaoo{2} & \ciaoo{3} & \ciaoo{4} & \ciaoo{5} & \ciaoo{6} & \ciaoo{7} & \ciaoo{8} & \ciaoo{9} & \ciaoo{10} & {\textbf{Avg.}} \\ \hline
			\multirow{4}{*}{$Best(M_g,\{m_i\})^d$}
			
			& \textbf{KS} & $0.955$ & $0.915$ & $0.961$ & $0.911$ & $0.868$ & {\cellcolor{green!25}$0.976$} & $0.943$ & $0.901$ & $0.906$ & {\cellcolor{green!25}$0.991$} & {\textbf{0.933}}\\ \cline{2-13}
			& \textbf{KR} & $0.934$ & $0.907$ & $0.955$ & $0.889$ & $0.885$ & $0.915$ & $0.912$ & $0.876$ & {\cellcolor{green!25}$0.924$} & $0.926$ & {\textbf{0.912}}\\ \cline{2-13}
			& \textbf{ES} & {\cellcolor{green!25}$0.982$} & $0.921$ & $0.971$ & $0.918$ & {\cellcolor{green!25}$0.914$} & $0.974$ & {\cellcolor{green!25}$0.980$} &{\cellcolor{green!25}$0.925$} & $0.904$ & $0.916$ & {\cellcolor{green!25}\textbf{0.940}}\\ \cline{2-13}
			& \textbf{JS} & $0.980$ & $0.916$ & {\cellcolor{green!25}$0.990$} & $0.876$ & $0.874$ & $0.949$ & $0.967$ & $0.865$ & $0.921$ & $0.890$ & {\textbf{0.923}}\\ \cline{2-13}
			& \textbf{RND} & $0.917$ & {\cellcolor{green!25}$0.939$} & $0.937$ & {\cellcolor{green!25}$0.947$} & $0.911$ & $0.927$ & $0.926$ & $0.908$ & $0.914$ & $0.926$ & {\textbf{0.925}}\\ \hline\hline
			
			\multirow{4}{*}{$Best(M_g,\{m_i\})$}			
			& \textbf{KS} & $0.963$ & $0.904$ & $0.974$ & $0.923$ & $0.899$ & {\cellcolor{green!25}$0.976$} & $0.952$ & {\cellcolor{green!25}$0.930$} & $0.917$ & {\cellcolor{green!25}$1.012$} & {\cellcolor{green!25}\textbf{0.945}}\\ \cline{2-13}
			& \textbf{KR} & $0.957$ & $0.903$ & $0.982$ & $0.891$ & $0.896$ & $0.917$ & $0.920$ & $0.886$ & {\cellcolor{green!25}$0.931$} & $0.964$ & {\textbf{0.925}}\\ \cline{2-13}
			& \textbf{ES} & $0.969$ & {\cellcolor{green!25}$0.925$} & $0.987$ & $0.928$ & $0.918$ & $0.970$ & $0.964$ & {\cellcolor{green!25}$0.930$} & $0.896$ & $0.940$ & {\textbf{0.943}}\\ \cline{2-13}
			& \textbf{JS} & {\cellcolor{green!25}$0.970$} & $0.904$ & {\cellcolor{green!25}$1.005$} & $0.854$ & $0.899$ & $0.936$ & $0.958$ & $0.865$ & $0.923$ & $0.926$ & {\textbf{0.924}}\\ \cline{2-13}
			& \textbf{RND} & $0.908$ & $0.912$ & $0.978$ & {\cellcolor{green!25}$0.941$} & {\cellcolor{green!25}$0.964$} & $0.931$ & {\cellcolor{green!25}$0.973$} & $0.922$ & $0.889$ & $0.912$ & {\textbf{0.933}}\\ \hline
		\end{tabular}
	\end{adjustbox}
	\label{ch5:tab:metricsRun}
\end{table*}

Table~\ref{ch5:tab:metricsRun} summarizes my evaluation by reporting the accuracy ratio obtained from the selected model based on the four distance metrics plus the baseline method (random selection denoted by ``RND'') across ten runs. 
Let us start with the dynamic inference, rows corresponding to $Best(M_g,\{m_i\})^d$ in Table~\ref{ch5:tab:metricsRun}. It can be seen that all metrics perform relatively well as their selected models are close to ideal (giving ratios greater than $0.9$). Given a run, the cell of the best-performing method is shaded in light green. ES seems to outperform the others by winning in $40$\% of the runs (\ciaoo{1}, \ciaoo{5}, \ciaoo{7}, and \ciaoo{8}), as well as by the average ratio of $0.940$ across the ten runs. The baseline RND wins in two runs \ciaoo{2} and \ciaoo{4}, but on average ($0.925$) ranks the third after ES and KS. Regarding the standard deviation of accuracy ratios, the four metrics plus the baseline give a measure of around $0.04$, indicating relatively consistent performance across ten runs.
In addition to the accuracy ratio, I measured the absolute delta of the overall accuracy $|A_S - A_I|$, which turns out to be about $0.05$, $0.07$, $0.04$, $0.06$, and $0.06$ for KS, KR, ES, JS, and RND, respectively, on average across ten runs. Again, by this measure, ES slightly outperforms its rivals, yielding a relatively minor loss of prediction accuracy compared to the upper bound (the ideal model) in practical settings. 

Moving to the static inference, rows corresponding to $Best(M_g,\{m_i\})$ in Table~\ref{ch5:tab:metricsRun}, it can be seen that the selected model sometimes outperforms the ideal model, giving a ratio of more than 1. 
We now observe that KS is the winner metric (marginally) by giving the highest ratio in three runs and yielding an average ratio of $0.945$.
Similar to what we saw for the dynamic inference, the baseline RND comes third after KS and ES metrics.
The standard deviation of accuracy ratios is about $0.05$, not much different from what was seen in the dynamic inference.  Finally, the absolute accuracy delta in the static inference is about $0.04$ for KS and ES, about $0.06$ for KR and JS metrics, and about $0.05$ for RND.

\textbf{F1-score vs. Accuracy:} Throughout this chapter I have been using accuracy as the performance metric of the inference models -- the best model was the one that yields the highest overall accuracy across IoT classes. Let us now evaluate the efficacy of inference models using two other well-known metrics, namely precision and recall. Precision is computed by the number of correctly predicted IoT classes (which indeed connect to the network) divided by the total number of predicted classes (correct and incorrect) in a given home. In other words, the precision metric indicates how precise the prediction of an inference model is, that decreases by false positives (those predicted IoT classes that are not present on the target home network). On the other hand, recall is the number of correctly predicted IoT classes divided by the total number of IoT classes connected to a home network. Recall decreases by false negatives (missing classes that indeed exist in the home network. Precision and recall are incomplete and cannot be reliable performance indicators individually. However, combining these two metrics gives a better picture of the model performance. $F_1$-score, the harmonic mean of precision and recall, is widely used as a performance measure in machine learning applications. $F_1$-score is computed by:

\vspace{-10mm}
\begin{equation}
F_1\mbox{-score} = 2\times\frac{\mbox{Precision} \times \mbox{Recall}}{\mbox{Precision} + \mbox{Recall}}
\end{equation}

\begin{table*}[t!]
	\caption{$F_1$-score ratio of a distance-/random-based selected model to its corresponding ideal model.}
	\renewcommand{\arraystretch}{1.4}
	\begin{adjustbox}{width=0.99\textwidth}
		\begin{tabular}{|c|c|c|c|c|c|c|c|c|c|c|c|c|}
			\hline
			\multicolumn{2}{|r|}{Runs} & \ciaoo{1} & \ciaoo{2} & \ciaoo{3} & \ciaoo{4} & \ciaoo{5} & \ciaoo{6} & \ciaoo{7} & \ciaoo{8} & \ciaoo{9} & \ciaoo{10} & {\textbf{Avg.}} \\ \hline
			\multirow{4}{*}{$Best(M_g,\{m_i\})^d$}
			
			& \textbf{KS} & $0.986$ & $0.988$ & $0.989$ & $0.992$ & {\cellcolor{green!25}$0.986$} & {\cellcolor{green!25}$0.995$} & $0.990$ & $0.994$ & $0.992$ & {\cellcolor{green!25}$0.9974$} & {\textbf{0.991}}\\ \cline{2-13}
			& \textbf{KR} & $0.991$ & $0.989$ & $0.989$ & $0.992$ & {\cellcolor{green!25}$0.986$} & $0.993$ & $0.989$ & $0.991$ & {\cellcolor{green!25}$0.994$} & $0.984$ & {\textbf{0.990}}\\ \cline{2-13}
			& \textbf{ES} & $0.992$ & {\cellcolor{green!25}$0.994$} & {\cellcolor{green!25}$0.993$} & $0.993$ & $0.984$ & {\cellcolor{green!25}$0.995$} & {\cellcolor{green!25}$0.996$} & {\cellcolor{green!25}$0.995$} & $0.987$ & $0.992$ & {\cellcolor{green!25}\textbf{0.992}}\\ \cline{2-13}
			& \textbf{JS} & {\cellcolor{green!25}$0.993$} & $0.991$ & {\cellcolor{green!25}$0.993$} & $0.989$ & $0.985$ & $0.994$ & $0.991$ & $0.987$ & $0.991$ & $0.981$ & {\textbf{0.990}}\\ \cline{2-13}
			& \textbf{RND} & $0.988$ & $0.991$ & $0.988$ & {\cellcolor{green!25}$0.995$} & $0.985$ & $0.988$ & $0.986$ & $0.989$ & $0.987$ & $0.983$ & {\textbf{0.988}}\\ \hline\hline
			
			\multirow{4}{*}{$Best(M_g,\{m_i\})$}			
			& \textbf{KS} & $1.000$ & $0.984$ & $1.003$ & {\cellcolor{green!25}$1.005$} & $0.990$ & $1.007$ & {\cellcolor{green!25}$1.008$} & $0.997$ & $0.997$ & $0.994$ & {\textbf{0.999}}\\ \cline{2-13}
			& \textbf{KR} & {\cellcolor{green!25}$1.007$} & $0.981$ & $1.008$ & $1.002$ & {\cellcolor{green!25}$0.999$} & {\cellcolor{green!25}$1.008$} & {\cellcolor{green!25}$1.008$} & {\cellcolor{green!25}$1.000$} & {\cellcolor{green!25}$1.003$} & {\cellcolor{green!25}$0.997$} & {\cellcolor{green!25}\textbf{1.001}}\\ \cline{2-13}
			& \textbf{ES} & $0.997$ & $0.991$ & {\cellcolor{green!25}$1.015$} & $1.002$ & $0.988$ & $1.004$ & {\cellcolor{green!25}$1.008$} & $0.997$ & $0.997$ & $0.987$ & {\textbf{0.998}}\\ \cline{2-13}
			& \textbf{JS} & {\cellcolor{green!25}$1.007$} & $0.984$ & {\cellcolor{green!25}$1.015$} & $0.992$ & $0.990$ & {\cellcolor{green!25}$1.008$} & $1.004$ & $0.991$ & $0.994$ & $0.984$ & {\textbf{0.997}}\\ \cline{2-13}
			& \textbf{RND} & $1.001$ & {\cellcolor{green!25}$0.994$} & $1.007$ & $1.004$ & $0.990$ & $0.994$ & {\cellcolor{green!25}$1.008$} & $0.981$ & $0.997$ & $0.981$ & {\textbf{0.996}}\\ \hline
		\end{tabular}
	\end{adjustbox}
	\label{ch5:tab:metricsF1Run}
	\vspace{-4mm}
\end{table*}

\vspace{-6mm}
Note that the objective is to select a model with the highest F1-score.
Table~\ref{ch5:tab:metricsF1Run} reports the ratio of $F_1$-score ratio of a distance-/random-based selected model to that of the ideal model across ten runs with dynamic and static inference strategies. Given a strategy, the best-performing model of each run is highlighted. Similar to the accuracy metric, the $F_1$-score ratio may be greater than 1 in some cases for the static inference. The reason for this is that the ideal model is selected based on the training dataset, but when the same model is applied to the testing dataset, its $F_1$-score might be less than the $F_1$-score of the model selected by the distance metric.
Though various metrics perform very closely, ES is the winner of the dynamic inference by giving the highest F1-score ratio in half of the runs and the average ratio of $0.992$. At the same time, KR outperforms other metrics with the static inference by giving the highest F1-score ratio in $70$\% of the runs and the average ratio of $1.001$. Compared with the metric-selected models, RND has the lowest $F_1$-score ratio in both dynamic and static strategies. It is important to note that distance-/random-based selected models yield a better F1-score ratio in Table~\ref{ch5:tab:metricsF1Run} compared to the accuracy ratio in Table~\ref{ch5:tab:metricsRun}.

\section{Conclusion}
Network operators of various sizes increasingly employ machine learning-based models to infer passively from their network traffic, identifying and characterizing connected assets. Vulnerable consumer IoT devices coming online in home networks are particularly interesting to residential ISPs who are tasked to manage large-scale networks. This chapter focused on the practical challenges of concept drifts in modeling the behavior of IoT devices in traffic flows from residential networks. I analyzed over 6 million IPFIX flow records from 12 homes, each serving 24 IoT device types. I quantified the impact of concept drifts in traffic data across time and space domains. Given concept drifts, I next quantitatively compared the performance of two broad inference strategies: global (one model trained on aggregate data from all seen homes) versus contextualized (one model per seen home) when predicting traffic data of unseen homes. I dynamically combined the capabilities of the two strategies, improving the baseline inference by 20\%. Finally, I developed a practical method to automatically and dynamically select the best model with overall accuracy and F1-score very close (94\% and 99\%, respectively) to those of the upper bound from the ideal model without needing labels of unseen data for a realistic inference.

In the next chapter, I will analyze and develop methods to combat data-driven adversarial attacks on machine learning inference models. My primary focus will be on decision tree-based models like Random Forest, which I have used so far throughout this thesis.

%% file: tex/Chapter6-Resiliency/main.tex
\chapter[Resiliency of Decision Tree-based Inference Models]{Resiliency of \\Decision Tree-based Inference Models}
\label{ch:secure infer}

%
%




Throughout the previous chapters, we witnessed the potential of machine learning models, specifically decision tree ensemble models like Random Forest, for IoT traffic inference. Despite their great power in finding network behavioral patterns, machine learning models are increasingly becoming targets of sophisticated data-driven adversarial attacks resulting in misprediction which downgrades their performance in traffic inference. Misprediction is more severe when it causes false negative \ie failing to detect a malicious network behavior. In this chapter, I develop AdIoTack, a system that highlights vulnerabilities of decision tree-based models against adversarial volumetric attacks, helping cyber security teams quantify and refine the resilience of their trained models for monitoring and protecting IoT networks. 

I begin by developing a white-box algorithm that takes a trained decision tree ensemble model and the profile of an intended network-based attack (\eg TCP/UDP reflection) on a victim device type as inputs and generates recipes that specify certain packets on top of the indented attack packets (less than 15\% overhead) that together can bypass the inference model unnoticed. For the evaluation, I validate the efficacy of AdIoTack on a testbed consisting of a handful of real IoT devices monitored by a trained inference model. Lastly, I develop systematic methods for applying patches to trained decision tree ensemble models, improving their resilience against 92\% adversarial volumetric attacks.
	
\section{Introduction}

	In chapters \ref{ch:smart infer}, \ref{ch:obscure infer}, and \ref{ch:scalable infer} we used Random Forest, a type of machine learning models based on decision trees, for identifying IoT device types with a fairly high accuracy. Decision tree-based models have also been used for monitoring purposes \cite{Doshi2018, Hasan2019, Alrashdi2019, Zhang2006}. Traditionally, the ``security of ML-based models'' has not been the main objective for designing algorithms and developing inference systems in various domains, especially in cyber security. Therefore, potential adversaries with certain incentives aim to subvert the ML model either during training or operation which is known as adversarial attack. Attackers may poison the training instances to influence the resulted model. They may attempt to carefully manipulate network traffic at run-time to flip predictions, yielding a poor performance of the inference model in distinguishing the malicious instances.
	
	Several countermeasures have been proposed, under the umbrella of Adversarial Machine Learning (AML), with practical successes in the area of machine vision and image recognition~\cite{Xie2019, Wang2019, Dhillon2018}. However, in the context of cyber security, it is in its nascent stage of development to the best of my knowledge. A recent report by McAfee~\cite{McAfee2019} highlights a few malware families that have bypassed machine learning engines in 2018. It predicts how cyber criminals will be increasingly employing artificial intelligence techniques to evade detection. The primary objective of adversarial cyber attackers is to find loopholes in ML-based network security models like traffic classifiers, anomaly detectors, or intrusion detection systems. To fortify inference models against these targeted attacks, cyber security teams need to quantify the resilience of their trained model and refine any loopholes. This chapter focuses on developing techniques for launching volumetric attacks (known as adversarial evasion attack) on IoT devices without being noticed by a decision tree ensemble inference model that continuously monitors the network traffic. To make these ideas more concrete, let us consider a scenario whereby an attacker aims to launch a TCP SYN reflection attack (common in large-scale DDoS attacks sourced from IoT devices \cite{synDDoS}) with 1000 attack packets. 
	
	\vspace{5mm}
	The attacker sends 1000 TCP SYN packets with spoofed source IP address to a target IoT device which replies (reflects) them to the victim destination (the source entity specified in the SYN packets) with 1000 TCP SYN-ACK packets. IoT devices typically generate a small amount of traffic volume on their limited set of TCP/UDP flows; thus, reflection attacks with high rates can be detected by inference models relatively easily. That said, attackers with sufficient \textit{prior knowledge} would have the ability to launch their intended attack (without being detected) by precisely generating and injecting some adversarial network traffic (say, 50 NTP packets along with 5 DNS packets and an SSDP packet) in addition to the intended attack packets (1000 TCP SYN packets). To humans, this additional traffic seems to be irrelevant and random; however, this well-crafted ``\textit{recipe}'' subverts the model's internal decision-making, leading it to accept this network traffic instance as benign. Finding these attack recipes would depend on the inference model's traffic features and detection algorithm, which requires a well-developed adversarial learning algorithm to generate them precisely.
	It is important to note that this chapter uses the term ``attack'' in two different contexts. First is the network-based volumetric attack (\eg TCP SYN reflection) on IoT devices, whereby target IoT devices reflect the incoming attack traffic towards an external victim. Second is the adversarial machine learning attack which is a technique that utilizes prior knowledge to subvert the ML-based inference model, and hence the volumetric attacks can go undetected.
	
	In the literature, researchers have studied adversarial attacks on IoT networks~\cite{Ferdowsi2019, Vahakainu2020, Ibitoye2019, Sagduyu2019, Caminero2019}. Prior works primarily focus on developing adversarial models that can learn how to bypass neural network-based attack detection models. In contrast, the literature has not devoted much attention to decision tree-based models. Unlike neural networks, decision tree-based models are not differentiable; thus, well-known approaches like~\cite{Goodfellow2014, Papernot2016, Croce2020} are not compatible with their structure. Additionally, given the dynamics of network traffic and some constraints on Internet Protocol (IP) packets, adversarial attacks in the context of network security (unlike image processing) become relatively more challenging. Another gap in the current literature is that adversarial attacks, to the best of my knowledge, have never been executed in a real network protected by an ML-based inference model. Lastly, no prior systematic attempt has been made to improve the resilience of decision trees against these sophisticated attacks without manipulating the training process which requires re-training the model.
	
	This chapter presents AdIoTack, a systematic approach for learning and launching adversarial attacks on IoT networks protected by decision tree ensemble models. This chapters has four contributions: \textbf{(1)} it develops a novel adversarial learning algorithm named offline learning that automatically generates adversarial attack ``recipes'' given an intended volumetric attack on a target class, subverting a trained decision tree ensemble model (\S\ref{sec:adiotack}). I consider two representative network-based volumetric attacks, namely TCP SYN and SSDP reflection, which are widely used on IoT devices for launching DoS/DDoS attacks \cite{IBMsec,synDDoS,Wisec2017,NETSCOUT}. Adversarial recipes specify certain overhead packets ($2$\% for TCP SYN reflection and $14$\% for SSDP reflection) to be injected for an intended volumetric attack to go unnoticed; \textbf{(2)} It develops an online execution method that launches the adversarial instances (learned from offline learning) on a real IoT device network monitored by a decision tree ensemble inference model (\S\ref{sec:online}); \textbf{(3)} I evaluate the performance of AdIoTack by applying it to a testbed comprising of nine consumer IoT devices. 
	I show that the online attack execution has at least a $95$\% chance of successfully launching an adversarial attack when the attacker has the prior knowledge of the inference model cycles (\S\ref{sec:eval}); and, \textbf{(4)} I develop a method for patching decision trees (without a need for re-training), making them more robust against adversarial volumetric attacks while maintaining the inference accuracy for benign traffic (\S\ref{sec:patching}). 
	The refined model can detect all adversarial SSDP reflection attacks with low (less than 200 attack packets per minute), medium (between 200 and 700 attack packets per minute), and high (greater than 700 attack packets per minute) impact rates. For adversarial SYN reflection attacks, the refined model detects $63\%$ of low, $75\%$ of medium, and $100\%$ of high impact attacks.

	\section{AdIoTack: System Architecture and Adversarial Learning}\label{sec:adiotack}
	This section describes the AdIoTack system, including major functional decisions and system components (\S\ref{sec:arc}), offline learning (\S\ref{sec:learning}), the core algorithms (\S\ref{sec:algorithms}) and consistency verification (\S\ref{sec:vercon}).
	
	\subsection[Threat Model: Functional Decisions and System Components]{Threat Model: Functional Decisions and \\System Components}\label{sec:arc}
	The main objective of AdIoTack is to help cyber-security teams quantify the resilience of their decision tree-based models against adversarial volumetric attacks that target \textit{integrity} of the inference models.
	Given a well-trained traffic inference model that is able to detect ``non-adversarial'' attacks relatively easily, AdIoTack generates adversarial instances that the model mispredicts as benign. Note that ML models may be tested for other aspects like their \textit{availability}, \ie whether they remain operational while being flooded by malicious requests \cite{Barreno2010}, which is beyond the scope of this chapter. In terms of influence, adversarial attacks may be either \textit{causative} or \textit{exploratory} \cite{Barreno2010}. In causative attacks, training data instances are poisoned in order to manipulate the inference logic. In exploratory attacks which happens post-training, well-crafted malicious traffic instances that resemble benign instances are used to bypass the model while launching an attack. My focus in AdIoTack is on exploratory attacks. 
	
	Exploratory attacks require a procedure for finding adversarial instances. For models like neural networks, logistic regression, and SVM, this procedure can be done by using gradient-based methods such as gradient descent, fast gradient sign \cite{Goodfellow2014}, Jacobian saliency map \cite{Papernot2016}, or auto projected gradient descent \cite{Croce2020}. While differentiable models like neural networks have been widely applied to computer vision tasks, non-differentiable models like decision trees have received less attention. In cyber security applications, instead, decision tree-based models are deemed more attractive because of some of their innate properties: (i) decision tree-based algorithms offer faster training time compared to neural networks \cite{Lim2000}; (ii) decision tree-based models require less computing resources at runtime (post-training)  \cite{Taghavi2019} (iii) better handling of imbalanced datasets \cite{Khoshgoftaar2007} without the need for data normalization or scaling prior to training; (iv) Random Forest algorithm is reasonably resilient to the overfitting problem \cite{Breiman2001}; (v) Decision tree-based models often yield good results in terms of accuracy and explainability, \ie their internal decision-making process is visible during testing, which is highly desirable for cyber-analysts (particularly for IoT network security \cite{Hasan2019, Doshi2018, Miettinen2017, Arunan2020a, arunan19, Arunan2017}) in taking remedial and/or preventive actions. In contrast, competitors like neural networks do not provide insights into their internal process of inference; (vi) Random Forest classifiers are more robust than other models studied in \cite{Katzir2018} against adversarial evasion attacks, hence, become more difficult to evade. Because of these distinct features available in decision tree-based models and fewer existing works compared to a rich literature for neural networks, this chapter focuses on decision tree-based models.
	
		One can quantify the resilience of an ML model in three different scenarios: (a)  White-box represents the ``worst-case'' scenario when the adversary comes with full knowledge of the model's structure, algorithm, and features; (b) Gray-box assumes that the adversary’s knowledge is relatively limited (say, partial access to the training data); (c) Black-box is when the inference model is targeted with no prior knowledge -- it can only be probed, say, via API calls. I choose the white-box approach because AdIoTack is primarily designed as a tool to empower defenders (network operators) in preparing for worst-case scenarios, identifying, and mitigating vulnerabilities of decision tree-based models against adversarial attacks. It is important to note that a white-box approach provides defenders with an upper-bound assessment by revealing all existing loopholes in their inference model that can be exploited (even with relatively lower probabilities) by adversarial attacks. Therefore, having an ultimate white-box resilience assessment is recommended as a best-practice \cite{Carlini2017}. Although gray/black box approaches highlight more realistic loopholes, they cannot serve as an ultimate resiliency check, \eg one model may be robust against gray-box attacks but falls short of expectations in white-box ones. It is important to note that although obtaining information about the properties of the model is practically challenging for adversaries, it is not impossible. Techniques (beyond the scope of this chapter) such as social engineering, penetration, and scanning can be employed for that purpose \cite{mcAfeeGTD}. Lastly, a white-box approach assesses the resilience of a model by its innate properties without making assumptions about how those properties are kept ``confidential''.


	\begin{figure}[t!]
		\centering
		\includegraphics[width=0.85\linewidth]{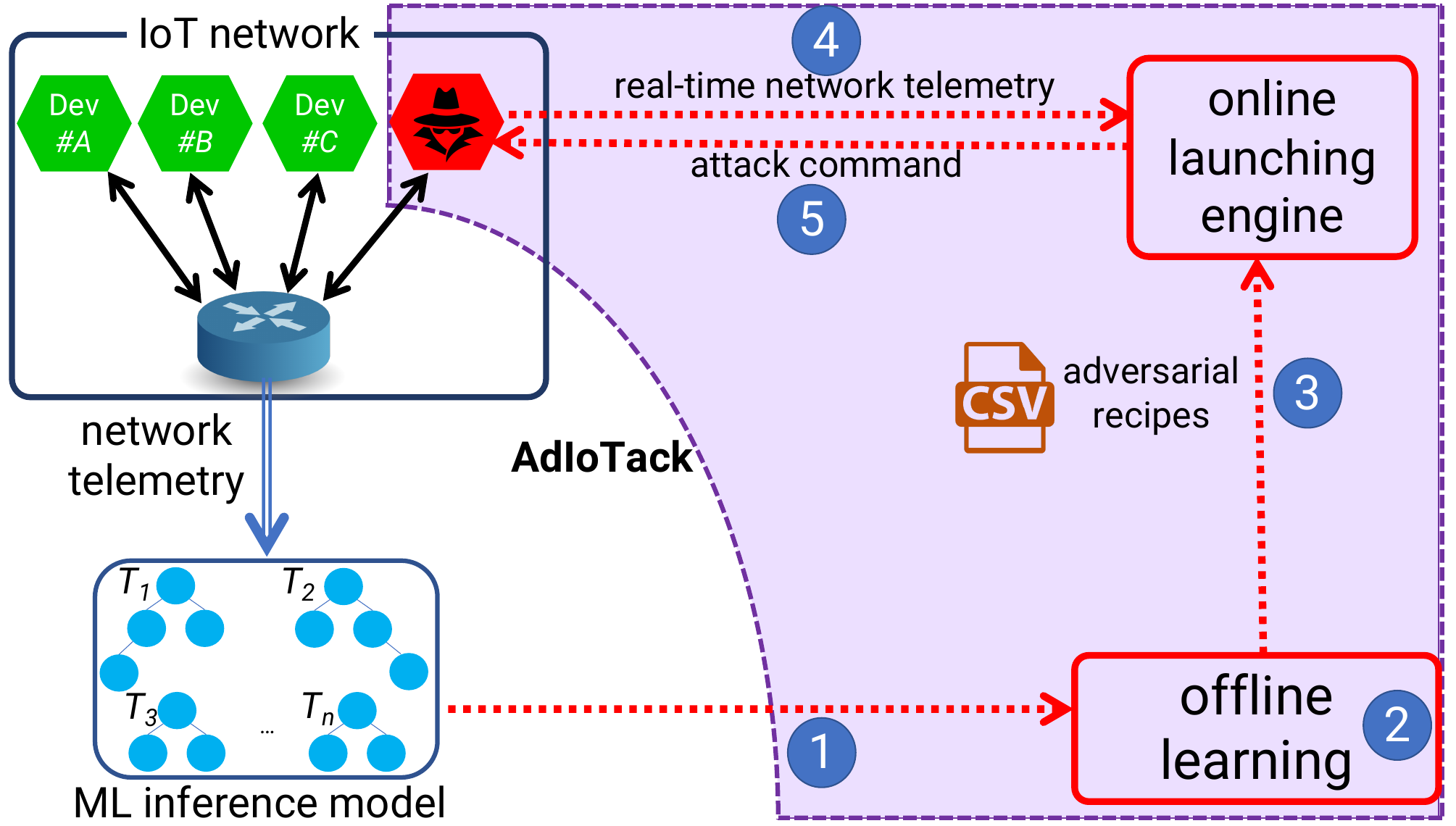}
		\caption{System architecture of AdIoTack.}
		\label{fig:arch}
	\end{figure}
	
	Fig.~\ref{fig:arch} shows the architecture of the AdIoTack system. On the top left, we see a network of IoT devices (green boxes) whose traffic is monitored periodically (\eg every one minute) by a decision tree ensemble model (\ie trees $T_1$ to $T_n$). The network router extracts the required telemetry (\eg packet/byte count of different traffic flows) from the network traffic of each IoT device and passes it to the ML inference model (bottom left). The model continuously infers the most probable class of the connected devices from the received telemetry. I use the classification score of the ML model to compute a device-specific threshold to distinguish benign behavior from malicious. That way, if the model yields a score below the given threshold, it is an indication for malicious behavior (details discussed in \S\ref{sec:offline_learning}). For AdIoTack (shown by shaded region), there are two essential phases, namely: (a) learning and (b) execution (launching). 
	
	\vspace{-4mm}
	In the learning phase (steps \ovalbox{1} and \ovalbox{2} in Fig.~\ref{fig:arch}), the offline learning module uses the inference model (note the white-box scenario) to learn blind spots of the model to generate adversarial recipes (step \ovalbox{3}). Each adversarial recipe consists of a number of conditions over features of the inference model (\eg $ 10 < f_1$, $f_2 \le 50$, and $20 < f_3 < 40$); thus, each recipe can generate several adversarial instances that conform to the recipe's conditions. For example, an adversarial instance for the mentioned recipe could be $f_1 = 15$, $f_2 = 30$, and $f_3 = 25$. For execution, network telemetry (step \ovalbox{4}) similar to inputs of the inference model must be collected, so that an ``appropriate'' adversarial recipe can be selected based on the current state of the network. To obtain real-time telemetry and execute the intended adversarial attack, network traffic of IoT devices is snooped. As a part of the threat model, I assume a ``malicious agent'' like an infected smartphone or a computer is already inside the local network (the red box inside the IoT network on the top left) for this purpose.
	
	\vspace{-4mm}
	
	To obtain network telemetry in real-time, the malicious agent can passively (\eg sniffing) or actively (\eg man-in-the-middle via ARP poisoning) monitor the behavior of victim IoT devices. The passive approach is stealthier to perform but could be practically challenging in some cases, like when the malicious agent and its victims are on different physical access mediums (wireless versus wired). For AdIoTack, I employ the active mode for adversarial network monitoring. Also, this method only analyzes packet metadata (headers), so it has no issue with encrypted payloads.

	In the next step, the online engine starts a search for feasible adversarial recipes, given the current state of the network. From those candidate recipes, the online engine selects the closest one to the current state of the network to minimize the amount of adversarial packets (overhead) to be injected (in real-time to the network traffic) on behalf of the victim. Eventually, the intended volumetric attack is launched (step \ovalbox{5}) by judiciously crafting network traffic matching the chosen recipe. Note that the local victim will reflect/amplify the attack traffic onto an ultimate victim on the Internet.
	It is important to note that the adversarial packets (overhead traffic) are injected on behalf of local victims for their reflected/amplified traffic to go undetected.
	

	\subsection{Offline Adversarial Learning}\label{sec:learning}
	\label{sec:offline_learning}
	Offline adversarial learning is a process of generating adversarial recipes which: (a) result in desired malicious impact, and (b) can go undetected \ie the traffic inference model raises no anomaly flag. The results of offline adversarial learning show to what extent the given model is vulnerable to adversarial volumetric attacks.
	
	I employ ML models' \textit{classification score} as a measure for detecting behavioral changes. I define a classification score threshold for each class (device type) according to the training results. At run-time, any instance with a classification score below the given threshold is flagged as an anomaly. The adversarial learning problem can be defined formally as below:
	\vspace{-14mm}
	
	\begin{gather*}
		\label{eq:adv}
		\forall{d \in \textbf{D}},~\{x~|~\pred(x) = d ~\aand~ \score(x) \ge T_d\} \\
		\textrm{s.t.}~~~\forall{f \in \textbf{F}},~x_f \ge f_{min}
	\end{gather*}
	\vspace{-12mm}
	
	This equation defines a set of adversarial instances $\textbf{X}$ for each IoT device type $d \in \textbf{D}$. The output of the inference model for each adversarial instance $x \in \textbf{X}$, $\pred(x)$, is $d$ and the classification score, $\score(x)$, is greater than or equal to the classification threshold of the corresponding class, $T_d$. Therefore, every adversarial instance $x$ can bypass the inference model's detection as it satisfies the minimal classification score. Also, I define a set of constraints to guarantee that $\textbf{X}$ fulfills the requirements of the desired attack. As the scope of this chapter is volumetric attacks, these constraints are defined as a set of lower bound conditions for each feature $f$ that belongs to the target feature set $\textbf{F}$. The set $\textbf{F}$ contains the relevant features that would be affected by the desired attack, \eg TCP packet and byte count features in TCP SYN reflection attack.
	
	
	\begin{figure}[t!]
		\centering
		\includegraphics[width=\linewidth]{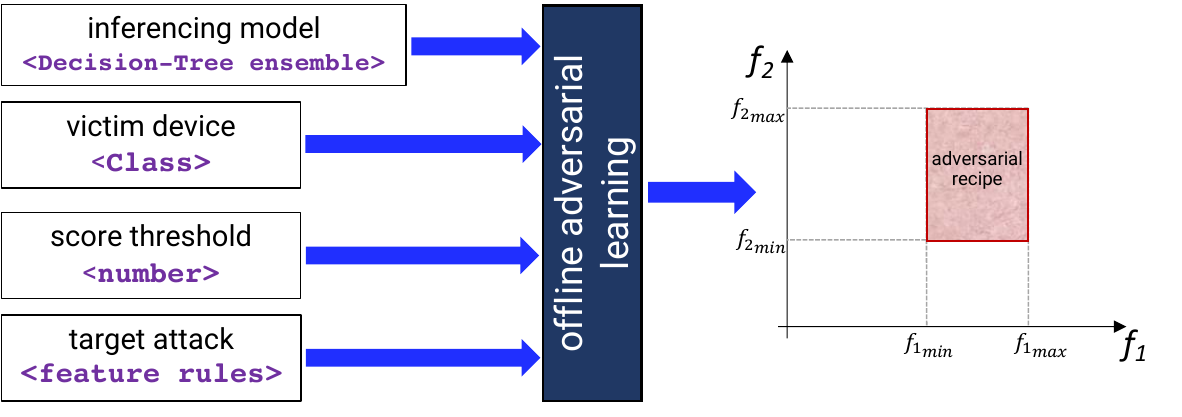}
		\caption{Inputs and outputs of the adversarial learning process.}
		\label{fig:adv_process}
	\end{figure}

	Fig.~\ref{fig:adv_process} illustrates inputs and outputs of the offline adversarial learning process. For illustration purpose I only show two features, $f_1$ and $f_2$. Suppose we want to test whether the model is vulnerable to a volumetric attack on a victim device (\eg Amazon Echo), where the attack offers more than 1000 packets over $f_1$. The desired attack impact is captured by a set of conditions for the target feature(s) (in our example, $f_1 > 1000$). The offline adversarial learning process then searches over a given tree ensemble model in order to determine recipes that the ensemble model predicts them as the class of the victim device with a classification score of greater than or equal to the given threshold. Adversarial recipes can be seen as a region in an $N$ dimension space, being $N$ the number of features required by the inference model. Conditions on target features ($f_1$ in our example) specified by the adversarial recipes must be consistent with those conditions given as input for the intended attack (\ie $f_1 > 1000$). In other words, any point in an adversarial recipe region is an adversarial instance that satisfies the conditions of the intended attack.
	
	
	\subsection{The AdIoTack Algorithms}
	\label{sec:algorithms}
	
	AdIoTack focuses on performing adversarial attacks on a given decision tree ensemble model. The model consists of several decision trees. Weighted or unweighted voting gives the final classification, \ie the fraction of trees yielding the final output determines the prediction and classification score.
	
	
	In order to explain how voting works, let us consider an example relevant to our device class inference problem. Let say, if a benign traffic instance $x$ belongs to IoT device class ``{\myverb{A}}'' (ground-truth), and say 95\% of the trees in the ensemble model classify it correctly, the final prediction would be ``{\myverb{A}}'' with the score of 0.95. Now, assume ``{\myverb{A}}'' is under a cyber attack, having a traffic instance $\hat{x}$. The model is expected to give a lower score \cite{Arunan2020a} for $\hat{x}$ since because of the deviation from benign instance $x$ there will be disagreement among various trees, each inferring from a specific set of features. Certain features of $\hat{x}$ will significantly deviate from their expected normal range, leading a portion of trees to classify $\hat{x}$ as a class other than ``{\myverb{A}}'', and other features may remain relatively unchanged hence some other trees predict the class of $\hat{x}$ as ``{\myverb{A}}''. Therefore, the attack can be flagged with a good chance. In adversarial attack, we aim to find a recipe which has the same malicious impact of $\hat{x}$ in such a way that the majority of decision trees predict it as ``{\myverb{A}}'', resulting in high score and implying it is still benign.
	
	\begin{figure}[t!]
		\begin{center}
			\includegraphics[width=0.95\linewidth]{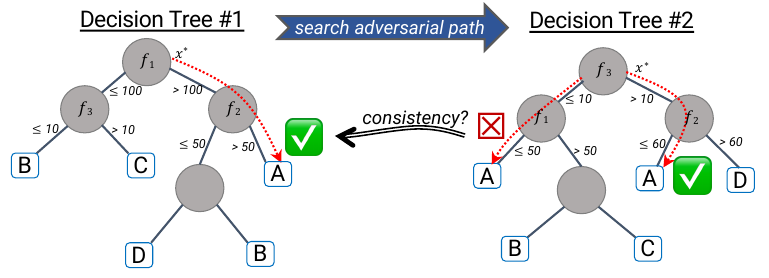}
			\caption[Searching adversarial paths in two decision trees]{Searching adversarial paths in two decision trees: The adversary starts traversing decision tree \#1 with no prior rule -- a path to target class ``{\myverb{A}}'' is found ($\checkmark$). Moving to decision tree \#2, the first path becomes inconsistent with rules of the chosen path from decision tree \#1 ($\times$), and thus an alternative path is sought and found ($\checkmark$).}
			\label{fig:RF_adv_path}
		\end{center}
	\end{figure}

	In the context of decision trees, we define an \textit{adversarial path} to be the path from the root node of a given decision tree to a leaf with the label of the victim device that we aim to launch an attack on it. Each adversarial path is associated with a set of conditions that are met across the path nodes. Fig.~\ref{fig:RF_adv_path}-\textit{left} shows how the adversarial path is found on an illustrative decision tree. Gray circles are decision nodes, each checking a certain condition (\eg $f_1 \le \tau_1$), proceeding to subsequent branch (left or right, depending on the check result). Adversarial path $x^*$ is a complete path from the root to a leaf node, landing on the target victim class ``{\myverb{A}}''. AdIoTack performs tree traversals using the pre-order method -- the root node is first visited, then recursively a pre-order traversal of the left subtree is performed, followed by a recursive pre-order traversal of the right subtree. 
	
	Note that searching the adversarial path needs to be extended to all decision trees inside the ensemble model where consistency of conditions across trees becomes a non-trivial challenge. Fig.~\ref{fig:RF_adv_path} shows an illustrative example of searching for an adversarial path across two decision trees while conditions (of paths in the two trees) need to be consistent. Assume we want to target class ``{\myverb{A}}'' with a cyber attack that requires $\{f_1 > 1000\}$ (\ie target rule). Below, I show the initial recipe to begin the attack:
	\[
	Recipe: f_1 > 1000 \Rightarrow A
	\]
	
	Starting from decision tree \#1 (on the left), AdIoTack finds the first path leading to a leaf with the target class ``{\myverb{A}}'' (by traversing the tree in pre-order mode). The adversarial path on decision tree \#1 consists of two conditions: $\{f_1 > 100$ and $f_2 > 50\}$, which are consistent with the initial recipe; therefore, we can add them to the recipe:
	\[
	Recipe: f_1 > 1000~\wedge~f_1 > 100~\wedge~f_2 > 50 \\ \Rightarrow A
	\]
	
	\noindent which can be simplified (\textit{merged}) to:
	\[
	Recipe: f_1 > 1000~\wedge~f_2 > 50 \Rightarrow A
	\]
	
	Moving on to decision tree \#2, the first possible path (in pre-order search) to the target class ``{\myverb{A}}'' requires $\{f_1 \le 50\}$, which is inconsistent with $\{f_1 > 1000\}$ in the current recipe -- this path cannot be taken, and hence an alternative path needs to be sought. The other path to the other leaf with class ``{\myverb{A}}'' (on decision tree \#2) results in two new conditions: $\{f_3 > 10$ and $f_2 \le 60\}$, not violating the current recipe determined by the path from decision tree \#1. The updated recipe is shown below:
	\[
	Recipe: f_1 > 1000~\wedge~f_3 > 10~\wedge~50 <f_2 \leq 60 \Rightarrow A
	\]
	
	To generate an adversarial recipe that drives the majority vote to the target class, AdIoTack needs to keep track of conditions and progressively augment them as new adversarial paths are found across individual decision trees. The search procedure visits each tree in a greedy manner, \ie it skips a tree if it could not find any consistent adversarial path on it. Also, the order of the trees influences the output of the method. I address this problem by running the algorithm multiple times with different tree ordering to generate new recipes.

	\begin{algorithm}[t!]
		\caption{Adversarial evasion attack on tree ensemble model.}\label{alg:adv}
		\footnotesize
		\begin{algorithmic}[1]
			\State {$\mathcal{IM} \gets$ inference model}
			\State {$TR \gets$ set of target rules}
			\State {$TD \gets$ target device class}
			\State {$T_d \gets$ classification score threshold of $TD$}
			\Function{FindRecipe}{$\mathcal{IM}$, $TR$, $TD$, $T_d$}
			\State {$R \gets (TR \Rightarrow TD)$} \Comment{Recipe}
			\For{$t~in~\mathcal{IM}.trees$}
			\State {$AdvPath \gets$ \Call{FindAdvPath}{$t.root$, $TD$, $R$, $[~]$}}
			\If{$AdvPath \ne \Null$}
			\State{$R \gets$ \Call{merge}{$R$, $AdvPath$}}
			\EndIf
			\EndFor
			\State{$x^* \leftarrow$ \Call{project}{$R$}}
			\If{\Call{$\mathcal{IM}$.pred}{$x^*$} = $TD$ \& \Call{$\mathcal{IM}$.score}{$x^*$} $\ge T_d$}
			\State \Return {$R$}
			\Else
			\State \Return \Null
			\EndIf
			\EndFunction
		\end{algorithmic}
	\end{algorithm}

	\vspace{5mm}
	Having these intuitive illustrations of the adversarial attack method, let us formally develop it by Algorithms~\ref{alg:adv} and \ref{alg:adv_leaf}. Algorithm \ref{alg:adv} searches for an adversarial recipe for a given target device. This algorithm expects the inference model ($\mathcal{IM}$), a set of target rules ($TR$) that are defined based on the target cyber attack, the target class of victim IoT device ($TD$), and the classification score threshold ($T_d$) as inputs.
	
	In essence, Algorithm~\ref{alg:adv} iterates over all trees in the ensemble model ($\mathcal{IM}$). Notice the tree order in a decision tree ensemble model is arbitrary, and different orders may give different results (new recipes or no recipe at all). In~\S\ref{sec:eval}, I evaluate how different orders can affect the number of generated recipes. For each tree, the algorithm calls the function {\myverb{FindAdvPath}} in Algorithm~\ref{alg:adv_leaf} to find an adversarial path consistent with the current recipe generated so far. If such a path is found, the corresponding adversarial path's rules are merged with the current recipe rules ({\myverb{Merge}}). The following examples show how {\myverb{Merge}} function works: 
	
	{\myverb{Merge}}$\left (f \le \tau_1,~f \le \tau_2,~\tau_1 \le \tau_2\right ) = f \le \tau_1$, or
	
	{\myverb{Merge}}$\left (f > \tau_1,~f > \tau_2,~\tau_1 > \tau_2 \right ) = f > \tau_1$

	\begin{algorithm}[t!]
		\caption{Searching for a consistent adversarial path.}
		\footnotesize
		\label{alg:adv_leaf}
		\begin{algorithmic}[1]
			\Function{FindAdvPath}{$n$, $TD$, $R$, $P$}
			\If{$n$ is $leaf$}
			\If{$n.label = target$}
			\State {\Return $P$} \Comment{Adversarial path}
			\Else
			\State \Return \Null
			\EndIf
			\EndIf
			\State {$r \gets$ \Null}
			\If{\Call{isConsistent}{$n.cond, R$} \& \Call{isConsistent}{$n.cond, P$}}
			\State {$r \gets$ \Call{FindAdvPath} {$n.left$, $TD$, $R$, $[P \wedge n.cond]$}}
			\If{$r \ne \Null$}
			\State \Return $r$
			\EndIf
			\EndIf
			\If{\Call{isConsistent}{$\lnot n.cond, R$} \& \Call{isConsistent}{$\lnot n.cond, P$}}
			\State {$r \gets$ \Call{FindAdvPath} {$n.right$, $TD$, $R$, $[P \wedge \lnot n.cond]$}}
			\EndIf
			\State \Return $r$
			\EndFunction
		\end{algorithmic}
	\end{algorithm}
	
	\vspace{8mm}
	
	As there is no guarantee that {\myverb{FindAdvPath}} will find an adversarial path for every tree, it is needed to validate the efficacy of the final recipe (whether it can bypass the model or not). We may find adversarial paths effective on only a small subset of the decision trees, or even no tree at all. In this case, the resulted adversarial instances will become ineffective in yielding the target class and/or the score threshold. I validate the efficacy of a recipe by generating a representative adversarial instance, $x^*$, using {\myverb{Project}} function. I note that every instance of a recipe would fall under the desired adversarial paths determined by Algorithm~\ref{alg:adv}; therefore, if $x^*$ can bypass the model, any other adversarial instance generated from the given recipe can do the same.
	The instance $x^*$ is presented to the model to obtain the prediction. If it is classified as the target class with a classification score greater than or equal to the specified threshold ($T_d$), the algorithm approves and returns the recipe. 
	
	Algorithm~\ref{alg:adv_leaf} (invoked from within Algorithm~\ref{alg:adv}) searches for a consistent adversarial path ($P$) to a leaf with the target class label ($TD$) on a given tree. The algorithm is a direct application of pre-order traversal for binary trees. During a search, I check for consistency ({\myverb{IsConsistent}}) between the nodes' condition and the recipe (explained in \ref{sec:vercon}). This helps to prune the search space, reducing the average execution time. Algorithm~\ref{alg:adv_leaf} assumes binary decision trees; therefore, three fields ({\myverb{left}}, {\myverb{right}} and {\myverb{cond}}) suffice to describe the nodes. The left subtree is associated with the condition {\myverb{cond}} and the right subtree with logical negation of it ({$\lnot$\myverb{cond}}).

	Note that an adversary may choose to stop searching when a certain level of majority across trees (\ie classification score) is obtained. My algorithm, instead, aggressively aims to maximize its chance of subverting the model by (greedily) searching all possible trees. Given this greedy nature of my offline adversarial learning, its time complexity is $\mathcal{O(}\mathcal{T}.\mathcal{N}.\mathcal{F})$, proportional to the number trees ($\mathcal{T}$), maximum number of nodes in the trees ($\mathcal{N}$), and total number of features ($\mathcal{F}$). It is important to note that the search across trees progresses only in a forward direction. Hence, in some cases, it may not yield any consistent adversarial path on a decision tree (in the middle of iterating over trees). In other words, if a consistent path on a tree is not found, then that tree is skipped without going backward, seeking alternative adversarial paths on the previous trees, and updating the recipe correspondingly.
	
	\subsection{Verifying Consistency of Conditions}\label{sec:vercon}
	My adversarial learning algorithm is general and can be applied to any decision tree ensemble model. However, certain constraints may be needed when employing the algorithm for specific domain problems. Given the primary use-case of this chapter is network-based volumetric attacks, three types of consistency checks are performed before taking a new branch (left or right subtree) while searching for an adversarial path:
	
	\textbf{Single-feature consistency:} This type of consistency is required when different conditions are imposed on a single feature across trees.
	Let us consider two illustrative conditions $f\le\tau_1$ and $f>\tau_2$. They are considered consistent if $\tau_2<\tau_1$, where $f$ denotes a feature that the inference model uses. 
	A single-feature consistency check is required in any problem of adversarial machine learning. However, in the context of cyber security, I identified two additional checks essential for launching attacks; otherwise, the offline learning algorithm may generate impractical adversarial recipes that cannot be launched during the online phase.

	\textbf{Frame size consistency:} This constraint is specific to the use-case of this chapter. Frame size is measured in bytes and has a minimum and maximum length, depending on the implemented technology. For example, an Ethernet frame must be at least 64 bytes for collision detection to work and can be a maximum of 1518 bytes to avoid IP fragmentation. This expected characteristic may get violated by a condition. For example, let us consider a set of conditions on two features of incoming ($\downarrow$) DNS traffic for a given adversarial recipe:
	
	\[
	\Bigg \{
	\begin{aligned}
		C_1: 9<\ovalbox{$\downarrow$DNS~packet~count} \\
		C_2: \ovalbox{$\downarrow$DNS~byte~count} \le 100
	\end{aligned}
	\]
	
	These translate into an average size of incoming DNS frames to be less than 10 bytes, which is impossible on Ethernet networks -- sending at least ten frames even with no payload would result in 640 bytes of traffic, violating the condition $C_2$ above. 
	
	\textbf{Boundary consistency:} Average frame size obtained by the ratio of byte count and packet count must be rounded to integer values (cannot be float values). There are certain corner cases where packet count and byte count conditions become inconsistent. The following conditions exemplify this situation:
	
	\[
	\Bigg \{
	\begin{aligned}
		C_3: 10<\ovalbox{$\downarrow$DNS~packet~count} \le 15 \\
		C_4: 998<\ovalbox{$\downarrow$DNS~byte~count} \le 1,000
	\end{aligned}
	\]
	
	Choosing the combination of \ovalbox{$\downarrow$DNS~packet~count} = 11  and \ovalbox{$\downarrow$DNS~byte~count} = 999 gives a minimum average frame size of 91 B (equally sized). At a high level, this combination would violate $C_4$ since 11 DNS incoming messages of size  91 B each will amount to 1001 B. Indeed, one may choose to solve this inequality problem by judiciously setting packet sizes (\eg nine packets of each 100 B and a packet of 99 B) -- this is beyond the scope of this chapter. I only consider recipes that allow adversarial attacks of equal packet size.  
	
	Note that in a single tree, conditions inside each path (from root to a leaf) satisfy single-feature consistency by default. Still, even conditions within a given path of a single tree may violate frame size and boundary consistencies. 
	
	\section{Launching Adversarial Volumetric Attack}\label{sec:online}
	This section describes how I use adversarial recipes to launch real volumetric attacks on an operational IoT network whose traffic is continuously monitored by a trained inference model. To clarify the method better, I explain this section with a set of the flow-based features that the model uses for inference. Inspired by \cite{Arunan2020a}, I choose packet count and byte count of certain network flows: $\downarrow${\myverb{DNS}}, $\uparrow${\myverb{DNS}}, $\downarrow${\myverb{NTP}}, $\uparrow${\myverb{NTP}}, $\uparrow${\myverb{SSDP}}, $\downarrow${\myverb{LAN}}, $\downarrow${\myverb{WAN}}, and $\uparrow${\myverb{WAN}}, proven to be reflective of IoT behavior, during fixed time windows (of length one minute); where $\downarrow$ and $\uparrow$ indicate incoming and outgoing directions, respectively.
	
	\begin{figure}[t!]
		\centering
		\includegraphics[width=0.85\linewidth]{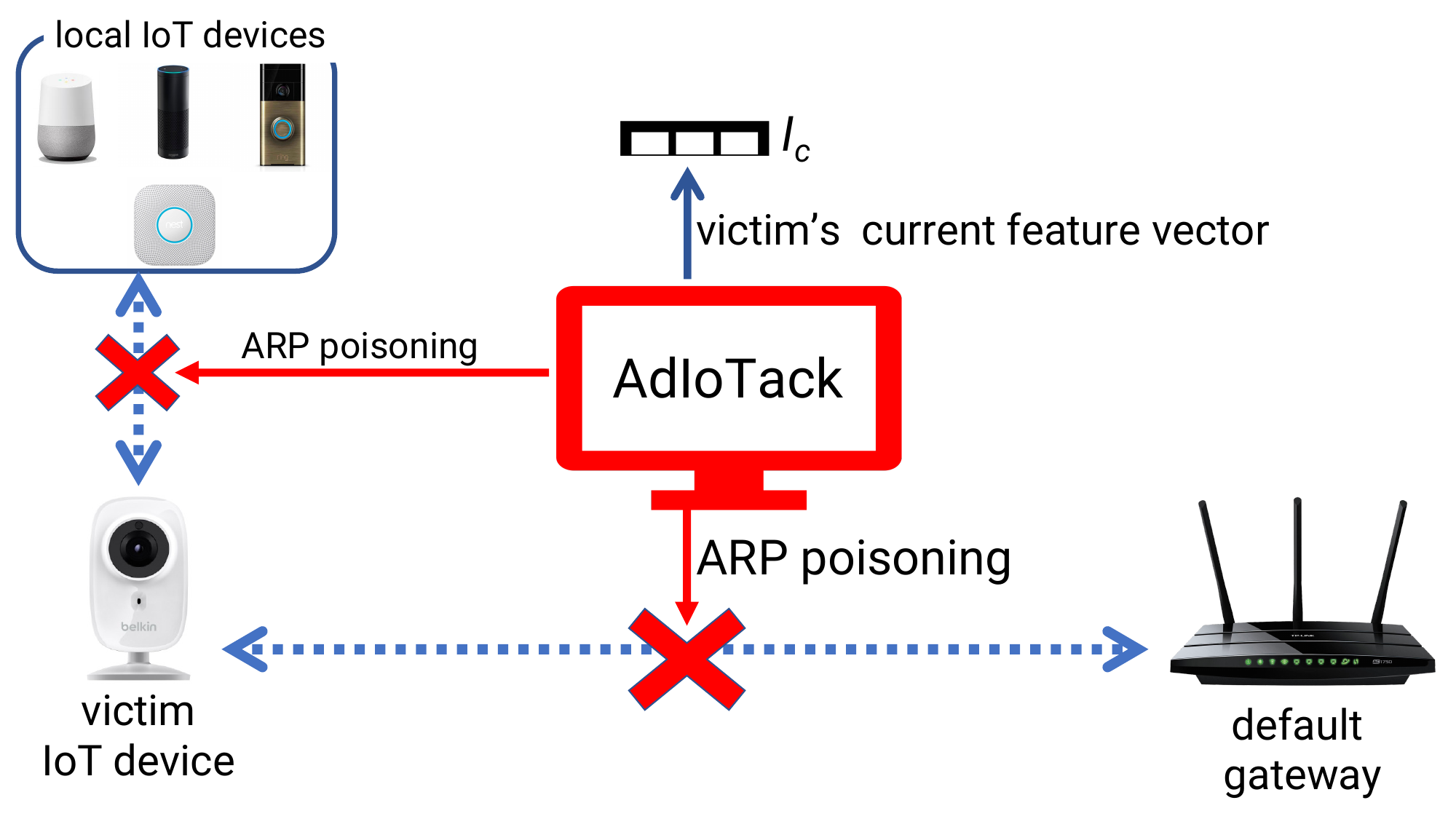}
		\caption{Local adversary (an infected machine) collects traffic features before launching an online attack.}
		\label{fig:attack_prep}
	\end{figure}

	This phase aims to project the current state (traffic feature vector) of a victim IoT device to an adversarial instance using one of the adversarial recipes obtained in the offline learning phase. A naive approach is to launch a cyber attack blindly based on a random recipe. This approach is not guaranteed to bypass the inference model given the variability in volume and frequency of traffic sent/received by the victim IoT devices during the last time window (\ie current state). It is important to note that malicious packets (pertinent to the intended network attack) get added to the benign traffic of the victim device. Hence, a more effective strategy is needed that considers the current state of the variable network.

\begin{figure}[t!]
		\centering
		\includegraphics[width=0.90\linewidth]{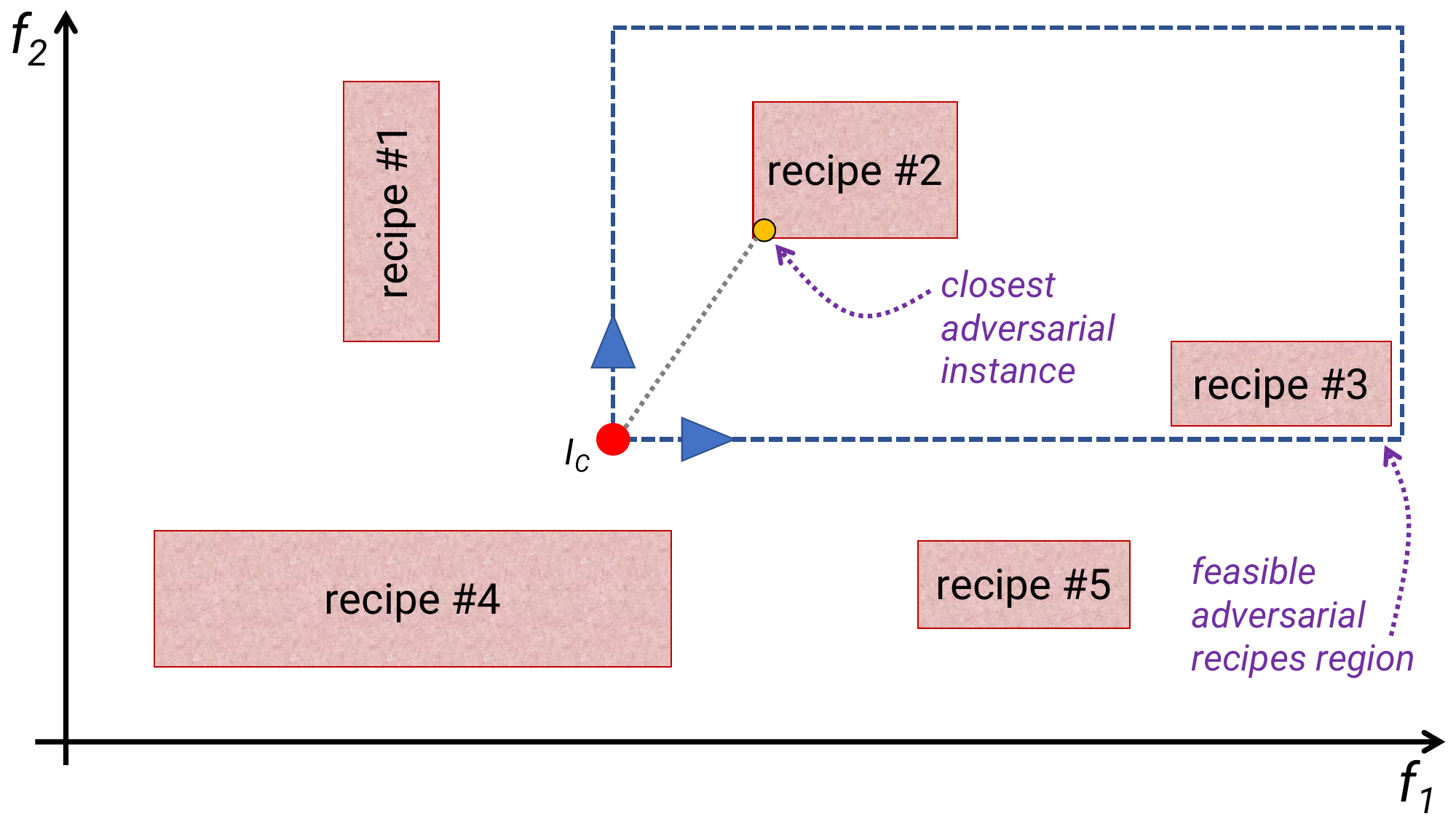}
		\vspace{-3mm}
		\caption[Finding feasible recipes' region and the closest adversarial instance given current state $I_c$]{Finding feasible recipes' region and the closest adversarial instance given current state $I_c$. For illustration purpose, only two features are shown.}
		\label{fig:closest_advrs}
	\end{figure}

	A well-known method to track network traffic and create a feature vector for the victim device is poisoning the ARP tables (discussed in \S\ref{sec:adiotack}) of local IoT devices and default gateway. Therefore, AdIoTack can sit in the middle of these entities by forwarding traffic between them. Fig.~\ref{fig:attack_prep} illustrates the monitoring phase before launching the attack. AdIoTack maintains a record (feature vector $I_c$) of the traffic features in the current epoch. Given $I_c$, AdIoTack would find feasible adversarial recipes towards the end of each epoch. A feasible recipe has to be an upper bound to vector $I_c$ (Fig.~\ref{fig:closest_advrs} shows it in a 2D space).
	If the victim displays network activities ($I_c$) exceeding a threshold ($f < \tau$) specified by a recipe, then $I_c$ cannot be projected to that recipe anymore simply because we cannot reduce the amount of traffic that already has been exchanged by the device but we can increase it.  
	Among all feasible recipes, AdIoTack selects the closest adversarial instance to $I_c$, minimizing the overhead packets to be injected. This strategy requires AdIoTack to be aware of the inference model's timing cycles, \ie when it is called for prediction. In the evaluation (\S \ref{sec:eval}), I will explain how AdIoTack has the chance to execute successful attacks even without syncing with cycles of the inference model.
	
	To construct recipes, I use linear search to find feasible instances with the time complexity of $\mathcal{O(}\mathcal{R}.\mathcal{F})$ in which $\mathcal{R}$ and $\mathcal{F}$ are the number of recipes and features, respectively. Though I do not encounter a challenge in terms of the time complexity of the linear search in the evaluation, there might be situations where a simple linear search becomes inefficient. 
	To achieve certain response time, one may attempt to optimize the search process and/or prune the generated recipes. Both of these objectives are beyond the scope of this chapter.

	I use an example in Table~\ref{tab:online_sample} to better clarify my method for executing a volumetric TCP SYN reflection attack on Google Chromecast. The top two rows indicate lower-bound and upper-bound feature values specified by the adversarial recipe for launching a successful attack. Note that the symbol ``*'' indicates unbounded features (no lower/upper limits). The third row shows the current feature vector of Google Chromecast over the last 60 seconds. The bottom row shows the adversarial instance comprising the intended volumetric attack traffic (bold cells) reflected by Google Chromecast and overhead packets (underlined cells) injected by AdIoTack.
	
	\begin{table*}[t!]
		\centering
		\small
		\caption{An example of a feasible SYN reflection attack on Google Chromecast.}
		\begin{adjustbox}{max width=\textwidth} 
			\renewcommand{\arraystretch}{1.4}
			\begin{tabular}{l|C{0.5cm} C{0.8cm} C{0.5cm} C{0.8cm} C{0.5cm} C{0.5cm} C{0.5cm} C{0.5cm} C{0.5cm} C{0.8cm} C{0.9cm} C{1cm} C{0.95cm} C{1.15cm} C{0.9cm} C{1.15cm}} \hline
				
				\hline
				\multicolumn{1}{L{1cm}|}{} & 
				\rotatebox{90}{$\downarrow$ {\myverb{DNS}} pkt} &
				\rotatebox{90}{$\downarrow$ {\myverb{DNS}} byte} &
				\rotatebox{90}{$\uparrow$ {\myverb{DNS}} pkt} &
				\rotatebox{90}{$\uparrow$ {\myverb{DNS}} byte} &
				\rotatebox{90}{$\downarrow$ {\myverb{NTP}} pkt} &
				\rotatebox{90}{$\downarrow$ {\myverb{NTP}} byte} &
				\rotatebox{90}{$\uparrow$ {\myverb{NTP}} pkt} &
				\rotatebox{90}{$\uparrow$ {\myverb{NTP}} byte} &
				\rotatebox{90}{$\uparrow$ {\myverb{SSDP}} pkt} &
				\rotatebox{90}{$\uparrow$ {\myverb{SSDP}} byte} &
				\rotatebox{90}{$\downarrow$ {\myverb{LAN}} pkt} &
				\rotatebox{90}{$\downarrow$ {\myverb{LAN}} byte} &
				\rotatebox{90}{$\downarrow$ {\myverb{WAN}}  pkt} &
				\rotatebox{90}{$\downarrow$ {\myverb{WAN}}  byte} &
				\rotatebox{90}{$\uparrow$ {\myverb{WAN}}  pkt} &
				\rotatebox{90}{$\uparrow$ {\myverb{WAN}}  byte}\\ \hline
				Condition ($>$) & 2 & 119 & 2 & 194 & * & 45 & * & * & 7 & 4353 & 94 & 9303 & 88 & 45568 & 10 & 5,650\\ 
				Condition ($\le$) & * & 1666 & * & * & * & * & 1 & * & 9 & 4553 & 148 & 13797 & * & * & * & *\\
				Current state ($I_c$) & 0 & 0 & 0 & 0 & 0 & 0 & 0 & 0 & 0 & 0 & 0 & 0 & 46 & 125857 & 51 & 5752\\
				Adversarial instance & \underline{3} & \underline{120} & \underline{3} & \underline{195} & \underline{1} & \underline{46} & \underline{0} & \underline{0} & \underline{8} & \underline{4354} & \underline{95} & \underline{9304} & \textbf{1000} & \textbf{74000} & \textbf{1000} & \textbf{74000}\\ \hline
				
				\hline
			\end{tabular}
		\end{adjustbox}
		\label{tab:online_sample}
	\end{table*}
	
	\begin{table}[t!]
		\centering
		\small
		\caption{Spoofed metadata for each feature.}
		\begin{adjustbox}{max width=0.80\textwidth}
			\renewcommand{\arraystretch}{1.5}
			\begin{tabular}{l|C{0.8cm}C{0.8cm}cccc} \hline
				
				\hline		
				Feature & src MAC & dst MAC & src IP & dst IP & src Port & dst Port \\ \hline
				$\downarrow$ DNS & GW & VIC & * & VIC & 53 & * \\
				$\uparrow$ DNS & VIC & GW & VIC & * & * & 53 \\ 
				$\downarrow$ NTP & GW & VIC & * & VIC & 123 & * \\ 
				$\uparrow$ NTP & VIC & GW & VIC & * & * & 123 \\ 
				$\uparrow$ SSDP & VIC & * & VIC & * & * & 1900 \\ 
				$\downarrow$ LAN & * & VIC & LAN IP & VIC & * & * \\ 
				$\downarrow$ WAN & GW & VIC & WAN IP & VIC & * & * \\
				$\uparrow$ WAN & VIC & GW & VIC & WAN IP & * & * \\ \hline
				
				\hline
			\end{tabular}
		\end{adjustbox}
		\label{tab:spoof}
	\end{table}
	
	To successfully launch the adversarial volumetric attack, AdIoTack needs to inject certain spoofed packets (the overhead traffic specified by the closest recipe) on behalf of the victim, so that the inference model does not detect any deviation in the behavior of the victim IoT device.
	Table~\ref{tab:spoof} summarizes certain packet metadata fields to be modified along with their spoofed value for a given set of traffic features in the specific model I studied. GW and VIC stand for gateway and victim, respectively. WAN IP and LAN IP respectively refer to external (public) and internal (private) IP address. Also, ``*'' highlights a wildcard value for the respective field.
	
	\textbf{Practical Challenges of Network-Based Attacks:} I discuss a few practical challenges to overcome for successful execution of adversarial attack instances in what follows:
	(i) in order to intercept bidirectional traffic exchanged between the victim and its local network, AdIoTack needs to employ Gratuitous ARP replies to send a broadcast spoofed ARP reply (on behalf of the victim device) to all local devices. That way, all local traffic destined to the victim device is forwarded to AdIoTack; and, 
	(ii) there might be cases whereby AdIoTack and victim may not share the same network interface (\eg one is wired and the other one is wireless). In that case, AdIoTack will have to send the spoofed traffic via the default gateway, which will disrupt the MAC-learning tables (wireless vs. wired) of the default gateway's interfaces due to the spoofed source MAC address in those packets (Table~\ref{tab:spoof}) -- this can lead to mis-forwarding of other packets to/from the victim (wireless LAN instead of wired LAN). Disrupting the default gateway's MAC tables results in dropping future packets with the same MAC address, this time as the destination field. To avoid this, AdIoTack sends a specific {\myverb{ICMP}} packet with destination broadcast MAC address ({\myverb{ff:ff:ff:ff:ff:ff}}) and victim's IP address. 
	The victim's reply will revert the MAC tables to their correct state.
	
	
	\section{Evaluation Results}
	\label{sec:eval}
	This section evaluates the performance of AdIoTack in its two phases, namely offline adversarial learning, and online adversarial execution. I begin by training three well-known decision tree ensemble models namely Random Forest, Gradient Boosting, and AdaBoost classifiers.
	
	
	

	\subsection{Multi-class Inference Models}
	\label{sec:infer model}
	Inspired by \cite{Arunan2020a}, I consider flow-level features from network traffic, periodically computed over one-minute windows. That way, no need for inspecting packet payloads.  
	Traffic features include statistics (packet and byte counts) of $\downarrow${\myverb{DNS}}, $\uparrow${\myverb{DNS}}, $\downarrow${\myverb{NTP}}, $\uparrow${\myverb{NTP}}, $\uparrow${\myverb{SSDP}}, $\downarrow${\myverb{LAN}}, $\downarrow${\myverb{WAN}}, and $\uparrow${\myverb{WAN}}; where $\uparrow$ indicates upstream traffic from each IoT device, and $\downarrow$ highlights downstream traffic to each IoT device. This means a total of 16 features.
	For training and testing, I use two datasets collected from the testbed (comprising nine IoT devices) during Feb-May 2020. The full dataset merely contains 954,384 benign instances with no trace of attacks (malicious instances). I use data from February, March, and the first half of April for training (i.e., 668,415 instances) and the second half of April and May for testing (i.e., 285,969 instances). Training based on “only benign” instances helps the model tighten and purify its acceptable boundaries to devices' expected behaviors (a finite set) without crippling the model by instances of certain known attacks (which can be infinite in practice). It has been shown \cite{Robin2010} that a model trained on known attacks will fall short in detecting unknown (``zero-day'') attacks.
	
	\begin{table}[t!]
		\centering
		\small
		\caption{Three ensemble models performance on the testing dataset.}
		\begin{tabular}{lcc} \hline
			\hline
			\textbf{Model} & \textbf{Accuracy} & \textbf{False positive} \\ \hline 
			Random Forest & 96\% & 9\% \\ 
			Gradient Boosting & 91\% & 0.1\% \\ 
			AdaBoost & 99\% & 5\%\\ \hline
			\hline
		\end{tabular}\label{tab:modelsPerformance}
	\end{table}

\begin{table}[t!]
	\centering
	\small
	\caption{Expected classification score ($\mu$ and $\sigma$) of the Random Forest model obtained from training.}
	\begin{tabular}{l|cc} \hline
		
		\hline
		& \multicolumn{2}{c}{Benign} \\ 
		IoT class & $\mu$ & $\sigma$ \\ \hline 
		Amazon Echo & 0.94 & 0.10 \\ 
		Belkin motion & 0.96 & 0.06\\ 
		Belkin switch & 0.86 & 0.18 \\ 
		Chromecast & 0.90 & 0.16 \\
		Hue bulb & 0.99 & 0.01 \\ 
		LiFX bulb & 0.72 & 0.15\\ 
		Netatmo camera & 0.88 & 0.14 \\ 
		Samsung camera & 1.00 & 0.002 \\ 
		TP-Link switch & 0.86 & 0.10\\ \hline
		\hline
	\end{tabular}	\label{tab:conf}
\end{table}
	
	I use the Python Scikit Learn library to train multi-class classifiers by three representative ensemble decision-tree algorithms, namely Random Forest, Gradient Boosting, and AdaBoost, to predict device class label. These models provide us with a classification score used as a baseline threshold for detecting misbehavior. 
	Table~\ref{tab:modelsPerformance} shows the accuracy and false positive rate of each model.
	For detecting misbehaviors, I use classification scores' $\mu$ and $\sigma$ for each class of IoT device (shown in Table~\ref{tab:conf} for Random Forest model) obtained from the training dataset in such a way that if for a given instance the model's classification score for its predicted label is below $\mu-\sigma$ (of that label), the instance is considered as malicious.
	Table~\ref{tab:conf} shows the Random Forest model's classification score for each class of IoT devices in the testbed obtained from the training dataset. I use $\mu$ and $\sigma$ values obtained from the training for detecting IoT devices' misbehavior in such a way that if for a given instance the model's classification score for its predicted label is below $\mu-\sigma$ (of that label), the instance is considered as malicious.

	%

	\subsection{AdIoTack Offline Instances}
	\begin{figure}[t!]
		\centering
		\includegraphics[width=0.80\linewidth]{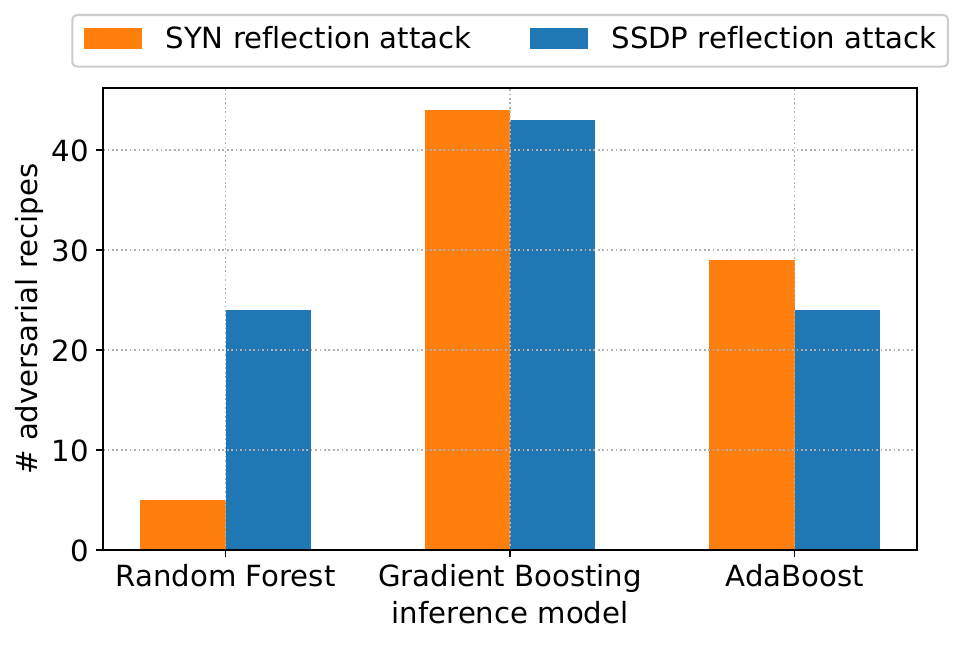}
		\caption{Number of generated adversarial recipes for three ensemble models for attack impact of 1000 packets.}
		\label{fig:recipe_count_model}
	\end{figure}

	In this part, I evaluate the performance of AdIoTack in offline adversarial learning with a view to launch two well-known network-based attacks, namely TCP SYN and SSDP reflection attacks.
	
	Let me begin by showing that my offline learning method is generalizable by applying it to the three popular ensemble models. Fig.~\ref{fig:recipe_count_model} indicates the number of adversarial recipes for SSDP and SYN reflection attacks per inference model. Random Forest seems to be more robust to adversarial attacks compared to the other two models, corroborating with the observations in \cite{Katzir2018}. Because of its relative robustness, I choose Random Forest for the rest of the experimental evaluation.

	\begin{figure}[t!]
		\centering
		\includegraphics[width=0.80\linewidth,height=0.73\linewidth]{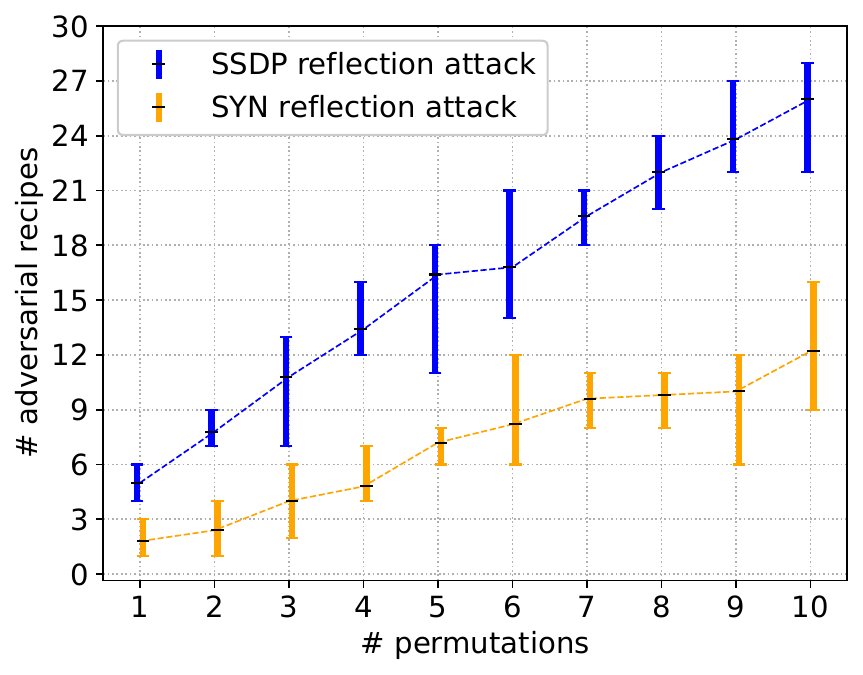}
		\caption{Number (min, max, avg) of generated adversarial recipes for different permutation counts for attack impact of 1000 packets.}
		\label{fig:recipe_permutation_count}
	\end{figure}
	
	{\vspace{5mm}}
	Now, I quantify the impact of the sequence order of trees in generating adversarial recipes. I define permutation as a random shuffle of the trees inside the ensemble model, which we need to iterate through them in the same order as the permutation suggests. I show that various permutations can result in different adversarial recipes. Fig.~\ref{fig:recipe_permutation_count} shows the min, max, and average number of unique recipes for different permutation counts by running the algorithm five times for each permutation value. For example, if the permutation count is three, I run the offline learning with three different permutations for each device over five rounds and count the number of unique recipes at each round. The plot shows an overall increasing trend which means as we try different permutations, there is a high chance of finding new recipes. Also, the plot suggests that SSDP reflection can generate more recipes than SYN reflection, in which the increasing trend seems to become smoother after three permutations. The average time taken for generating adversarial recipes per permutation is about 1.5 minutes for SSDP reflection and 2.2 minutes for SYN reflection attack. This is probably because four features (packet and byte count of flows $\downarrow${\myverb{WAN}} and $\uparrow${\myverb{WAN}}) are affected by SYN reflection attack, whereas only two features (packet and byte count for $\uparrow${\myverb{SSDP}}) are affected by the SSDP reflection attack. 

	\begin{figure}[t!]
		\centering
		\includegraphics[width=0.80\linewidth,height=0.73\linewidth]{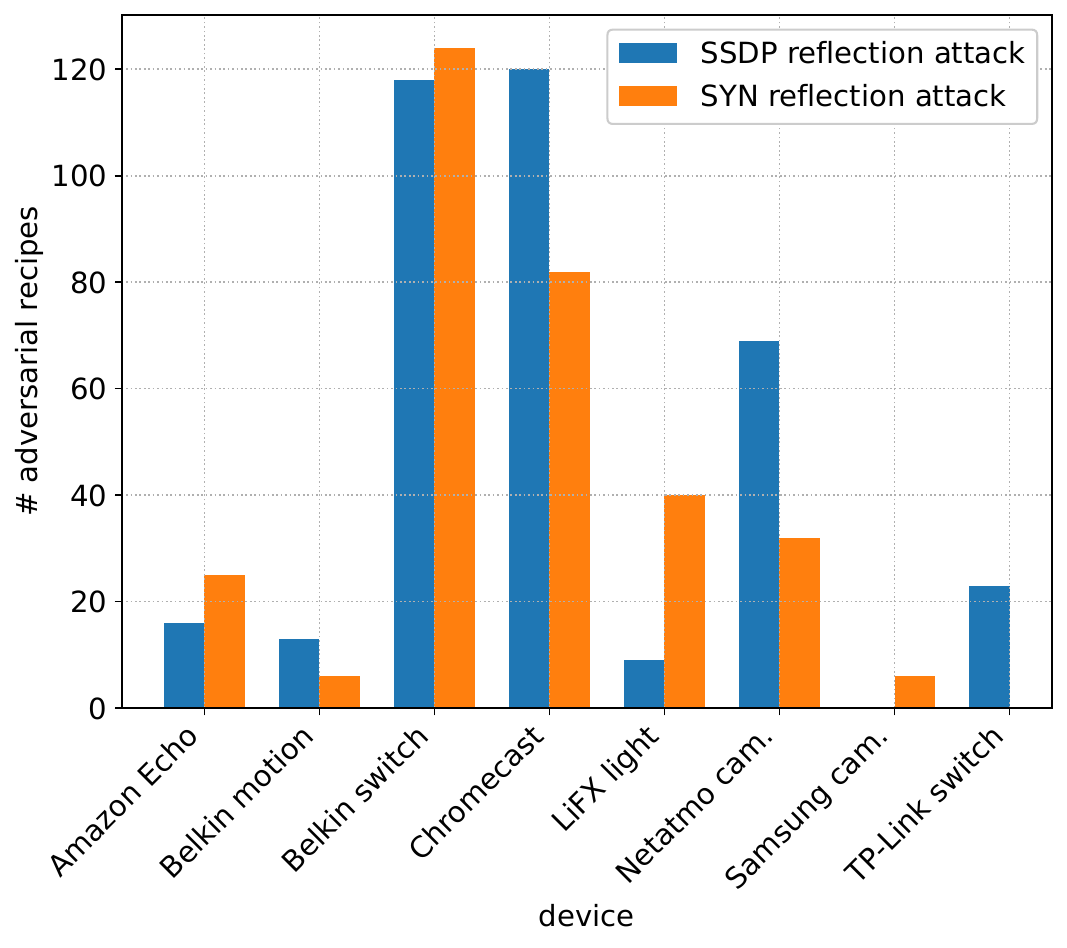}
		\\
		\caption{Number of generated adversarial recipes per IoT device for two representative attacks (result of a 20-permutation run).}
		\label{fig:dev_recipe_count}
	\end{figure}


	\begin{figure}[t!]
		\begin{center}
			\mbox{
				\subfigure[Amazon Echo.]{
					{\includegraphics[width=0.75\textwidth]{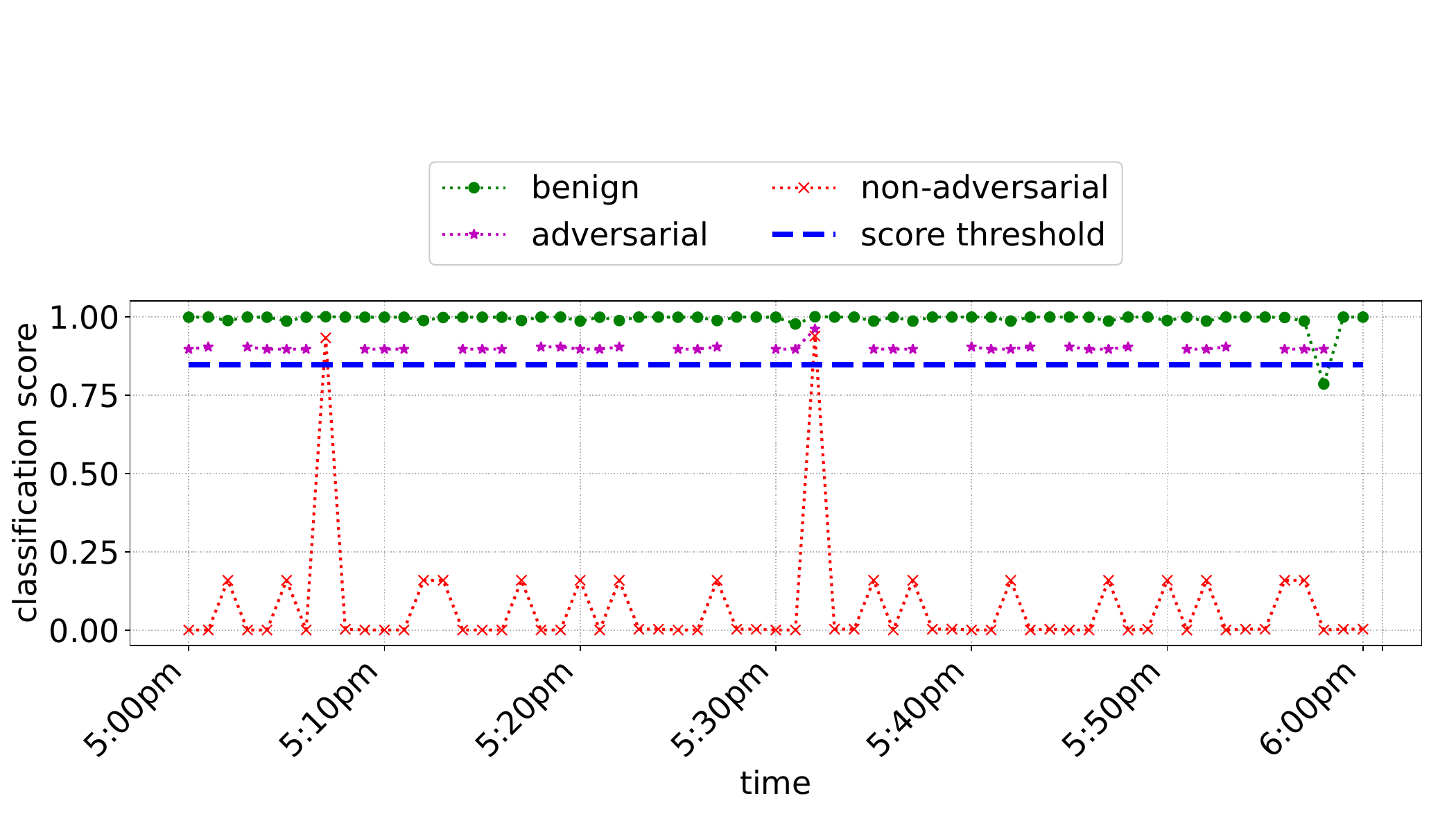}}\quad
				}
			}
			
			\mbox{
				\subfigure[Belkin switch.]{
					{\includegraphics[width=0.75\textwidth]{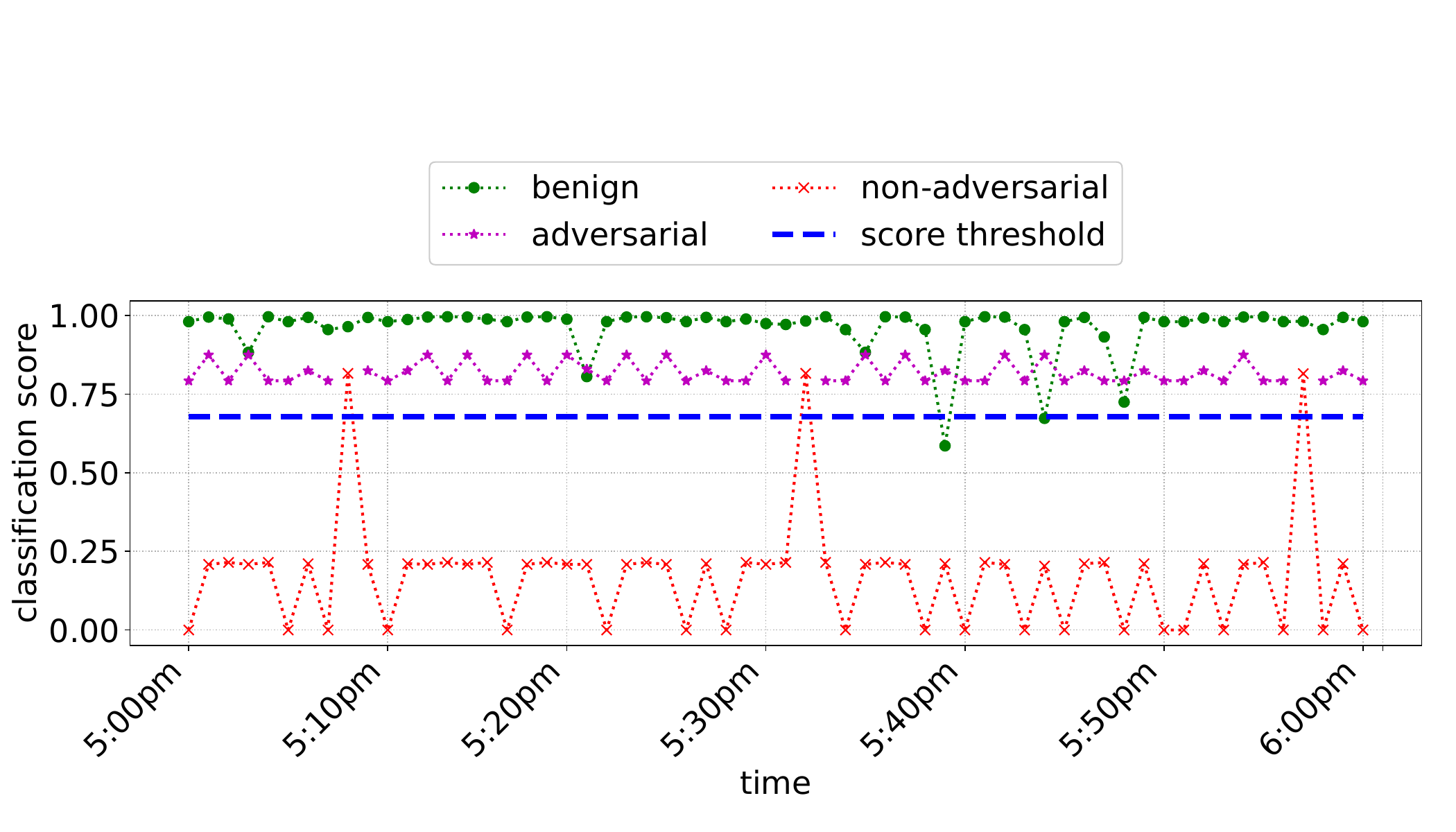}}\quad
				}
			}
			\caption[Time trace of model performance against replayed instances of two representative IoT devices]{Time trace of model performance against replayed instances of two representative IoT devices: (a) Amazon Echo, and (b) Belkin switch, in three scenarios: benign (green), synthetic adversarial SYN reflection attack (purple) and synthetic non-adversarial SYN reflection attack (red).}
			\label{fig:replay_attack}
		\end{center}
	\end{figure}

	Fig.~\ref{fig:dev_recipe_count} illustrates the number of generated recipes for each IoT device in the testbed by running the offline learning algorithm with 20 permutations. Belkin switch and Chromecast in both types the attacks. Samsung smart camera has no recipe for SSDP reflection, TP-Link has no recipe for SYN reflection attack, and Hue lightbulb has no recipe for both of the attacks, indicating tight constraints for those classes.

	Before experimenting with the adversarial recipes in an operational network of IoT devices, I evaluate the performance of my inference model and efficacy of offline adversarial instances by replaying various unseen datasets in three scenarios, namely: (1) a purely benign dataset is expected to receive high classification scores from the model, (2) syntactically changed features of the benign instances (to represent non-adversarial SYN reflection attack) are expected to receive low scores from the model (highlighting a detection), and (3) synthetically changed features of the benign instances (to represent adversarial SYN reflection attack) are expected to receive high scores from the model (highlighting a miss).

	
	Fig.~\ref{fig:replay_attack} shows the results for two representative IoT devices over an hour. As expected, almost all benign instances are classified with a high score (\ie above the threshold shown by dashed blue line), highlighting the expected performance under normal situations (benign traffic). The dotted red curve shows the model's score for non-adversarial attack instances, which the model detects a majority of them (receiving scores below the threshold). Finally, the purple curve highlights the model's inability to detect adversarial attacks (\ie giving scores above the threshold, meaning benign). In some epochs like at 5:03pm for Amazon Echo and 5:08pm for Belkin switch, there is no recipe in the feasible region to project the current instance, thus missing data points.

	\subsection{AdIoTack Online Evaluation}
	
	\begin{figure}[t!]
		\centering
		\includegraphics[width=0.95\linewidth]{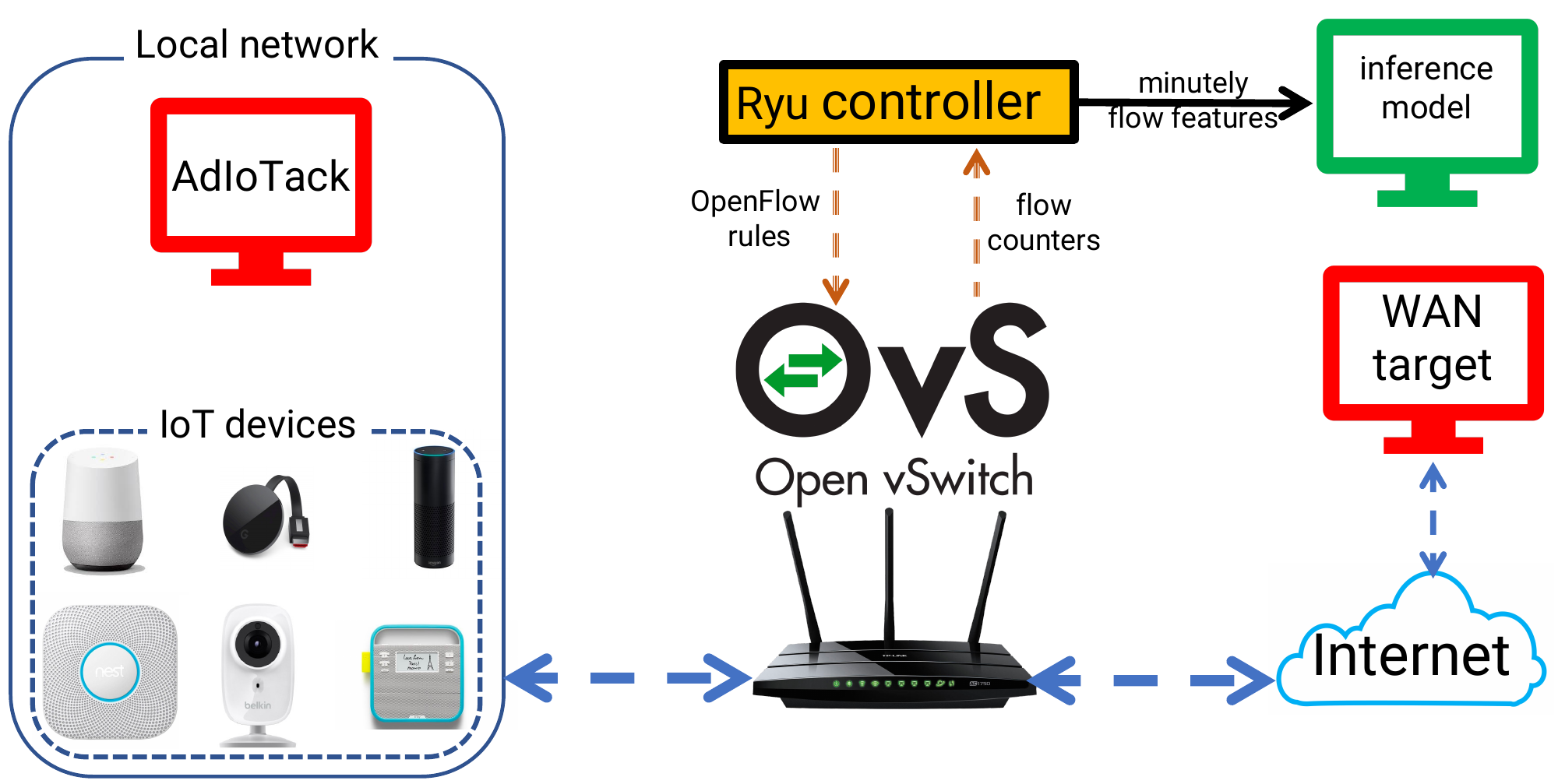}
		\\
		\caption{Prototype of AdIoTack.}
		\label{fig:setup}
	\end{figure}
	
	To demonstrate adversarial attacks in a realistic scenario, I deploy the inference model in a system that can classify traffic in real-time. It receives a stream of flow-based telemetry and predicts traffic features of individual connected devices every minute. To collect real-time telemetry, I use an SDN-enabled home gateway in my setup, shown in Fig.~\ref{fig:setup}. I installed Open Virtual Switch (OVS) on the gateway of the IoT network. Ryu, an open-source SDN controller, is installed on another machine that communicates with this OVS. The Ryu controller initializes the setup by sending a fixed set of flow rules specific to each IoT device on the network, of which the inference model needs the corresponding flow counters as features (features described in \S \ref{sec:infer model}). Every minute, the model receives flow counters from the controller and classifies individual devices. The model also detects unexpected behavior by way of classification score (explained in \S \ref{sec:infer model}).
		
	\begin{figure}[t!]
		\begin{center}
			\mbox{
				\subfigure[Belkin switch.]{
					{\includegraphics[width=0.43\textwidth]{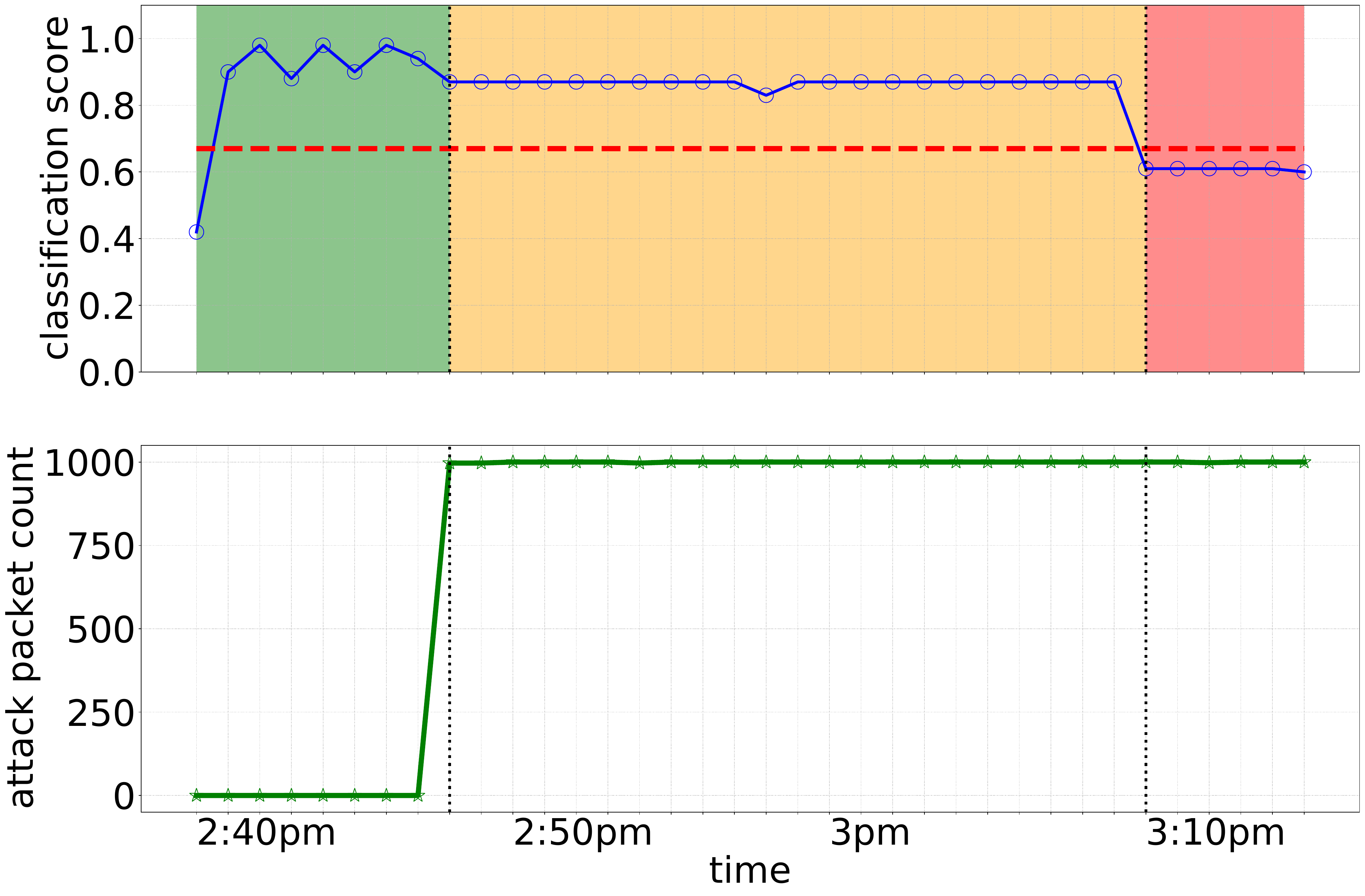}}\quad
				}
			}
			\hspace{-8mm}
			\mbox{
				\subfigure[Chromecast.]{
					{\includegraphics[width=0.43\textwidth]{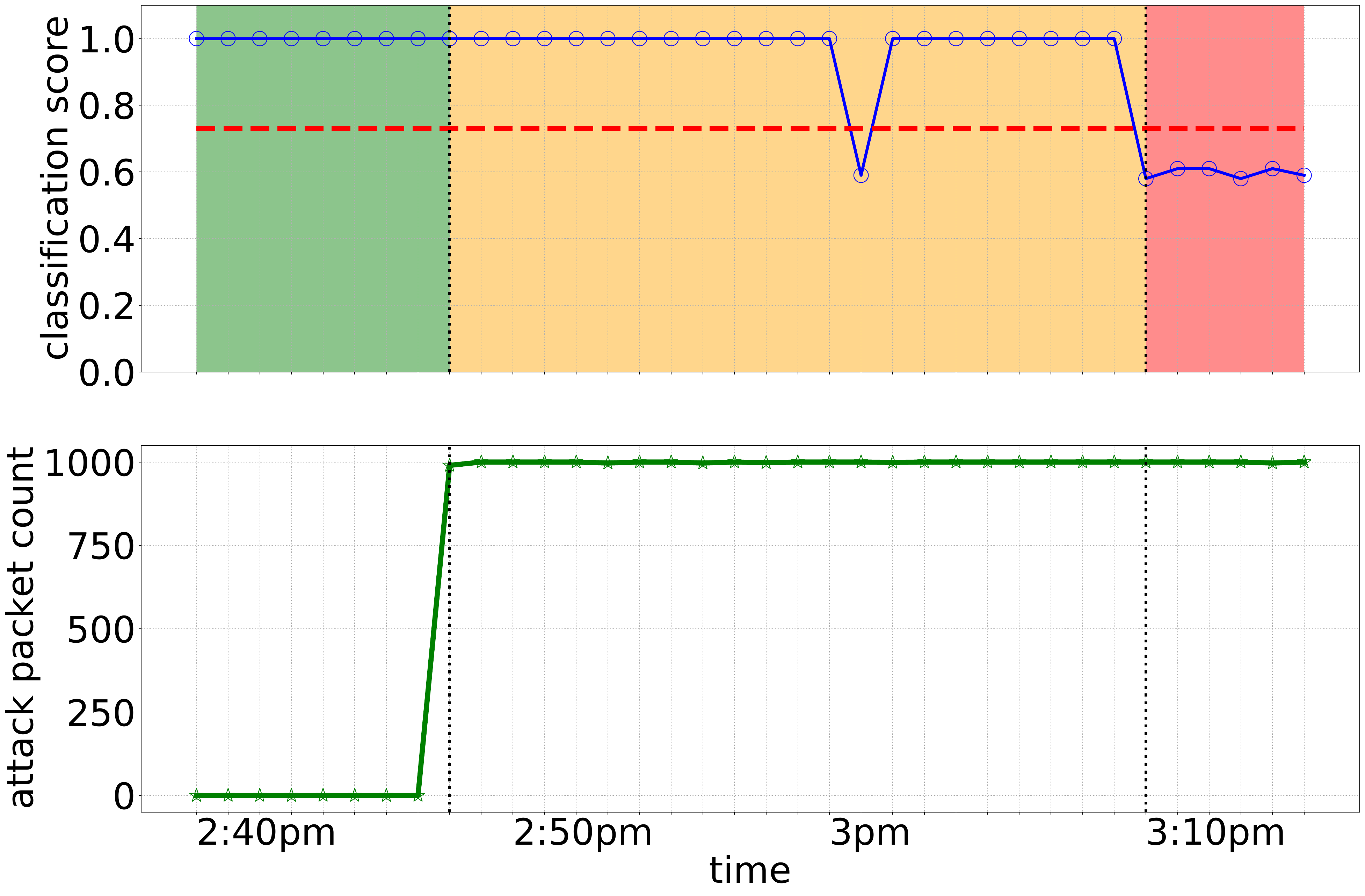}}\label{fig:ExecChrom}\quad
				}
			}
			\mbox{
				\subfigure[Netatmo camera.]{
					{\includegraphics[width=0.43\textwidth]{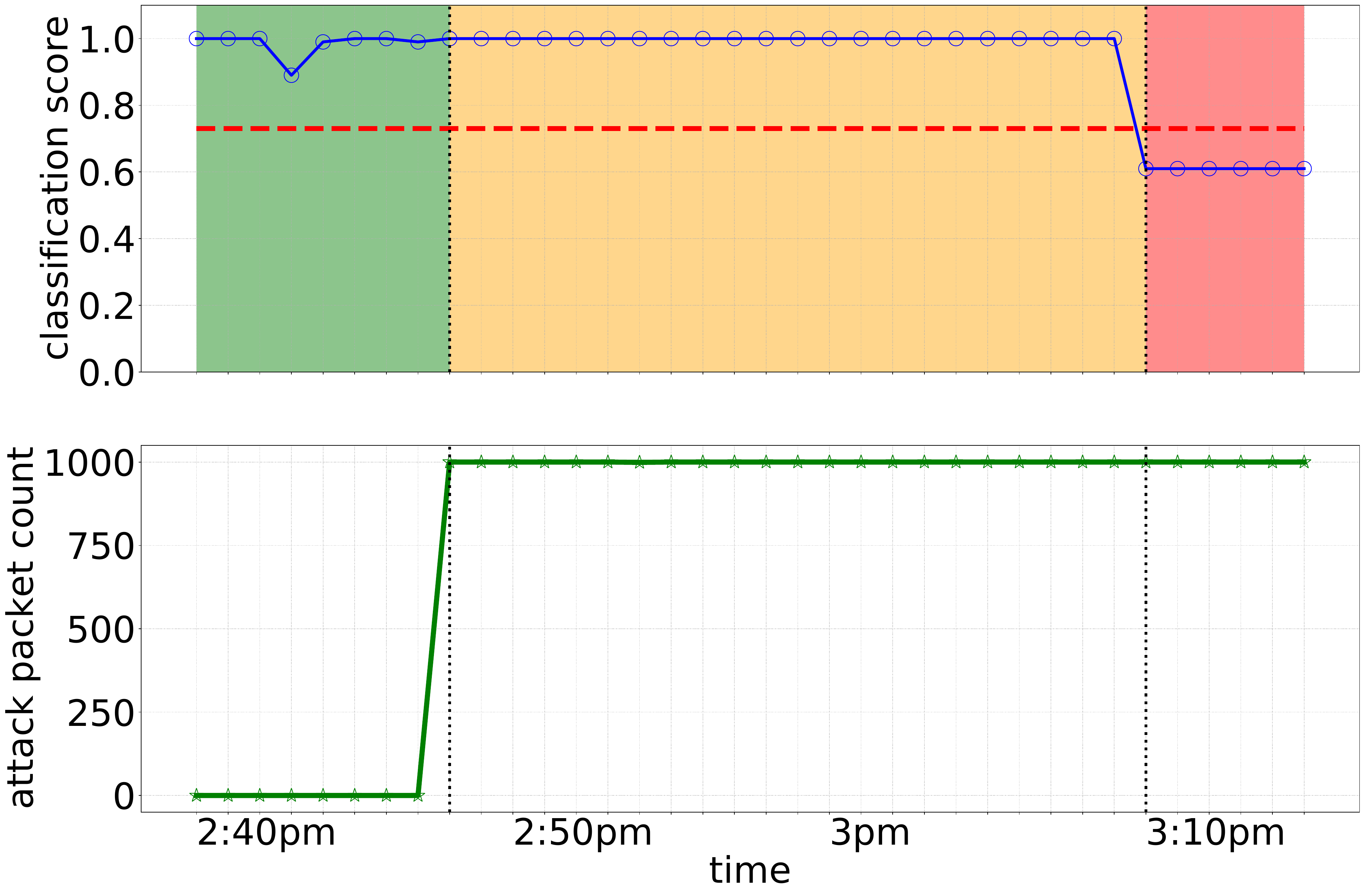}}\quad
				}
			}
			\hspace{-8mm}
			\caption{Time trace of model performance and attack traffic on three IoT devices in three scenarios: normal operation (green), adversarial attack (orange) and non-adversarial attack (red).}
			\label{fig:time_trace}
		\end{center}
	\end{figure}
	
	Fig.~\ref{fig:time_trace} shows the performance of AdIoTack in the online execution phase for three representative devices with SYN reflection attack. 
	The entire evaluation lasted for 35 minutes with three stages: between 2:40pm and 2:48pm (highlighted in green), no attack was executed to observe the behavior of the inference model in normal operation; between 2:48pm and 3:10pm (highlighted in orange), I performed adversarial SYN reflection attacks in each epoch; lastly, between 3:10pm and 3:15pm (highlighted in red), I performed non-adversarial SYN reflection attack with the same impact of the adversarial ones launched in the middle stage. The upper subplots show the inference model's classification score at the end of each epoch. The score threshold of each class is shown by a constant dashed red line. The lower subplots show the number of attack packets received by the ultimate target machine (victim server) residing on the WAN interface of the gateway (outside the local IoT network).

		It can be seen that the inference model displays an acceptable performance under normal operation (\ie pure benign traffic, without any attack). Moving to the orange regions, we observe that all adversarial reflection attacks are successful (classification scores above the threshold), except during an epoch (at 3:01 pm) for Chromecast in Fig.~\ref{fig:ExecChrom} -- this sudden drop in the score was due to a short network disruption that caused packet loss in the traffic sent by AdIoTack.
		Lastly, focusing on the red regions, the model gives scores lower than the expected threshold to non-adversarial instances across the three representative devices.
	
	
	During the online evaluation, I found that it is impossible to execute any adversarial attack on three devices, namely Samsung smart camera, Philips Hue lightbulb, and Amazon Echo. For Samsung smart camera, the reason is that no feasible adversarial recipe was found for $I_c$ (device's current feature vector) to be projected on as $I_c$ during my experiment violated the upper bound of the rule on $\uparrow${\myverb{SSDP}} feature. For the other two devices, the reason relies on the network aspect. Philips Hue lightbulb responds to the corrective ping message sourced by AdIoTack. However, the response does not revert the gateway's MAC table; probably because Philips Hue is not a standalone device and relies on a separate bridge for communication. Amazon Echo did not respond to the ping message, hence the disrupted MAC table issue stopped us from launching the attack.
	
	\begin{figure}[t]
		\begin{center}
			\mbox{
				\subfigure[Belkin motion sensor.]{
					{\includegraphics[width=0.445\textwidth]{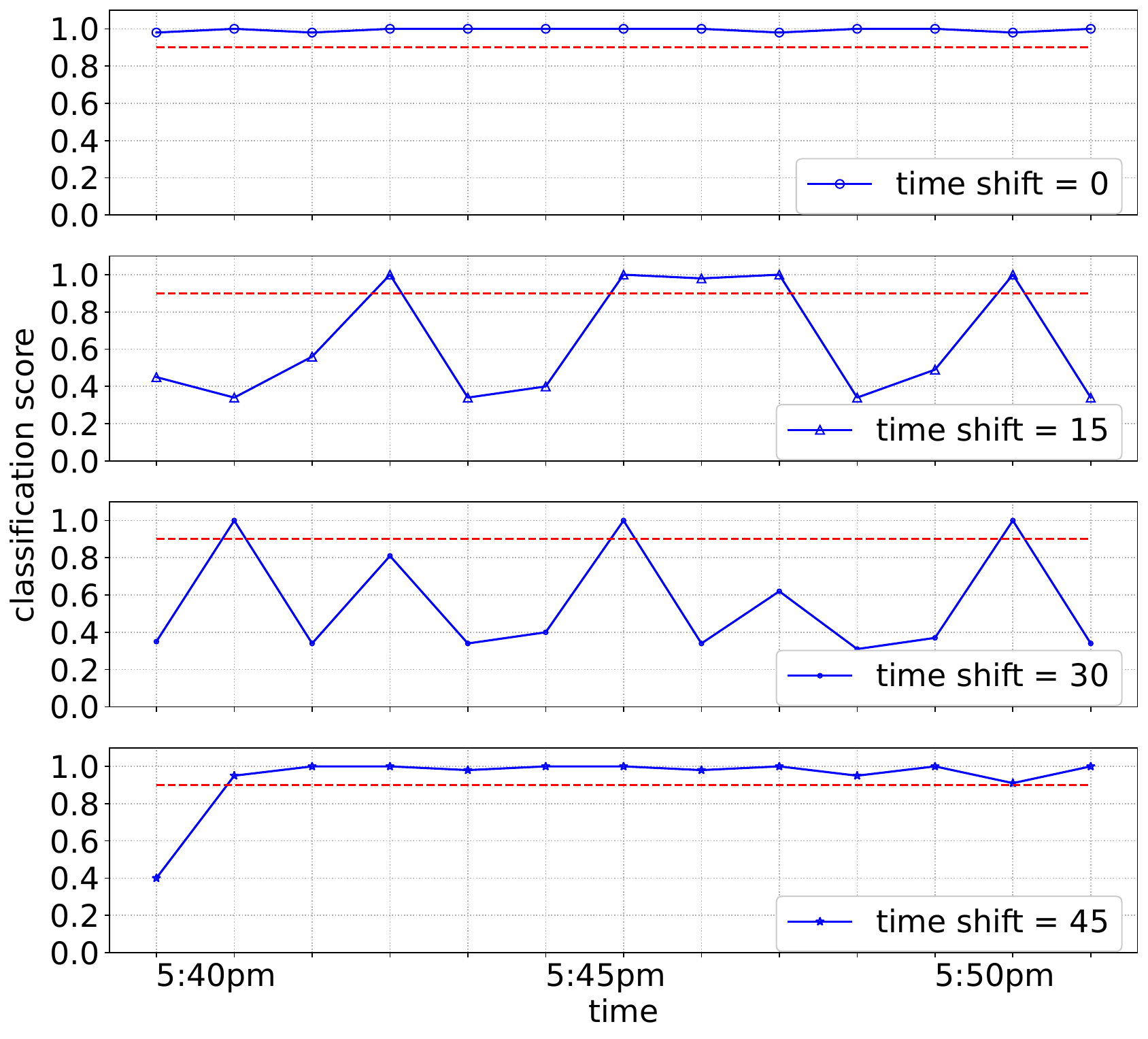}}\quad
				}
			}
			\hspace{-8mm}
			\mbox{
				\subfigure[Chromecast.]{
					{\includegraphics[width=0.445\textwidth]{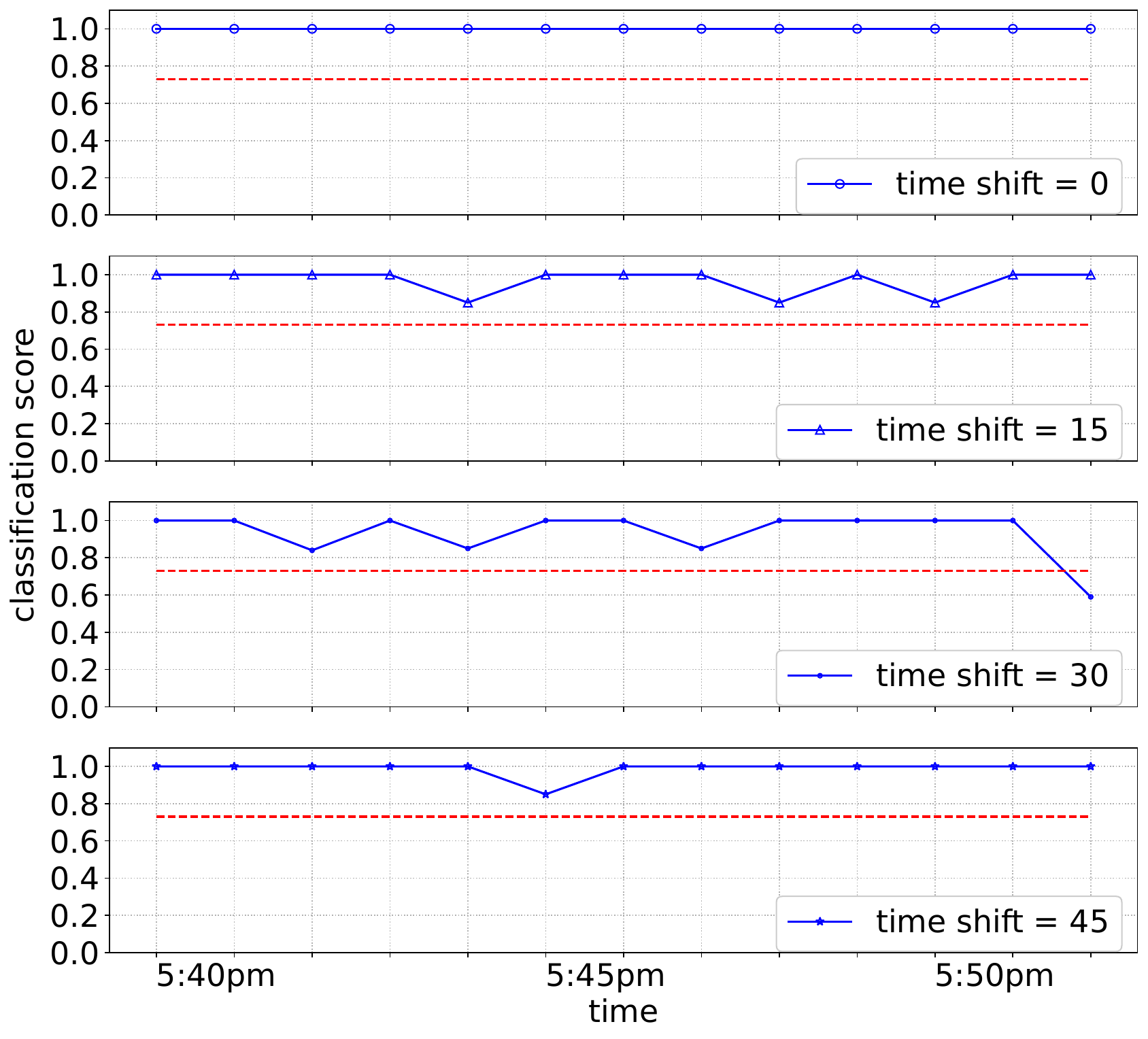}}\quad
				}
			}
			\caption{Effect of time-shift for two IoT devices.}
			\label{fig:time_shift}
		\end{center}
	\end{figure}
	
	During this evaluation, I assumed that AdIoTack knows the inference cycles of the model \ie when the model fetches the flow counters from the SDN controller and classifies the traffic. This is a valid assumption as I already declared my work to be a white-box approach. Still, I further evaluate AdIoTack online execution in scenarios where inference cycles are unknown. Fig.~\ref{fig:time_shift} illustrates this evaluation for Belkin motion sensor and Chromecast. I launched adversarial attacks on these devices with four different timings. Time-shift of $T$ means that the attacker is $T$ seconds ahead of the inference model. I need to note that the time window in this evaluation is 60 seconds meaning that the model becomes online every 60 seconds.
	
	Unsurprisingly, Fig.~\ref{fig:time_shift} shows that adversarial attacks with the time shift of zero are $100$\% successful in bypassing the model. Fig.~\ref{fig:time_shift} shows that for Chromecast, time-shift almost has no effect on the success of adversarial attacks; however, for Belkin motion sensor, it has significant effects. The results suggest that a time shift of 30 seconds leads to the minimum success rate ($23$\%), but when it gets closer to the inference model's time cycle, \ie time shift of 45 seconds, the success rate significantly rises ($92$\%). The key takeaway is that regardless of when AdIoTack launches the attack, there is always a chance to bypass the model.

	\section{Refining Resilience of Decision Trees Against Adversarial Volumetric Attacks}\label{sec:patching}
	
	I have so far highlighted the vulnerability of decision tree-based models against systematically crafted adversarial attacks. In this section, I develop a method called \textit{patching} to refine the model post-training, making decision trees robust against adversarial volumetric attacks.
	
\begin{figure}[t!]
		\centering
		\includegraphics[width=0.92\linewidth]{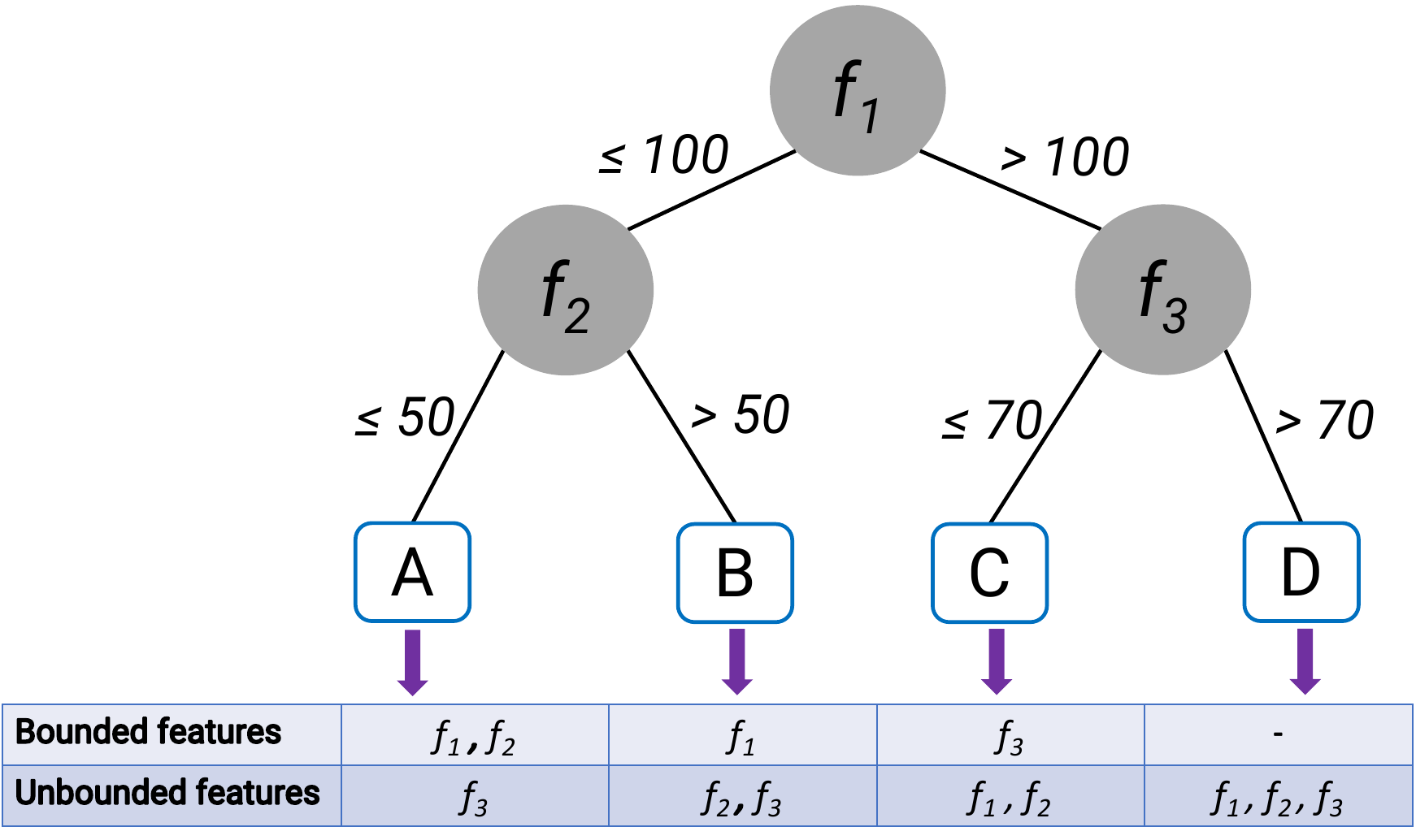}
		\\
		\caption{Bounded and unbounded features for each leaf in a decision tree.}
		\label{fig:vulnerable_tree}
	\end{figure}

	{\vspace{5mm}}
	Decision trees trained purely by benign traffic instances are inherently vulnerable (given their structure) to volumetric attacks. Fig.~\ref{fig:vulnerable_tree} illustrates an illustrative decision tree with four leaves and three decision nodes on three features. For each leaf, I define \textit{bounded} and \textit{unbounded} feature sets. For a given leaf $l$, its bounded feature set includes feature $f$, if inside the path from the tree's root to $l$, there is at least one decision node on $f$ which must take the left branch (\ie $f \le \tau$) to reach $l$. Otherwise, if there is no such a decision node for $f$, I consider $f$ as an unbounded feature for $l$. For example, in Fig.~\ref{fig:vulnerable_tree}, consider the leaf with label `{\myverb{C}}'. To reach this leaf from the root, the following conditions must be met: $f_1>100$ and $f_3\le70$. Thus if an instance has $f_1=+\infty$, $f_2=+\infty$, and $f_3=70$, it still reaches the leaf with no issue as there is no upper boundary check for $f_1$ and there is no upper or lower boundary check for $f_2$ at all.

	In the literature, there exist methods \cite{Chen2019, Calzavara2020} for developing robust decision trees against adversarial attacks. The current techniques manipulate the training process of decision trees by changing the original model to make sure if an adversarial attack happens, the model still gives the correct output label. This is desirable in image classification which requires a given set of adversarial instances to be provided before refining the model, \ie the model still might be vulnerable to another set of adversarial instances. That way, they sacrifice the prediction accuracy for benign instances, in order to make decision trees robust in adversarial scenarios. Also, given their primary focus is on image recognition, they do not have a notion of volumetric attacks, and thus, unbounded features remain loose. My method, however, focuses on a different problem (\ie volumetric cyber attacks), and aims to refine a trained model. Therefore, it does not interfere with the training process and does not require re-training. My method receives a trained decision tree-based model and then patches the unbounded features with no dependency on any adversarial recipes. The patching process is carried out offline, but the refined model is capable of detecting adversarial attacks in real-time (online detection). If the training dataset gives a correct representation of devices' traffic characteristics by at least capturing the maximum traffic volume for each class correctly, the patched model will have the exact same accuracy/false positive over benign instances as the original model. However, if there are data instances in the testing dataset that have higher traffic features' value than the limits which were captured from the training dataset, my method would flag them an as malicious (\ie higher false positives than the original unpatched model).

	\begin{figure}[t!]
		\centering
		\includegraphics[width=0.85\linewidth]{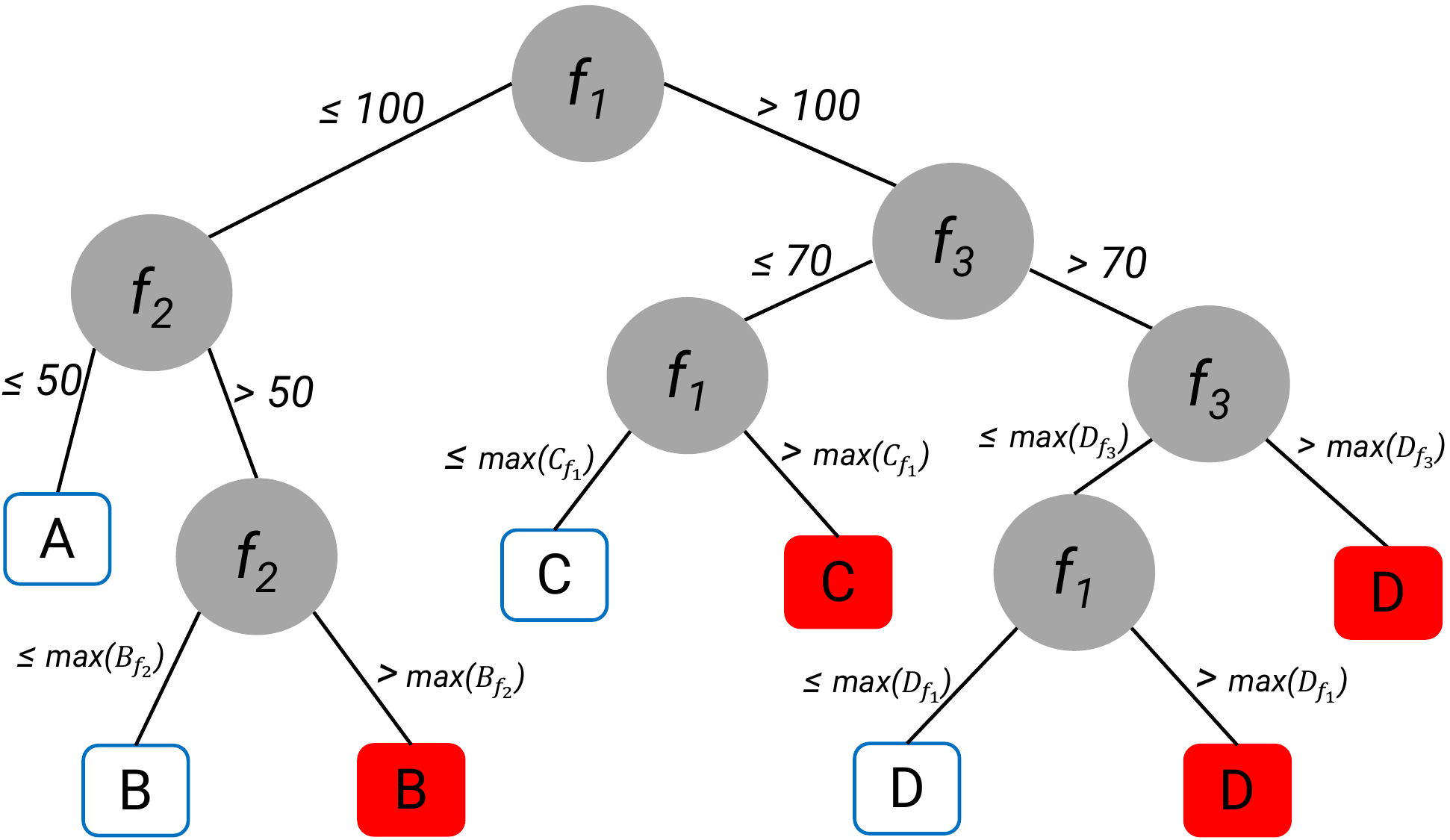}
		\caption{Essential patching performed on individual leaves of the vulnerable decision tree from Fig.~\ref{fig:vulnerable_tree}.}\label{fig:patched_tree}
	\end{figure}
	
	I define a patch for feature $f$ on leaf $l$ as follows:
	\begin{gather*}
		\label{eq:patch}
		patch_{l, f} = f \le max(l_f)
	\end{gather*}
	
	This patch can be seen as a new decision node added just before reaching the leaf, that checks the upper bound of $f$ for $l$. The upper bound is calculated based on the training dataset used for training the original model. As an example, based on the training dataset, Amazon Echo sends a maximum of four DNS packets every minute; thus, $max(Amazon~Echo_{\uparrow DNS~pkt}) = 4$. Any instance $x$ reaching leaf $l$, which $x_f>max(l_f)$, goes to a new leaf that yields $x$ as a malicious instance.
	
	There are two noteworthy points: (1) Because IoT devices send/receive a low volume of traffic and have repetitive behavioral patterns, individual devices' maximum traffic volume can be captured \cite{Arunan2017}. However, this could be challenging for personal computers and smartphones which their traffic volume highly depends on users' activities. (2) My method reduces the impact of adversarial volumetric attacks significantly. However, low-impact attacks (\ie within the normal behavior range of IoT devices) can still bypass the model.

	There are two scenarios where feature $f$ can become unbounded for leaf $l$: (1) When there is a decision node on $f$, but its right branch (\ie $f > \tau$) is taken on the path from the tree's root to $l$ (\eg $f_1$ for leaf `{\myverb{C}}' in Fig.~\ref{fig:vulnerable_tree}); and (2) When there is no decision node on $f$ through the path at all (\eg $f_2$ for leaf `{\myverb{C}}' in Fig.~\ref{fig:vulnerable_tree}). I develop two patching techniques to address these two scenarios. \textit{Essential Patching} solves the first problem by traversing a given vulnerable decision tree and patching leaves with unbounded features through paths from the root. Fig.~\ref{fig:patched_tree} shows the result of essential patching performed on the tree from Fig.~\ref{fig:vulnerable_tree}. Adversarial instances will be routed to the red leaves, which results in true detection of the attack.
	
\begin{figure}[t!]
		\centering
		\includegraphics[width=0.9\linewidth,height=0.75\linewidth]{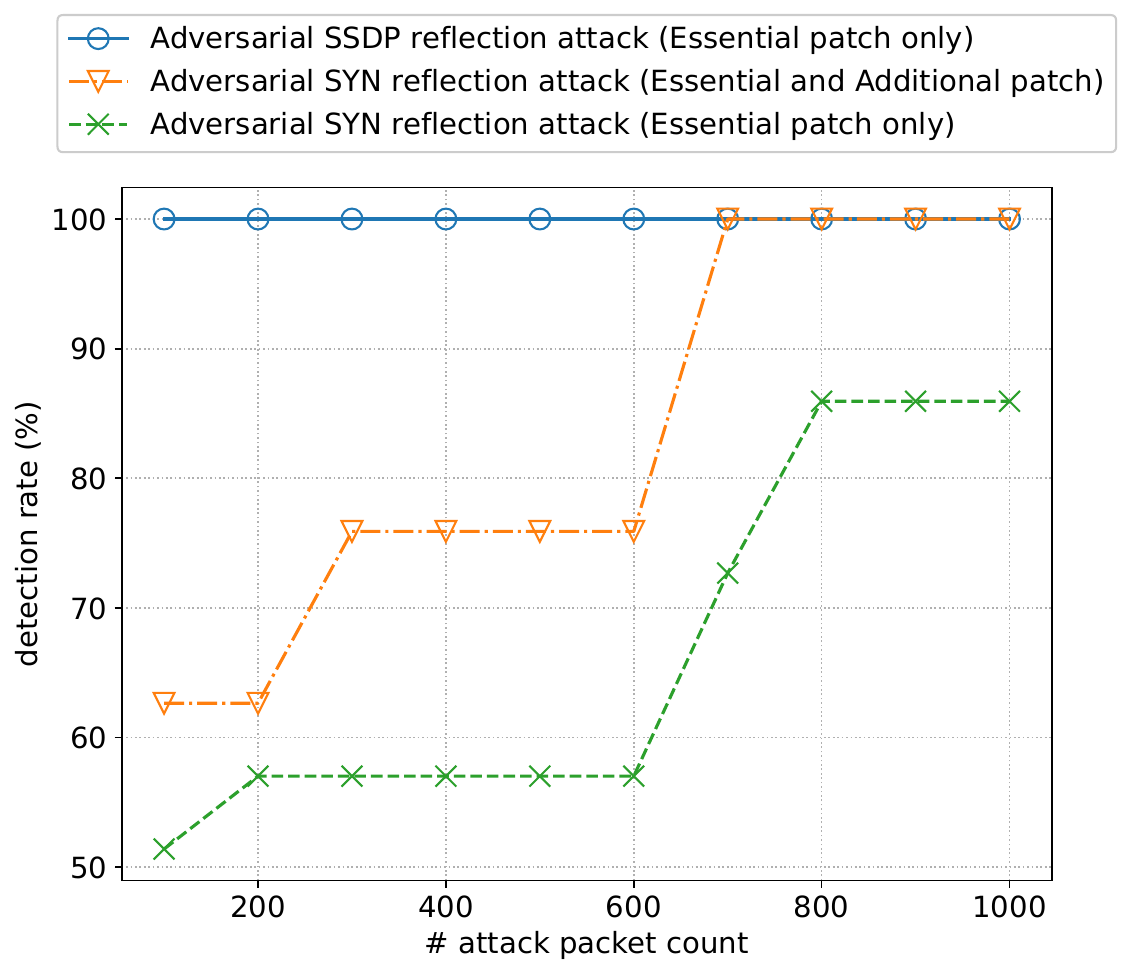}
		\caption{Detection rate of Essential and Additional patching for adversarial SYN and SSDP reflection attacks.}
		\label{fig:detection_rate}
	\end{figure}

	Although essential patching is necessary to fix some unbounded features, it does not cover the second scenario where some features do not even exist on some paths. For example, leaf ``{\myverb{A}}'' in Fig.~\ref{fig:patched_tree}, still can be targeted with volumetric attack on $f_3 = +\infty$. In fact, some adversarial attacks can be captured solely with essential patching (\eg SSDP reflection attack in Fig.~\ref{fig:detection_rate}). To overcome this problem, I define \textit{Additional Patching} which patches all leaves for a given feature. This way, regardless of the trees' structure, I make sure all leaves are patched against adversarial volumetric attacks on the desired features. The time complexity of the patching technique (essential and additional) is $\mathcal{O(}\mathcal{T}.\mathcal{N}.\mathcal{F})$, which is proportional to the number trees ($\mathcal{T}$), maximum number of nodes in the trees ($\mathcal{N}$), and total number of features ($\mathcal{F}$).
	
	Additional patching can be done in different ways; for example, one may choose to patch all leaves for all possible features to create maximum robustness,  making the model complex and slow. Another one may use expert knowledge to select specific vulnerable features to limit the complexity of the model while making it robust against attacks over those features that are more likely to be targeted by attackers.
	After applying the essential patching to the original model, I quantify the accuracy and false-positive over the testing dataset and they are 96\% and 0.09\%, respectively, which are exactly the same as the original model.
	
	Fig.~\ref{fig:detection_rate} shows the performance of my patching methods over adversarial SSDP and SYN reflection attacks with a range of attack impacts (100 to 1,000 packet count). Note that these attack instances are created based on adversarial recipes, bypassing the unpatched model, as shown in the previous section. Essential patching is sufficient to detect all SSDP reflection attacks while detecting $86$\% of the high-impact SYN reflection attacks. Having an additional patch over $\uparrow$ {\myverb{WAN}} packet count feature increases the detection rate for low, medium, and high impact attacks; and detects all attacks with an impact greater than $636$ packets. The average impact of SYN reflection attacks that can still bypass the model is $117$ packets across the IoT devices, with the maximum impact of $636$ packets for Chromecast. Compared to the vulnerable model, which has no limit over the attack packet count (\ie unlimited impact), this impact shows a significant improvement in terms of robustness.
	
	\begin{figure}[t!]
		\centering
		\includegraphics[width=0.99\linewidth]{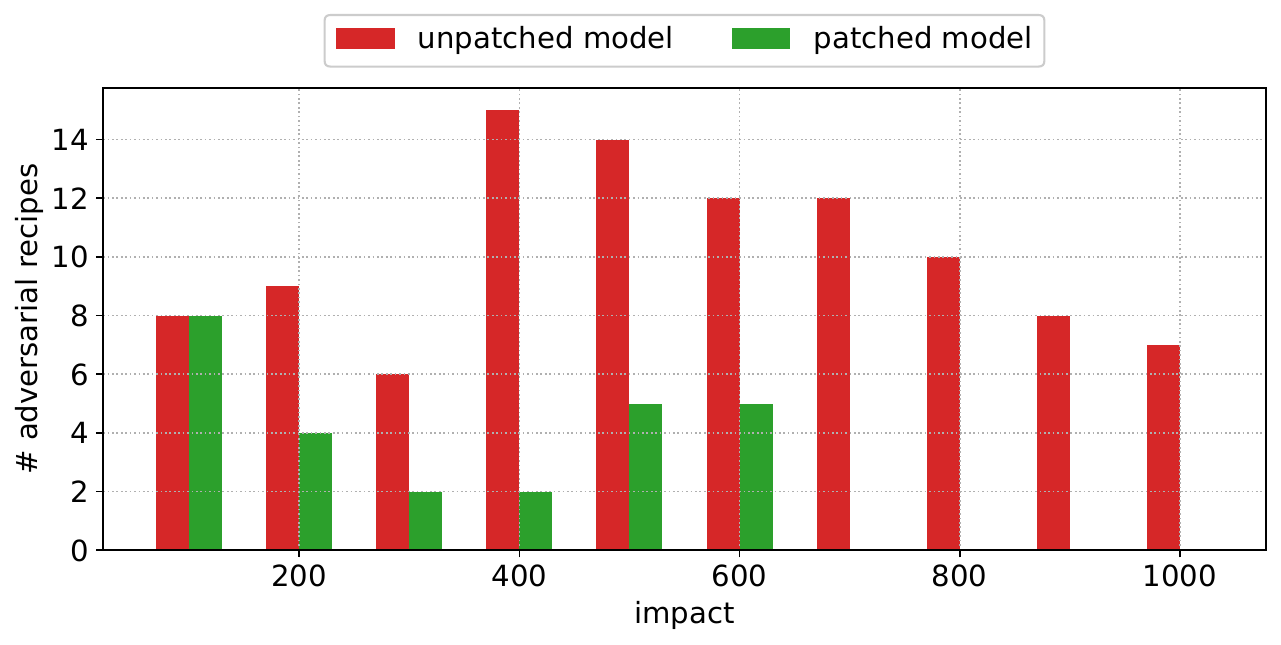}
		\caption{Number of recipes generated for SYN reflection attack before and after patching the model.}
		\label{fig:patched_vs_unpatched}
	\end{figure}
	
	Finally, I validate the efficacy of my patched model by re-running the adversarial offline learning function on the refined model. Fig.~\ref{fig:patched_vs_unpatched} compares the number of generated recipes in the patched model versus the original (unpatched) model. The patched model (green bars) has become more robust against SYN reflection attacks with any impact above 200 attack packets and the improvement becomes more evident as the impact increases. More importantly, no recipe exists for volumetric attacks with an impact above 600 packets. We observe that the number of generated adversarial recipes starts to increase from the impact value equal to 400 packets per second. It is important to note that an adversarial recipe is solely a function of the given decision tree and hence has no correlation with the attack impact. This is because the decision tree is generated from the benign training dataset with no knowledge and/or assumption about possible attacks. Therefore, the number of adversarial recipes is not expected to increase or decrease by the impact of attacks.

		\subsection{Impact of Hyper-Parameters}
		Decision tree ensemble models come with two key hyper-parameters, namely the number of trees and the maximum depth of each tree. These parameters can be tuned to gain the best accuracy results. One may wonder about their role in the model resilience against adversarial attacks. It seems more decision trees or deeper trees can introduce additional constraints to the process of generating adversarial recipes, thus, potentially limiting those recipes. Let us investigate the impact of these two parameters on adversarial recipes. I first train six different models with 50, 100, 150, 200, 250, and 300 trees (all share a depth of 10) and thereafter generate adversarial recipes for each of these models. Similarly, I train seven different models (with 100 trees), varying the depth from 8 to 20 in steps of 2 units.
		
		My experiments revealed that the number of decision trees has almost no significant impact on adversarial recipes. In other words, incorporating more trees in an ensemble model does not necessarily make the model robust against adversarial attacks. However, increasing the depth of trees would noticeably limit the recipes. For example, for SSDP reflection attacks, the number of recipes generated for the seven models are 37, 33, 9, 4, 2, 1, and 0 as the max depth increases from 8 to 20, respectively. Similar patterns in SYN reflection attacks are observed.
		
		Different impacts of these two hyper-parameters could be attributed to the ``bootstrap'' technique used by ensemble algorithms. This technique helps to decrease the variance of model accuracy for unseen data (testing phase) by training each decision tree with a random sub-sample of the training dataset \cite{Leo1996}. Therefore, in a given decision tree, as it gets deeper, new constraints apply to the same sub-sample space, resulting in smaller recipe regions, increasing the chance for inconsistency with other constraints in the recipe. Moving to a new decision tree, on the other hand, is built on a different sub-sample of the training dataset compared to other trees. Therefore, its constraints are likely to be on a different scope, more likely not to violate the ones found in other trees.

	\section{Conclusion}
	
	In this chapter, I developed \textit{AdIoTack}, a systematic way of quantifying and refining the resilience of decision tree ensemble models against data-driven adversarial attacks.
	I first developed a white-box algorithm that automatically generates recipes of volumetric network-based attacks that can bypass the inference model unnoticed. My algorithm takes the intended attack on a victim class and the model as inputs. I developed a systematic method to successfully launch the intended attack on real networks. Next. I prototyped AdIoTack and validated its efficacy on a real testbed of real IoT devices monitored by a trained Random-Forest model. I demonstrated how the model detects all non-adversarial volumetric attacks on IoT devices while missing many adversarial ones. Finally, I developed systematic methods to patch loopholes in trained decision tree ensemble models. I demonstrated how my refined model detects 92\% of adversarial volumetric attacks.

%% file: tex/Chapter7-Conclusion/main.tex
\chapter{Conclusions and Future Work}\label{chapter:conclusion}


\vspace{-4mm}
\section{Conclusions}
IoT devices are being increasingly deployed in different domains, from home networks to smart manufacturing and critical infrastructures. The imbalanced maturity of IoT device manufacturers coupled with economic factors resulted in a lack of embedded device-level security measures, making IoTs vulnerable to various cyber-attacks. These vulnerabilities impose risks to user privacy as well as the security of the entire Internet ecosystem. Network-level solutions have emerged as viable methods to manage these risks. Passive traffic inference is a network-level technique that consumes network traffic of connected assets to draw insights into individual networked devices (\eg determining their make/model, type, role, cluster) and/or tracking their behavior to detect known malicious patterns or deviations from the norm. IoT traffic inference has been studied for the past several years with various objectives and strategies. However, existing works do not fully address some of the real challenges of IoT traffic inference which deter them from being deployed in large-scale production settings. In this thesis, I used real IoT traffic data collected from different sources like laboratory testbeds and residential networks to demonstrate the existence and quantify the impact of these challenges. I then developed some possible solutions to address them. My contributions are summarized as follows.

(1) In Chapter~\ref{ch:smart infer}, I studied the challenge of balancing the computation costs against the accuracy of inference systems. I started with distinguishing the two tasks of classification and monitoring, which were brought together as separate stages into a generic inference architecture. For the classification, I compared the characteristics of various packet/flow-based features and demonstrated how specialized models outperformed a single combined-view model. Then, I developed an optimization problem that selects an optimal set of specialized models and consolidates their outputs in a multi-view classification scheme. For the monitoring, I developed a cost-effective and explainable progressive model with three levels of granularity using one-class models trained by a refined version of convex hulls.  
I publicly released data (training and testing instances of classification and monitoring tasks) and enhancement code for the convex hull algorithm to the research community.

(2) In Chapter~\ref{ch:obscure infer}, I studied the challenges of network traffic measurement cost and the lack of high-quality data, which leaves us with partial and/or coarse-grained data for performing the inference. For this chapter, I analyzed over nine million IPFIX records of 28 IoT devices, each containing aggregate flow information. The dataset is released as open data to the public. I developed and compared two inference models for classifying IoT devices based on features of IPFIX flow records. The first model was a stochastic machine learning-based model that uses 28 flow features extracted from IPFIX records, yielding an average accuracy of $96$\% in classifying the device types. The second model I developed was a deterministic model that creates a fingerprint for each device type based on the cloud services it consumes, giving an average prediction accuracy of $92$\%. Eventually, I developed a method for combining the outputs of the stochastic and the deterministic models that mitigates false positives in $75$\% of cases.

(3) In Chapter~\ref{ch:scalable infer}, I investigated how managing concept drifts at scale can be challenging. I analyzed over six million IPFIX records collected from 12 real home networks and demonstrated the existence of concept drift in time and space domains. Then, to manage concept drifts, I developed and evaluated the performance of two inference strategies: global (a single model trained with all data) and contextualized (a separate model trained for each home) modeling. I showed that incorporating the global model into contextualized modeling makes our inference system more resilient against concept drifts. As contextualized modeling requires a model to be selected for inference from data of an unseen home, I developed a model selection strategy based on classification score distribution. My method selects the best model without the need for true labels of unseen data during the testing phase. Evaluation results show that the models selected by this method can achieve $94$\% and $99$\% of the accuracy and F1-score of the ideal model, respectively.

(4) In Chapter~\ref{ch:secure infer}, I focused on the resiliency of decision tree-based machine learning models against adversarial volumetric attacks. I started by developing a system called AdIoTack that generates adversarial recipes for a given decision tree and an intended volumetric attack on a target IoT class. Then, if possible, AdIoTack launches real volumetric attacks based on previously found recipes without getting flagged by the decision tree-based inference model. I evaluated AdIoTack in a testbed of nine IoT device classes. It found at least an adversarial recipe for eight devices (out of nine), with an almost perfect chance of bypassing the model while launching attacks. Finally, to overcome the vulnerability of decision trees against adversarial volumetric attacks, I developed a patching technique that refines models, making them robust against $92$\% of the generated adversarial attacks.

\section{Future Work}

Contributions made by this thesis set a foundation for developing and evaluating the efficacy of IoT traffic inference from a practical viewpoint. Inference models can be further improved in different ways, like increasing the prediction accuracy, reducing computation costs, or broadening the scope and coverage. In what follows, I briefly highlight a few directions that can be taken for improvement in future work.

(1) In Chapter~\ref{ch:smart infer}, I showed how the optimal multi-view scheme detects almost all device classes correctly. However, it might miss a few device classes at certain resolution periods. In order to further improve this optimal inference, one may want to continue classifying the traffic of  assets for several resolution periods instead of relying on a single period (my evaluation). By doing that, various techniques (\eg majority voting, consecutive/consistent predictions) can be employed to resolve possible conflicts across multiple predictions.

(2) Developing models to infer from opaque and coarse-grained data is indeed a challenging task. In Chapter~\ref{ch:obscure infer}, my main focus was determining the connected device types from their IPFIX flow records. I also developed a method with metrics to perform a monitoring task (\S\ref{ch4:sec:trust}), extending the classification ability of the stochastic model. An important direction could be developing more sophisticated monitoring models to track the behavior of IoT devices by analyzing coarse-grained/opaque data. Additionally, I developed a deterministic model based on cloud services consumed by individual IoT device classes for the classification task. Another piece of information contained in IPFIX records is the IP address of cloud servers that can be directly used or converted to domain names for expanding and enhancing the performance of deterministic models.

(3) Concept drifts are known and well-studied problems in machine learning for various application domains.
In Chapter~\ref{ch:scalable infer}, I studied the challenges of concept drifts in network traffic data across temporal and spatial domains for a distributed and scalable inference. I mainly focused on methods to manage concept drifts, maximizing the performance of prediction. However, one can target the occurrence of concept drift itself in order to gain a better understanding of the underlying challenges. For instance, future work can develop methods to quantify the severity of concept drifts and determine affected traffic features. Another area to explore is developing model selection techniques. Although the method developed in this thesis gives reasonably acceptable results, there are more sophisticated methods \cite{Reis2018b} for this purpose that might be able to select a model closer (in terms of accuracy) to the ideal upper bound.

(4) Adversarial machine learning has been a popular topic, especially in computer vision. Chapter~\ref{ch:secure infer} of this thesis developed a system for generating and launching adversarial volumetric attacks on networked IoT devices, whose traffic activity is continuously monitored by a model. Although the algorithms developed in Chapter~\ref{ch:secure infer} were specific to decision tree-based models (\eg Random Forest and AdaBoost), the same concept (with variations in actual techniques) can be applied to other machine learning-based models (\eg neural networks, SVM, Naive Bayes). In this thesis, I performed the adversarial attack study with a white-box approach (a worst-case scenario) which is desirable for a resiliency assessment of network operators or cyber defenders. However, studying more sophisticated and challenging attacks (gray-box and black-box approaches) will reveal the realistic capabilities of adversaries who aim to subvert the efficacy of inference models in production networks.